\documentclass[11pt,letterpaper,oneside]{book} 

\usepackage{fullpage} 
\usepackage{amsmath,amssymb,amsthm} 
\usepackage{graphicx}
\usepackage{imakeidx} 
\usepackage{xr-hyper} 
\usepackage{physics}
\usepackage{titlesec}
\usepackage{appendix}
\usepackage{tensor}
\usepackage{framed}
\usepackage{bm}
\usepackage{mdframed}
\usepackage{xcolor}
\usepackage{scalerel} 
\usepackage{cite}
\usepackage[bb=pazo]{mathalpha}
\usepackage{bm}
\usepackage{fontawesome5}

\usepackage[colorlinks=true]{hyperref}

\newcommand*\xref[1]{%
    \hyperref[#1]{SM.\ref{#1}}}
\newcommand*\xeqref[1]{%
    \hyperref[#1]{\textup{\tagform@{SM.\ref{#1}}}}}

\usepackage{centernot}

\numberwithin{equation}{section}  

\usepackage[nottoc]{tocbibind}

\usepackage[automake,symbols,nogroupskip,nonumberlist]{glossaries-extra}

\makeglossaries

\newglossaryentry{speedc}{
    name = \ensuremath{c},
    description={The speed of light.},
    type = main
}
\newglossaryentry{beta}{
    name = \ensuremath{\beta},
    description={Velocity in units of \ensuremath{c} (i.e., \ensuremath{v/c})},
    type = main
}
\newglossaryentry{eta}{
    name = \ensuremath{\eta_{\mu\nu}},
    description={The Minkowski metric diag\ensuremath{(-1, 1, 1, 1)}},
    type = main
}
\newglossaryentry{g}{
    name = \ensuremath{g_{\mu\nu}},
    description={A general metric tensor},
    type = main
}
\newglossaryentry{vec}{
    name = \ensuremath{\vec{v}},
    description={Three-vectors have arrows},
    type = main
}
\newglossaryentry{4vec}{
    name = \ensuremath{v},
    description={Four-vectors do not have arrows},
    type = main
}
\newglossaryentry{veccomp}{
    name = \ensuremath{v^\mu/v^i},
    description={Components of a vector. Greek indices start from 0, Latin indices start from 1.},
    type = main
}
\newglossaryentry{cocomp}{
    name = \ensuremath{v_\mu/v_i},
    description={Components of a co-vector. Greek indices start from 0, Latin indices start from 1.},
    type = main
}
\newglossaryentry{defeq}{
    name = \ensuremath{:=},
    description={Left side ``is defined as'' right side. The \ensuremath{=} indicates equality, and the colon tells you which side is being defined.},
    type = main 
}
\newglossaryentry{R}{
    name = \ensuremath{\mathbb{R}},
    description={The real numbers. This includes all negative and positive numbers, all integers, all fractions (rational numbers), and even the irrational numbers ($e$,  $\pi$,  $\sqrt{2}$, etc.).},
    type = main 
}
\newglossaryentry{Rn}{
    name = \ensuremath{\mathbb{R}^n},
    description={The \ensuremath{n^\text{th}} Cartesian product of \ensuremath{\mathbb{R}}. Elements take the form \ensuremath{(r_1, r_2, \dots, r_n).}},
    type = main 
}
\newglossaryentry{1}{
    name = \ensuremath{\mathbb{1}},
    description={The identity matrix, consisting of only 1 along the diagonal and 0 in every other entry. In components, $\mathbb{1} = \delta_{ij}$. We use the same symbol for the identity matrix in every dimension.},
    type = main 
}
\newglossaryentry{delta}{
    name = \ensuremath{\delta_{ij}},
    description={The Kronecker delta, equal to $1$ if $i = j$ and $0$ otherwise. Equivalently, the components of the identity matrix.},
    type = main 
}
\newglossaryentry{fMtoN}{
    name = \ensuremath{f: M\to N},
    description={Mathematical mapping notation. This is a handy way of writing functions that makes it crystal-clear what kinds of things the inputs and outputs are. Read this as, \ensuremath{f} is a machine that eats an element of \ensuremath{M} and spits out one or more elements of \ensuremath{N}.},
    type = main 
}
\newglossaryentry{fmton}{
    name = \ensuremath{f: m\to n},
    description={Mathematical element mapping notation. The local version of the mapping notation; this tells you explicitly how each element \ensuremath{m} in \ensuremath{M} turns into one or more elements \ensuremath{n} in \ensuremath{N}.},
    type = main 
}
\newglossaryentry{lhs}{
    name = LHS,
    description = {The left-hand side of an equation.},
    type = main
}
\newglossaryentry{rhs}{
    name = RHS,
    description = {The right-hand side of an equation.},
    type = main
}

\newcommand{\email}[1]{\href{mailto:#1}{\nolinkurl{#1}}}

\titleformat
{\chapter}
[display]
{\bfseries\Large\itshape}
{Chapter \thechapter}
{0.5ex}
{
	\rule{\textwidth}{1pt}
	\vspace{0.5ex}
}
[
\vspace{-1.5ex}%
\rule{\textwidth}{0.3pt}
]

\makeindex

\usepackage{chngcntr}
\usepackage{etoolbox}

\AtBeginEnvironment{subappendices}{%
  \chapter*{\thechapter A\quad\appendixname}
  \addcontentsline{toc}{chapter}{%
    \protect\numberline{\thechapter A}\appendixname}
  \counterwithin{figure}{section}
  \counterwithin{table}{section}
}

\apptocmd{\subappendices}{%
  }{}{}

\makeatletter
\patchcmd{\l@chapter}{1.5em}{2em}{}{}
\renewcommand*\l@section{\@dottedtocline{1}{2em}{3em}}

\renewcommand*\l@figure{\@dottedtocline{1}{1.5em}{3.3em}}
\let\l@table\l@figure

\makeatother

\usepackage{authblk}

\title{A Lean and Mean Introduction to Modern General Relativity}
\author[*]{Peter Hayman\thanks{\email{peter@haymanphysics.com}}}
\affil[*]{Department of Physics, The University of Auckland}
\date{}

\mdfdefinestyle{postulatestyle}{%
	linecolor=purple,
	linewidth=1pt,%
	frametitlerule=true,%
	frametitlebackgroundcolor=gray!20,
	innertopmargin=\topskip,
}
\mdfdefinestyle{definitionstyle}{%
	linecolor=blue,
	linewidth=1pt,%
	frametitlerule=true,%
	frametitlebackgroundcolor=gray!20,
	innertopmargin=\topskip,
}
\mdfdefinestyle{asidestyle}{%
	linecolor=gray,
	linewidth=1pt,%
	frametitlerule=true,%
	frametitlebackgroundcolor=gray!20,
	innertopmargin=\topskip,
}
\mdtheorem[style=postulatestyle]{postulate}{Postulate}
\mdtheorem[style=definitionstyle]{definition}{Definition}
\mdtheorem[style=asidestyle]{aside}{Aside}

\begin{document}
\maketitle
\tableofcontents
\listoffigures

\printunsrtglossary[type=main,title={List of Symbols and Math}]

\chapter*{Forward}
\addcontentsline{toc}{chapter}{Forward}

Think of these notes more as a novella than a textbook. They are intended to cover a single semester introductory course on general relativity at the upper undergraduate level, and to do so from a modern differential geometric perspective while maintaining a strong, logically consistent and physically-oriented narrative structure. The aim is to tell the story of how a reader freshly familiar with the Lorentz transformations of special relativity must eventually discover the full structure of modern general relativity, and then explore those major applications (gravitational waves, black holes, and cosmology) that one would naturally consider. Deviations from the narrative are confined to boxed asides and a few appendices (and perhaps someday separate supplementary materials). This is written for physicists, so the main takeaways of every section are those that enable practical work in general relativity, but as much as possible I also introduce differential geometric language and notation so the reader will have seen them when they encounter them in the wild. For example, vectorfields on manifolds are defined and even motivated by the concept of sections of tangent bundles, but calculations are all set up and carried out in (coordinate basis) components.

Materially, these notes draw mainly from Carroll \cite{carroll_2019} (one of the best modern graduate texts, takes a geometric approach well-adapted to physicists) and Schutz \cite{schutz_2009} (an excellent undergraduate introduction), with occasional calculations from Weinberg \cite{weinberg_gravitation_1972} (follows the old, coordinate-based approach, but contains almost every calculation you could ask for), all adapted to my own personal take with some topics not covered in standard texts (in particular the affine structure of Minkowski space, Rindler coordinates, and the proper near-source boundary condition for Schwarzschild black holes). Additional resources one might find useful include Wald \cite{wald2010general} (a comprehensive, modern text for physicists, but heavier on the math); Misner, Thorne, and Wheeler \cite{Misner:1973prb} (a legendary omnibus); and Spivak \cite{spivak1999comprehensive} (the epic, definitive five-volume treatise on pure differential geometry). As background, I assume at least some introductory special relativity, a comfortable handle on Newtonian mechanics, and basic electrodynamics and quantum mechanics. 

Throughout, I use the only sensible choice of units in which the speed of light is numerically 1, and so it will be omitted almost everywhere (except at the very beginning before natural units are introduced). I also use the correct (mostly-plus) metric signature. In keeping with standard practice, spacetime vectors are written with unaltered letters and are indexed by Greek letters, while spatial vectors are written with an arrow and are indexed by Latin letters. 

\chapter*{Introduction}
\addcontentsline{toc}{chapter}{Introduction}

General Relativity is easily in contention for the title of ``most elegant theory'' in Physics. The entire theory rests on two compact, yet extremely powerful and meaningful sets of equations. The first set are the Einstein field equations:
\begin{equation*}
	R_{\mu\nu} - \frac 12 R g_{\mu\nu} = \kappa T_{\mu\nu}.
\end{equation*}
What makes this set of equations so elegant is that without knowing the details of either side, the equation graphically represents the core idea of GR:
\begin{equation*}
	\text{Geometry} = \text{Matter}.
\end{equation*}
Spacetime---the substrate of all physics---is not set in stone, it is dynamic and its dynamics are intimately tied to the matter that lives within it. 

Another way of looking at the Einstein field equations is that they are the equations of motion for the geometry of spacetime. Closely related to that is the second crucial set of equations in this course, the geodesic equations:
\begin{equation*}
    \frac{D}{ds} U^\mu = U^\nu \nabla_\nu U^\mu = 0.
\end{equation*}
The geodesic equations are the equations of motion of point-like matter in curved space, and are the relativistic version of Newton's second law\footnote{Of course as written the equations only represent motion in a gravitational field, there are additional terms present when other forces are involved.}. These two sets of equations of motion are coupled to each other, and of course they are. The geometry of space reacts to the properties of the matter within it, but the behaviour of the matter in space reacts to the shape of the space it lives in, which influences the geometry of space, etc. This recursive relationship just reinforces the identity of spacetime and matter. 

These extraordinary phenomena are in the public consciousness by this point, but understanding and accepting them scientifically means going in skeptically and seeing that they are essentially an \emph{inevitable} consequence of the experimental facts of special relativity and the equivalence principle (the equality of inertial and gravitational mass). Starting with years of Newtonian mechanics under our belt (and at least a touch of electromagnetism and quantum mechanics), we will build the mathematical language that special relativity demands we use, paying close attention to the aspects that make it special in chapter \ref{ch:geomRel}, and following the consequences of breaking those properties when we need to incorporate the equivalence principle in chapters \ref{ch:manifolds} and \ref{ch:curve} (i.e.~we will naturally develop the language of differential geometry). After establishing the theory of general relativity in chapter \ref{ch:EFEs}, we will proceed to apply it to gravitating sources of increasing complexity, first source-free and perturbative solutions in chapter \ref{ch:AppsI}, and then to point-sources and (highly symmetric) fluids in chapter \ref{ch:AppsII}. A single course is nowhere near enough time to fully explore the enormous landscape of this theory, but hopefully this work will sufficiently develop for you the fundamental tools required to be able to learn any further topics with relative ease (many of which are teased but not explored throughout), and to do so in a way that is simultaneously intuitive and up-to-date.

\chapter{The Geometry of Special Relativity}
\label{ch:geomRel}

Act I, scene I: the physicist-in-training has spent many years studying Newtonian mechanics, has started to tackle electrodynamics and quantum mechanics, and has recently been introduced to the special theory of relativity. We begin by recapping the basics of the latter while introducing a little linguistic precision that will be important in the long run.

\section{Minkowski Space}
\label{ch:geomRel:sec:mink}

Special relativity is a beautiful thing. Two postulates, empirically determined, combine to break our hard earned intuition about how to mathematically represent the laws of physics. It turns out that time is in fact not a universal parameter but instead just a place, and every observer has their own notion of what it means to move along through time, and consequently their own personal notion of the shapes and sizes of extended objects. Having been convinced of the veracity of these principles and effects in your previous introduction to special relativity, it is now time to make precise the mathematical language this forces upon us, and to begin the inescapable task of re-writing all the laws of physics to comply with this new framework (and as you can probably foresee, the only physical theory that will really cause us problems in this endeavour will be gravity). 

\subsection*{Lightning SR}

Special relativity is built upon two fundamental postulates. The first postulate goes back as far as Galileo\index{Special Relativity Postulates!Inertial Frames}, a convenient framing of which is:
\begin{postulate}[The Principle of Relativity]
	There exists a special class of coordinate systems (called \textbf{inertial coordinate systems}) in which the laws of physics take the same mathematical form. 	
\end{postulate}
This is clearly true from experience; everything you know about how Nature works is equally true when you're at home as it is when you're at the library, at the park, on a mountain, etc. We call these systems \textbf{inertial} because they are the systems in which Newton's first law is manifest. They are special in that they are \emph{clean}, there are no confounding external influences. The cockpit of a plane performing aerial acrobatics, for example, is good example of a \emph{non}-inertial frame, as any experiment you conduct in that environment would be very hard to reproduce indeed. Perhaps a less intuitive non-inertial frame is the one you're in now; the ground pushing up on you is an external force that augments your measurements of the fundamental laws of physics.  

The first postulate applies equally well to Newton's laws as it does to Einstein's, the difference comes with the second postulate:
\begin{postulate}[Invariance of the Speed of Light]
	There exists a finite speed\index{Special Relativity Postulates!Finite Speed of Light} $\mathbf{c}$ whose value is measured to be the same in every inertial coordinate system (numerically, this is the speed of light in a classical vacuum).
\end{postulate}
This is the unshakable conclusion drawn from the failure of experiments to detect the luminiferous aether, and it is this second postulate that completely changes the mathematical landscape of physics. You'll recall from your earlier introduction to special relativity that these postulates lead to the famous \textbf{Lorentz transformations}\index{Lorentz Transforms} or \textbf{Lorentz Boosts}. For any two collinear inertial observers, their coordinate systems $\mathcal{O},$ and $\mathcal{O}^\prime$, are related by:
\begin{align}
	\label{eq:ch1:Lorentz}
	ct^\prime &= \gamma ct - \gamma\beta x, \quad \text{and} \\
	x^\prime &= \gamma x - \gamma\beta ct,
\end{align}
where we take the collinear axis to be the $x$ axis. The relativistic factor is defined as:
 \begin{equation}
	\label{eq:ch1:gamma}
	\gamma := \frac {1}{\sqrt{1 - \beta^2}},
\end{equation}
where $\beta := v/c$ is the normalized relative velocity of the frames (and in $\gamma$ we see the limit on $\beta$ built in, since $\beta \ge 1$ would lead to imaginary coordinates). 
\begin{aside}[Natural Units]
    In fact, in relativistic contexts it is always only useful to talk of velocities relative to $c$, to the point where it is really just clumsy to keep around the factor of $c$ that is always there. Instead, what we'll do is work in a more \emph{natural} system of units\index{Natural Units} where the numerical value of $c$ ``happens'' to be 1. For instance, we may define a unit of time, the du, such that $1 \, \text{du} = 1 \, m / (299 792 458 \, m/s)$, and then measure all time intervals in du\footnote{Common practice in the field, see e.g. \url{http://www.peebleslab.com/20}}. Of course, this strategy would work equally well for units of spatial separation---the important thing is that we work with a system of units that makes the relation between distances in time and space absurdly simple, so we never have to think about it again. 
\end{aside}
The Lorentz transformations stand in stark contrast to the \textbf{Galilean boosts} of old:
\begin{align}
    \label{eq:ch1:Galileo}
    t^\prime &= t, \quad\qquad \text{and} \\
    x^\prime &= x - vt,
\end{align}
which clearly preserve a universal choice of time coordinate and come with no limit to allowed velocity parameters.

You will have seen all the quirky business with length contraction and time dilation, and various and sundry silly paradoxes that arise from forcing our naive animal viewpoint on a special relativistic world, but here we are only focusing on the important mathematical properties implied by the Lorentz transformations. There are two keys lessons here. The first is of course that time is not a parameter, it is a \emph{place}. Instead of writing physical systems as functions of time, we must parameterize physics that takes place in the combined space+time = ``spacetime'' landscape by some arbitrary external parameter (typically using something physically meaningful, like a particle's personal clock---i.e.~proper time). The second lesson is that as a place, time is a little bit unusual. We can make this statement more concrete, but first we have to look a little closer at the context of the Lorentz transformations just described, and the landscape of special relativity we call \textbf{Minkowski space}\index{Minkowski Space} (as opposed to the \textbf{Euclidean space}\index{Euclidean Space} of Newton's world).

\subsection{Minkowski Coordinates and Vectors}

\begin{figure}[ht]
    \centering
    \includegraphics[width=\textwidth]{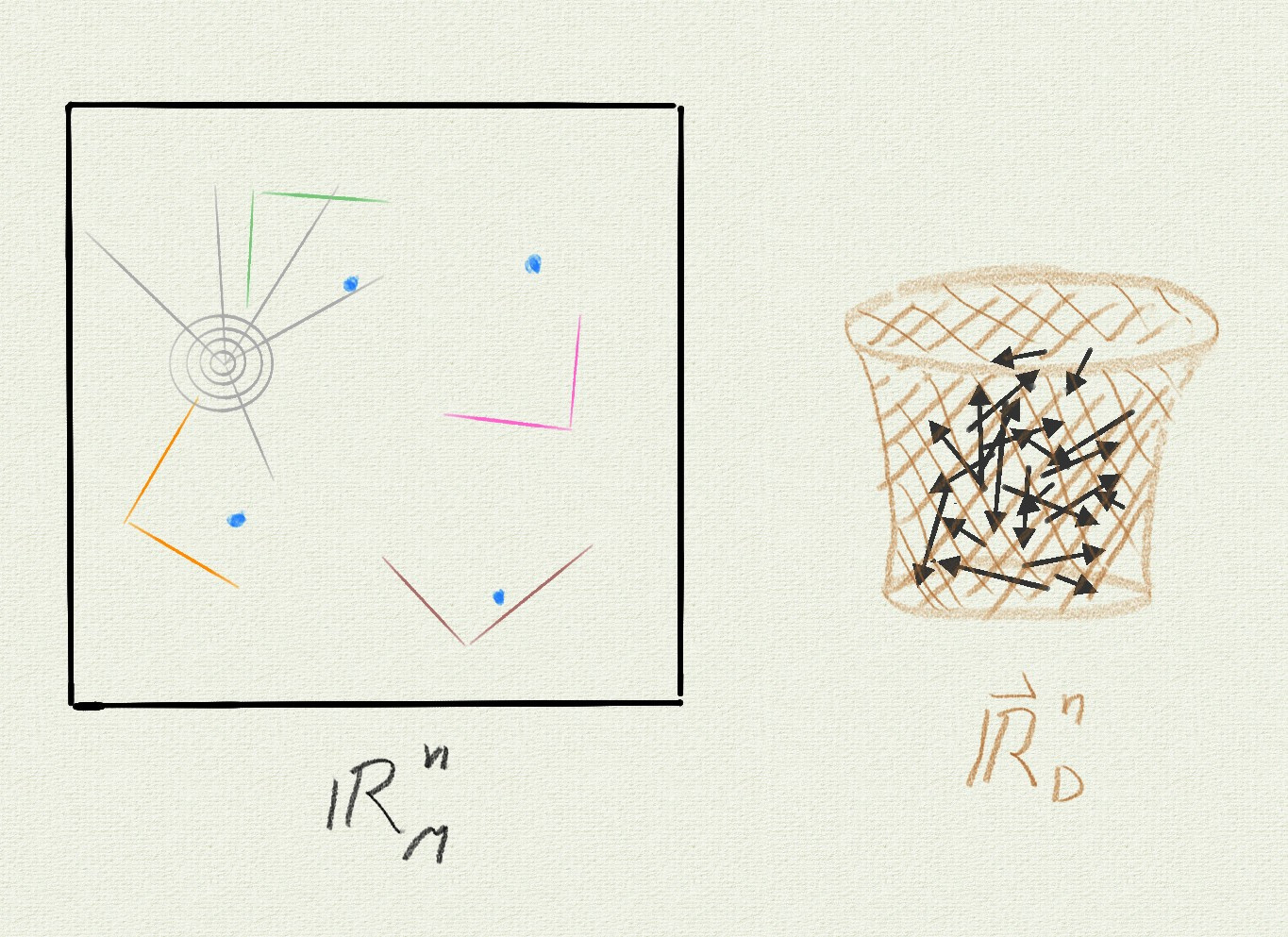}
    \caption[The Affine structure of Minkowski space.]{The Affine structure of Minkowski space. Left: the abstract points of spacetime are the canvas itself (some specific points plotted in blue), with arbitrary coordinate systems providing systematic labels to each point. The companion displacement vectors represent physically meaningful---and linearly operable---vectors that span the space. Cartesian coordinates are the priviledged coordinate systems that speak directly to displacement vectors.}
    \label{fig:minkowski}
\end{figure}

As written, the Lorentz transformation above describes a change of coordinates between one observer's coordinate system $(t,x)$ and another's $(t^\prime, x^\prime)$. The keen---possibly a bit impatient---reader has probably noticed that the system of equations \eqref{eq:ch1:Lorentz} resembles a system of linear transformations, and so is probably eager to start talking about vectors, but here it is important to take a step back. In fact I have a confession to make: you've been lied to. At some time in your life, someone has pointed to a set of equations like \eqref{eq:ch1:Lorentz} and told you it is a linear system relating so-called ``position vectors,'' but this is a fiction. Even when they look like it, \emph{positions and coordinates are not vectors}. Let's be clear about what they are (see figure \ref{fig:minkowski} for a graphical depiction). 

\subsubsection*{Points and Coordinates}
Picture an empty sheet of paper. You know this serves as a canvas you can use to describe physics, and you're comfortable using it to draw free-body diagrams or other physical pictures. You know those things you've drawn occupy places in spacetime, you know you are going to represent those places with two numbers (or more if you're projecting higher dimensions onto the sheet), but you can't actually write those numbers down until you fix a set of axes (i.e.~coordinates). The numbers (coordinates) you ascribe to individual places in space are a mathematical crutch in that the points of spacetime are abstract objects, but to do math with them we need to turn them into numbers and exactly how we do that is virtually arbitrary. You already know this of course, you know that you can put the origin of your coordinate system anywhere you please, and you know you can use polar, or cylindrical, or any number of different types of coordinate systems to dissect the exact same piece of paper. Phrased in this way, it's obvious that positions can't be vectors. What is result of adding the position of the Eiffel tower to the position of the Sydney opera house? Obviously a nonsense question, but one that must be answerable for vector quantities (by definition). In special relativity, the points of spacetime are geometrically the same as $\mathbb{R}^n$ (the set of tuples of $n$ real numbers) but with \emph{no} additional structure, \emph{no} vectors. To emphasize this detail, when referring to the points of spacetime we'll call it $\mathbb{R}^n_M$. 

\subsubsection*{A Happy but Confusing Coincidence: Affine Structure}
The confusion with vectors arises because of a happy coincidence relating to $\mathbb{R}^n$. It just so happens that $\mathbb{R}^n$ is also the prototypical example of a structure that can be made into a vectorspace, with the obvious rules of vector addition and scalar multiplication, and the existence of the various identity vectors. We are very used to writing vectors as columns of numbers just like elements of $\mathbb{R}^n$, and we are similarly very used to thinking of them as arrows that can be drawn on a sheet of paper. There is a coincidental relationship between points and arrows on a piece of paper in that any two points can be used to define an arrow, and any arrow can be attached to any point to define a second point. This special relationship between points and vectors defines a special structure called an \textbf{Affine space}\index{Affine space}, and it is a fluke of nature that both the Euclidean landscape of Newtonian mechanics \emph{and} the Lorentzian landscape of special relativity have this remarkable structure. 

The arrows we draw on spacetime are precisely the \textbf{displacement vectors}\index{Displacement vectors} that represent physically meaningful distances between points and they are the objects we can do linear operations with. The associated displacement vectors are so important that we'll give them their own name $\vec{\mathbb{R}}_D^n$ to denote all the additional linear structure they contain that the points of spacetime do not, and we'll define Minkowski space as the \emph{pairing} $M := \{ \mathbb{R}^n_M, \vec{\mathbb{R}}^n_D\}$. In most choices of coordinate systems, the number associated with the tip of a displacement vector arrow is not simply related to the number associated with its tail, but there exists a priviledged class of coordinate systems in which that relationship is as simple as can be---those are the \textbf{Cartesian coordinates}\index{Cartesian Coordinates} (also sometimes called \textbf{Affine coordinates}\index{Affine Coordinates} for this reason). Cartesian coordinate systems align with our notion of \emph{inertial} coordinate systems, they are the coordinates in which inertial particle worldlines are simple linear functions, and in which their velocity vectors are constants. They are so special (and inertial frames so important) that in discussing special relativity we will exclusively restrict our attention to that class of coordinates, but it is crucial to remember we are making this choice because eventually we will be forced to abandon it.

\subsubsection*{Minkowski Vectors}
With those important but perhaps a bit dry semantics out of the way, let's return to the Lorentz transformations and follow up on the claim that time is a weird place. First, we can extend the coordinate transformation \eqref{eq:ch1:Lorentz} into an operation on vectors by using that the Cartesian (Affine) coordinates are aligned with displacement vectors and writing
\begin{align}
    \label{eq:ch1:Lorentz1D}
    \Delta t^\prime &= \gamma \Delta t - \gamma\beta \Delta x, \quad \text{and} \\
    \Delta x^\prime &= -\gamma\beta\Delta t + \gamma \Delta x.
\end{align}
This is now a system of linear equations, so it's better to write it in matrix (and later index) form:
\begin{equation}
	\label{eq:ch1:Lorentz2}
	\Delta x^\prime = \Lambda \Delta x,
\end{equation}
where in a minor abuse of notation we'll use $\Delta x$ for both the displacement in the $x$-direction and to name the entire displacement vector $ \Delta x = (\Delta t, \Delta x)^T$, but the meaning will be obvious from context (in fact we'll rarely need to talk about the $x$-component on its own at all coming up). The Lorentz transformation matrix is then clearly (restricted to 1+1 dimensions) 
\begin{equation}
	\label{eq:ch1:lambda}
	\Lambda = \mqty[ \gamma & -\gamma\beta \\
	-\gamma\beta & \gamma].
\end{equation}
\begin{aside}[A Note on Notation]
    You'll note we just defined a spacetime displacement vector $\Delta x = (\Delta t, \Delta x)^T$ without any arrow or underline. This will be the standard practice in relativistic theories; relativistic vector quantities will be denoted without decorations while purely spatial vector quantities will carry a vector sign, so $\Delta x$ is a vector over spacetime, but $\vec \Delta = (\Delta x, \Delta y, \Delta z)^T$ is an old-timey spatial vector. We also sometimes talk about the different types of vectors with different names as well, calling relativistic vectors ``Lorentz vectors'' or ``four-vectors'' when dealing with 3+1 dimensional Minkowski space. 

    While we're talking notation, we also just used the standard $(\# space)+(\# time)$ format for spacetime dimensions. Usually this means 3+1 dimensions meaning 3 spatial dimensions and 1 time dimension, but as above it's sometimes helpful to restrict to fewer spatial dimensions temporarily (though some more adventurous theorists also like to tinker with the number of time dimensions).
\end{aside}

The Lorentz transformation is a statement about how the components of a single physically meaningful spacetime vector relate between inertial coordinate systems. That the coordinate systems are inertial means that \emph{all} physical laws must take the same exact mathematical form in each frame. In particular, this means that any operation that turns a vector into a scalar quantity must be \emph{invariant} under the action of a Lorentz boost. This is where Minkowski space finally makes a name for itself, for now the old simple inner product $\langle v, w \rangle := v^T w$ is no longer a meaningful operation. A simple calculation shows
\begin{equation}
    \label{eq:ch1:breakEuclid}
    (v^\prime)^T w^\prime = v^T \Lambda^T \Lambda w \neq v^T w,
\end{equation}
since $\Lambda^T\Lambda \neq \mathbb{1}$, so this is no way to form scalars from vectors in a relativistic theory. The general structure still has to work though (combinatorially ``row-matrix-column'' is the only way to turn $n$-dimensional vectors into a single number), so let's broaden it a little by searching for a matrix $\eta$ such that  $\Lambda^T\eta\Lambda = \eta$, for then
 \begin{equation}
    \label{eq:ch1:fixNorm}
    (v^\prime)^T \eta w^\prime = v^T \Lambda^T \eta \Lambda w \neq v^T \eta w,
\end{equation}
would be true. Sure enough, it is easy to see that any matrix proportional to 
 \begin{equation}
	\label{eq:ch1:defEta}
	\eta := \mqty[-1 & 0 \\ 0 & 1] \qquad \text{(1+1 dim)}
\end{equation}
satisfies this relation (and so we'll be sensible human beings and choose the proportionality factor to be 1. Note that some peculiar folks---notably in the world of particle physics---choose a factor of -1 for some reason). This matrix goes by the name of the \textbf{Minkowski metric tensor}\index{Minkowski Metric}, and of course is similarly defined for the full 3+1 dimensions of the real world as
 \begin{equation}
	\label{eq:ch1:defEtaFul}
	\eta := \mqty[-1 & 0 & 0 & 0 \\ 0 & 1 & 0 & 0 \\ 0 & 0 & 1 & 0 \\ 0 & 0 & 0 & 1] \qquad \text{(3+1 dim)}. 
\end{equation}

I realize this re-discovery of the Minkowski spacetime interval might seem a bit of an odd choice given the usual tale of rods and cones in a special relativity course has a lot to say about invariant proper times and proper distances, but I assure you there's a method to the madness. Of course the generalized length of Lorentz vectors just defined applies exactly to the spacetime intervals of Einstein's thought experiments, but they also apply much more broadly to \emph{any} physically meaningful vectors in a Minkowski spacetime. This means momentum vectors have well-defined invariant lengths (masses in that case), relativistic currents of all sorts have well-defined invariant lengths, and much more as we shall soon see. To reflect this generality, we'll keep using generic Lorentz vectors like $v$ and $w$, and only specialize to coordinate displacement vectors when the need arises. 

\subsection*{\texorpdfstring{A Closer Look at $\eta$}{A Closer Look at the Metric Tensor}}

The metric tensor $\eta$ is a crucial component of Minkowski space, and fundamentally what makes it different from Euclidean space. We defined it as a matrix, and while that will continue to be a useful way to look at it at times, it is more useful in general to remember our motivation was to define an invariant measure of lengths of spacetime vectors. Mathematically, Newton's Euclidean space equipped with the usual dot product is called an \emph{inner product} space. The Euclidean inner product is a true inner product in that it satisfies the following properties (ex: check for yourself that they are all true).
$\langle \cdot, \cdot \rangle$:
\begin{itemize}
	\item Bilinear: The Euclidean inner product is linear in ``both slots,'' e.g., $\langle \vec{v} + \vec{w}, \vec{r} \rangle = \langle \vec{v}, \vec{r} \rangle + \langle \vec{w}, \vec{r} \rangle$, and $\langle \vec{v}, \vec{w} + \vec{r} \rangle = \langle \vec{v}, \vec{w} \rangle + \langle \vec{v}, \vec{r} \rangle$, etc. 
	\item Non-Degenerate: The Euclidean product satisfies that only the zero vector is the zero vector. That is, if $\langle \vec{v}, \vec{w} \rangle = 0$ for all $\vec{w}$, then $\vec{v} = \vec{0}$ (and obviously the same for the second slot).
	\item Positive-Definite: The overlap between two non-zero Euclidean vectors is always positive, $\langle \vec{v}, \vec{w} \rangle > 0$ for all non-zero $\vec{v}$ and $\vec{w}$. 
\end{itemize}

The metric tensor $\eta$ is the corresponding way we measure vector lengths and overlaps in Minkowski space. As an operator $\eta(\cdot, \cdot)$ acting on Lorentz vectors $v = (v^t, v^x, v^y, v^z)^T$, \\$w = (w^t, w^x, w^y, w^z)^T$, we see from the matrix form \eqref{eq:ch1:fixNorm} that $\eta(\vec{v}, \vec{w}) := -v^tw^t + v^xw^x + v^yw^y + v^zw^z$. That minus sign makes all the difference, it ensures the Minkowski metric tensor is \emph{not} a proper inner product. Instead, it satisfies the following properties (ex: verify these properties). $\eta( \cdot, \cdot )$:
\begin{itemize}
	\item Bilinear: The Minkowski metric tensor is linear in ``both slots,'' e.g., $\eta(u + v, w) = \eta(u, w) + \eta(v, w)$; $\eta(u, v + w) = \eta(u, v) + \eta(u, w)$; etc.
	\item Non-Degenerate: The Minkowski metric tensor satisfies that only the zero vector is the zero vector. That is if $\eta(v, w) = 0$ for all $w$, then $v = 0$ (and obviously the same for the second slot).
	\item \textbf{NOT Positive-Definite}: The overlap between two non-zero Minkowski vectors is \emph{not necessarily} positive; $\eta(v, w)$ can be negative, positive, or vanish for all non-zero $v$ and $w$. 
\end{itemize}

\subsection*{Dividing Spacetime}

The fact that the overlap of two Minkowski vectors can be non-positive is responsible for a lot of physics. We can classify vectors as follows:

\begin{enumerate}
	\item \textbf{Timelike}: If $\eta(v,v) < 0$
	\item \textbf{Spacelike}: If $\eta(v,v) > 0$
	\item \textbf{Lightlike (or Null)}: If $\eta(v,v) = 0$
\end{enumerate}

Causally-connected events must be either timelike or lightlike, it must be possible for something to connect the spatial distances in \emph{more than or equal to} the amount of time it would take a light beam ($\Delta t^2 \ge \Delta x^2 + \Delta y^2 + \Delta z^2$ means any $\beta \le 1$). Also notice that since Lorentz transformations preserve the form of the Minkowski metric tensor, it is impossible to change perspective from one inertial frame to another in which a vector changes its nature (e.g., there is no way to boost to a frame where a timelike, causally connected sequence becomes spacelike and acausal). 

\section{Vector Variance}
\label{ch:geomRel:sec:vec}
Nope, not done with vectors yet---in fact we're just getting started. We identified displacement vectors $\vec{\mathbb{R}}_D^n$  as the fundamental vectors associated with Minkowski spacetime, but as mentioned above, pure displacement vectors are not the most useful by themselves, physical vectors tend to be built \emph{upon} them. This can be as simple as a scaling, for example inertial relativistic velocity is just displacement scaled by change in proper time:
\begin{equation}
    \label{eq:ch1:avFourVel}
    U = \mqty(\Delta t/\Delta\tau \\\Delta x/\Delta\tau \\\Delta y/\Delta\tau \\\Delta z/\Delta\tau ). 
\end{equation}
(This is of course for inertial particles for whom $\Delta x^\mu \sim \Delta \tau$). Or it can be a more complicated relationship like the electromagnetic four-potential, in which the components are essentially arbitrary functions of spacetime. To keep things general, we will use generic components for our four-vectors, just remembering they are built on displacement vectors. 

We've made a big deal about vectors being abstract objects, this was even an important part of how we (re)discovered the Minkowski metric above. But the vector/matrix notation as in \eqref{eq:ch1:avFourVel} actually obscures this fact since as we know the components change under coordinate transformations. To make our language more precise, we'll introduce \textbf{component notation}\index{Component Notation}:
\begin{equation}
	\label{eq:ch1:delCoords}
	v = \sum_{\mu=0}^3\, v^\mu \hat{e}_\mu := v^\mu \hat{e}_\mu.
\end{equation}
\begin{aside}[Einstein Notation]
    And straight away we'll invoke Einstein notation as well. We see sums of the form \eqref{eq:ch1:delCoords} a \emph{lot} in relativity, so it's more convenient to assume a sum and drop the symbol. For our safety though, we establish a couple of rules with this. First, the terms are only summed if there is an index that is repeated, once as a subscript and once as a superscript. Second, when the index is a Greek letter the sum is over space and time indices (0 to 3), but when it is a Latin letter, the sum is only over the spatial indices (1 to 3---trust me, this will be handy). We're also using the notation that $(t, x, y, z) := (x^0, x^1, x^2, x^3)$, and that $\hat{e}_\mu$ denotes a unit vector in the $x^\mu$ direction\footnote{Well, really the $x^\mu$ Cartesian axis is defined by the $\hat{e}_\mu$ basis vector, but that kind of pedantry gets a physicist called rude names, like ``mathematician.''}. It may be more familiar to see these basis vectors as $ \hat{x}, \hat{y}$, etc., but $\hat{e}_\mu$ is the standard notation in the field since it generalizes easily to higher dimensions. 
\end{aside}

Let's dissect this notation \eqref{eq:ch1:delCoords}. $v$ is a vector, an abstract object. It is a single unique element of $\vec{\mathbb{R}}^n_D$ and it doesn't care how we represent it. The set $\{\hat{e}_\mu\}$ is a particular \emph{set} of vectors (abstract objects) that are linearly independent, orthogonal, and of unit length, meaning no $\hat{e}_\mu$ can be written as a linear combination of the other vectors, and that $\eta(\hat{e}_\mu, \hat{e}_\nu) = \eta_{\mu\nu}$ (this is how we \emph{define} orthonormality in Minkowski space). The set $\{\hat{e}_\mu\}$ form a \textbf{basis}\index{Vectorspace Basis} for the displacement vectorspace that can be used to label every other vector in the vectorspace. The \textbf{components}\index{Vector Components} $\{v^\mu\}$ are simply numbers---by themselves they're meaningless, but taken together they amount to a \emph{basis-specific} nametag for $v$, and as such the set of components together needs to follow some rules. Specifically, if we grow dissatisfied with our current inertial frame and wish to relocate to another, the new basis vectors will be related to the old by means of a Lorentz transformation\footnote{\faThumbtack: Put a pin in this; almost all of this notation ports over exactly to the non-Cartesian coordinates we'll use later, but getting there takes a little conceptual effort.} (which includes spatial rotations) $\Lambda$, 
\begin{equation}
	\label{eq:ch1:changeBasis}
	\overline{e}_\mu := \Lambda_\mu^\nu\hat{e}_\nu,
\end{equation}
but since the abstract vector $v$ doesn't care how we look at it, we must impose that the components in the new frame be related by the \emph{inverse} map to their old names, 
\begin{equation}
	\label{eq:ch1:changeComps}
	\overline{v}^\mu :=  (\Lambda^{-1})_\nu^\mu v^\nu.
\end{equation}
In this way, we build all the important information about a vector into the index that labels its components. Any vector whose components transform the same way as the components of $v$ under a coordinate transformation is said\footnote{By physicists. Some mathematicians use the opposite convention.} to be \textbf{contravariant}\index{Vectors!Contravariant}, because its components \emph{contra}vary with changes of basis.
%
%
\begin{aside}[A Note on Indices]
	Normally indices refer to components of vectors (and similar objects) as just described, but the unit vectors $\hat{e}_\mu$ are an exception. Each $\hat{e}_\mu$ is independently a complete abstract vector, so for each of them the $\mu$ is just an inert label, the \emph{name} of the vector. As they are complete vectors, we can even write an individual basis vector in components, e.g.~$\hat{e}_\mu = \delta_\mu^\nu \hat{e}_\nu$, where we identify $[\delta_\mu]^\nu$ as the components of the abstract vector $\hat{e}_\mu$. 
\end{aside}
Besides the examples mentioned already, there's one very important class of contravariant vectors we should mention, since we'll see them a lot later.

\subsection*{Tangents from Displacement Vectors}

Doing physics in a spacetime very often involves looking at paths through the world. Paths are the accumulation of life experience of observers, and can be as simple as particle worldlines (which just accumulate proper time), or they can be combined with scalar fields to tell more colourful stories (a particle's experience travelling through a temperature field, for instance). Whatever they represent, where there's a path there's a tangent, and in the language of special relativity, tangents to curves are related to a special class of displacement vectors, the \emph{infinitessimal} displacement vectors.

Consider the curve $\gamma : [0,1] \to \mathbb{R}^4_M$, where  $\gamma : s \mapsto (x^0(s), x^1(s), x^2(s), x^3(s))$ in some inertial coordinate system $x^\mu$. The tangent to this curve can be evaluated at any point $s$ in the usual way, 
\begin{equation}
    \label{eq:ch1:derivGamma}
    \dot\gamma(s) := \lim_{h\to 0} \frac{\gamma(s + h) - \gamma(s)}{h}. 
\end{equation}
But since $\gamma(s+h)$ and $\gamma(s)$ are just points in spacetime, the difference between them defines a displacement vector. In components,
\begin{equation}
	\label{eq:ch1:defDeriv}
	\dot\gamma(s) = \dot{x}^\mu(s)\hat{e}_\mu := \left(\lim_{h\to 0} \frac{x^\mu(s + h) - x^\mu(s)}{h}\right)\hat{e}_\mu.
\end{equation}
This is a re-scaled displacement vector, so clearly a contravariant vector. It is a bit of a special re-scaling though, it turns the subset of displacement vectors of infinitessimal length about a point into a proper vectorspace by dividing through by an infinitessimal. Interestingly, the discovery of a proper vectorspace from the renormalization of an infinitessimal displacement vectorspace is a very generic procedure; what makes the Affine Minkowski space so special is that the resulting tangent vectorspace is at every point fully mathematically equivalent (isomorphic) to the global displacement vectorspace\footnote{Put a pin \faThumbtack\, in this, this identification (or potential lack thereof) will be very important later on.}. We'll have more to say about this later, but first, all this discussion about contravariant vectors is probably making you suspect there are non-contravariant vectors, and indeed there are.

\subsection*{Covectors}

The natural vectors on Minkowski space are displacement vectors. They may be used directly to represent inertial motion or in more complicated ways to represent the propagation of physically meaningful quantities, like energy-momentum, or potential or force fields. Whatever they represent, at the end of the day Minkowski vectors are still abstract objects and we do not measure abstract objects in the lab, we only measure numbers, so physical predictions always involve turning vectors into numbers. We already know one consistent tool for this, the Minkowski metric $\eta$ measures the lengths of vectors in a frame-independent way. This is a good start, but any vector contains more information than just its length, so we still need more independent ways to extract information from physical vectors. Vectors being linear creatures, the most general consistent\footnote{More precisely, almost all fundamental physical laws are linear systems of equations, hence why we represent physical properties by vectors in the first place. } way to turn one into a number is through a linear operator $\omega : v \mapsto C$. The requirement that $\omega$ be a linear operator is equivalent to the statement that $C$ must be a polynomial linear in the components of $v$, so 
 \begin{equation}
    \label{eq:ch1:linOp}
    \omega(v) = \omega_0 v^0 + \omega_1 v^1 + \omega_2 v^2 + \omega_3 v^3 = C.
\end{equation}
The form \eqref{eq:ch1:linOp} makes it clear that the linear operators themselves form a vectorspace in their own right, and so the coefficients $\omega_\mu$ can be seen to be components of some sort of vector $\omega = \omega_\mu \hat{e}^\mu$, but exactly what sort takes a moment of thought.

Our experience defining the Minkowski metric should be enough to see that $\omega$ is not simply a displacement-type vector on its side, $\omega \neq w^T$ ($w \in \vec{\mathbb{R}}_D^n$), since if it were, then under a change of inertial coordinates the ``constant'' $C$ would very much not be so, $w^T v = C \neq w^T \Lambda^T \Lambda v$ . For the resulting number to be a true number---a \textbf{Lorentz scalar}\index{Lorentz Scalar}---the components of $\omega$ must transform in the \emph{opposite} direction to those of  $v$, that is we must have
 \begin{equation}
    \label{eq:ch1:covary}
    \overline{\omega}_\mu = \omega_\nu \Lambda^\nu_\mu.
\end{equation}
This is the same direction as the displacement basis vectors, so we call $\omega$ a \textbf{covariant vector}\index{Vectors!Covariant}, or a \textbf{covector}\index{Covector}, but note that this also means we define the basis covectors to transform in the opposite way, $\overline{e}^\mu = \hat{e}^\nu (\Lambda^{-1})^\mu_\nu$.

\subsection*{Gradients as Covectors}

Just as displacement vectors have an infinitessimal relative in tangent vectors, so too do covectors have an infinitessimal cousin in the gradient operator. To see this, consider a particle cutting a path through a scalar field on spacetime. That is, if $x^\mu(\tau)$ is the worldline of some particle (inertial or not) and $\phi : \mathbb{R}_M^4 \to \mathbb{R}$  is a scalar field on Minkowski space, then the field felt by this particle throughout its life is the composition $\phi(x^\mu(\tau))$. Now look to its infinitessimal---how does  $\phi$ incrementally change along the particle's life?  Easy:
\begin{equation}
    \label{eq:ch1:gradCov}
    \dv[]{\phi(x^\mu(\tau))}{\tau} = \frac{\partial\phi(x^\mu)}{\partial x^\mu} \frac{\dd x^\mu(\tau)}{\dd \tau} = \partial_\mu \phi \, \dot x^\mu,
\end{equation} 
which defines the very useful notation $\partial_\mu := \frac{\partial}{\partial x^\mu}$. Here the gradient $\partial_\mu \phi$ is doing exactly what a covector should do, it is (in a linear way) turning the tangent displacement vector $\dot x^\mu$ into a simple finite scalar $\dot \phi := \dv[]{\phi}{\tau}$. Moreover, it clearly tranforms in the correct way; schematically, if $x^\mu$  goes one direction, one must expect $1/x^\mu$ to go the opposite way. 

It is hard to imagine disentangling vectors and covectors, so we're not even going to try. Instead we'll dive even deeper into their relationship and eventually find both physical intuition and calculational efficiency. 

\subsection*{Covectors as Dual Vectors}

Covectors formally belong to the \textbf{dual space}\index{Dual Space} to displacement vectors. For any (finite-dimensional) vectorspace $V$, one finds there exists a complementary isomorphic vectorspace $V^*$ (sometimes $\widetilde{V}$) made up of all linear functions $f \in V^*$ such that $f : V \to \mathbb{R}$. $V^*$ is called the dual space to $V$ and elements $f$ are called \textbf{dual vectors}\index{Vectors!Dual}, and when $V$ has the additional structure of a bilinear form (such as our metric tensor $\eta$) there is a \emph{natural} isomorphism between the two. This mapping is easy enough to see; $\eta$ is a very hungry bilinear operator, it wants to eat two vectors and turn them into a real number. If we only partially feed the metric by giving it a single vector $\eta(v, \cdot)$ then what remains is a perfectly linear operator that is still hungry for one more vector to turn into a real number. In this way, for every vector $v \in \vec{\mathbb{R}}_D^n$, we define a dual covector $\omega_v := \eta(v, \cdot) \in \vec{\mathbb{R}}_D^{n*}$. This can be represented as a row vector if necessary:
\begin{equation}
	\omega_v = \eta(v, \cdot) = v^T\eta = \mqty(-v_0 & v_1 & v_2 & v_3),
\end{equation}
or better yet, in terms of a basis of orthonormal covariant vectors:
\begin{equation}
	\omega_v = \eta_{\mu\nu}v^\nu \hat{e}^\mu =: (\omega_v)_\mu \hat{e}^\mu,
\end{equation}
where the basis covectors $ \hat{e}^\mu$ are now uniquely defined by the condition $ \hat{e}^\mu(\hat{e}_\nu) = \delta^\mu_\nu$ (n.b., here again, the superscript on $\hat{e}^\mu$ is the \emph{name} of the covector, not an index, just like for the contravariant basis vectors). As we're starting to get comfortable using index notation, it is a helpful exercise to ensure the action of $\omega_v$ on a vector $w$ is as expected: 
\begin{equation}
	\label{eq:ch1:covecActionComps}
	\omega_v(w) =  (\omega_v)_\mu \hat{e}^\mu (w^\nu \hat{e}_\nu) = (\omega_v)_\mu w^\nu \hat{e}^\mu (\hat{e}_\nu) = (\omega_v)_\mu w^\nu \delta^\mu_\nu = (\omega_v)_\mu w^\mu = \eta_{\mu\rho}v^\rho w^\mu = \eta(v, w)
\end{equation}
as it should be. 

An important corollary of this duality is that our covectorspace is a pseudo-inner product space like our displacement vectors, except that here the pseudo-inner product is the \emph{inverse} matrix $\eta^{-1}$. To see this, consider that if we want covectors to really be dual to vectors, we should want it to be the case that the notion of lengths is preserved. That is, if we have two covectors  $\omega_v$ and $\omega_w$ dual to the vectors $v$ and $w$ respectively, then we want it to be the case that the scalar product between $\omega_v$ and $\omega_w$ satisfies:
\begin{equation}
	\label{eq:ch1:findDualProduct}
	\omega_v\cdot\omega_w =\, ?^{\mu\nu}(\omega_v)_\mu(\omega_w)_\nu = v\cdot w. 
\end{equation}
We can find the mystery matrix $?^{\mu\nu}$ by following our nose:
\begin{align}
	\label{eq:ch1:findInvEta}
	?^{\mu\nu}(\omega_v)_\mu(\omega_w)_\nu &=\, ?^{\mu\nu}\eta_{\mu\rho}v^\rho\eta_{\nu\sigma}w^\sigma \notag \\
										   &=\, ?^{\mu\nu}\eta_{\mu\rho}\eta_{\nu\sigma}v^\rho w ^\sigma.
\end{align}
Then if the mystery matrix satisfies $?^{\mu\nu} = (\eta^{-1})^{\mu\nu}$, we have\footnote{Note that this is the definition of an inverse matrix in component form, $M^{-1}M = \mathbb{1} \to (M^{-1})^{\mu\nu}M_{\nu\rho} = \delta^\mu_\rho$. See \ref{ch:geomRel:apps:matrices} for more detail. } $(\eta^{-1})^{\mu\nu}\eta_{\mu\rho} = \delta^\nu_\rho$ and
\begin{equation}
	\omega_v\cdot\omega_w = \delta^\nu_\rho \eta_{\nu\sigma}v^\rho w^\sigma = \eta_{\rho\sigma}v^\rho w^\sigma = v\cdot w,
\end{equation}
so the natural metric tensor associated with the dual space is the inverse of the Minkowski metric, $\eta^{-1}$.

To help drive home the idea of dual vectors, here are a few examples you're likely familiar with already. First, ordinary row vectors are already dual vectors, just for Euclidean space. The metric on Euclidean space is just the identity matrix, so if we take a Euclidean column vector $v$, then the dual vector is $\omega_v = \mathbb{1}(v, \cdot) = v^T \mathbb{1} = v^T$. This is the more natural way we would shmush two column vectors together, Minkowski space just amounts to throwing a minus sign onto the first component (general curved spaces will be more complicated). A less obvious example you will likely have seen comes from quantum mechanics. In Dirac notation, kets are the vectors, and bras are the dual vectors. Pictorially, this is pretty obvious---$\bra{\psi}$ shmushed with  $\ket{\phi}$ is a (complex) number. More concretely, the vectors in quantum mechanics are square-integrable functions, and (mod-)squared integration is the inner product:  $\braket{\psi}{\phi} = \int \dd V \psi^*\phi$, so if $\phi$ is the vector $\ket{\phi}$, then $\int \dd V \psi^* \cdot$ is the dual vector $\bra{\psi}$. Finally---and a bit more visually---topographical maps are just gradient maps, and we already know gradients are covectors. This provides a good conceptual way of picturing them though, covectors are like custom rulers for vectors. Place a vector across a steep gradient on a topographical map and you interpret its length very differently from when you place the very same arrow across a shallower part of the map (see figure 3.3 in \cite{schutz_2009}).

\subsection*{Vectors as Dual Dual Vectors}

Fix a displacement vector and feed it to one slot of the metric tensor and you get a covector. But the covectors have their own metric tensor, the inverse $\eta^{-1}$, so what if we feed that just one covector? Indeed, we can just as easily define the dual space to the covectors space in this way, but if we do we find ourselves in a familiar land. In components, the dual to an already dual vector $\omega_v$ is
 \begin{equation}
    \label{eq:ch1:checkDualDual}
    \omega_v^{*\mu} = (\eta^{-1})^{\mu\nu}(\omega_v)_\nu = (\eta^{-1})^{\mu\nu} \eta_{\nu\alpha}v^\alpha = \delta^\mu_\alpha v^\alpha = v^\mu,
\end{equation}
which is just the original displacement vector, hence the dual space to the dual space of displacement vectors is itself, it's a round trip. But of course this is the case, any displacement vector can easily be thought of as a function on linear maps from vectors to numbers as being the function that feeds itself to the linear map---i.e. $v(\omega) := \omega(v)$. This ``duals all the way down'' way of thinking is not especially helpful for physics, but mathematicians love it so you may well come across it in your travels (thought it will briefly be useful in discussing tangent vectors on manifolds, section \ref{ch:manifolds:vecs}). 

\subsection*{Physical Units}

Seeing that covectors are duals of vectors are duals of covectors might lead one to be suspcious 
of the importance of our choice of which direction is covariant and which contravariant. Specifically in Minkowski space we have a good justification in choosing variance along coordinate lines since we have this special relationship between displacement vectors and coordinate axes, but in general it is arbitrary, and in fact we already know this intuitively as physicists. The abstract nature of vectors is usually built into the concept of physical units, and covectors are quantities that are \emph{rates} with respect to preferred units. For instance, when we define the unit of displacement to be the metre m we measure the rate of change of something with respect to a displacement in inverse metres m$^{-1}$. Maybe more intuitively, we think of distances in time as measured in seconds s, and rates of change (with respect to displacements in time) as being measured in s$^{-1}$. But in principle no one threatened us and told us to base our system on displacement. There's nothing \emph{a priori} wrong with choosing gradients to be the fundamental quantity of interest, say orienting things in terms of Hz and then thinking about displacements in time as rates of change with respect to gradients---i.e., measuring time in Hz$^{-1}$. The important thing about conventions is that once you set them, you stick to them for ever and ever, so since our whole journey started with displacement vectors, we will respect them.


\subsection*{Index Gymnastics}

With an understanding of how vectors and covectors are related by the metric, the practical calculational consequence for the working physicist is that we now have a way of switching between upper and lower indices. That is, the dual vector with components $\omega_\mu$ can be understood as the dual to the vector with components  $\omega^\mu$ by defining $\omega_\mu := \eta_{\mu\nu}\omega^\nu$. To go the other way, we define the components of the inverse metric $\eta^{-1}$ to be $\eta^{\mu\nu}$, so that we may identify the components of a vector $v^\mu$ with the components of a dual vector $v_\mu$ by $v^\mu := \eta^{\mu\nu}v_\nu$. These mappings\index{Raising and Lowering Indices} are known as ``\textbf{raising}'' and ``\textbf{lowering}'' indices, and will become as natural as addition and multiplication when we get into the meat of general relativity. Incidentally, mathematicians refer to index raising and lowering as the musical isomorphisms, and represent the process with the musical ``sharp'' $\#$ and ``flat'' $\flat$ symbols respectively. 

\begin{aside}[Passive and Active Transformations]
    We have just spent a lot of time talking about how \emph{not} to change a vector with a linear transformation. In physics this is the common understanding of a basis transformation, but in some semantic circles, people like to refer to this as a \textbf{passive transformation}\index{Coordinate Transformations!Passive} as opposed to an \textbf{active transformation}\index{Coordinate Transformations!Active} where the vector itself is transformed into a different vector. This can be done with any type of linear map, but the Lorentz boost provides an excellent example of the difference. Imagine an inertial particle in its rest frame, so that its four-velocity vector is simply the unit vector in the time direction. A passive Lorentz boost corresponds to changing \emph{perspective} to that of a different observer passing by at some relative velocity $\beta$. In the passer-by's frame, the original particle's four-velocity is now $(U^\prime)^\mu = U^\nu(\Lambda^{-1})^\mu_\nu = (\Lambda^{-1})^\mu_0$. The vector is unchanged, but its coordinates in a different reference frame are notably different. In contrast, an active Lorentz boost would be like instantaneously accelerating that original inertial observer up to a velocity $\beta$ \emph{within its original reference frame}, so in this case the components of its four-velocity vector have changed from $U^\mu = \delta^\mu_0$ to $U^\mu = \Lambda^\mu_\nu \delta^\nu_0 = \Lambda^\mu_0$, both using the \emph{same} basis vectors. 

    The takeaway is that passive transformations are just changes in perspective and must not affect the physics, while active transformations \emph{are} the physics.
\end{aside}



\section{A Little Physics}

The aim is to get to gravitational physics as soon as possible, but to avoid getting bogged down in abstract math, it helps to place some of this in the context of actual physics. 

\subsection*{\emph{Heavy}locities\footnote{Heavy (i.e., non-relativistic) velocities. Let's make this happen.}}

Let's start with the simplest possible physical thing in Minkowski space: a single massive particle in its rest frame. Recall that we \emph{define} a particle's proper time $\tau$ to be the time elapsed in its rest frame. Turning that around, we can say that the parameter that governs a particle's motion is its proper time, and its trajectory in its rest frame is simply $x^\mu(\tau) = (\tau, 0, 0, 0)$. Then the particle's \textbf{four-velocity}\index{Four-velocity} is the tangent to this curve, given by $U(\tau) = \dv[]{}{\tau}x(\tau) = (1, 0, 0, 0)^T$ (notice the length of this vector is $U\cdot U = \eta(U, U) = -1$, a property of massive particles). Need some convincing that this makes sense as a velocity? Easy, we'll just boost out of the particle's rest frame. Write the general abstract vector $U$ in terms of the basis vectors associated with the particle's rest frame:
\begin{equation}
	\label{eq:ch1:fourvelRest}
	U = \delta^\mu_0\, \hat{e}_\mu.
\end{equation}
Here $\delta^\mu_0$ is the usual Kronecker delta symbol, and we are denoting the unit basis vectors in the particle's rest frame as $\hat{e}_\mu$. To look at this vector from the perspective of a frame that is moving at velocity $\beta$ in say the $x$-direction with respect to this one, the basis vectors of the boosted frame will be related to $\hat{e}_\mu$ by a Lorentz transformation $\Lambda(\beta)$ (where we make the velocity-dependence of the transformation explicit). So we write:
\begin{equation}
	\label{eq:ch1:boostBasis}
	\overline{e}_\mu = (\Lambda(\beta))^\nu_\mu\hat{e}_\nu,
\end{equation}
where
\begin{equation}
	\label{eq:ch1:Lorentz3}
	\Lambda(\beta) = \mqty[\gamma & -\gamma\beta & 0 & 0 \\ -\gamma\beta & \gamma & 0 & 0 \\ 0 & 0 & 1 & 0 \\ 0 & 0 & 0 & 1 ].
\end{equation}
Then the components must \emph{contra}vary with this boost, so we write:
\begin{equation}
	\label{eq:ch1:boostComps}
	\overline{U}^\mu = (\Lambda(-\beta))^\mu_\nu \hat{U}^\nu = (\Lambda(-\beta))^\mu_\nu\delta^\nu_0 = \mqty[\gamma \\ \gamma\beta \\ 0 \\ 0].
\end{equation}
(Ex: Verify that the \emph{vector} $U$ is unchanged by this transformation. That is, verify explicitly that $ \overline{U}^\mu \overline{e}_\mu = \hat{U}^\nu \hat{e}_\nu$). Recall that $\gamma := 1/\sqrt{1 - \beta^2}$, so that for $\beta \ll 1$, we have $\gamma \sim 1$ and $\gamma\beta \sim \beta$, so the components of the four-velocity $U$ in the boosted frame are $ \overline{U}^\mu \sim (1, \beta, 0, 0)^T$, justifying the name.

In particle mechanics, the incredibly important quantity \emph{momentum} is related to velocity by $p = m\beta$. We therefore define the \textbf{four-momentum}\index{Four-momentum} to be the four-vector $p = mU$. In the particle's rest frame, this has components $\hat{p}^\mu = (m, 0, 0, 0)^T$, while in a boosted frame (along the $x$-axis, say), it has components $ \overline{p}^\mu = (m\gamma, m\gamma\beta, 0, 0)^T$. Notice that to leading order in $\beta$, we have in the boosted frame that $\overline{p}^0 = m\gamma \sim m + \frac 12 m\beta^2 + \ldots$ which has the form ``rest mass + kinetic energy'' (ex: verify this). This leads to the identification of the energy $E := p^0$, and linear momentum $\vec{p}^{\,i} := p^i$. Notice that the length of the vector $p^2 = -m^2 = -E^2 + \vec{p}^{\,2}$ is independent of inertial reference frame (by definition), hence why it is often called a particle's \emph{invariant mass}.

\subsection*{Light Velocities}

Speaking of invariant mass, what about invariant massless things, like photons? Our discussion above started with a timelike vector ($U\cdot U = -1 < 0$), so it's worth asking, what would lightlike and spacelike four-velocities and momenta represent? Well in the first case, the name is kind of a spoiler. Lightlike vectors satisfy $U\cdot U = 0$, or in the case of momentum, $p\cdot p = -m^2 = 0$---a simple example of such a vector would be $U = (1, 1, 0, 0)$ (ex.~show that the norm of this vector vanishes). What would \emph{not} be an example of such a vector? Obviously the one we started with, $U = (1, 0, 0, 0)$. Recall that these are \emph{disjoint} classes of vectors, there is no Lorentz transformation that will take you from a lightlike to a timelike vector and vice versa. This leads to the important following statement: \emph{there is no rest frame for massless particles}. The curse of the photon; to maintain its velocity in everyone else's frame it must give up a rest frame of its own. Fortunately, since all of us studying physics have mass, it won't be important that we grapple with this fact of life, just keep it handy when looking at vectors.

\subsection*{Imaginary Velocities}

Having explored the meaningful cases, there's one more class of vectors to think about, spacelike vectors. These must satisfy $U\cdot U > 0$, or in the case of momentum,  $p\cdot p = -m^2 > 0 \implies m^2 < 0$. While this class of vectors is not so relevant when talking about physical particles (for the same reason as the null vectors above, no perspective shift can take us from physically relevant $m^2 > 0$ vectors to these), it is interesting to note that they do have some sort of meaning in particle physics. Weinberg has a good short discussion of why (see chapter 1, section 13 in \cite{weinberg_gravitation_1972}), the takeaway of which is that special relativity demands the existence of anti-particles (particles of the same mass but opposite charge of a given particle) because the uncertainty principle allows particles to traverse spacelike displacements.

\subsection*{Accelerating Ahead}

The four-velocity and four-momentum just defined are as applied to \emph{unaccelerated} or \emph{inertial} bodies, hence why we were able to find an inertial rest frame. In fact, while we defined them as tangents to a curve, we didn't need to take that infinitessimal limit, the four-velocity and four-momentum of inertial particles are just as easily defined in terms of displacement vectors. For any two proper times $\tau_1$ and $\tau_2$ (where $\tau_2 > \tau_1)$, we have in the particle's rest frame $\Delta x = (\tau_2 - \tau_1, 0, 0, 0)$, and so the velocity defined by $\Delta x / \Delta \tau = (1, 0, 0, 0)$. This is \emph{unique} to inertial observers in Minkowski space. 

In general, particles may be subject to external forces (e.g., an electron in a magnetic field) in which case the tangent to the particle's trajectory will \emph{vary} along its path through spacetime. Now you can't find a globally defined (displacement) vector that represents the particle's velocity, but you can do the next best thing, you can find a vector-\emph{field} for the particle's velocity. A vectorfield is as it sounds, a function that takes a point in spacetime (just a place not a vector, remember) and turns it into a vector---mathematically\footnote{\faThumbtack: Put a pin in this definition too. Here the vectorfield maps points in $\mathbb{R}_M^n$ to the same vectorspace $V$, but in GR it will map different points to \emph{different} vectorspaces. }, $v : \mathbb{R}_M^n \to V$ (where $V$ is a possibly re-scaling of displacement vectors). In practice, this means the components of the vector take on a dependence on position. For example, if $U = (1, 0, 0, 0)$ is a vector, then $U(x^\mu) = (t^2, 0, 0, 0)$ is a vectorfield, or if we were dealing with a particle tracing a path through spacetime, we would write $U(x^\mu(s)) = (x^0(s), x^1(s), x^2(s), x^3(s))$.

The quintessential vectorfields in classical mechanics are again velocity and momentum, and now also acceleration and force. These are defined as one would expect, $a = \frac{d}{d\tau}U(x^\mu(\tau))$ and  $f = \frac{d}{d\tau}p(x^\mu(\tau))$. Another four-vectorfield you are familiar with comes from electromagnetism. EM has a special history with relativity, it was discovered accidentally to already be completely compatible with special relativity. In fact, it turns out that the electric scalar potential and the magnetic vector potential naturally fit together into a four-vector  $A = (\phi, \vec{A})^T$ known as the electromagnetic four-potential. In four-vector (and tensor) notation, Maxwell's equations simplify greatly, but that is beyond the scope of this course.

\subsection*{ICRFs}

One important note to make about the four-velocity field of an accelerated particle. While it is not possible to find a single inertial ``rest'' frame for an accelerated particle, it is important to remember that accelerated particles are still massive particles, and are good observers of Nature. As a result, it must always be possible to \emph{instantaneously} find a rest frame, even for an accelerated particle. This frame, only defined for a specific value of $\tau$, is called the \textbf{Instantaneously Co-Moving Rest Frame}\index{Instantaneously Co-Moving Rest Frame!(ICRF)} of the particle\footnote{Schutz \cite{schutz_2009}, as well as some others, calls this a ``momentarily'' co-moving rest frame. I choose ``instantaneous'' because I feel it is more descriptive. Neither of these are usually mentioned in the literature, so use whatever makes the most sense to you.}. This actually imposes a fairly stringent restriction on the types of paths a massive particle can trace out in spacetime. A general result is that at every proper time $\tau$, there exists a frame such that $U(\tau)\cdot U(\tau) = -1$ (these frames are not the same from $\tau$ to $\tau$, but the dot product is Lorentz-invariant, so you can move between frames with impunity). The four-acceleration therefore satisfies:
\begin{equation}
	\label{eq:ch1:fourAcel}
	\dv[]{}{\tau}\left(U(\tau)\cdot U(\tau)\right) = 2 U(\tau)\cdot \dv[]{U(\tau)}{\tau} = 2 U(\tau)\cdot a(\tau) = 0.
\end{equation}
That is, the four-acceleration of physical particles is always perpendicular to their instantaneous four-velocity. (Ex.~prove the first equality).

\section{Tensors}
\label{ch:geomRel:tensors}

\index{Tensors}
The four-momentum $p$ of a particle is sufficient to describe everything about its energy and momentum. However, what we're working towards is a theory of gravity, and that necessarily requires thinking about more than one thing. In fact, for typical gravitational scenarios, we're interested in very macroscopic quantities of things, so knowing a single particle's four-momentum is not extraordinarily helpful. Instead, we'll need to construct more complicated objects to describe $N$ particles' energy-momenta, as well as their interactions. 

In fact, this need to expand from one-dimensional vectors should be familiar. Look no further than quantum mechanics for an example why: The quantum state of a system of more than one particle is a weird, giant, coupled Hilbert space, $\mathcal{H}_1 \otimes \mathcal{H}_2 \otimes \ldots$. For instance, a three-dimensional particle-in-a-box is described by three independent particle-in-a-boxes shmushed together. What we need is a general way to build up more complicated structures out of the relatively simple ones we've come to know and love\footnote{Love not necessarily required.}, and the most general (nice) way to do that is through the \textbf{tensor product}\index{Tensors!Tensor Product}.

Take two vectorspaces $V$ and $W$, and define an operator $\bigotimes$ that maps the Cartesian product (pairs) $V\times W$ to a brand new vectorspace $V\otimes W$. All we really need from this operator is that it preserves the linearity of the original vectorspaces, so it must be ``linear in both slots'', like an inner product, only this time it maps to a new vector instead of a number. It's easiest to see how this works by first looking at basis vectors. Say $V = \text{span}\{\hat{e}_i\}$ and $W = \text{span}\{\hat{f}_i\}$, then $V\otimes W = \text{span}\{\hat{e}_i\otimes\hat{f}_j\}$. The objects $ \hat{e}_i\otimes \hat{f}_j$ are just individual vectors with a long (descriptive) name, and there are $\text{dim}(V)\times\text{dim}(W)$ of them. A general vector in the tensor product space $V\otimes W$ therefore looks like:
\begin{equation}
	\label{eq:ch1:defGenTens}
	T = T^{\mu \nu} \hat{e}_\mu\otimes\hat{f}_\nu.
\end{equation}
A couple of important points here: 1.~any two vectors $v$ and $w$ can be tensored together, the result being a tensor\index{Tensors!Tensor} $T = v\otimes w := v^\mu w^\nu  \hat{e}_\mu\otimes \hat{f}_\nu$, and 2.~the converse is \emph{not} true, \emph{not} every tensor is the tensor product of two vectors (ex. prove $T = \hat{e}^0\otimes \hat{f}^0 + \hat{e}^1\otimes \hat{f}^1 \neq v\otimes w$ for any $v \in V\, , w \in W$). This second point is maybe less intuitive, but is extremely important, it is the reason the tensor product is so powerful. The tensor product space is much larger than the sum of its parts and in that extra space is room for a lot of physics. Again quantum mechanics furnishes an excellent example of this point. When two or more quantum particles are studied together, the Hilbert space of the whole system is the tensor product of their individual Hilbert spaces. When you can think of a state as being $\ket{\psi}\otimes\ket{\phi}$, it is possible to think of the two particles individually, but the full (tensor product) space has room for more complicated states---\emph{entangled} states---that don't look like that, and force us to lose our ability to understand the particles as existing separately. 

So now that we have some sense of what tensor products are, and what the tensor product space is, we can write down some things to do with them, and conventions we will follow. First, our tensors will be defined in terms of a select pair of vectorspaces. For obvious reasons we'll use spaces related to displacement vectors $D \propto \vec{\mathbb{R}}_D^n$ and their duals $D^* \propto \vec{\mathbb{R}}_D^{n*}$. We then define the following:
\begin{itemize}
    \item The \textbf{rank}\index{Tensors!Rank} of a tensor is the number of copies of $D$ and $D^*$ that make up the tensor product space to which it belongs. For example, if $T$ is a tensor in the space $D\otimes D\otimes D^*\otimes D^*$ the it is a rank-4 tensor\footnote{For physicists. There are some areas of mathematics where the rank of a tensor is defined as the fewest number of basis vectors it takes to express it.}. Equivalently, if we write it in component form:
		\begin{equation}
			T = T\indices{^\mu^\nu_\rho_\lambda} \hat{e}_\mu\otimes \hat{e}_\nu\otimes \hat{e}^\rho\otimes \hat{e}^\lambda,
		\end{equation}
	    then the rank of the tensor is the number of indices its components carry.

        \item The \textbf{type}\index{Tensors!Type} of a tensor is more descriptive. If a tensor lives in a space that is $M$ copies of $D$ and $N$ copies of $D^*$, we call it a type-$\mqty(M \\ N)$ tensor. As above, we can get this information from the components as well. According to convention, we write indices for contravariant vectors in superscript, and indices for covariant vectors in subscript, so a type-$\mqty(M \\ N)$ tensor has $M$ upper indices and $N$ lower indices. Notice that in this way, we can consider contravariant vectors to be type-$\mqty(1 \\ 0)$ rank-1 tensors, while covariant vectors are type-$\mqty(0 \\ 1)$ rank-1 tensors. It is also convenient to define scalars to be rank-0 tensors.
\end{itemize}

Next, let's go over some of the things we can do with tensors. 
\begin{itemize}
	\item Since tensor product spaces are vectorspaces, it follows that we can shmush any two tensors together into a bigger tensor. If $T$ is a type-$\mqty(M \\ N)$ tensor, and $R$ is a type-$\mqty(P \\ Q)$ tensor, then $T\otimes R$ is a rank-$(M+N+P+Q)$, type-$\mqty(M+P \\ N+Q)$ tensor.
        \item Interestingly, we can also find a natural way to \emph{reduce} the rank of a tensor, as long as it has one of each type of index. It might have always seemed a bit precarious having vectors live so close to things that eat vectors, and we can exploit that. Whenever we so choose, we can take a type-$\mqty(M \\ N)$ tensor and turn it into a type-$\mqty(M-1 \\ N-1)$ tensor by letting one of the dual vectors eat one of the vectors. For example, if we have $T = T\indices{^\mu^\nu^\ldots_\rho_\sigma_\ldots}\hat{e}_\mu\otimes\hat{e}_\nu\otimes\cdots\hat{f}^\rho\otimes\hat{f}^\sigma\otimes\cdots$, we can ``contract''\index{Tensors!Contraction} say the $\nu$ and $\sigma$ indices by using $\hat{f}^\sigma(\hat{e}_\nu) = \delta^\sigma_\nu$ to get a new tensor $T^\prime = T\indices{^\mu^\nu^\ldots_\rho_\nu_\ldots}\hat{e}_\mu\otimes\cdots\hat{f}^\rho\otimes\cdots$. Note: this is also sometimes referred to as taking the trace across the indices $\nu$ and $\sigma$, since this is a generalized notion of the trace from linear algebra (to see why, notice that a rank-2 type-$\mqty(1 \\ 1)$ tensor can be represented as a matrix, and the contraction of its two indices is $M\indices{^i_i}$, exactly the usual trace).
	\item As alluded to earlier, we can also play a little gymnastics with our indices. The multilinearity of the tensor product is really a beautiful thing, any linear thing we could do with our separate vectorspaces we can \emph{also} do to the vectorspaces sitting inside a tensor product. In particular, we can also use the metric tensor to raise or lower indices on a general tensor. In index notation, for $T\indices{^\mu^\nu_\rho_\sigma}$ we ``lower the index $\nu$'' by defining a new tensor by with components $T\indices{^\mu_\nu_\rho_\sigma} := \eta_{\nu\lambda}T\indices{^\mu^\lambda_\rho_\sigma}$. Similarly, for the same tensor, we can ``raise the index $\rho$'' by defining a new tensor with components $T\indices{^\mu^\nu^\rho_\sigma} := \eta^{\rho\lambda}T\indices{^\mu^\nu_\lambda_\sigma}$. (Note, if it helps you can think of this as a two-step operation: 1.~take the tensor product with the metric, 2.~contract the desired index with one of the metric's indices).
\end{itemize}

Lastly, a word of caution: order matters. The object $v\otimes w$ is \emph{not} the same thing as $w\otimes v$. In fact, this is such an important point, that we use special symbols when there is a symmetry or anti-symmetry. In the mathematical literature, you will find the symmetric tensor product denoted $\odot$, while the anti-symmetric product is $\wedge$ (the anti-symmetric ``wedge'' product is actually very useful in physics, and while not so important for this course, it does come in handy in appendix \ref{ch:curve:app:int}). In index notation, no special symbols are employed, it's generally just stated that something is (anti)symmetric on two specific indices. Sometimes, however, it is handy to symmetrize or anti-symmetrize a given tensor, in which case we use a special notation around the indices. For example,  $T_{(\mu\nu)} = \frac 12 (T_{\mu\nu} + T_{\nu\mu})$ represents the symmetrization of the tensor $T_{\mu\nu}$, while $T_{[\mu\nu]} = \frac 12 (T_{\mu\nu} - T_{\nu\mu})$ represents the anti-symmetrization of the same tensor.

\begin{aside}[Tensors as operators]
\label{ch1:aside1}
    \index{Tensors!As Operators}
	In some more modern circles, it is common to think of all tensors as functions that map vectors and covectors to a single number. Recall that we found even the original displacement vectors can be thought of as functions on their dual vectors, so everything in a tensor can (if one wishes) be interpreted as a linear map, and so tensors can be defined as multi-linear maps. This is often how they are introduced in the mathematical literature, but here it is more important to think about them as built up from physically meaningful quantities. 
\end{aside}

\subsection*{The Metric Tensor Really is a Tensor}

By the way, we've been calling $\eta$ the metric ``tensor,'' it is time to justify the name. Recall that one of the ways we have been thinking about $\eta$ is as a bilinear form on contravariant vectors, that is that $\eta(v, w) \in \mathbb{R}$ and is linear in both slots. Knowing what we now know, another way to express something like that would be as a type-$\mqty(0 \\ 2)$ tensor:
\begin{equation}
	\label{eq:ch1:defMetTens}
	\eta = \eta_{\mu\nu}\hat{e}^\mu\otimes\hat{e}^\nu.
\end{equation}
When presented with two vectors $v = v^\mu \hat{e}_\mu$ and $w = w^\mu \hat{e}_\mu$, we find:
\begin{align}
	\label{eq:ch1:metTensEat}
	\eta(v, w) &= \eta_{\mu\nu}(\hat{e}^\mu\otimes \hat{e}^\nu)(v^\rho \hat{e}_\rho, w^\sigma \hat{e}_\sigma), \notag \\
			   &= \eta_{\mu\nu}v^\rho w^\sigma \hat{e}^\mu(\hat{e}_\rho)\hat{e}^\nu \hat{e}_\sigma, \notag \\
			   &= \eta_{\mu\nu}v^\rho w^\sigma \delta^\mu_\rho \delta^\nu_\sigma, \notag \\
			   &= \eta_{\mu\nu}v^\mu w^\nu,
\end{align}
exactly as we wanted. Similarly, we can define the metric inverse as a type-$\mqty(2 \\ 0)$ tensor,  $\eta^{-1} := \eta^{\mu\nu} \hat{e}_\mu \hat{e}_\nu$ (show that this also leads to $v^*\cdot w^* = \eta^{\mu\nu}v_\mu w_\nu$). It is also a useful calculation to show that the mixed metric $\eta\indices{^\mu_\nu}$ is just the identity $\delta^\mu_\nu$ (hint: use the inverse metric to raise an index on the metric).

\subsection*{Tensor Equations}

And this brings us to the point of tensors, and why it's so important that we phrase physics in terms of them. Tensors are geometric objects, so when we write equations in the form $T = P$, where $T$ and $P$ are tensors (including scalars, vectors, and covectors), the equations are valid in \emph{all} equivalent coordinate systems. This is also why it's so important that we specify what vectorspaces we are using to construct our tensors, and what constitutes equivalent coordinate systems. For example, consider a rank-3 tensor given by $T = U\otimes U^* \otimes U^*$, or in components\footnote{Here Wald \cite{wald2010general} would use his ``abstract index notation'' and use Latin letters for the indices to indicate that this is a proper tensor equation in index form, as opposed to say \eqref{eq:ch1:evalExTens}, which is only valid in a particular coordinate system. I feel it is just as good to make clear notes in the margin, or in the variable names, so I will follow the more common convention of always using Greek letters, and let the meaning be clear from context.}:
\begin{equation}
	\label{eq:ch1:exTensEq}
	T\indices{^\mu_\nu_\lambda} = U^\mu U_\nu U_\lambda,
\end{equation}
where $U$ is the four-velocity of an unaccelerated particle. This equation might seem intimidating---a rank-3 tensor in 3+1 dimensions has $4^3 = 64$ independent components!---but knowing it is valid in \emph{all} inertial coordinates greatly simplifies the calculation. In the rest frame of the unaccelerated particle, its four-velocity is simply $U^\mu = (1, 0, 0, 0)^T$ and the covariant form $U_\mu = (-1, 0, 0, 0)$, so we can say that:
 \begin{equation}
	\label{eq:ch1:evalExTens}
	T\indices{^\mu_\nu_\lambda} = \begin{cases}
		1, \quad \text{if} \quad \mu = \nu = \lambda = 0, \\
		0, \quad \text{else}
	\end{cases}, \qquad \text{(Rest frame)}
\end{equation}
and from there, the components can be retrieved in any other inertial frame by appropriate Lorentz transformations:
\begin{equation}
	\label{eq:ch1:TransExTens}
	 \overline{T}\indices{^\mu_\nu_\lambda} = \Lambda\indices{^\mu_\alpha}\Lambda\indices{_\nu^\beta}\Lambda\indices{_\lambda^\gamma} T\indices{^\alpha_\beta_\gamma} = \Lambda\indices{^\mu_0}\Lambda\indices{_\nu^0}\Lambda\indices{_\lambda^0}. 
\end{equation}
Here again we'll emphasize that the appearance of Lorentz transformations on the RHS of \eqref{eq:ch1:TransExTens} is a direct consequence of our choice to work exclusively in flat, Cartesian Minkowski space. Had we been working in Cartesian Euclidean space, for example, the matrices would have been orthogonal instead. 




\section{A Little More Physics: The Stress-Energy Tensor}
\label{ch:geomRel:stressEnergy}

While we've mostly discussed a single particle so far, most of physics is concerned with multiple objects---even macroscopic quantities of objects---and as we're moving towards a theory of gravity this becomes ever more important. Fundamentally, it is of course possible to describe a macroscopic system in terms of a collection $\{(p_n)^\mu\}$ of four-momentum vectors, but this very quickly becomes intractible as $n \to \infty$. Fortunately, Nature is kind enough to ``decouple,'' and allow a qualitatively different treatment of phenomena at different scales. Consequently, at macroscopic numbers of particles, an effective strategy becomes to think about four-momentum \emph{flux}, the density of four-momentum vectors $p^\mu$ flowing through a given 3-surface, defined by a vector $v^\nu$ (rather, via its covector $v_\nu$). As we've just shown, the most general way to track both of these quantities while retaining their linearity is through a tensor product, so this must be described not by a four-vector, but by a rank-2 tensor $T^{\mu\nu}$, known as the \textbf{Stress-Energy Tensor}\index{Stress-Energy Tensor}. The components refer to the flux of the $\mu$-component of four-momentum through a surface defined by the dual vector $\hat{e}^\nu$.

Later we will see an algorithmic way of constructing a Stress-Energy Tensor for a particular physical system through an action principle, but for now we will argue our way to a specific example that is particularly useful in a course on gravity: the \textbf{perfect fluid}. A fluid is the simplest kind of macroscopic object, since without rigidity it can be made as symmetric and structureless as possible, and a \emph{perfect} fluid is the limiting case of that exercise. A perfect fluid is one without heat conduction or shear or viscous forces, which boils down to a distribution of essentially inertial particles, with no appreciable local torques, long-range forces, binding energies, etc. We'll start with yet more simplicity, and consider even the non-relativistic limit of a perfect fluid, which in cosmological circles is called \textbf{dust}.

\subsection*{Dust}

\begin{figure}[ht]
    \centering
    \includegraphics[width=\textwidth]{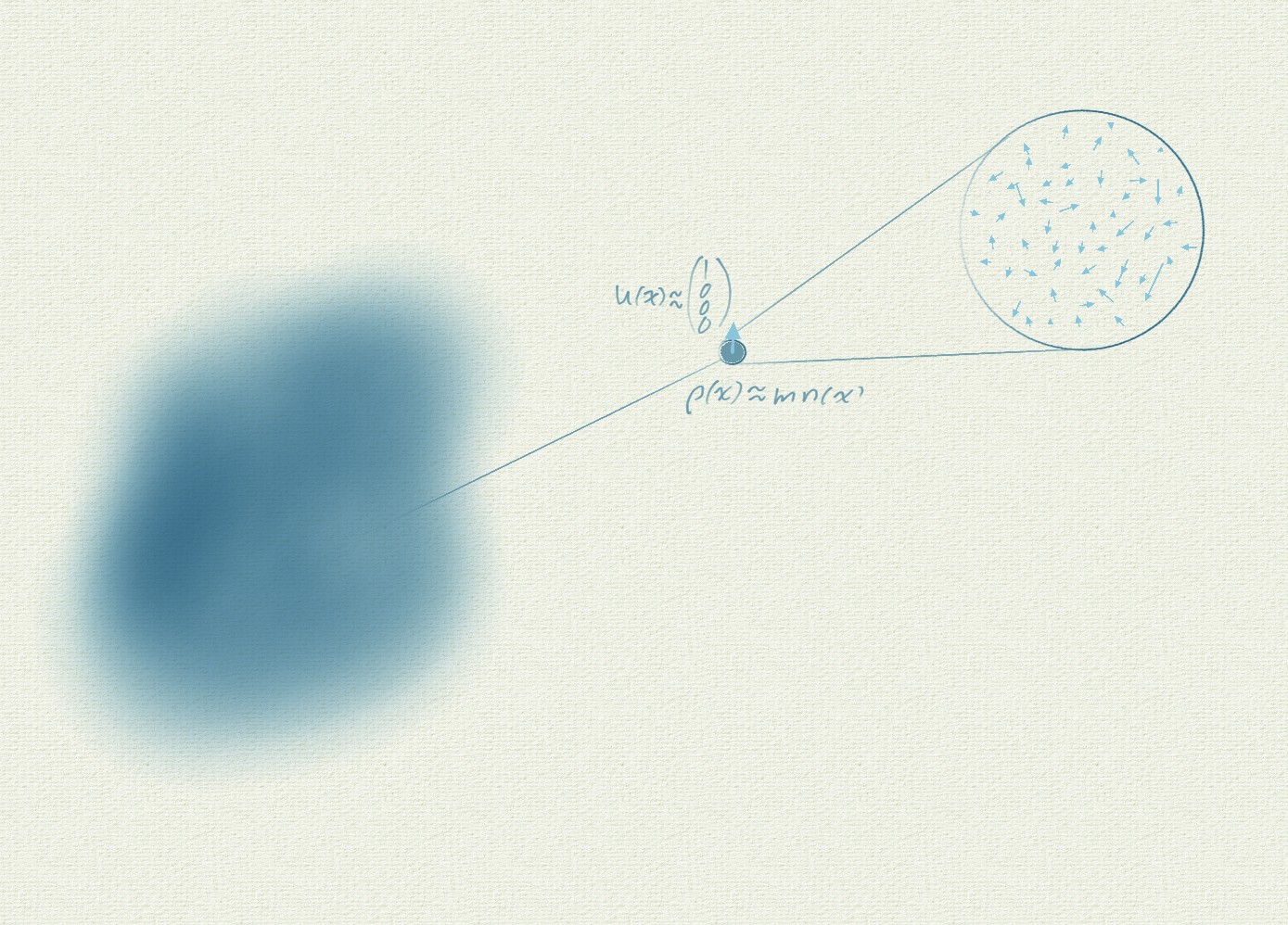}
    \caption[A non-relativistic perfect fluid.]{A non-relativistic perfect fluid. A macroscopic number of randomly distributed particles, each with their own worldlines and four-velocities, can be smoothed out by averaging over regions that are sufficiently large to wash out random relative velocities, but sufficiently small to be approximately continuous on the scale of the fluid as a whole. }
    \label{fig:ch1:dust}
\end{figure}

Imagine a field of microscopic non-interacting particles, all travelling randomly relative to each other at non-relativistic speeds (basically a gas---see figure \ref{fig:ch1:dust}). With enough constituent particles, we can zoom out to the point that any given region contains sufficiently many particles that their random motions on average cancel out, and the region is well described by an average energy-momentum density $p^\mu_i \sim (\rho_i, 0, 0, 0)^T$, computed in an inertial frame of reference centred on that region. 

Next, we would like to collect together all of these point-wise calculations of energy-momentum densities into an overall field, but we face the problem that each element was computed in its own reference frame, so gathering them together requires a choice of global reference frame. Here we make use of our simplifying assumptions, in particular that there are no long-range or external forces that induce \emph{macroscopic} motion which would require additional degrees of freedom to quantify. In that case, the motion of any one region relative to another must be small, so we may optimize that difference by finding the frame of reference that best approximates $u^\mu(x) \approx (1, 0, 0, 0)^T$, where $u : \mathbb{R}^4_M \to \vec{\mathbb{R}}^4_D$ is a \emph{vectorfield} representing the averaged four-velocity of each local region of the fluid. This reference frame is called the fluid's \textbf{rest frame}\index{Stress-Energy Tensor!Rest Frame}.

\begin{aside}[Thinking Globally]
    Defining the rest frame of the fluid is a good point to discuss the fundamental shift in perspective it takes to go from individual particles to macroscopic bodies. Any given observer lives life according to its own proper time, and when only a few observers exist, it is well worth keeping track of everyone's proper time, and calculating effects like time dilation, and playing around with apparent paradoxes. But once we graduate to working with macroscopic systems, it becomes sensible to pick one observer and just interpret the world through their eyes. In the case of a fluid, we have identified one special observer whose personal rest frame is ideally suited to monitoring the global structure of the fluid. This ``comoving'' rest frame observer will not see the world exactly the same way as any individual particle in the fluid, but it doesn't matter, they will still be able to do science, make predictions, and model the universe. This way of thinking will become more useful later on, when choosing the right reference observer becomes a critical part of modelling a system.
\end{aside}

With the special rest frame defined, we are now in a position to put together a consistent parameterization of the fluid. We define an energy density field $\rho(x)$ that consists of the average local energy density in every region $x$ \emph{as measured in the rest frame of the fluid}. Energy density alone is not an invariant quantity (see appendix \ref{ch:manifolds:app:tensorDensities}), and while $\rho(x)$ measured in the fluid's rest frame will differ numerically from $\rho_i$ measured in the local rest frame of each region, choosing the global fluid rest frame as our point of reference minimizes that difference, and what remains is a small price to pay for putting everyone on the same playing field. 

Staying here in the non-relativistic limit, the energy of each particle is approximately its mass, so the energy density is equivalent to a mass density, and for simplicity considering only a single species of particle with mass $m$, we can also write this in terms of a number density, so that $\rho(x) = mn(x)$. With no appreciable local momentum flow, and working in the rest frame of the fluid, we are now in a position to construct a stress-energy tensor field. In this frame, where all energies are inertial particle masses and all four-velocities are approximately time-directed, the only measurable energy-momentum flux is $p^0$ in the direction $x^0$, so we must have
\begin{equation}
	\label{eq:ch1:restFrameT00}
	T^{00}(x) = mn(x)u^0(x) \qquad \text{(rest frame)}.
\end{equation}

Now we get to see some immediate return on our investment of defining tensors as geometric, coordinate-free objects. Equation \eqref{eq:ch1:restFrameT00} is \emph{not} quite a tensor equation, there's a rank-2 tensor on the left-hand side, but only a rank-1 tensor on the right-hand side, so so far it's only an equation for a single component in a single reference frame. However, if we could find a second-rank tensor whose 00-component is equivalent to the right-hand side in the rest-frame of the dust, that \emph{would} be a tensor equation, and it would \emph{have} to be equal to the stress-energy tensor in all frames. While this kind of guess-work is not easy as a general rule, it is shockingly easy here. We have only one rank-1 tensor field available to us, $u^\mu(x)$, and with its simple form in the rest frame, we observe $T^{00}(x) = mn(x)u^0(x)u^0(x)$ is a proper tensor equation with the correct value in the fluid's rest frame, so define:
\begin{equation}
	\label{eq:ch1:dustT}
	T^{\mu\nu}_{\text{dust}} = mn u\otimes u = \rho u\otimes u,
\end{equation}
where we drop the depence on $x$ and take it as implied from here on.

\subsection*{Perfect Fluids}

\begin{figure}[ht]
    \centering
    \includegraphics[width=\textwidth]{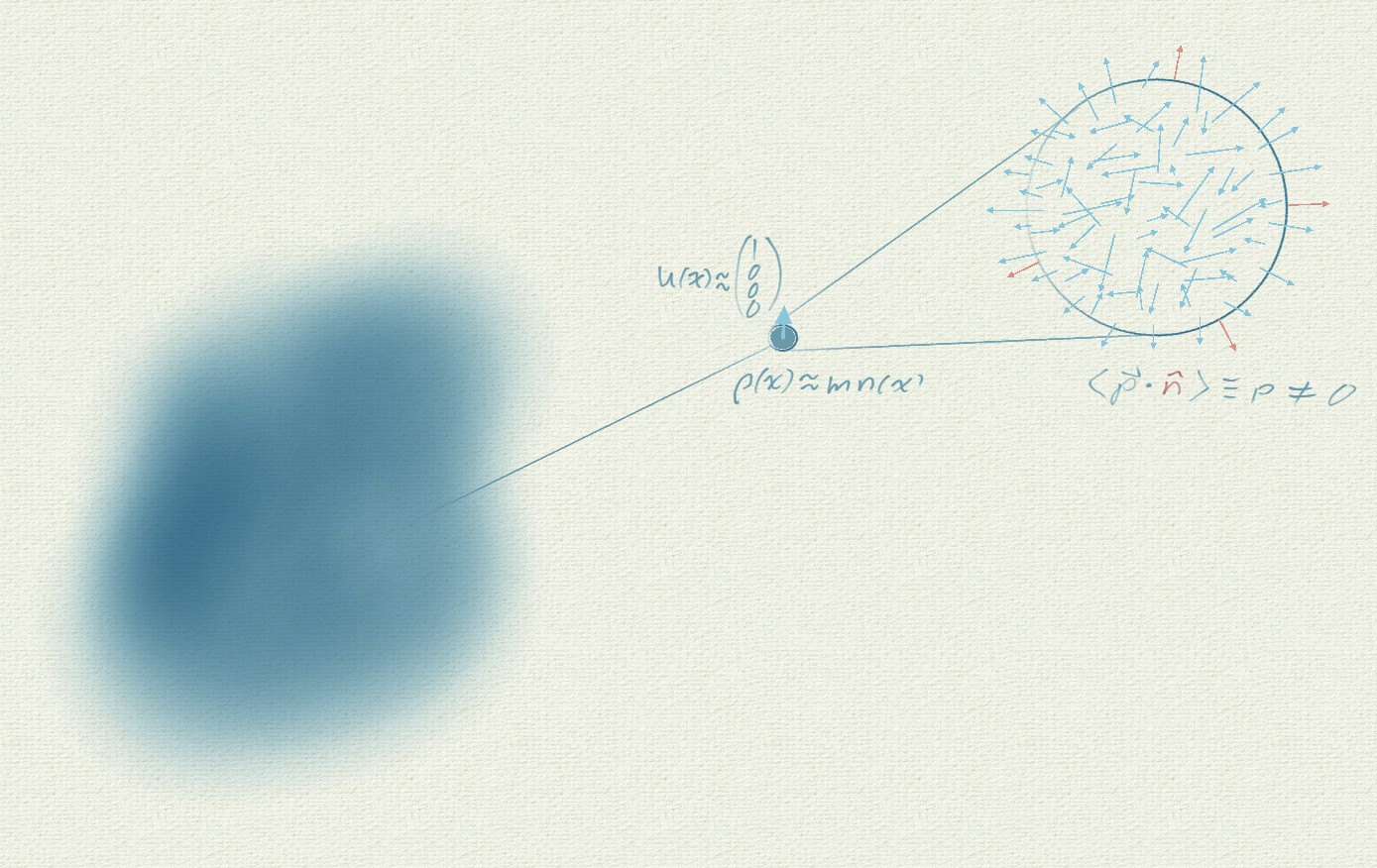}
    \caption[A relativistic perfect fluid.]{A relativistic perfect fluid. The construction here is the same as the non-relativitstic case, but now that individual particles move with relativitstic four-velocities, there is a non-zero average overlap between particle three-momenta and surface normal vectors in any bounded region (i.e., a pressure). In the simplest case (the perfect fluid), this overlap must be isotropic, otherwise anisotropies would accumulate, resulting in overall macroscopic motion in the structure of the fluid.}
    \label{fig:ch1:radFluid}
\end{figure}

Back again in the rest frame of the fluid, now we simply relax the requirement that the microscopic particle motion be non-relativistic. We do not change any other restriction, and as a result can carry over the whole discussion from above with one minor change: at the local level although motion is still random and so averages out to being overall stationary, individual particles now carry enough kinetic energy that their motion in and out of the region matters on a macroscopic scale (see figure \ref{fig:ch1:radFluid}). Particle motion still must be inertial, so particle momenta must flow linearly in and out of every local region (i.e., there must be no stress, which also implies the motion must on average be isotropic), but now in addition to a local energy density, we may also characterize each region with an average local flow of momentum parallel to coordinate directions---i.e., a local \emph{pressure} $p_i$.

As above, we may write this in components in the rest frame, and infer the appropriate tensor structure. We must have:
\begin{equation}
	\label{eq:ch1:restFrameFluid}
	T^{\mu\nu} = \mqty[\rho & 0 & 0 & 0 \\ 0 & p & 0 & 0 \\ 0 & 0 & p & 0 \\ 0 & 0 & 0 & p] \qquad \text{(rest frame)}.
\end{equation}
(Recall we have dropped the explicit field notation, but note that the pressure $p(x)$ is indeed a field as well). Now we do the same exercise as in the non-relativistic case, we have to use the tensors we know to put together a tensor expression that takes the form \eqref{eq:ch1:restFrameFluid} in the rest frame of the fluid. This time it's a little bit trickier, but one way to make the solution more apparent is write the matrix as a sum:
\begin{align}
	\label{eq:ch1:matSum}
	\mqty[\rho & 0 & 0 & 0 \\ 0 & p & 0 & 0 \\ 0 & 0 & p & 0 \\ 0 & 0 & 0 & p] &= \rho\mqty[1 & 0 & 0 & 0 \\ 0 & 0 & 0 & 0 \\ 0 & 0 & 0 & 0 \\ 0 & 0 & 0 & 0] + p\mqty[0 & 0 & 0 & 0 \\ 0 & 1 & 0 & 0 \\ 0 & 0 & 1 & 0 \\ 0 & 0 & 0 & 1], \notag \\
			   &= (\rho + p)\mqty[1 & 0 & 0 & 0 \\ 0 & 0 & 0 & 0 \\ 0 & 0 & 0 & 0 \\ 0 & 0 & 0 & 0] + p\mqty[-1 & 0 & 0 & 0 \\ 0 & 1 & 0 & 0 \\ 0 & 0 & 1 & 0 \\ 0 & 0 & 0 & 1],
\end{align}
where we used the magic of $0 = 1 + (-1)$. Now we have recognizable tensors in both terms ($u\otimes u$ in the first, and $\eta$ in the second), so we can immediately write:
\begin{equation}
	\label{eq:ch1:fluidT}
	T_{\text{Perfect fluid}}^{\mu\nu} = (\rho + p)u\otimes u + p\eta^{-1},
\end{equation}
which again is valid in all inertial frames because it is a \emph{tensor} equation. (Recall that tensors form a vectorspace so the sum of two tensors is indeed a valid tensor, and note that we wrote $\eta^{-1}$ since we need contravariant indicies to match)\index{Stress-Energy Tensor!Perfect Fluid}. 

\subsection*{Symmetry of the Stress-Energy Tensor}

It is always important to track whether or not a tensor is symmetric or anti-symmetric on any of its indices. Take the metric tensor, for example: it is very important that $\eta$ be symmetric in order to reasonably be considered a sort of inner product, and this imposes strong constraints on the possible values for its components. The stress-energy tensor also turns out to be symmetric. A detailed, general proof of this statement is somewhat orthogonal to this course, so we refer the reader to the nice explanation in Schutz \cite[\S 4.5]{schutz_2009}. In lieu of a full explanation, we offer a few short observations: a) the stress-energies of dust and perfect fluids given above are both clearly symmetric tensors, b) $T^{0i} = T^{i0}$ because energy and momentum are the same thing (energy flowing through time is  $\rho u = nmu = n p$), and c) $T^{ij} = T^{ji}$ must be true so that there is no local torque (e.g., shear in $x-y$ plane induces a torque in the $z$-plane, so it must be balanced by exactly the same shear in the $y-x$ plane, or else the material would be very odd indeed).

\subsection*{Conservation of Stress-Energy}

The stress-energy tensor also satisfies one more property, it is conserved. This is encoded in the statement:
\begin{equation}
	\label{eq:ch1:consT}
	\partial_\nu T^{\mu\nu} = 0.
\end{equation}
Notice that this is essentially a 3+1-dimensional divergence, and is the statement that energy and momentum can flow in one spacetime dimension as long as that flow \emph{came from} or is \emph{going to} another spacetime dimension. Conservation of stress-energy is not a property of the tensor as a geometric object, it has to be built in by construction, or as a consequence of an underlying symmetry (as in the case when it is derived from an action). This is the property that generalizes the notion of conservation of energy. 

To understand the conservation equation, first we have to understand how to define the derivative of a tensor. In Minkowski space this is fairly straightforward (spoiler: this will be more complicated very soon). Take the gradient of the entire tensor:
\begin{align}
	\label{eq:ch1:DvT}
	\partial T = \partial\left( T^{\mu\nu} \hat{e}_\mu\otimes \hat{e}_\nu \right) &:=  (\partial_\lambda T^{\mu\nu})\hat{e}_\mu\otimes \hat{e}_\nu\otimes\hat{e}^\lambda + T^{\mu\nu}\partial\left( \hat{e}_\mu\otimes \hat{e}_\nu \right), \notag \\
																				  &=  (\partial_\lambda T^{\mu\nu})\hat{e}_\mu\otimes \hat{e}_\nu\otimes \hat{e}^\lambda.
\end{align}
The second term in the first line vanishes because of the affine structure of Minkowski space; displacement vectors are defined throughout \emph{all} of spacetime, and as such we choose to describe displacement vectors in terms of constant, orthogonal, Cartesian bases. The end result is that the gradient of a tensor is the tensor product of the gradient of the components of that tensor (the gradient of scalar fields, which we know is a well-defined covector in Minkowski space) with the associated tensor basis vectors. For a general type-$\mqty(M \\ N)$ Minkowski tensor, the gradient is a well-defined type-$\mqty(M \\ N+1)$ Minkowski tensor. The conservation equation \eqref{eq:ch1:consT} is then just the trace of the gradient of the stress-energy tensor.

\begin{aside}[Notation]
	Derivatives of tensors are going to be pretty common in the next chapter, so for convenience, we define the following notation for tensor components. The simple partial derivative of a tensor is written $\partial_\mu T\indices{^\nu_\rho} =: T\indices{^\nu_\rho_,_\mu}$.
\end{aside}

\subsection*{Looking Forward}

Fun as that exercise was, in general it is not trivial to guess the form of meaningful tensors, and later on we will define the stress-energy tensor more generally by means of an action. For now though, we have all that we need to describe the physical situations we will study in this course. The three main topics we'll look at are: gravitational waves, black holes, and cosmology. These situations surprisingly parallel what one would find in a course on electromagnetism: sources are respectively nothing (gravitational radiation $\leftrightarrow$ EM radiation), a delta function (black hole $\leftrightarrow$ point-charge), and full distributions (perfect fluid matter $\leftrightarrow$ dielectric media). But first, we need to ``covariantize'' gravity, and that is the task we tackle in the next chapters.

\section{Recap}

Let's recap the most important points, and lay out the rules for how to use what we've learned.

\begin{itemize}
	\item Time is a \emph{place} not a parameter
	\item Minkowski space is like Euclidean space, but the inner product is not a proper inner product, it is the Minkowski metric tensor $\eta$ and it is \emph{not} positive-definite
	\item Displacement vectors are \emph{not} guaranteed to be a thing, they're a peculiarity of Minkowski/Euclidean space
	\item Vectors are abstract objects, not tied to a particular choice of basis. Think of them as carrying units, so that if we write $v = v^\mu \hat{e}_\mu$, then the vector quantities are $v$ and $\{\hat{e}_\mu\}$ so the dimensionalities are $[v] = L$, $[\hat{e}_\mu] = L$, and $[v^\mu] = 0$
	\item Despite the fact that their existence is fragile, we base our whole lives around displacement vectors, so we say that if the basis vectors of displacement vectors are transformed by a Lorentz transformation, then objects whose components are corrected by the inverse matrix are \emph{contravariant}, while objects whose components are corrected by the original transformation are \emph{covariant}
	\item Metric induces natural duality, quantities $\leftrightarrow$ rates, vectors $\leftrightarrow$ covectors
	\item Tensor product to build up more complicated structures while retaining all the linearity we can.
	\item Macroscopic theory means macroscopic description of matter-energy $\longrightarrow$ the stress-energy tensor $T_{\mu\nu}$.
\end{itemize}

\begin{subappendices}

\section{Matrices and Rank-2 Tensors}
\label{ch:geomRel:apps:matrices}

In the main text, we make a lot of use of index notation, describing vectors/covectors, tensors, and non-tensorial matrices. Vectors, covectors, and tensors have a very clear and defined structure to their index notation, but the non-tensorial matrices---especially Lorentz matrices---can be a little confusing, so here we will clarify a few things about matrices and rank-2 tensors. 

\subsection{Matrices as Notation}
Very simply, matrices are orderly boxes of numbers; they generalize the concept of vectors as tubes of numbers to $m\times n$ grids of numbers. For example, the following is a matrix:
 \begin{equation}
    \label{eq:ch1:app1:matrixA}
    A = \mqty[1 & 2 \\ 3 & 4 \\ 5 & 6].
\end{equation}
It is trivial to bestow upon these grids the powers of matrix addition and scalar multiplication, which elevates them to the status of $nm$-dimensional vectors in the abstract sense. The matrix $A$ above could, for example, be written as any other vector,  $A = 1 \hat{e}_1 + 2 \hat{e}_2 + 3 \hat{e}_3 + 4\hat{e}_4 + 5\hat{e}_5 + 6\hat{e}_6$. Typically however, it is more convenient to build into the notation a memory of the grid structure of the matrix, so we may write instead, for example:
\begin{equation}
    \label{eq:ch1:app1:matrixAComps}
A = \sum_{i = 1}^{3}\sum_{j = 1}^{2} A(i,j) \hat{e}(i,j).
\end{equation}
Alternatively, we may write this still more compactly using indices, such as $A_{ij}$ or $A^{ij}$ or $A^i_j$. In standard linear algebra uses, one would typically choose the convention $A_{rc}$ ($r$ for row and $c$ for column).

Matrix addition and scalar multiplication are very easy to represent in components, we simply arrange for the indices to line up: $C = kA + B$ is as simple as  $C_{ij} = kA_{ij} + B_{ij}$ (the component form of element-wise addition and scalar multiplication). There are, however, a few other operations that are natural (and useful) to define on matrices, the most important being \emph{matrix multiplication}. This is an exercise in matching dimensions and playing with columns and rows as individual vectors. Matrix multiplication is defined\footnote{In fact there are many possible constructions of matrix products, but the one we describe here is the universal standard.} as an operation $\cdot : M_{m\times n} \times M_{n\times p} \to M_{m\times p}$ by matching the rows of the first matrix to the columns of the second (hence why those dimensions must match, but the others need not). For example:
\begin{equation}
    \label{eq:ch1:app1:exMult}
    \mqty[1 & 2 \\ 3 & 4 \\ 5 & 6]\mqty[v_1 & w_1 \\ v_2 & w_2] = \mqty[(1v_1 + 2v_2) & (1w_1 + 2w_2) \\ (3v_1 + 4v_2) & (3w_1 + 4w_2) \\ (5v_1 + 6v_2) & (5w_1 + 6w_2)].
\end{equation}
Consistency and care allow the construction of a component representation of this product. If $C = AB$, we write  $C_{ij} = \sum_k^n A_{ik}B_{kj}$. Note that any other choice of index placement in $A$ and  $B$ is also perfectly acceptable as long as we know we are summing over the \emph{columns of A} and  \emph{rows of B}.

To finish up, the following matrix operations/properties are also very natural to define, and it is helpful to spell out their component forms. The \emph{transpose} operation interchanges rows and columns, so naturally $(A^T)_{rc} = A_{cr}$, and the \emph{trace} operation is the sum of the diagonal elements, so easily written $\Trace{A} = \sum_i A_{ii}$.

\subsection{Matrices as (Bi)Linear Maps}
Fundamentally, matrices are just notation like column and row vectors, and the principle confusion between matrices and tensors lies in simple errors in the use of matrix notation in representing tensors. A secondary (related) confusion, however, also arises due to preconceptions from linear algebra. Matrices of real numbers are also the natural representation of linear maps on row and column vectors, so we are very used to thinking of transformations on columns and rows of real numbers in matrix form using the matrix multiplication defined above, e.g. $\vec{w} = A\vec{v}$ for  $\vec{v} \in \mathbb{R}^n$ and $\vec{w} \in \mathbb{R}^m$. This naturally leads to prescriptions for how matrices transform under changes of bases. For instance, we might recall that linear maps are meant to change as $A \to B^{-1}AB$ under a change of basis operation $B$, but also that bilinear maps (e.g., $\vec{w}^T A \vec{v} \in \mathbb{R}$) must instead transform as $A \to B^T AB$. In linear algebra, we learn these rules by rote memorization, but with tensors, we now have a systematic way to relate (bi)linear maps and matrices.

\subsection{Rank-2 Tensors Represented as Matrices}
Rank-2 tensors are $nm$-dimensional vectors, and any two vectorspaces of equal dimension are isomorphic, so all rank-2 tensors (and the linear operations involving them) can be represented as  $n\times m$ matrices. What's more, the notation used to define rank-2 tensors is eerily similar to that used to define matrices, so the mapping is trivial,
 \begin{equation}
    \label{eq:ch1:app1:mapTensMat}
    T_{ij} \hat{e}_i\otimes \hat{e}_j \to \sum_{ij} T_{ij} \hat{e}_{ij} = \mqty[ T_{11} & T_{12} & \ldots \\ T_{21} & T_{22} & \ldots \\ \ldots & \ldots & \ldots].
\end{equation}

Each of these notations has advantages and disadvantages. Matrix notation is very visual, and can make calculations with specific tensors very quick to carry out. However, the tensor index notation carries much more information than its matrix representation; phrased another way, the map from tensors to matrices is not unique so the same matrix can represent any number of different tensors. Most prone to confusion is the fact that $T_{ij}$, $T^{ij}$, $T\indices{^i_j}$ and $T\indices{_i^j}$ are all represented by the same matrix. 

Although confusing at first, starting with tensors can help make sense of matrix equations and their behaviours. For instance, linear maps on vectors can now be thought of as elements of $V\otimes V^*$. Consider an element $T\indices{^i_j} \hat{e}_i\otimes\hat{e}^j$ acting on a single vector $v^k \hat{e}_k$:
\begin{align}
    \label{eq:ch1:app1:tensLinOp}
    T(\cdot, v) &= T\indices{^i_j} \hat{e}_i\otimes \hat{e}^j\left( v^k \hat{e}_k \right), \notag \\
                &= T\indices{^i_j}v^k \delta^j_k \hat{e}_i \notag \\
                &= T\indices{^i_j}v^j \hat{e}_i.
\end{align}
Representing $T\indices{^i_j} \hat{e}_i\otimes \hat{e}^j$ in matrix form as $\sum_{ij} T_{ij} \hat{e}_{ij}$, we have simply $T(v) = \sum_{j} T_{ij} v_j$. But now we \emph{also} know immediately how $T$ transforms under a change of basis: 
\begin{align}
    \label{eq:ch1:app1:tensChangeB}
    (T^\prime)\indices{^i_j} = T\indices{^k_l} (B^{-1})_k^i B^l_j.
\end{align}

To represent the change of basis calculation in matrix form, recall the pairing of rows and columns in matrix multiplication (columns of the first matrix, rows of the second). If we choose the first index of $T$ to be rows and the second to be columns, then the index that sums over the columns of $T$ must be the rows of $B$, and the index that sums over the rows of $T$ must be the columns of $B^{-1}$: 
\begin{align}
    \label{eq:ch1:app1:matChangeB}
    T^\prime &= \sum_{kl} T_{kl}B^{-1}_{ik}B_{lj}, \notag \\
             &= \sum_{kl} B^{-1}_{ik}T_{kl}B_{lj}, \notag \\
             &= B^{-1} T B,
\end{align}
where in the second line, we rearranged the factors to align the summed indices for visual ease. Given a specific form for $T$, and $B$, the matrix form \eqref{eq:ch1:app1:matChangeB} can be much easier to evaluate directly than the component-wise tensor expression, but as you can see, it takes care to set up the matrix equation correctly. A good exercise is to repeat this procedure for an element of $V\otimes V$ acting on two vectors (i.e., a bilinear form, so you should recover the linear algebra expression $B^T T B$).


\subsection{Some Examples}
Practice makes perfect. Consider the following: 
\begin{equation}
    \label{eq:ch1:app1:defC}
    C\indices{^a_b} = A^{ac}B_{cb}.
\end{equation}
Here, $A, B,$ and $C$ are all elements of different tensor product spaces and $C$ is constructed from $A$ and $B$ through a two-step process---first $A$ and  $B$ are tensored together, and then two indices are contracted. This calculation looks almost desperate to be in matrix form, so starting with $A$, choose to represent the first index as its rows and the second as its columns, so naturally the columns and rows of $B$ must similarly be its first and second indices, and the matrix representation of \eqref{eq:ch1:app1:defC} is as simple as $C = AB$. 

The second step in the construction of $C$ required a choice of indices to contract. Here we chose the second of $A$ with the first of $B$, but other choices were possible. We could instead have chosen to define:
 \begin{equation}
    \label{eq:ch1:app1:defD}
    D\indices{^a_b} = A^{ac}B_{bc}.
\end{equation}
Now the matrix form is a little less obvious, since the indices are not aligned. Suppose we choose $A$ as our reference again, so that its first and second indices are repsectively its rows and columns. Then to represent this process as matrix multiplication, the indices of $B$ must be inverted, its first and second representing its columns and rows. While we are legally allowed to do this, it is a bit perverse to represent $A$ and $B$ in opposite ways---much better to represent $A$ and $B$ in the same way, and simply add more detail to the representation of their product. In that case, start with matrix components:
\begin{align}
    \label{eq:ch1:app1:repD}
    D_{ab} &= \sum_{c} A_{ac}B_{bc}, \notag \\
           &= \sum_{c} A_{ac} (B^T)_{cb}, 
\end{align}
so that the easiest way to represent the calculation of $D$ in matrix notation is as $D = AB^T$. As an exercise, do the same for the contractions based on the first index of $A$.

\subsection{When a Matrix is Not a Tensor}

Finally, it bears repeating that matrices are a \emph{notation}, and not all matrices represent rank-2 tensors. A priori, no matrix represents a tensor until its tensor structure is spelled out, but even then some matrices are never meant to represent tensors. We have already encountered two examples of this: 1) components of a tensor specified in a specific frame, and 2) change-of-basis matrices (i.e., Lorentz matrices at this point). 

Components of a tensor in a specific frame are just numbers. Typically we see this when trying to construct a tensor from physical principles, like when constructing the perfect fluid stress-energy tensor; the expressions
\begin{equation}
    \label{eq:ch1:app1:TcompsDef}
    T^{\mu\nu} = \begin{cases}
        \rho u^0, &\quad \mu = \nu = 0, \\
        0, &\quad \text{else}
    \end{cases}, 
\end{equation}
is easily represented as
\begin{equation}
    \label{eq:ch1:app1:TcompsMat}
    T^{\mu\nu} = \mqty[\rho & 0 & 0 & 0 \\
    0 & 0 & 0 & 0  \\
    0 & 0 & 0 & 0  \\
    0 & 0 & 0 & 0  
    ],
\end{equation}
but the first equation is not a tensor equation, so neither is the second.

And change-of-basis matrices are not tensors. They are not vectors (one does not add, subtract, and scale changes-of-bases) and their indices do not represent components of vectors (one would not perform a change-of-basis on a change-of-basis matrix). More technically, a change-of-basis is a \emph{passive} operation, it is equivalent to the identity map if acting on vectors. This is in stark contrast to the view of tensors as (multi-)linear operators on vectors, where each individual tensor represents an \emph{active} transformation that turns one vector into a different one. The confusion here arises because every change-of-basis matrix is defined in reference to the action of an active linear transformation applied individually to the original basis vectors, but when used passively on any \emph{other} vector, it does not substantively change that vector, and hence is no longer a (non-trivial) linear map.

\section{The Universal Property of Tensors}

Mathematicians like to be a bit more technical in their definition of tensors, typically defining them in terms of a ``universal property.'' It may seem obtuse, but their definition really does coincide with our notion of the ``most general multi-linear way to combine vectors.'' We'll go through this correspondence explicitly here.

Formally, define the tensor product as follows. For any two vectorspaces $V$ and $W$, the \textbf{tensor product space} $V\otimes W$, together with the \textbf{tensor product}  $\otimes : V\times W \to  V\otimes W$, are the unique vectorspace and bilinear product (up to isomorphism) that satisfy the following \textbf{universal property}: for any bilinear map $f : V\times W \to B$ that takes elements of $V$ and $W$ to the vectorspace $B$, there exists a \emph{unique} linear map $F : V\otimes W \to B$ such that $f(v, w) = F(v\otimes w)$.

I won't prove here that the tensor product defined above satisfies this definition (that's a good exercise for you), but I just want to draw your attention to this universal property as a mathematical statement of our intuition of the tensor product. The universal property states that \emph{any} other way you can think of to combine two vectors in a bilinear way is equivalent to a \emph{linear} map from the tensor product space. It is in this sense that the tensor product encodes the \emph{multi}-linearity of any combination of vectorspaces, and its uniqueness (up to isomorphism) is what makes it the most general way to do so. This is actually a very powerful statement; in a world in which the tensor product was not unique, physicists would have to construct their theories with all types of objects that could be constructed from basic vector spaces and experiments would need to be used to determine which objects represented elements of the real world. If some objects were excluded, it would raise questions as to why Nature preferred some structures to others, and if none were excluded, physical theories could be intractably complex! So next time you find yourself writing down a nice, succinct physical law, thank the tensor product.

\end{subappendices}

\chapter{Manifolds, Tangent Vectors, Cotangent Vectors, and Tensors}
\label{ch:manifolds}

Special relativity is all well and good, but what's this we keep hearing about gravity? What could that possibly have to do with anything we've just seen? Well of course, the setting of Minkowski space, and removal of time as a parameter, requires a refactoring of our other physical laws, all of which are typically written in what some call a ``non-covariant'' way. In principle though, that should be a separate topic, a separate course of relativistic electromagnetism, relativistic thermodynamics, relativistic quantum mechanics, etc. Why is the theory of relativistic gravity called the \emph{general} theory of relativity? The answer boils down to a very simple but profound physical fact that serves as the first of two new postulates we need to add to special relativity to make it a theory of gravitation:
\begin{postulate}[The Equivalence Principle]
   Inertial mass is equal to gravitational mass, $m_I = m_G$. 
\end{postulate}
This is (a phrasing of) \textbf{the Equivalence Principle}\index{Equivalence Principle}: inertial mass (the one in Newton's second law, $\vec{F} = m_I \vec{a}$) is equivalent to gravitational mass (the reactive mass in Newtonian gravity \\$F_G = -G m_G M/r^2$). It directly follows from the equivalence principle that no two bodies can exist unaccelerated, and that \emph{the gravitational acceleration of every body is identical, regardless of its gravitational charge ($m_G$)}. That is to say, while it's not theoretically impossible to construct a relativistic version of Newtonian gravity\footnote{And in some practical cases, this is actually useful, cf. Gravitoelectromagnetism, \cite{maartens_gravito-electromagnetism_1998}.}\ldots why would you? If everything responds the same way to everything else's gravitational field, why bother with a gravitational field at all, why not just map out the geometry that everything will follow? 

\begin{aside}[Different Types of Equivalence Principles]
    Out in the wild, you will find different phrasings and versions of the equivalence principle. What we have just defined (the postulated equivalence of inertial and gravitational mass) tends to go by the name of the \textbf{Weak Equivalence Principle}. If we can take it a step further and postulate that all other physical laws take the same mathematical (and numerical) form in every local inertial reference frame, we arrive at the \textbf{Einstein Equivalence Principle}. Finally, we can be very strict and require that \emph{even gravitational binding energy} behaves exactly the same way in every local inertial reference frame. This is known as the \textbf{Strong Equivalence Principle}, and it postulates the \emph{complete} equality of gravity with geometry. It won't come up in this course, but technically we'll be assuming at least the Einstein equivalence principle.
\end{aside}

The absence of an impartial observer really is a fundamental point here as well. How do we construct every other physical theory? Who can impartially observe the electromagentic interaction? Surely someone without electric charge. Who can observe the nuclear interactions? Again, someone with no nuclear charge. Even thermodynamic ensembles can safely be observed by someone not involved in the ensemble (e.g.~the comoving observer that defines the coordinates of the rest frame of the perfect fluid from section \ref{ch:geomRel:stressEnergy}). But in gravity, every potential observer has mass and must partake in the festivities, and the only objects without mass (photons, gluons, maybe gravitons) are cursed to live without an inertial frame of reference.

This is essentially the argument Einstein used to derive the general theory of relativity. Today, it's very common to be introduced to this line of thinking through the \emph{gedanken} experiment of a person in an accelerating rocket in space (under a constant acceleration, and with no windows in the craft, any experiment the astronaut performs in the rocket will return precisely the same results as it would in a lab on Earth---an isolated, non-rotating Earth, of course). Whichever way you slice it, the statement is that gravity is identical to non-trivial spacetime geometry, so we have two tasks before us: 1. How do we understand motion in curved spacetime? and 2. What equation describes the geometrical response to matter? The first question just requires us to learn some math, and is the topic of the next couple of chapters. The second question is answered by the Einstein Field Equations. Strictly speaking, the EFEs are an axiom of the theory, but we will at least motivate them as the equations of motion of spacetime geometry in chapter \ref{ch:EFEs}, and with the rest of the material explore some of their physical applications.

\section{Differentiable Manifolds}
\label{ch:manifolds:manifolds}

When two or more massive bodies exist in spacetime, it is impossible for any of them not to accelerate. Phrased another way, it is impossible to associate a global inertial frame with any object when there exists more than one object. Think about it like this, if I try to use the (isolated, non-rotating) Earth as my global coordinate system, I will immediately fail, as no object that I let go of will stay put, they'll all start moving towards the ground (violating Newton's first law). And from the perspective of the ball I drop, the Earth will rapidly approach it, no matter how much it asks it not to. There was a crucial word there though: \emph{global}. Einstein's brilliant insight that led to GR is completely analogous to the brilliant insight that led Newton to his laws of motion. When Newton wanted to study the complicated curved trajectories followed by physical bodies, he broke their motion down into small line segments and invented calculus; when Einstein wanted to study the curved geometry that results from universal gravitation and the equivalence principle, he broke the big curved universe down into small, flat patches, and (co-)invented differential geometry. Where Newton realized all we really know how to work with are lines, Einstein realized all we really know how to live in is a Cartesian space, so the general strategy of GR is to describe physics as \textbf{locally Minkowski}, even in situations where it can't be Minkowski everywhere. The statement that spacetime looks like something \emph{locally} actually has two meanings, one in terms of the global structure (that of a manifold) and another in terms of the setting for vectors (tangent spaces). We'll treat them in turn; first the universe.

Step 1 of GR is constructing spacetime as a \emph{differentiable manifold}. We'll start with the definition, and pick it apart. 
\begin{definition}[Manifolds]
    A smooth \textbf{differentiable manifold}\index{Differentiable Manifold} is a topological space\footnote{Strictly speaking also Hausdorff (ensuring the ability to isolate points from each other) and second-countable (just making sure you don't need infinite coordinate patches).} $M$ together with an atlas of charts $A = \cup_i (U_i, \varphi_i)$ such that $M$ is fully contained in $A$, such that the coordinate functions $\varphi_i$ are $C^\infty$ and invertible, and such that the transition functions are also $C^\infty$.  
\end{definition}
That's a lot, I know, but it's really just mathematical lingo for a very intuitive idea: we only understand how to think about points as numbers, and while not everything in Nature is as simple as $\mathbb{R}^n$, everything important can at least be broken up into segments that can be viewed through the lens of $\mathbb{R}^n$. Maybe the best part of this lingo is that the overly-generic mathematical language handily represents the enormous freedom one has to say exactly \emph{how} to cover these lovely things in patches of $\mathbb{R}^n$. Let's break it down.

\begin{figure}[ht]
	\centering
	\includegraphics[width=\linewidth]{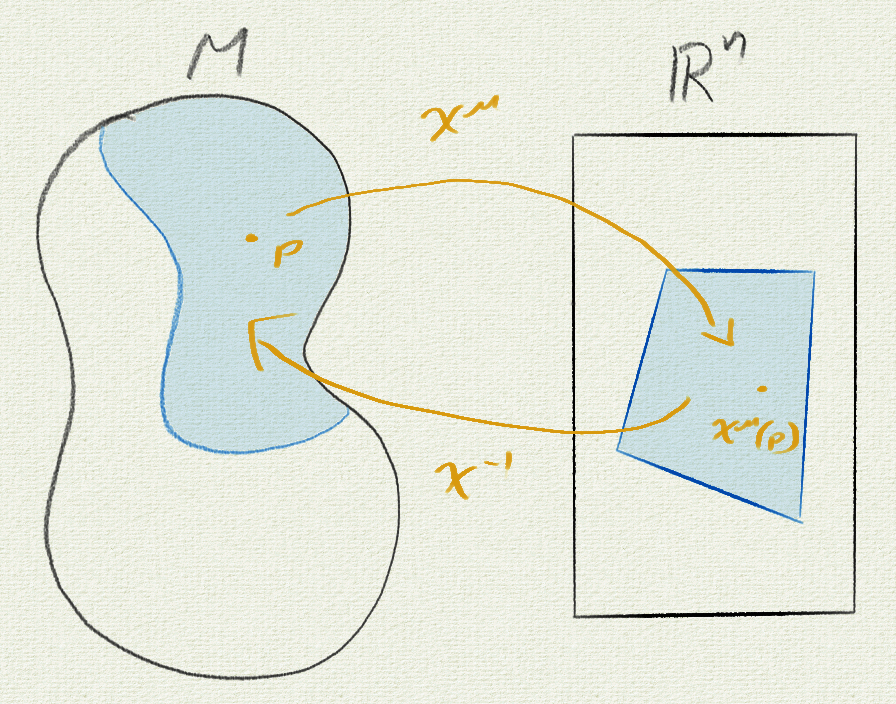}
        \caption[The coordinate functions on a manifold.]{The coordinate functions on a manifold. The coordinate charts formalize the intuitive notion of coordinates for a given point, $x^\mu(p)$.}%
	\label{fig:manifold_coords}
\end{figure}

\begin{itemize}
	\item A topological space $M$ is a bunch of points with a minimum of structure (an ability to define limits, group points, and whatnot).
	\item A $C^n$ function is a function whose $n$th derivative is continuous, so a $C^0$ function is just a continuous function, while a $C^\infty$---or smooth---function is infinitely differentiable. It is important we work with smooth functions because we have good empirical evidence that everything physical is smooth, there are no singularities anywhere in Nature. 
        \item A coordinate chart is a map $\varphi$ from an open subset $U \subset M$ to an open subset of the \emph{topological space} $\mathbb{R}^n$ (\emph{not} Euclidean space, we do \emph{not} include a vectorspace structure here), see \ref{fig:manifold_coords}. It is important the sets be \emph{open} (so do not include boundaries) so that derivatives can be well defined, and it is important that $\varphi$ is both smooth (because Nature is) and invertible because \ldots
	\item Whenever two coordinate charts $(U, \varphi)$ and $(V, \psi)$ overlap, they have to agree on what they're mapping, so in the overlap $U \cap V$, it better be the case that  $\psi(\varphi^{-1})$ and $\varphi(\psi^{-1})$ are completely well defined and well behaved.
\end{itemize}

Always best to put things in context. Consider the very simple manifold $\mathbb{R}^2 - L$, where $L$\footnote{Well, $L$ is a semi-infinite line, but a good place to put a positive $x$-axis.} is the positive $x$-axis (including the origin). There are loads of ways to cover this manifold in charts, but one way we're very familiar with is polar coordinates, so we can say the ``coordinate map'' $x_{\text{pol}}^\mu : M \to V \subset \mathbb{R}^2$ looks like $x_\text{pol}^\mu(p) = (r, \theta)$. As physicists, we're maybe more familiar with going the other way; we visualize the ``polar'' manifold in terms of $x_\text{pol}^{-1} : V \to M$ with $x_\text{pol}^{-1}(r, \theta) = (r\cos\theta, r\sin\theta)$, called a \textbf{paramaterization}\index{Coordinate Charts!Parameterization} of the manifold patch. In practice, we need both directions: the fundamental things we need to work with are coordinate charts, but we need to know what they look like to use them, and to know what they look like we need to know how to represent points on the manifold, but a parameterization of points on the manifold is just the inverse of a coordinate chart. Confused? Good. An example should sort us out, lets see what a simple coordinate transformation means in this language. 

\begin{figure}[ht]
	\centering
	\includegraphics[width=\linewidth]{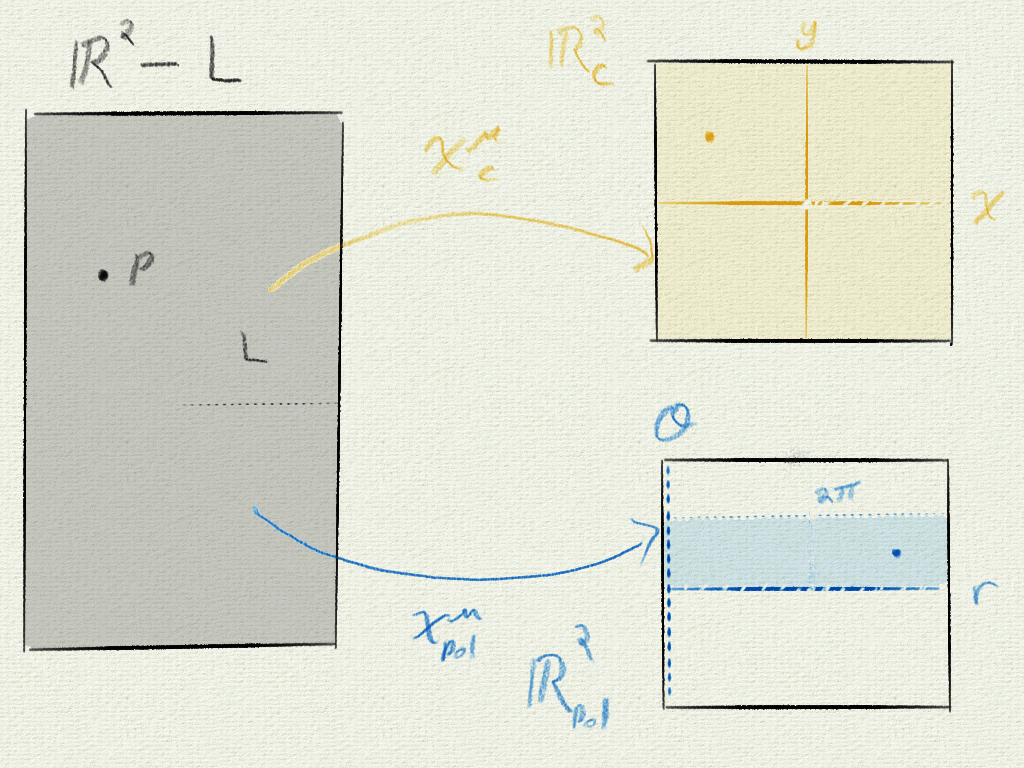}
        \caption[The manifoldy way of changing from Cartesian to polar coordinates.]{The manifoldy way of changing from Cartesian to polar coordinates on $\mathbb{R}^2 - L$. The dotted line on the manifold is the removed line $L$, the shaded regions indicate the support and range of the coordinate maps (dashed lines not included), and the points in each plane represent respectively $p$, $x_\text{C}^\mu(p)$, and $ x_\text{pol}^\mu(p)$.}%
	\label{fig:manifold_coordChange}
\end{figure}

Again considering $\mathbb{R}^2 - L$, this time we'll look at two charts, the polar coordinates $x_\text{pol}^\mu(p)$ and Cartesian coordinates, $x_\text{C}^\mu(p)$, and we'll imagine taking a point $p$ in Cartesian coordinates and changing it into polar coordinates (see figure \ref{fig:manifold_coordChange}). The Cartesian map is exactly what it sounds like, $x_\text{C}^\mu(x,y) = (x,y) \leftrightarrow x_\text{C}^{-1}(x,y) = (x,y)$, while the polar map is as given above. A change of coordinates is a map between $\mathbb{R}^2_{\text{Cart}}$ and $\mathbb{R}^2_{\text{Polar}}$ given by function composition (and in general only defined on the overlap of the patches on the manifold, but they coincide in this example). That is, changing our outlook on the manifold means taking all the points we had mapped with Cartesian coordinates, un-mapping them, then mapping them with the new function, polar coordinates: $ x_\text{pol}^\mu(x_\text{C}^{-1}): \mathbb{R}_{\text{Cart}} \to \mathbb{R}_\text{Polar}$. Explicitly, for a point $(x, y)$ in $\mathbb{R}_{\text{Cart}}$, we have $x_\text{pol}^\mu(x_\text{C}^{-1}(x, y)) = x_\text{pol}^\mu(x, y) = (r, \theta)$ where $r$ and $\theta$ satisfy:
\begin{align}
	\label{eq:ch2:cartToPol}
	x = r\cos\theta, \quad &\text{and} \quad y = r\sin\theta, \notag \\
	\implies r = \sqrt{x^2 + y^2}, \quad &\text{and} \quad \theta = \atan(y/x).
\end{align}
so we have $x_\text{pol}^\mu(x_\text{C}^{-1}(x,y)) = (\sqrt{x^2 + y^2}, \atan(y/x))$. Very importantly, notice that the coordinate transformations $x_\text{pol}^\mu(x^{-1})$ are \emph{not} usually linear functions---a major departure from the lovely linear Lorentz transformations from chapter \ref{ch:geomRel}! (Ex: perform this calculation again, but now instead of regular Cartesian coordinates, use the rotated coordinates $x^\mu(x + y, x - y) = (x,y)$).
\begin{aside}[No Vectors!]
	Here is a good place to really emphasize the point that so far there are \emph{no} vectors in sight. Sure $\mathbb{R}_\text{Cart}$ again can easily be given additional vectorspace structure, but $\mathbb{R}_\text{Polar}$ \emph{definitely} cannot (try it out, take a putative ``vector'' $(0, 3\pi/2)$ and add it to itself, you'll get $(0, 3\pi)$, and since $3\pi > 2\pi$ it is \emph{not} in the space). This is the \emph{norm} for manifolds, not the exception, so we simply have to give up any notion of a general displacement vector.	
\end{aside}

In differential geometry, the best non-trivial example is always a sphere. Consider an ordinary 2D sphere (called the two-sphere, or $S^2$). Clearly it is most convenient to label points on the sphere with polar coordinates, $(\theta, \phi)$, but this is actually harder than it looks. First, there is the problem of open sets; charts on manifolds need to be defined on open sets, and need to be one-to-one, so for angular coordinates, we always have to exclude a whole line of longitude in order to keep the coordinates well-defined, so it takes a \emph{minimum} of two charts to cover the sphere. The other issue is in finding a good mapping to $\mathbb{R}^2$ (and back). Even though the coordinates $(\theta, \phi)$ look nice, we have to remember they represent points on a sphere, so the maps to and from $\mathbb{R}^2$ must reflect that. We won't dwell on it here, but it is a good idea to look at some texts (e.g., \cite{spivak1999comprehensive}) for the ``stereographic projection'' representation of the 2-sphere, and for a simpler example on the circle see appendix \ref{ch:manifolds:app:coordExamples}. We'll just observe two important lessons the 2-sphere tells us about general manifolds: a) a complete description of the two-sphere requires at least two coordinate patches, and b) the simplest coordinates are not always the simplest coordinates.

This is the first meaning of spacetime being ``locally Minkowski,'' that you know how to translate whatever weird coordinates you're using into sensible sets of numbers and back again (even if you can't do it the same way everywhere). The next step is to figure out what vectors mean on a manifold, and this is why we so strongly emphasized the unique affine structure of Minkowski space earlier. Again, a general collection of points is \emph{not} an affine space by default, there do not big displacement vectors like in Minkowski space that span the whole world. Think about the two-sphere again; what happens if I take the point $(\pi/2, \pi/2)$ and add it to itself? Well that would be a pole, and we said poles can't be contained in the simple $(\theta, \phi)$ coordinate chart. Try as you might, you can't come up with a consistent vector space structure on just the coordinate charts of a manifold. However, since vectors are crucial to physics, we're going to have to find some way of adapting them to curvy spaces. 

\section{Tangent Vectors, Cotangent Vectors, and Tensors}
\label{ch:manifolds:vecs}

To figure out how to construct some sort of vector on general spacetime manifolds, we have to invoke a second notion of ``locally Minkowski.'' On the one hand, spacetime is locally Minkowski if we can understand its points in terms of mappings to $\mathbb{R}^n$ (the topological space), but on the other hand we can also think about a space being approximately Minkowski if it looks like Minkowski space when we zoom in \emph{really, really} close. That is, maybe we can't define general displacement vectors in a general spacetime, but if we look at really, really, infinitessimally small displacement vectors? Well surely those must be valid---after all, we were able to study physics for a long time before GR came around.

\begin{figure}[htpb]
	\centering
	\includegraphics[width=\linewidth]{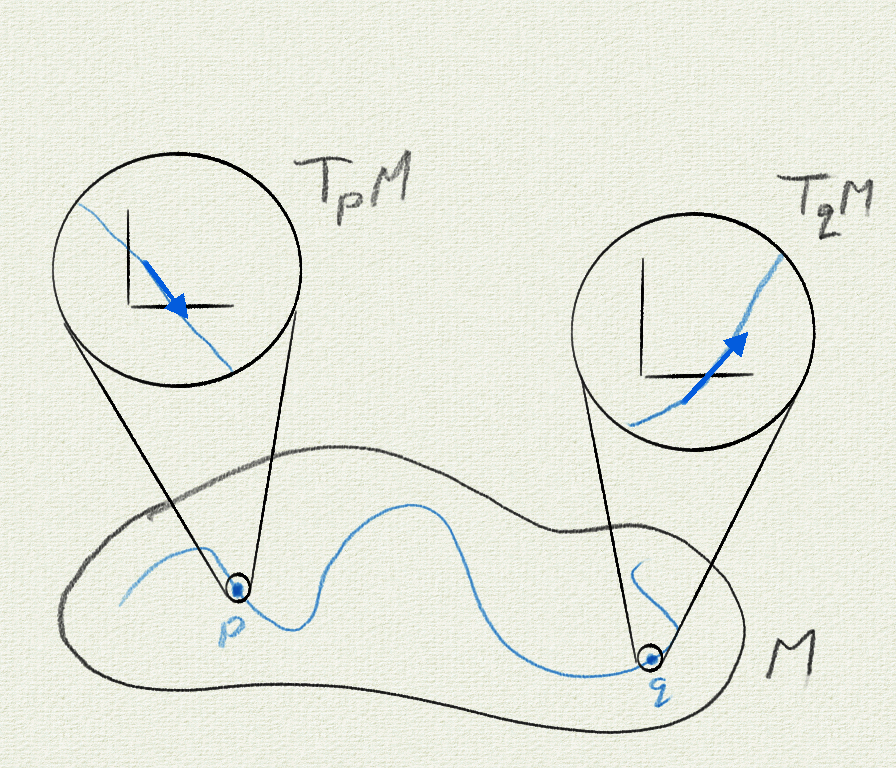}
	\caption{A visualization of a path traced out in a manifold, and the tangent spaces associated with different points.}%
	\label{fig:manifold-tangents}
\end{figure}

At any point on the manifold of spacetime, a human-sized ideal observer can be placed and that observer can establish what they think of as a Minkowski coordinate system, complete with its affine structure of displacement vectors. What this tiny observer sees as global displacement vectors are infinitessimal in the grand scheme of things, but as we know from section \ref{ch:geomRel:sec:vec}, infinitessimal displacement vectors correspond to tangents of worldlines. In this way, the only vectors that survive the transition from SR to GR are the tangent vectors to worldlines, e.g.: $v_p = \frac{d}{ds}\gamma(s)\vert_p$ is a valid vector (where $\gamma : \mathbb{R} \to M$ is a curve along the manifold, $s$ is the parameter of the curve, and $p$ is a point on that manifold). This leads us to define:
\begin{definition}[Tangent Space]
    At any point $p$ in the spacetime manifold $M$, the set of all tangent vectors to curves forms a vectorspace called the \textbf{Tangent Space at p}\index{Tangent Space}, denoted $T_pM$. In general, the tangent spaces at different points $p$ and $q$ are entirely different vectorspaces $T_pM$ and $T_qM$ (see figure \ref{fig:manifold-tangents}), and vectors in one cannot interact with vectors in the other.
\end{definition}
%

%

\subsection*{Vectors and Vectorfields}

\begin{figure}[htpb]
	\centering
	\includegraphics[width=\linewidth]{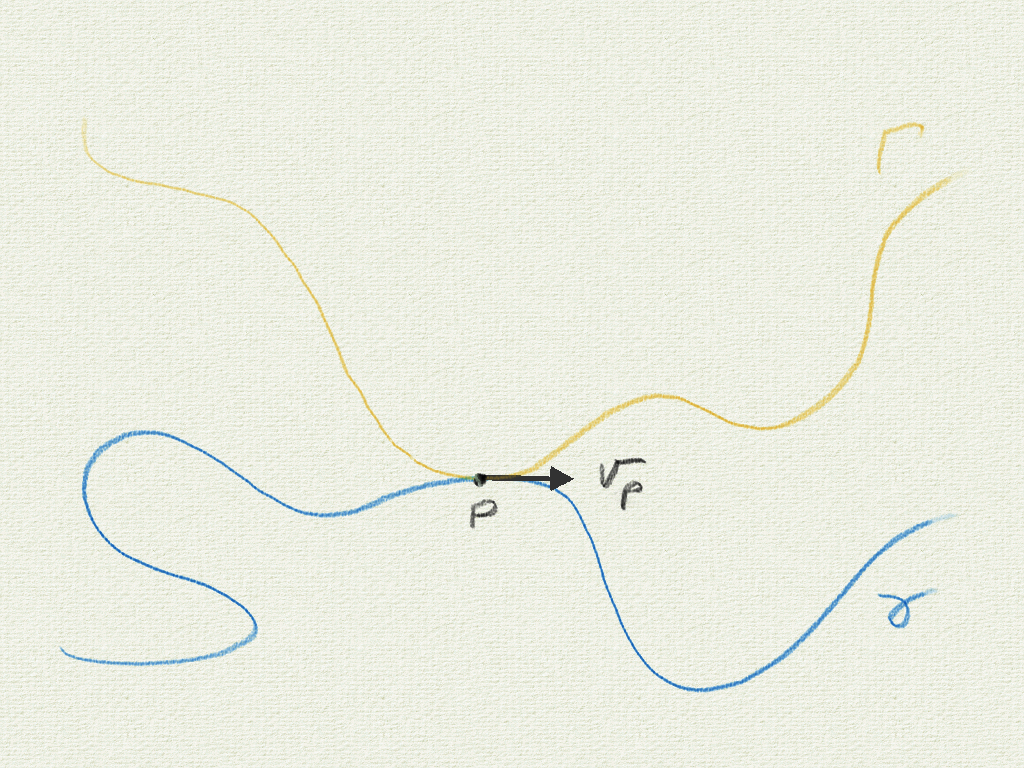}
	\caption{Two very different curves with the same tangent at a point $p$.}%
	\label{fig:manifold-tanvecs}
\end{figure}

In fact, we can do even a bit better than to think about tangents to specific paths. The tangent space is tangents to \emph{all} paths, so if a vector can be defined by a given path: $v_p = \frac{\dd}{\dd s}\vert_{s_0}\gamma(s) = \frac{x_\gamma(s)^\mu}{\dd s}\vert_{s_0}\hat{e}_\mu(p)$, then surely the same vector can just as well be defined by another path (see figure \ref{fig:manifold-tanvecs}) that just \emph{happens} to have the same tangent at $p$: $v_p = \frac{\dd}{\dd r}\vert_{r_0}\Gamma(r) = \frac{\dd x^\mu(\Gamma(r))}{\dd r}\vert_{r_0}\hat{e}_\mu(p)$ (where $\gamma(s_0) = \Gamma(r_0) = p$), which suggests that the important thing isn't the derivative of a specific path, but rather the derivative itself. That is, we can equally well define the tangent space as the set of all $\frac{\dd}{\dd s}\vert_{s_0} = \frac{\dd x^\mu}{\dd s}\vert_{s_0}\partial_\mu\vert_{p}$. This even lets us pick out the handy dandy basis vectors $\{\hat{e}_\mu(p)\} = \{\partial_\mu\vert_{p}\}$. This identification is so important it warrants its own definition.
\begin{definition}[Tangent Vectors]
	An element (a \textbf{tangent vector}\index{Vectors!On Manifolds}) of the tangent space $T_pM$ at point $p \in M$ is an object $\dv[]{}{s}\vert_{s_0}$ that generates tangents to curves (where the parameterization is chosen such that $\gamma(s_0) = p$ for all curves). If $p$ lives in a coordinate chart $x^\mu : M \to R^n$, then $\dv[]{}{s}\vert_{s_0} = \dv[]{x^\mu}{s}\vert_{s_0}\partial_{\mu}\vert_{p}$, and the coordinate partial derivatives $\{\partial_\mu\vert_{p}\}$ form a basis for the tangent space.
\end{definition}
Moreover, the expression $\partial_\mu\vert_{p}$ is begging to have that evaluation symbol plucked right off, and in fact we can immediately extend the definition of a vector to a \emph{vectorfield}:
\begin{definition}[Tangent Vectorfields]
    A \textbf{tangent vectorfield} is an object that returns a tangent vector at every point in space: $v = v^\mu\partial_\mu$ and $v(p) = v^\mu(p)\partial_\mu\vert_p$. Mathematically, we say that vectorfields live in the \textbf{tangent bundle}\index{Tangent Bundle} of the manifold, $TM := \bigsqcup_{p \in M}T_p M$ (some even call vectorfields ``sections'' of vector bundles).	
\end{definition}

\begin{aside}[Evaluating Functions on Manifolds]
	We've now come across terms like $v^\mu(p)$ and it is worth taking a second to be clear about what we mean. A function on a manifold is formally just some map $f: M \to \mathbb{R}$, but to actually \emph{calculate} anything, we need a representation of that function in terms of some parameters, and that is what the coordinate maps do, they are are nice mappings from the manifold to $\mathbb{R}^n$ that mean we can identify any function on the manifold with an equivalent function $F: \mathbb{R}^n \to \mathbb{R}$ given by composition of $f$ with the coordinate map $x^\mu$: $f(p) = F(x^\mu(p))$, and of course, normally we are sloppy with notation and just write $f$ for both. This is what we mean when we say we write or evaluate things in ``local coordinates.'' Note that these are ``local'' under the first sense of manifolds being ``locally'' Minkowski, it just uses the fact that we can parameters patches of points on the manifold with patches of points in $\mathbb{R}^n$.
\end{aside}

\begin{aside}[Differential Geometry Vector Fields]
	The identification of vectors with derivatives is even deeper than this. In modern differential geometry, vectors are \emph{defined} as derivatives, so that vectorfields on manifolds are defined as linear operators on smooth functions that satisfy the Leibniz law (i.e., the product rule). This definition coincides with the definition in terms of partial derivatives when coordinates are chosen, but coordinates aren't necessary, the definition is geometric and coordinate-independent in nature. 	

        This way of thinking about vectors is intimately tied to the idea of directional derivatives, and actually gives us a great way to justify the identification of the basis vectors with partial derivatives. Recall from section \ref{ch:geomRel:sec:vec} that the directional derivative of a scalar field along a path was a team effort of a covector (in the form of the gradient) and a vector (in the form of a tangent). In that section we wrote it as the covector acting on the vector, but we are perfectly within our rights to phrase it the other way, treating the tangent vector as an operator on a gradient:
        \begin{equation}
            \label{eq:ch2:dirDeriv}
            \frac {\dd \phi(x^\mu(s))}{\dd s}\eval_{s_0} = v_p(\dd \phi) = \frac {\dd x^\mu(s)}{\dd s}\eval_{s_0} \hat{e}_\mu(\dd \phi)\eval_{s_0} \leftrightarrow \frac {\dd x^\mu(s)}{\dd s}\eval_{s_0} \partial_\mu\phi\eval_{s_0}
        \end{equation}
so it must be the case that $\hat{e}_\mu$ is the operator that selects the $\partial_\mu$ component from gradients of scalar fields. But why stop there, why not let $\hat{e}_\mu$ do the work itself? Separately define vectors to act on scalar fields on their own and you can directly use $\hat{e}_\mu = \partial_\mu$ to get the job done without any knowledge of covectors at all, simply define $v_p(\phi) := v_p(\dd\phi)$.  
\end{aside}

\subsection*{Tensors}

One quick diversion: now that we know the vectors and vectorspaces involved, we can immediately port over our whole discussion of tensors from \ref{ch:geomRel:tensors}, we define tensors at a point $p$ on a manifold to be elements of the tensor product space $T_pM\otimes T_pM \otimes \ldots T^*_pM \otimes T^*_pM\otimes\ldots$, where $T_pM$ is the tangent space as above, and $T_p^*M$ is the dual of the tangent space at a point (we haven't discussed the Minkowski metric yet but the dual space can be defined without a bilinear form, the form is only necessary to provide a \emph{natural} mapping between the two). More generally, tensor fields on manifolds are elements of the tensor product space $TM\otimes TM\otimes \ldots T^*M \otimes T^*M\otimes \ldots$, where $TM$ is the tangent bundle (the collection of all tangent spaces on the manifold) and $T^*M$ is the cotangent bundle (the collection of all dual tangent spaces on the manifold).

Notice that this justifies the effort we put in defining things in terms of displacements! The tangent space is just the space of infinitessimal displacements, so while this might look big and new, it's really the same stuff we've been looking at, just done on a point-by-point basis.

\subsection*{The Metric Tensor}

At this point, we have one very important similarity to the content of chapter \ref{ch:geomRel}, and one very important divergence. The divergence is that the coordinates used for the basis vectors are \emph{no longer necessarily Cartesian}. Coordinate charts on manifolds can be all sorts of things, there's definitely no restriction to make them be Cartesian (otherwise why bother curving space?), so for example we may have $\partial_1 = \partial_r$ and $\partial_2 = \partial_\theta$---i.e., polar coordinates. The similarity though, is that the Minkowski spaces our tiny observers see have \emph{all} of the structure we discussed in chapter \ref{ch:geomRel}, in particular the Minkowski metric tensor may be used to measure lengths of vectors, and may be used to define covectors. 

However, there is a giant catch to this similarity---as the gods of this spacetime manifold, we may be kind to tiny Bob at point $p$ and choose our coordinates such that the coordinate vectors at Bob's location $\partial_\mu\vert_p$ coincide with Bob's notion of a Cartesian coordinate system: e.g., $\partial_r\vert_{p} = \partial_y$ and $\partial_\theta\vert_{p} = \partial_x$. In this way, Bob understands how to measure lengths of vectors and define covectors, he just uses the Minkowski metric. However, in those \emph{same} coordinates, tiny Alice, observing her little world from point $q$ is not so lucky! In her world, we have $\partial_r\vert_q \neq \partial_y$ and $\partial_\theta\vert_q \neq \partial_x$. Rather, it is generally true that $\partial_r\vert_q = a\partial_x + b\partial_y$ and so on. (And of course, Bob is not special, we could have chosen to bless Alice with orthogonal bases instead, or in a more traditional deity fashion, we may have chosen to smite them both). It is a pain to change coordinates every time we want to look at a different point\footnote{Well, with the help of some auxiliary fields (vielbeins) we technically can, but it's a nuisance}, so we instead have to live in a world where our basis vectors are in general \emph{not} orthogonal. Still, even if our coordinate basis vectors are not orthogonal, we can always establish some notion of lengths of vectors at a point, we just have to take into account non-trivial overlaps of the basis vectors, where in general $\hat{e}_\mu\cdot \hat{e}_\nu \neq \delta_{\mu\nu}$ and instead varies from point to point. Then just as we always do with point-wise defined objects, we may collect together all of these independent pseudo-inner products into a pseudo-inner product \emph{field}, the metric tensor:
\begin{definition}[The metric tensor]
    Collecting together the pseudo-inner products on all tangent spaces in the manifold, we define the \textbf{metric tensor}\index{Metric Tensor!General Manifold} $g : TM\times TM \to \mathbb{R}$ as a tensor field that at each point $p$ in $M$ defines the notion of lengths of vectors. As a pseudo-inner product it is symmetric, and as a physical thing it is continuous and infinitely differentiable. In distinct contrast to the Minkowski metric $\eta$, the metric tensor $g$ generally has off-diagonal elements. Notice that according to our definition, $g \in T^*M\otimes T^*M$.
\end{definition}
A manifold equipped with a metric tensor is called a \textbf{Riemannian manifold}\index{(pseudo)-Riemannian Manifold} if $g$ is positive-definite, and a \textbf{pseudo-Riemannian manifold} otherwise (so in GR we will be concerned with the latter).

\subsection*{Cotangent vectors}

Now that we have a metric tensor, we have a natural relationship between vectors (elements of $T_pM$) and covectors (elements of $T^*_pM$), induced in just the same way as in \ref{ch:geomRel:sec:vec}. That is, for every tangent vector $v_p \in T_pM$, there is a natural dual vector given by $v_p^* = g(v, \cdot)\vert_p \in T^*_pM$. Recall that even in Minkowski space, the \emph{basis} dual vectors satisfied a smoother relationship with the basis displacement vectors, $\hat{e}^\mu(\hat{e}_\nu) = \delta^\mu_\nu$. In fact, the same thing applies here (as it should), we can find a lovely basis given by the gradient\footnote{And here we use $\dd$ to represent the gradient instead of $\partial$ or $\nabla$ both because it is the standard notation in the literature, and to avoid confusion with the tangent vector bases, and later the covariant derivative.} of the coordinate functions: $\dd(x^\mu)(p) = \frac{\dd x^\mu}{\dd x^\nu}\vert_p(\hat{e}_p)^\nu = \delta^\mu_\nu\vert_p (\hat{e}_p)^\nu \, \implies \dd x^\mu\vert_p = (\hat{e}_p)^\mu$. We then have a wonderful shorthand for the action of a dual basis vector on a tangent basis vector: $\dd x^\mu(\partial_\nu) = \partial_\nu(x^\mu) = \delta^\mu_\nu$.
\begin{aside}[Exterior derivative]
    In fact, this shorthand can be generalized: the gradient of any scalar field $\phi: M \to \mathbb{R}$ is a covector $\dd \phi_p = \frac{\partial \phi}{\partial x^\mu}\vert_p \dd x^\mu\vert_p$, and the action of this dual vector on any tangent vector is $\dd \phi_p(v_p) = v_p(\phi) = v^\mu_p\partial_\mu\vert_p(\phi)$. This is the most basic example of the exterior derivative $\dd$, which in general maps differential forms, $\dd : \bigwedge^nT^*M \to \bigwedge^{n+1}T^*M$ (for more on differential forms, see appendix \ref{ch:curve:app:int:forms}).
\end{aside}

\subsection*{General Covariance}

In Minkowski space, there existed priviledged global coordinate systems, the inertial coordinate systems. These coordinates had a special relationship with vectors, so we restricted our attention to them when defining covariance and contravariance and therefore restricted coordinate transformations to Lorentz transformations. With the loss of global inertial coordinates, we lose the special status of Lorentz transformations (except in a few places) as well, but the importance of vectors, covectors, and tensors as purely geometric objects remains, where it now goes by the name of \textbf{general covariance}\index{General Covariance}\footnote{Here is a point of great division in the community. You will hear this type of thing variously referred to as ``general coordinate invariance,'' ``general covariance,'' ``diffeomorphism invariance,'' and others. There are subtle differences between them all, but the truth is that the matter is simply not settled. In these notes I aim for the simplest take, which I understand to be general covariance. A good discussion of the matter can be found in \cite{Giulini2007}.}. Given two overlapping coordinate patches $U, V$, $U\cap V \neq \emptyset$, defined respectively by the coordinate functions $x^\mu$ and $y^\mu$, the coordinate transformation $x\circ y^{-1} : \mathbb{R}^n \to \mathbb{R}^n$ given by $x^\mu(y^{-1})$ induces a transformation on vectorfield coordinate bases
\begin{equation}
	\label{eq:ch2:vectorLaw}
	\partial_{y^\mu} = \frac{\partial x^\nu}{\partial y^\mu}\partial_{x^\nu}.
\end{equation}
(By the way, we might remember in previous courses defining $\frac{\partial x^\nu}{\partial y^\mu}$ as the ``Jacobian'' of the coordinate transformation. We will continue to use that naming convention). The components of geometric objects---vectors, covectors, and tensors---then transform in the opposite way, just as we saw with Lorentz transformations:
\begin{align}
	\label{eq:ch2:compLaws}
	v^\mu \to \frac{\partial y^\mu}{\partial x^\nu}v^\nu, \\
	v_\mu \to \frac{\partial x^\nu}{\partial y^\mu}v_\nu, 
\end{align}
and tensors follow trivially by sticking one copy of the appropriate vector/covector law onto each contravariant/covariant index. This relation between tensor components in different general coordinates is trivial to implement in the notation we've developed with one caveat: the metric tensor changes as well. 

This last point holds both a great deal of freedom and complexity. On the one hand we can now use any coordinates best suited to a given problem, but on the other we now carry around and handle a generally non-trivial tensorfield in the metric tensor (and it's now non-trivial inverse). As always an example helps: consider the Minkowski metric tensor in polar coordinates (in 2+1 dimensions, for simplicity). Simply turning the crank, we write:
\begin{align}
	\label{eq:ch2:MinkPolar}
        (\eta_{\text{pol}})_{\mu\nu} = \frac{\partial x_\text{C}^\rho}{\partial x_\text{pol}^\mu}\frac{\partial x_\text{C}^\sigma}{\partial x_\text{pol}^\nu}(\eta_{\text{C}})_{\rho\sigma}.
\end{align}
Now obviously the switch from Cartesian to polar coordinates does not mess with the time component, so we can pretty easily read off that $(\eta_\text{pol})_{0\nu} = (\eta_\text{C})_{0\nu} = -\delta_{0\nu}$. For the rest, we have to use the inverse of the coordinate transformation (so the first line of \eqref{eq:ch2:cartToPol}). We have:
\begin{equation}
	\label{eq:ch2:MinkPolar2}
	\begin{aligned}
		\frac{\partial x_\text{C}^1}{\partial x_\text{pol}^1} = \cos\theta, &\qquad \frac{\partial x_\text{C}^1}{\partial x_\text{pol}^2} = -r\sin\theta, \\	
		\frac{\partial x_\text{C}^2}{\partial x_\text{pol}^1} = \sin\theta, &\qquad \frac{\partial x_\text{C}^2}{\partial x_\text{pol}^2} = r\cos\theta
	\end{aligned}.
\end{equation}
Then to chug this into \eqref{eq:ch2:MinkPolar}, use that $(\eta_\text{C})_{ij} = \delta_{ij}$, so we find:
\begin{equation}
	\label{eq:ch2:MinkPolar3}
	\begin{aligned}
		(\eta_{\text{pol}})_{11} = \cos^2\theta + \sin^2\theta, &\qquad (\eta_{\text{pol}})_{12} = -r\sin\theta\cos\theta + r\sin\theta\cos\theta, \\	
		(\eta_{\text{pol}})_{22} = r^2\sin^2\theta + r^2\cos^2\theta, &\qquad (\eta_{\text{pol}})_{21} = r\sin\theta\cos\theta - r\sin\theta\cos\theta 	
	\end{aligned}.
\end{equation}
Or in a more conventional form:
\begin{equation}
	\label{eq:ch2:MinkPolar4}
	\eta_\text{pol} = -\dd t \otimes \dd t + \dd r \otimes \dd r + r^2 \dd\theta\otimes \dd\theta.
\end{equation}

\subsection*{Next up}

And there we have it, we now have generalized everything we had cared about in Minkowski space to general manifolds\ldots almost. We haven't talked about differentiation of tensors yet, and for good reason. So far, we've resigned ourselves to living in a world of infinitely many unrelated tangent spaces and cotangent spaces, none of which can talk to each other. But physics takes place \emph{across} tangent spaces, so next we need to learn how to get tangent spaces to talk to each other, how to take physical things along paths through different tangent spaces, and ultimately tease out measures of the underlying curvature of the manifold, and connect that in a physical way to the matter content of spacetime. 

\section{Recap}

\begin{itemize}
	\item The equivalence principle $m_I = m_G$ destroys the idea that there can exist a global inertial reference frame for any physical observer $\longrightarrow$ everything is affected by gravity in the same way, so think of gravitational motion as a \emph{property} of spacetime, not something that happens on top of it.
	\item No global Minkowski space leads to general manifolds instead
	\item No global Minkowski space means no displacement vectors
	\item \emph{Local} observers still exist though, so \emph{infinitessimal} displacement vectors still exist $\longrightarrow$ tangent vectors (and cotangent vectors)
	\item Basis for tangent vectors is $\partial_\mu$ 
	\item Basis for cotangent vectors is $\dd x^\mu$ 
	\item At most one point in general can have $\partial_\mu\vert_p\cdot\partial\nu\vert_p = \eta_{\mu\nu}\vert_p$, everywhere else will be generally non-orthogonal: $\partial_\mu\vert_q\cdot \partial_\nu\vert_q = g_{\mu\nu}\vert_q$ $\longrightarrow$ metric tensor field $g_{\mu\nu}$ (symmetric but not diagonal).
\end{itemize}

\begin{subappendices}

\section{An Example of Coordinate Functions and Transformations}
\label{ch:manifolds:app:coordExamples}

\begin{figure}[ht]
	\centering
	\includegraphics[width=0.8\linewidth]{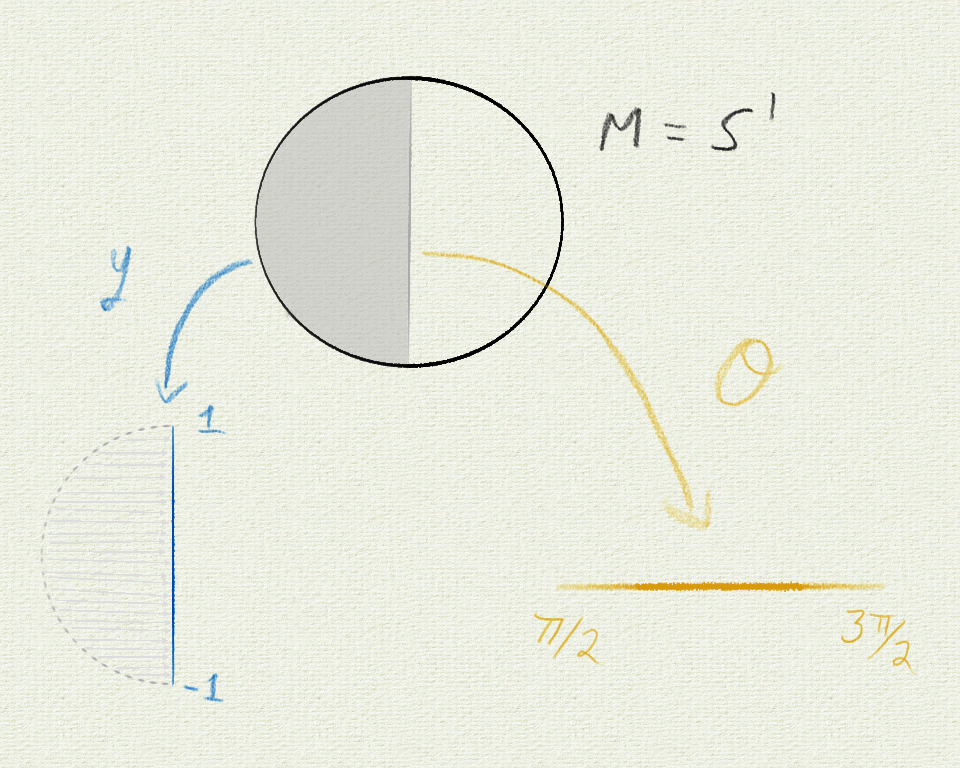}
        \caption[The coordinate maps of the circle under study.]{The coordinate maps of $S^1$ under study. The shaded region indicates the overlap in domain between the polar coordinate and the projection coordinate.}%
	\label{fig:manifolds_coordS1}
\end{figure}

To really drive home how to think about coordinates and transformations, it helps to do some examples. First, let's treat some patches on the 1-sphere, the circle. For our maps, let's take the usual angle $\theta \in (0, 2\pi)$, and let's take projection from the left hemisphere onto the $y$-axis (see figure \ref{fig:manifolds_coordS1}, and note that throughout I will engage in a systematic abuse of notation, using $\theta$ and $y$ as both the coordinate maps and the points in the images of those maps, and I will use them in both different senses in the same expression. Madness). 
The polar angle induces a parameterization on the manifold $\theta^{-1}(\theta) = (\cos\theta, \sin\theta)$, while the projection induces the parameterization $y^{-1}(y) = (\sqrt{1 - y^2}, y)$. A change of coordinates is then effected by composition, either from polar coordinates to the projection: $y(\theta^{-1})$ or the reverse, $\theta(y^{-1})$. Explicitly, we have  $y(\theta^{-1}(\theta)) = y(\cos\theta, \sin\theta) = \sin\theta$, and $\theta(y^{-1}(y)) = \theta(\sqrt{1 - y^2}, y) = \asin(y)$.

\section{Tensor Densities}
\label{ch:manifolds:app:tensorDensities}

We have made a big point about defining everything in terms of invariant geometric (tensorial) objects, but there are still times when we will really want to use objects that we just made up and definitely aren't tensors. Even so, the only objects with physical meaning are still tensors, so anything else we just have to view as a \emph{piece} of a tensor. When we're able to break things up such that the quantity we're interested in carries the indices of the complete tensor, we call that structure a \textbf{tensor density}\index{Tensor Density}. Of specific relevance for us will be two examples: physical scalar densities, and the Levi-Civita tensor. 

Probably the most commonly used tensor density on the mathematical side of our material is the Levi-Civita tensor density. The point is that we would really like to be able to work with the totally anti-symmetric Levi-Civita symbol in our equations, the object (in $n$ spacetime dimensions):
\begin{equation}
	\label{eq:ch2:app2:leviSymbol}
	\tilde\epsilon_{\mu_1\mu_2\ldots\mu_n} = \begin{cases}
		+1, \quad \text{if even permutation of indices} \\
		-1, \quad \text{if odd permutation of indices} \\
		0, \quad \text{else}
	\end{cases},
\end{equation}
where these components are the same \emph{everywhere and in all coordinate systems}. Clearly that does \emph{not} represent the components of a tensor, so we need to find something \emph{like} it that does. This turns out not to be too hard, but takes just a bit of foresight. 

First, consider the defining formula for a matrix determinant (one of the most common places to find the Levi-Civita symbol in linear algebra, and so also in physics):
\begin{equation}
    \label{eq:ch2:app2:detForm}
    \abs{M} = \tilde\epsilon_{\mu_1\mu_2\ldots\mu_n}M^{\mu_1}_1 M^{\mu_2}_2\ldots M^{\mu_n}_n,
\end{equation}
for some matrix $M$. Now if we swap any two of the lower indices, say $M_2^{\mu_1}M_1^{\mu_2}\ldots M_n^{\mu_n}$ , then to recover the formula for the determinant we have to swap the corresponding indices on the Levi-Civita symbol as well:
\begin{equation}
    \label{eq:ch2:app2:detForm2}
    \abs{M} = \tilde\epsilon_{\mu_1\mu_2\ldots\mu_n}M^{\mu_2}_1 M^{\mu_1}_2\ldots M^{\mu_n}_n = -\epsilon_{\mu_2\mu_1\ldots\mu_n}M^{\mu_2}_1 M^{\mu_1}_2\ldots M^{\mu_n}_n.
\end{equation}
Repeating the same process for any permutation of the lower indices shows we pick up a minus sign for any odd permutation and no sign for an even permutation. This information is what the Levi-Civita symbol was built for, so we can encode all permutations as:
\begin{equation}
    \label{eq:ch2:app2:detForm3}
    \tilde\epsilon_{\nu_1\nu_2\ldots\nu_n}\abs{M} = \tilde\epsilon_{\mu_1\mu_2\ldots\mu_n}M^{\mu_1}_{\nu_1} M^{\mu_2}_{\nu_1}\ldots M^{\mu_n}_{\nu_n}. 
\end{equation}
To see how this relates to tensors just look at the case that $M = J$, the Jacobian of a coordinate transformation. In that case, the RHS of \eqref{eq:ch2:app2:detForm3} is exactly the transformation law for a tensor, but the LHS is not quite there. In other words, if we constructed a tensor in a particular coordinate system with components $T_{\mu\nu\ldots} = \tilde\epsilon_{\mu\nu\ldots}$, then in a different coordinate system the components would be $T_{\alpha\beta\ldots} = \abs{J}\tilde\epsilon_{\alpha\beta\ldots}$. If only we had a separate scalar function $f(x)$ that transformed as $f(y) = \abs{J}f(x)$, then we could define components $\epsilon_{\mu\nu\ldots} = f(x)\tilde \epsilon_{\mu\nu\ldots}$ which really would be tensorial. 
%
%
Fortunately there is a natural example of just such a function already built into our manifolds: the metric determinant. Consider a coordinate transformation of the metric, $g_{\alpha\beta} = J^\mu_\alpha J^\nu_\beta g_{\mu\nu}$ . Treating it as a matrix equation, take the determinant of both sides and find:
\begin{equation}
    \label{eq:ch2:app2:transfG}
    g(y) = \abs{J}^2 g(x)
\end{equation}
(definine the metric determinant as $g$) so $\sqrt{\abs{g}}$ transforms as $\abs{J}$ under a coordinate transformation. We therefore define the \textbf{Levi-Civita Tensor}\index{Levi-Civita Tensor} as 
\begin{equation}
    \label{eq:ch2:app2:defEpsTensor}
    \epsilon_{\mu\nu\ldots} := \sqrt{\abs{g}}\tilde \epsilon_{\mu\nu\ldots}.
\end{equation}
The Levi-Civita tensor serves the role of the Levi-Civita symbol but in a proper tensorial way, so in principle one should only ever use the tensor form when working on a differentiable manifold. When necessary, the exact same process serves to define a fully contravariant form as well, $\epsilon^{\mu\nu\ldots} := (\abs{g})^{-1/2}\tilde \epsilon^{\mu\nu\ldots}$ (though note constructing this by raising each index individually returns an overall $\text{sgn}(g)$ that we are incorporating in the definition of $\tilde \epsilon^{\mu\nu\ldots}$). 

In fact, it turns out that the only things we like to think of as ``pieces'' of tensors also end up transforming non-tensorially only by picking up factors of the Jacobian (or inverse) determinant, so we can more precisely define tensor densities as indexed objects $\mathfrak{T}\indices{^\mu^\nu^\cdots_\rho_\sigma_\cdots}$ related to proper tensors by
\begin{equation}
    \label{eq:}
    T\indices{^\mu^\nu^\cdots_\rho_\sigma_\cdots} = \abs{g}^{w/2}\mathfrak{T}\indices{^\mu^\nu^\cdots_\rho_\sigma_\cdots},
\end{equation}
where the power $w$ is called the \textbf{weight}\index{Tensor Density!Weight} of the tensor density\footnote{Note that some authors use the opposite convention, defining tensor density weight by putting the $\sqrt{\abs{g}}$ on the other side, so their weights will be the negative of ours.}. By this definition, the Levi-Civita symbol is a tensor density of weight $+1$, while the contravariant form is a tensor density of weight $-1$. The metric determinant itself is trivially a scalar density of weight $-2$ ($g/g = 1$ is a perfectly invariant scalar).

Findally, the term ``density'' already speaks to the second example of tensor densities commonly seen in physics: physical densities. Here we are thinking about things like charge, energy, currents, etc., physical quantities that we treat in terms of their densities in space (or spacetime). But a physical density is a ratio of a quantity with respect to a volume, and the calculation of volumes becomes a very coordinate-dependent thing on a general manifold. We will see this in more detail when discussing integration on manifolds in \ref{ch:curve:app:int:forms}, but the takeaway is that volumes introduce a cumbersome factor of $\abs{J}$ to physical densities under a change of coordinates, so when handling physical densities in curved geometries we need to consider them as having weight $-1$, so for example energy density $\rho \to \rho/\sqrt{\abs{g}}$, some current density $j^\mu = \mathfrak{j}^\mu/\sqrt{\abs{g}}$, etc. Note that densities defined with respect to a submanifold of the full spacetime should be scaled by the determined of the metric restricted to the appropriate hypersurface on which the densities are calculated (e.g., spatial densities in 3+1 dimensional spacetime should be scaled by $\sqrt{g_{3}}$, the determinant of the spatial metric).

\end{subappendices}

\chapter{Covariant Differentiation and Curvature}
\label{ch:curve}

Now we can get to the task of \emph{quantifying} curvature. Our heuristic argument about the equivalence principle and the curvature of spacetime is all well and good, but it doesn't make for a mathematical theory until we can find a mathematical expression to represent some notion of the curvature of a manifold, and a way to connect it to the matter content of the universe. Ultimately, curvature is a property of the manifold as a whole, so quantifying it necessarily means drawing information from multiple tangent spaces. 

The first thing we need to do then is get tangent spaces to talk to each other. Normally this is first handled by looking at parallel transport (a consistent idea of ``moving'' a vector along some path in the manifold). While that's a great tool and the reader is encouraged to look it up themselves (see e.g.~\cite[\S 3.3]{carroll_2019}), in the interests of expedience we'll skip ahead a little and go straight to the idea of defining a derivative operator on vectorfields.

\section{The Covariant Derivative}
\label{ch:curve:sec:cov}

Our goal is to study physics in curved space, so at some point we need to set up physical equations of motion, which invariably involve derivatives. At first glance, this seems to be straightforward, we defined a derivative of tensors in \ref{ch:geomRel:stressEnergy}, when discussing the conservation of the stress-energy tensor. Looking closer though, we'll see that the definition \eqref{eq:ch1:DvT} applies \emph{only} to Minkowski space (note: not just flat space, but specifically \emph{Cartesian} flat space). At the time, we noted that the derivative acting trivially on the basis vectors was something to keep an eye on, and indeed, we'll get to that shortly, but an easy way to see why the Minkowski tensor \eqref{eq:ch1:DvT} does not remain a tensor on a general manifold is to see how it transforms under a general coordinate transformation. 

\subsection*{The failure of the partial derivative to turn vectors into tensors}

For a simple vectorfield, the components of the derivative as given by \eqref{eq:ch1:DvT} have the form $\partial_\mu V^\nu$ and should transform as a type-$\mqty(1 \\ 1)$ tensor under a general coordinate transformation:
\begin{equation}
	\partial_\mu V^\nu \to \frac{\partial x^\rho}{\partial y^\mu}\frac{\partial y^\nu}{\partial x^\sigma}\partial_\rho V^\sigma \qquad \text{(If a tensor)}.
\end{equation}
However, since $V^\nu$ definitely are the components of a tensor, and we know how partial derivatives transform under a coordinate transformation (the chain rule, i.e., covariantly), we can perform the computation explicitly:
\begin{equation}
	\label{eq:ch3:badPartial}
	\partial_\mu V^\nu \to \frac{\partial x^\rho}{\partial y^\mu}\partial_\rho\left( \frac{\partial y^\nu}{\partial x^\sigma} V^\sigma \right) = \frac{\partial x^\rho}{\partial y^\mu} \frac{\partial^2 y^\nu}{\partial x^\rho \partial x^\sigma} V^\sigma + \frac{\partial x^\rho}{\partial y^\mu} \frac{\partial y^\nu}{\partial x^\sigma} \partial_\rho V^\sigma,
\end{equation}
so for a \emph{general} coordinate transformation, where the Jacobian of the transformation is \emph{not} a constant matrix (like the Lorentz transformations), the naive derivative of a vectorfield does \emph{not} transform as a tensor. 

So what's really the problem here? We've suggested the derivatives of the basis vectors are throwing a wrench in the works, and while that's true, it's instructive to see why. Recall that formally, a derivative is just a difference, $\frac{\dd f}{\dd x} := \lim_{h\to 0}\frac{f(x + h) - f(x)}{h}$. For scalar functions on a manifold, this leads to the definition of tangent spaces; we assert that over infinitessimally small distances, life is Minkowski enough and we can construct an entire vectorspace out of the derivatives of fields and paths at a point. But when it comes time to take the derivative \emph{of those derivatives}, we can no longer pretend we're in a flat Cartesian world, now we have to compute $\frac{\dd^2 f}{\dd x^2} = \lim_{s\to 0} \frac{f^\prime(x+s) - f^\prime(x)}{s} = \lim_{s,h\to 0} \frac{f(x+h+s) - f(x+s) - f(x+h) + f(x)}{sh}$, which stretches across too much of the manifold, it necessarily involves two distinct tangent spaces. 
Fundamentally, that's why the derivative of the basis vectors can't be neglected, $\partial_\mu(\hat{e}_\nu)$ is an object in a new tangent space, and in general cannot always be taken to vanish (note that we temporarily restore the notation $\hat{e}_\mu$ for the tangent space basis vectors in order to avoid confusion with the ordinary partial derivative, which is an operator in this context, not a basis vector).  

\subsection*{Defining a new derivative}
So if we can't just use the partial derivative willy nilly to construct tensorial derivatives of tensors, what can we do? In the long-standing tradition of mathematics, the way out of this jam is to define a new thing to do what we want, and see if we can represent it in terms of things we know. So let's say there exists something like a derivative that really does turn tensors into tensors---we'll get ahead of ourselves and call it a \textbf{covariant derivative}\index{Covariant Derivative}\footnote{In some mathy circles, this is called a ``connection,'' since it connects different tangent spaces.} $\nabla$. 

Well it better be linear, that's the first thing; but linearity's not so hard, we're dealing with linear things left right and centre. More informatively, if it's going to be some sort of a derivative, it also needs to satisfy the Leibnitz law (product rule), so we need:
\begin{equation}
	\label{eq:ch3:covLeibnitz}
	\nabla(T\otimes S) = \nabla(T)\otimes S + T \otimes \nabla(S).
\end{equation}
We also really want this thing to represent what we think of as a gradient operator, so we will require that it \emph{is} the gradient when it acts on scalar fields (since the gradient lives in the cotangent space, and that's well-defined already):
\begin{equation}
	\label{eq:ch3:defCovScalar}
	\nabla(\phi) = \dd\phi = \partial_\mu\phi\, \dd x^\mu,
\end{equation}
where the second equality specializes to a coordinate system. Notice that just like the gradient and our naive derivative of vectors in Minkowski space, it must be the case that the covariant derivative turns tensors of type $\mqty(M \\ N)$ into tensors of type $\mqty(M \\ N + 1)$.

Consider the action of this new derivative on a vector, $V = V^\mu \partial_\mu$. We can think of each term in the coordinate representation of a vector as the tensor product between a scalar field ($V^\mu$) and a contravariant vector $(\partial_\mu)$, so the action of the covariant derivative must follow the product rule and read:
 \begin{equation}
	\label{eq:ch3:actCovVec}
	\nabla V = (\nabla_\nu V^\mu) \partial_\mu \otimes \dd x^\nu + V^\mu \nabla_\nu(\partial_\mu)\otimes \dd x^\nu = (\partial_\nu V^\mu) \partial_\mu \otimes \dd x^\nu + V^\mu \nabla_\nu(\partial_\mu)\otimes \dd x^\nu.
\end{equation}
In the second equality, we use that the covariant derivative of a scalar field is just the ordinary gradient. The question remains about what to do with the second term. In general, there is no way to derive what the result of $\nabla_\nu(\partial_\mu)$ must be, but we can at least parameterize our ignorance. For instance, it must be \emph{a vector}, since the equation is tensorial, so we know we can write it in a basis, and we might as well just give it a name,
\begin{equation}
	\label{eq:ch3:actCovVecDef}
	\nabla V = (\partial_\nu V^\mu) \partial_\mu \otimes \dd x^\nu + V^\mu (\Gamma_{\nu\mu})^\lambda \partial_\lambda\otimes \dd x^\nu.
\end{equation}
Note that the indices on $\Gamma$ are \emph{not} tensorial, they are a \emph{naming convention}. The components $\Gamma$, often called the \textbf{connection coefficients}\index{Covariant Derivative!Connection Coefficients}, are explicitly \emph{not} the components of a rank-3 tensor, so we do not space the indices, and typically we write them without the brackets as $\Gamma^\lambda_{\nu\mu}$. In components then, we say the covariant derivative takes the form:
\begin{equation}
	\label{eq:ch3:covCoords}
	\nabla_\nu V^\mu = \partial_\nu V^\mu + \Gamma^\mu_{\nu\lambda}V^\lambda =: V\indices{^\mu_;_\nu}.
\end{equation}
(Notice the re-labelling of dummy indices compared to equation \eqref{eq:ch3:actCovVecDef}). In the second equality, we define an often handy shorthand for the covariant derivative. Similar to how we defined the ordinary partial derivative shorthand notation with a comma, here we use a semi-colon to denote a covariant derivative. 

If $\nabla V$ really is a tensor, it must be the case that the coefficients \eqref{eq:ch3:covCoords} transform as a type-$\mqty(1 \\ 1)$ tensor. However, we already know what happens to the first term, it's \eqref{eq:ch3:badPartial}---the derivative product rules up an extra term out of the derivative of the Jacobian. So in order to keep things tensorial we have to define that the components of $\nabla_n (\partial_\mu)$ transform in just the right way to cancel off that extra term. That is, we need the connection coefficients to correct for the over-eager ordinary derivative. Some simple algebra shows the connection coefficients must therefore satisfy:
\begin{equation}
	\label{eq:ch3:gammaCoordCh}
	\Gamma^\mu_{\nu\lambda} \to \frac{\partial x^\alpha}{\partial y^\nu}\frac{\partial x^\beta}{\partial y^\lambda}\frac{\partial y^\mu}{\partial x^\gamma}\Gamma^\gamma_{\alpha\beta} - \frac{\partial x^\alpha}{\partial y^\nu}\frac{\partial x^\beta}{\partial y^\lambda} \frac{\partial^2 y^\mu}{\partial x^\alpha \partial x^\beta}.
\end{equation}
(Ex.~show that this transformation cancels the extra term from \eqref{eq:ch3:badPartial}).

One quick follow-up: by repeatedly applying the product rule, what we've derived here for a rank-1 vector field generalizes immediately to completely contravariant tensors of any rank. Each contravariant index comes with a basis vector, $T = T^{\mu\nu\rho\ldots}\partial_\mu\otimes\partial_\nu\otimes\partial_\rho\otimes\cdots$, so when we take the covariant derivative of $T$, the product rule gives us a different factor of $\Gamma$ for each index. E.g.:
\begin{equation}
	\label{eq:ch3:exRank3}
	\nabla_\nu T^{\rho\sigma\ldots} = \partial_\nu T^{\rho\sigma\ldots} + \Gamma^\rho_{\nu\alpha}T^{\alpha\sigma\ldots} + \Gamma^\sigma_{\nu\beta}T^{\rho\beta\ldots} + \ldots.
\end{equation}

So far so good, and in general any choice of $\Gamma$ that satisfies the properties above is a good covariant derivative, or connection (actually, you can even squeak by without the reduction to the partial derivative, if you have no respect for your own wellbeing). This means that in \emph{addition} to the metric being an external object we slap on to a manifold to give it structure, we \emph{also} have to adorn our manifolds with a \emph{choice} of connection. With so much flexibility, it helps to narrow down to a useful choice of connection by asking for a wishlist of nice properties. First up, compatibility with contractions.

It may seem obvious that the covariant derivative should be compatible with contractions---i.e. $\nabla_\nu T\indices{^\lambda_\lambda_\beta} = (\nabla T)\indices{^\lambda_\lambda_\beta_\nu}$---but we're kind of winging the transformation properties of the connection coefficients here, so there are no guarantees a priori. If we assume this holds though, we get a great relation between the covariant derivatives of vectors and covectors. Generically, there would be no relation between the vector object $ \nabla_\nu(\partial_\mu) =: \Gamma^\lambda_{\nu\mu}\partial_\lambda$ and the covector object $\nabla _\nu(\dd x^\mu) =: \widetilde \Gamma^\mu_{\nu\lambda} \dd x^\lambda$. However, if we can freely commute contractions with the covariant derivative, something fun happens. Consider the scalar $V^\mu W_\mu$. It's a scalar field, so if we take its covariant derivative we should get  $ \nabla_\rho (V^\mu W_\mu) = \partial_\rho( V^\mu W_\mu )  = \partial_\rho(V^\mu)W_\mu + V^\mu \partial_\rho(W_\mu)$. But if we contract after taking the covariant derivative, we need to bring in the connection coefficients:
\begin{align}
	\label{eq:ch3:connectVecCovec}
	\nabla _\rho \left( V^\mu W_\mu \right) &= (\nabla V)\indices{^\mu_\rho}W_\mu + V^\mu(\nabla W)\indices{_\mu_\rho}, \notag \\
											&= (\partial_\rho V^\mu)W_\mu + \partial_\rho(W_\mu)V^\mu + \Gamma^\mu_{\rho\lambda}V^\lambda W_\mu + \widetilde \Gamma^\lambda_{\rho\mu} V^\mu W_\lambda.
\end{align}
For the last line to equal $\partial_\rho(V^\mu W_\mu)$, it must be the case that (re-labelling some dummy indices):
\begin{equation}
	\label{eq:ch3:gammaTildeGamma}
	\Gamma^\mu_{\rho\lambda} = -\widetilde\Gamma^\mu_{\rho\lambda}.
\end{equation}
And with this (and the product rule), we arrive at a way to take the covariant derivative of any rank of mixed tensor: one copy of $+\Gamma$ for each contravariant index, and one copy of $-\Gamma$ for each covariant index. 
\begin{equation}
	\label{eq:ch3:genExpDeriv}
	\nabla_\rho T\indices{^\mu^\ldots _\nu_\ldots} = \partial_\rho T\indices{^\mu^\ldots _\nu_\ldots} + \Gamma^\mu_{\rho\lambda}T\indices{^\lambda^\ldots _\nu_\ldots} + \ldots - \Gamma^\lambda_{\rho\nu}T\indices{^\mu^\ldots _\lambda_\ldots} - \ldots .
\end{equation}

One last quick observation before we move on to the specific choice of connection coefficients we'll be using for the rest of the course. Although the connection coefficients $\Gamma$ are \emph{not} the coefficients of tensors (they transform funny), it is easy to show that the \emph{difference} between any two connection coefficients \emph{is} a tensor. That is, define two different covariant derivatives $ \nabla $ and $ \widehat \nabla$ made out of two different choices for connection coefficients $\Gamma$ and $\widehat \Gamma$. The difference between their actions on a vector must be a tensor (of course, they're proper tensor operators after all), so
\begin{align}
	\label{eq:ch3:difGams}
	\nabla _\nu V^\mu - \widehat\nabla_\nu V^\mu &= \partial_\nu V^\mu - \partial_\nu V^\mu + \Gamma^\mu_{\nu\lambda}V^\lambda - \widehat\Gamma^\mu_{\nu\lambda}V^\lambda, \notag \\
												 &= \Gamma^\mu_{\nu\lambda}V^\lambda - \widehat\Gamma^\mu_{\nu\lambda}V^\lambda
\end{align}
is a proper tensor. This is particularly useful in one specific case. When working out the components for $ \nabla _\nu$ above, we somewhat arbitrarily decided the order of the indices on $\Gamma$. Had we chosen another convention, for example $\widehat \Gamma^\mu_{\nu\lambda} := \Gamma^\mu_{\nu\lambda}$, we would have found another perfectly distinct, perfectly valid connection, so we can use this to construct an important tensor by taking the difference:
\begin{equation}
	\label{eq:ch3:defTor}
        T\indices{^\mu_{\nu\lambda}} := \Gamma^\mu_{\nu\lambda} - \Gamma^\mu_{\lambda\nu},
\end{equation}
named the \textbf{torsion}\index{Torsion} tensor for a particular connection.

\subsection*{The Levi-Civita Connection}

Finally, we'll impose a couple more handy properties on our covariant derivative to really nail it down to a particular form. First, let's get rid of torsion, it's really ugly. From now on, we'll only work with a connection that is ``torsion-free,'' so satisfies $\Gamma^\rho_{\mu\nu} = \Gamma^\rho_{\nu\mu}$. Next, we'll impose what will turn out to be the last condition we need to uniquely define a connection: metric compatibility.

Having attached a metric to our manifold, it is extremely useful to know how the covariant derivative we are trying to add relates to it (if at all). As it happens, we generally have the freedom to choose the best possible relation, called \textbf{metric compatibility}\index{Metric Compatibility}, the property of a covariant derivative acting trivially on the metric. That is, we choose $ \nabla $ to satisfy:
\begin{equation}
	\label{eq:ch3:defMetComp}
	\nabla_\rho g_{\mu\nu} = 0,
\end{equation}
for all indices. It turns out that all these conditions are finally sufficient to actually once-and-for-all uniquely define the connection and its coefficients, which in this case gets the name the \textbf{Levi-Civita connection}\index{Levi-Civita Connection}, or the \textbf{Christoffel connection}. 

To see how we get an expression for the connection coefficients, consider the permutations of \eqref{eq:ch3:defMetComp}\footnote{The following is mostly repeated from \cite[\S 3.2]{carroll_2019}.}:
\begin{align}
	\label{eq:ch3:metCompPerms}
	\nabla_\rho g_{\mu\nu} &= \partial_\rho g_{\mu\nu} - \Gamma^\lambda_{\rho\mu}g_{\lambda\nu} - \Gamma^\lambda_{\rho\nu}g_{\mu\lambda} = 0 \notag \\
	\nabla_\mu g_{\nu\rho} &= \partial_\mu g_{\nu\rho} - \Gamma^\lambda_{\mu\nu}g_{\lambda\rho} - \Gamma^\lambda_{\mu\rho}g_{\nu\lambda} = 0 \notag \\
	\nabla_\nu g_{\rho\mu} &= \partial_\nu g_{\rho\mu} - \Gamma^\lambda_{\nu\rho}g_{\lambda\mu} - \Gamma^\lambda_{\nu\mu}g_{\rho\lambda} = 0. \notag 
\end{align}
Using the symmetry of the metric, and the newly minted symmetry of the connection coefficients, we can subtract the second and third equations to find:
\begin{equation}
	\label{eq:ch3:connectLinComb}
	\partial_\rho g_{\mu\nu} - \partial_\mu g_{\nu\rho} - \partial_\nu g_{\rho\mu} + 2\Gamma^\lambda_{\mu\nu}g_{\lambda\rho} = 0,		
\end{equation}
which can be solved for the coefficients to find the unique expression
 \begin{equation}
	\label{eq:ch3:christofs}
	\Gamma^\lambda_{\mu\nu} = -\frac 12 g^{\lambda\rho} \left(\partial_\rho g_{\mu\nu} - \partial_\mu g_{\nu\rho} - \partial_\nu g_{\rho\mu}\right).
\end{equation}
This is an incredibly useful relation because (apart from all of the lovely conditions we've already imposed) we no longer have to worry about the connection coefficients as independent quantities, and can instead think of them as just complicated functions of the metric, which is enough to worry about on its own. In the form \eqref{eq:ch3:christofs}, the connection coefficients are often called the \textbf{Christoffel symbols}\index{Christoffel Symbols} (``symbols'' to reflect that they are \emph{not} tensors), or the \textbf{affine connection} (so called because it generalizes the affine structure of Minkowski space that let us play with vectors at different places on the same footing), and in older texts have been given the slightly odd notation $\Gamma^\lambda_{\mu\nu} =: \begin{Bmatrix} \lambda \\ \mu\nu\end{Bmatrix}$ (we will not employ this notation).

Finally, another handy formula that falls out of the definition of the Christoffel connection is the expression
\begin{equation}
	\label{eq:ch3:metDetCovDeriv}
	\Gamma^\mu_{\mu\nu} = \frac{1}{\sqrt{\abs{g}}}\partial_\nu\left( \sqrt{\abs{g}}  \right),  
\end{equation}
where $\abs{g}$ is the determinant of the metric. We will not derive this here, and will have little use for it, but it is good to know, particularly as a convenient way to write the divergence of a vectorfield $\nabla_\mu V^\mu = (\abs{g})^{-1/2}\partial_\mu\left( \sqrt{\abs{g}}V^\mu  \right) $.

\section{Geodesics}
\label{ch:curve:sec:geod}

Now that we have a consistent notion of a derivative on tensors, we can start to think about things to do with it. As physicists, the first thing we might want to do is use this tool to put physics on general non-Minkowskian manifolds. For example, if we look at (two of) the Maxwell equations in their covariant special relativistic form:
\begin{equation}
	\label{eq:ch3:maxwelSR}
	\partial_\nu F^{\mu\nu} = J^\mu \qquad \text{(Minkowski Space)},
\end{equation}
where $F^{\mu\nu} = \partial_\mu A_\nu - \partial_\nu A_\mu$ is the electromagnetic field strength tensor, and $J^\mu$ is a four-current, containing both electric and magnetic currents. Don't worry too much about the form here, the important part is just that since the ordinary partial derivative is \emph{not} a tensor operator, this physical law in the form \eqref{eq:ch3:maxwelSR} is \emph{not} a tensor equation on a general manifold. However, we can easily see what \emph{would} be a tensor equation, and indeed we make the bold assumption that the laws of electromagnetism in curved space actually are governed by the re-covariantized equation of motion\footnote{Interestingly, the electromagnetic field strength tensor needs no re-definition, since the difference of partial derivatives cancels out their non-conforming transformation properties. This is an example of the ``exterior derivative'' acting on one-forms---see appendix \ref{ch:curve:app:int:forms}.}:
\begin{equation}
	\label{eq:ch3:maxwelGR}
	\nabla_\nu F^{\mu\nu} = J^\mu \qquad \text{(General Space)}.
\end{equation}
We say ``general space`` but of course the physical interpretation here is that \eqref{eq:ch3:maxwelGR} completely describes the behaviour of electromagnetic particles and materials in a relativistic \emph{and} gravitational system (though does not yet incorporate the response of the gravitational field---for that we need chapter \ref{ch:EFEs}).

A much less obvious generalization can be carried out for another crucial physical law: Newton's second law in the absence of an external force:
\begin{equation}
	\label{eq:ch3:Newton2}
	\vec{F} = m\vec{a} = 0 \qquad \text{(Flat space, no external force)}.
\end{equation}
Here the statement is that unaccelerated particles follow straight lines. The generalization of this notion is the idea of \textbf{geodesics}\index{Geodesic}, the ``straightest'' paths on a manifold. We'll get into more details, but the simplest way to carry out this generalization is by following your nose. Newtonian acceleration is a 3-vector, $\vec{a} = \dv[2]{\vec{x}}{t}$, so the first thing we do is swap out the three-vector for a four-vector $\vec{x} \to x^\mu$. Next, recall we identify the formerly universal parameter of time $t$ with an external physical variable, usually a particle's proper time $\tau$, so ideally we would like to write Newton's second law as (dropping the mass because it falls out for unaccelerated trajectories) $\frac{\dd^2 x^\mu}{\dd \tau^2} = 0$. Unfortunately, this doesn't quite work! Of course $\frac{\dd x^\mu}{\dd \tau} =: U^\mu$ is a good tensor, it's the particle's four-velocity. However, the second derivative, $\frac{\dd}{\dd\tau} = U^\mu \partial_\mu$ turns the putative equation into $U^\mu \partial_\mu U^\nu = 0$, which is \emph{not} tensorial because of the presence of the lone ordinary partial derivative. There is an obvious way to remedy this deficiency though, we simply swap out the partial derivative for a covariant one---then by a simple algorithm, we find the unaccelerated version of Newton's second law becomes 
\begin{equation}
	\label{eq:ch3:defGeod}
	\frac{\dd x^\mu}{\dd s}\nabla _\mu \frac{\dd x^\nu}{\dd s} = 0
\end{equation}
on a general manifold. Easy peasy!

It is great to have arrived at this incredibly important equation from a simple algorithm and an old physical law. For later use though, it is beneficial to think about what this equation means geometrically, and how to derive it just from considerations of the properties of straight lines. Here's the basic premise. When we take the covariant derivative of a tensor, we get an object $\nabla T$ that will tell you how that tensor changes in the tangent space next door in any given direction. This is the reason the covariant derivative adds a covariant index, the extra covector is waiting for you to give it a vector to tell it in which direction you want to compute the change in the tensor. Mathematically, this follows very similarly to the notion of the directional derivative from \ref{ch:geomRel:sec:vec}, the object $V^\mu \nabla_\mu T\indices{^\rho^\ldots_\sigma_\ldots}$ is the covariant directional derivative, the (infinitessimal) change in the tensor $T$ along the direction of $V$.

Now thinking about paths---especially paths of things, physical things---we recall there is a natural vector associated with any path $\gamma(s)$ along a manifold, which is its tangent vector, $v_s(\gamma) = \frac{\dd}{\dd s}\gamma(s) = \frac{\dd x^\mu}{\dd s}\partial_\mu\vert_{s}\gamma$. Newton's second law states that an unaccelerated particle's trajectory is a straight line, never deviating from the direction of the line. Following the manifold philosophy, we can zoom in on our path $\gamma$ and ask is it basically a straight line? That is, does the tangent vector \emph{covariantly} stay the same in the direction it was headed? If it does stay the same, we say that the path is a \textbf{geodesic}\index{Geodesic}, an unaccelerated path on the manifold. Mathematically, we can write this exactly as \eqref{eq:ch3:defGeod}. Try to read that equation as: ``the change in the tangent vector vanishes in the direction it's pointing.'' ($\nabla \dot x$ is the gradient of the vector $\dot x$, so the inner product $\dot x \cdot \nabla \dot x$ vanishes along $\dot x$ if there is no change in $\dot x$ in that direction). 

Using the chain rule, $\frac{\dd}{\dd s} = \frac{\dd x^\mu}{\dd s}\partial_\mu$, the geodesic equation \eqref{eq:ch3:defGeod} can neatly and conveniently be written:
\begin{equation}
	\label{eq:ch3:defGeod2}
	\boxed{
		\frac{\dd^2 x^\mu}{\dd s^2} + \Gamma^\mu_{\nu\rho}\frac{\dd x^\nu}{\dd s}\frac{\dd x^\rho}{\dd s} = 0.
	}
\end{equation}
This is the equation that generalizes the notion of straight lines in curved spaces. To see this explicitly, notice that the Christoffels are all derivatives of the metric, and in Minkowski space (in Cartesian coordinates), the components of the metric are all constants, so the Christoffel symbols all vanish, and \eqref{eq:ch3:defGeod2} reduces to $\frac{\dd^2 x^\mu}{\dd s^2} = 0$, which for $s = \tau$ (proper time) yields $a = 0$, or $x^\mu \propto \tau$. Geodesics are extremely important in GR as the fact that some unaccelerated trajectories are \emph{not} straight lines is the mathematical expression of the statement that gravity is geometry. That is, an unaccelerated particle moving along a curved trajectory is simply moving in a non-trivial gravitational field.

\begin{aside}[Affine parameters]
	Here again we run into the word ``affine.'' Strictly speaking, the geodesic equation \eqref{eq:ch3:defGeod2} only defines geodesics with a specific parameterization, called an affine parameterization. To see this, take a curve that solves the geodesic equation, and change the parameter $s \to s(t)$, where $s(t)$ is some wild, crazy function of some other number $t$. If you stick this change of parameter into the geodesic equation, the derivatives change by the chain rule, $\dd/\dd s \to (\dd t/ \dd s)\dd/\dd t$, but the first term is a \emph{second} derivative, so it picks up an extra term, and in general you would have to solve:
	\begin{equation}
		\frac {\dd^2 x^\mu(t)}{\dd t^2} + \Gamma^\mu_{\nu\rho}\frac{\dd x^\nu(t)}{\dd t} \frac{\dd x^\rho(t)}{\dd t} = -\left( \frac{\dd t}{\dd s} \right)^{-2} \frac{\dd^2 t}{\dd s^2} \frac{\dd x^\mu(t)}{\dd t}. 
	\end{equation}
	Fortunately though, we have a great deal of freedom in how we choose to parameterize curves, so to make our lives easier we always choose to parameterize geodesics in such a way that the right hand side of the geodesic equation vanishes. This still leaves us with a certain latitude in choosing parameters, and you can verify easily that if $s$ parameterizes a geodesic (with the RHS vanishing), then so does $t = as + b$, where $a$ and $b$ are constants. This relation defines $s$ and $t$ as elements of a one-dimensional affine space (points can be identified with displacement vectors).
\end{aside}

\section{Curvature}
\label{ch:curve:sec:curve}

At long last we can finally start to talk about curvature---what it means, how to describe it, and soon how it relates to the matter content of space. There are a number of different ways of thinking about curvature. Intuitively when we think about curved geometries, we usually think about two-dimensional surfaces embedded in three spatial dimensions, like spheres. Here, the embedding carries with it a natural notion of how something like a sphere deviates from three-dimensional Cartesian space in terms of confining 3D objects to the 2D surface (think of stretching out a triangle to fit it onto a sphere, or trying to flatten out the peel of an orange). These intuitive notions of curvature coming from the external embedding are called \emph{extrinsic curvature}, and while that can have its use, it's much more helpful to have an internal, \emph{intrinsic} notion of curvature of a manifold. So far, we've done very well constructing spacetime as a single manifold with the additional structure of a metric and a connection, it would be a real shame if we could only understand the curvature of our spacetime manifold by \emph{also} having to carry around the entire structure of a larger Cartesian manifold as well. Phrased another way, we live in spacetime, so we'd better be able to characterize the curvature of our world while living in it.

Fortunately there exist such intrinsic ways of quantifying curvature. A great overview can be found in \cite[\S 1]{weinberg_gravitation_1972}, but the gist is that Euclidean space defines what we mean by ``flat'' space. Euclidean space is based on a set of axioms, and if you can violate any one of them (or any of the myriad equivalent formulations of them), then you have yourself a curved, non-Euclidean space. One example of (a phrasing of) these axioms is the statement that the internal angles of triangles add up to $\pi$, a statement it is easy to verify is contradicted on the sphere: consider the triangle formed by connecting the North pole to the equator, running along the equator over an angle of $\pi/2$ longitude, then running back up to the North pole. Each corner of the triangle is at an angle of $\pi/2$, adding up to a total of $3\pi/2$. Notice that the way we described this triangle did \emph{not} require us to picture it embedded in three dimensions. All we have to do is imagine we're a tiny person living on the sphere who starts walking in a straight line from a particular point, turns to their left when they feel like it, does the same after walking the other direction, and eventually arrives back home. However long this tiny person walks in either direction\footnote{Barring circumnavigations, of course.}, they will form a triangle with at least two right angles, so a sum of internal angles that is always greater than $\pi$.

More convenient for calculations is the requirement that parallel lines remain parallel. One of the ways to define Euclidean space is to include in it the notion of infinitely parallel lines, lines that start parallel and never meet in either direction. A space where lines that were initially parallel ever deviate (either meet or shoot off from each other) is inherently a non-Euclidean, or \textbf{curved} space.

\subsection{Geodesic Deviation}

\begin{figure}[htp!]
	\centering
	\includegraphics[width=\linewidth]{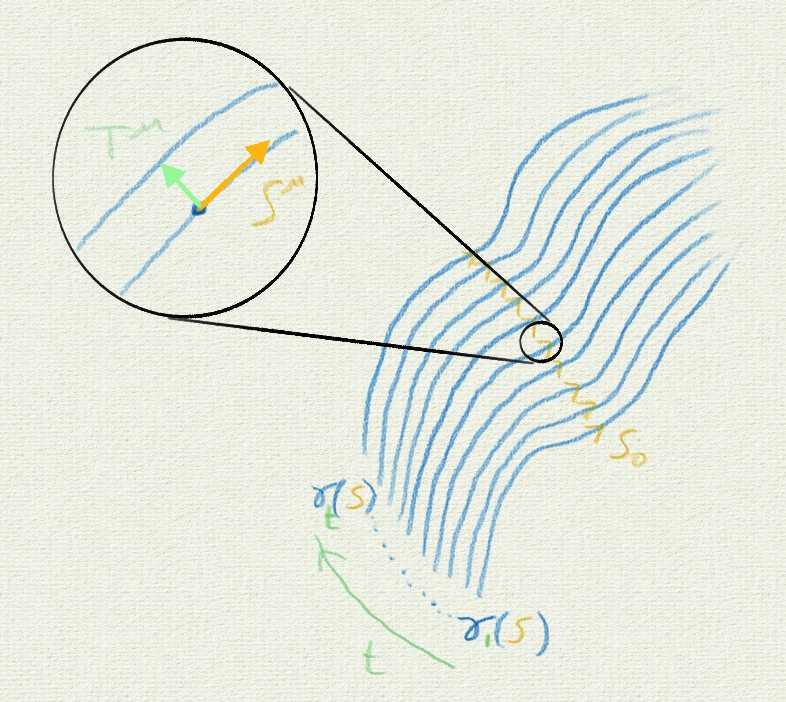}
        \caption[A smooth field of geodesics used to examine local geodesic deviation.]{A smooth field of geodesics. The blue lines are geodesics on the manifold, all conveniently parameterized by the same $s$, and there is a smooth field of them, so inbetween all of the blue lines are more geodesics. The parameter $t$ is a continuous variable that labels each of the geodesics. Together, $s$ and $t$ map out a two-dimensional surface in the manifold, each point on the surface corresponding to a point on a single geodesic curve. This time it's not just my bad drawing, the lines really can be non-parallel, and even cross at some points if they want to.}%
	\label{fig:geodesics}
\end{figure}

To find a way of mathematically describing curvature by means of the deviation of parallel lines, we start with the generalization of lines, geodesics. This time we'll need more than one of them, so let's go wild and define a whole \emph{field} of geodesics, $\gamma_t(s)$, where $t$ is a continuous parameter (see figure \ref{fig:geodesics}) and all of the geodesics are parameterized by the same $s$. 

Each of the $\gamma_t(s)$ is the generalization of a straight line, so if we zoom in really close on one, we can find a neighbour that is essentially parallel. Mathematically, ``looking really close'' means looking at tangent vectors, so consider the tangent to a single one of these curves, call it $S^\mu := \frac{\dd x^\mu(\gamma_{t_0}(s))}{\dd s}\vert_{s_0}$. Each $\gamma_t$ is a geodesic, so we must have:
\begin{equation}
	\label{eq:ch3:geodGam}
	S^\nu\nabla_\nu S^\mu = 0.
\end{equation}
Next, since our field of geodesics is continuous, we can define a direction in the manifold given by the parameter $t$, call it $T^\mu := \frac{\dd x^\mu(\gamma_{t}(s_0))}{\dd t}\vert_{t_0}$. The vector $T$ represents the direction toward the nearest geodesic from the one at $s_0, t_0$, so it gives us a way to define a sort of ``neighbour'' to the curve we're studying---graphically, picture this as the perpendicular connecting two parallel lines. In fact, we have a great deal of latitude here, so we can go ahead and parameterize things just right so that $T$ really is the perpendicular to $S$, $T^\mu S_\mu = 0$. Note that this is a condition that eliminates a free parameter, so where we had $2n$ for the two vectors, now we have $2n - 1$ free numbers to work with. Remember that, it'll come in handy shortly.

So $S^\mu$ is the tangent to a geodesic, and $T^\mu$ is the perpendicular infinitessimal distance to a neighbouring geodesics. The next thing we can ask is: does $T^\mu$ change if we move along the geodesic that defines $S^\mu$? This can be computed the same way we defined geodesics, take the gradient of  $T^\mu$ and project out the  $S^\nu$ component (i.e.~find the directional derivative of $T^\mu$ in the direction of the geodesic flow, $S^\mu$):
 \begin{equation}
	\label{eq:ch3:geodVel}
	v^\mu := S^\nu \nabla_\nu T^\mu.
\end{equation}
The quantity $v^\mu$ can be thought of as the \emph{geodesic separation velocity}, it tells us whether the geodesic and its neighbour are moving towards each other, away from each other, or parallel to each other. Our interest is in checking Euclid's Parallel Postulate for the manifold, so we'd like to consider a setup where $v^\mu = 0$, which we have more than enough degrees of freedom to do. Setting the $n$ components of the geodesic separation velocity to zero cuts our free parameters down to $n-1$.

Now we have $S^\mu$ the tangent to a geodesic, $T^\mu$ the perpendicular to this tangent connecting it to another geodesics, and we have chosen initial conditions and coordinates such that the separation $T^\mu$ is initially stationary in the direction of the geodesics. Finally we can ask, will that separation \emph{always} be stationary? To find out, just take the derivative again to find the \emph{geodesic separation acceleration}:
\begin{equation}
	\label{eq:ch3:geodAcc}
	a^\mu := S^\nu \nabla_\nu v^\mu = S^\nu \nabla_\nu \left( S^\rho \nabla_\rho T^\mu \right).
\end{equation}
Now remember, we're only working with $n-1$ free parameters at this point, and the geodesic separation acceleration is an $n$-component vector, so we no longer have enough degrees of freedom to \emph{force} the acceleration to vanish, so if it is \emph{non-zero}, then we know the manifold did is indeed gravitating and we're living in a non-Euclidean world. The only trouble is, this expression at the moment is highly dependent on our particular choices of $S$ and $T$. What we need is to massage it into a form where we can yank out the components of the tangent vector and its perpendicular, and have something that depends only on the properties of the manifold, its metric and connection (and for the Levi-Civita connection that's \emph{just} the metric).

At this point, it helps to temporarily choose our coordinates wisely. Let's pick the $x$-direction to be along $S$ and the $y$-direction to be along $T$, so $S = \delta^\mu_1 \partial_\mu$ and  $T = \delta^\mu_2 \partial_\mu$. Then the covariant derivative is fairly easy to evaluate: the components of $S$ and $T$ are constant so we have $ \nabla_\nu S^\mu = \Gamma^\mu_{\nu\rho} S^\rho$ and $ \nabla_\nu T^\mu = \Gamma^\mu_{\nu\rho}T^\rho$, so in particular, $T^\nu \nabla_\nu S^\mu = T^\nu \Gamma^\mu_{\nu\rho} S^\rho = S^\rho \Gamma^\mu_{\rho\nu} T^\nu = S^\rho \nabla_\rho T^\mu$. In other words, in these coordinates we can swap $S$ and $T$ in the directional derivative, but since that is a tensor equation it is valid in \emph{any} coordinate system, so we have in general:
\begin{align}
	\label{eq:ch3:geodAcc2}
	a^\mu &= S^\nu \nabla_\nu \left( S^\rho \nabla_\rho T^\mu \right), \notag \\
		  &= S^\nu \nabla_\nu \left( T^\rho \nabla_\rho S^\mu \right), \notag \\
		  &= S^\nu \left(\nabla_\nu  T^\rho \right)\nabla_\rho S^\mu  + S^\nu T^\rho \nabla_\nu \nabla_\rho S^\mu , \notag \\
		  &= T^\rho \left(\nabla_\rho  S^\nu \right)\nabla_\nu S^\mu  + S^\nu T^\rho \nabla_\rho \nabla_\nu S^\mu  + S^\nu T^\rho \comm{\nabla_\nu}{\nabla_\rho} S^\mu , \notag \\
		  &= T^\rho \nabla_\rho  \left(S^\nu \nabla_\nu S^\mu\right) + S^\nu T^\rho \comm{\nabla_\nu}{\nabla_\rho} S^\mu,
\end{align}
but since $S$ is the tangent to a geodesic, the first term vanishes by the geodesic equation. (Note: there was a lot of re-labelling dummy indices and swapping $S$ and $T$ in there, read it carefully). We are therefore left with the following expression for the geodesic separation acceleration:
\begin{equation}
	\label{eq:ch3:geodAcc3}
	a^\mu = S^\nu T^\rho\comm{\nabla_\nu}{\nabla_\rho} S^\mu.
\end{equation}
This may not look like much of an improvement, but it's actually great---the commutator of the covariant derivatives is a tensor operator, and astoundingly its components do \emph{not} depend on $S$. The easy way to see why is to note that the partial derivatives commute, so the only non-commuting parts of the covariant derivative operators comes from the Christoffel symbols and their derivatives, all depending only on the metric. In components, we have:
\begin{equation}
	\label{eq:ch3:defRiem}
	\boxed{
	\comm{\nabla_\nu}{\nabla_\rho} S^\mu =: R\indices{^\mu_\lambda_\nu_\rho}S^\lambda = \left( \partial_\nu \Gamma^\mu_{\rho\lambda} - \partial_\rho \Gamma^\mu_{\nu\lambda} + \Gamma^\mu_{\nu\sigma}\Gamma^\sigma_{\rho\lambda} - \Gamma^\mu_{\rho\sigma}\Gamma^\sigma_{\nu\lambda}  \right)S^\lambda .
}
\end{equation}
This defines the incredibly important \textbf{Riemann Curvature Tensor}\index{Riemann Curvature Tensor} $R\indices{^\mu_\lambda_\nu_\rho}$, which encodes the intrinsic notion of curvature for Riemannian manifolds equipped with the Levi-Civita connection (the definition can be extended with little difficulty to more general connections). In terms of the curvature tensor, the geodesic separation acceleration is
\begin{equation}
	\label{eq:ch3:geodAcc4}
	a^\mu = R\indices{^\mu_\lambda_\nu_\rho}S^\nu T^\rho S^\lambda,
\end{equation}
and now we know what it means: if initially parallel lines remain parallel, then $R\indices{^\mu_\lambda_\nu_\rho} = 0$ and the manifold is Euclidean, otherwise the spacetime is \emph{curved}. 

In its final form, \eqref{eq:ch3:geodAcc4} is known as the \emph{geodesic deviation equation}, and represents a very important physical effect. What are two parallel lines you made extensive use of eons ago when deriving special relativity? How about the worldlines of the ends of some extended object, like a rod? The ultimate divergence or convergence of parallel lines is the same thing as a force that twists, stretches, and contracts macroscopic objects---a \textbf{tidal}\index{Tidal Forces} force. Tidal forces arise from \emph{non-homogeneous} gravitational fields, and in general relativity, that manifests as intrinsically \emph{curved} spacetimes. It is also very interesting to note that this is a \emph{local} expression, so tidal forces are observable locally. This does not contradict the equivalence principle though, the elevator thought-experiment still works if you give it a non-uniform acceleration (but in general, don't be a jerk to people in elevators).

\subsection{Properties of the Curvature Tensor}

The Riemann tensor is such an important object it will be very helpful later for us to elaborate on some of its key properties now. For listing symmetries, it's easiest if all the indices are on the same level, so to start we'll consider the purely covariant form, $R_{\mu\lambda\nu\rho} := g_{\mu\sigma}R\indices{^\sigma_\lambda_\nu_\rho}$, which satisfies the following symmetries:
\begin{enumerate}
	\item Anti-symmetry in pairs: 
		\begin{align}
			\label{eq:ch3:Rsymm1}
			R_{\mu\lambda\nu\rho} &= -R_{\lambda\mu\nu\rho}, \quad \text{and} \\
			R_{\mu\lambda\nu\rho} &= -R_{\mu\lambda\rho\nu}.
		\end{align}

	\item Symmetry of pairs:
		\begin{equation}
			\label{eq:ch3:Rsymm2}
			R_{\mu\lambda\nu\rho} = R_{\nu\rho\mu\lambda}.
		\end{equation}
	\item Cyclic permutations (aka the Algebraic Bianchi identity):
		\begin{equation}
			\label{eq:ch3:Rsymm3}
			R_{\mu[\lambda\nu\rho]} = R_{\mu\lambda\nu\rho} + R_{\mu\nu\rho\lambda} + R_{\mu\rho\lambda\nu} = 0.
		\end{equation}
	\item The Bianchi identity:
		\begin{equation}
			\label{eq:ch3:Rsymm4}
			\nabla_{[\sigma}R_{\mu\lambda]\nu\rho} = \nabla_{\sigma}R_{\mu\lambda\nu\rho} + \nabla_\mu R_{\lambda\sigma\nu\rho} + \nabla_\lambda R_{\sigma\mu\nu\rho} = 0.
		\end{equation}
\end{enumerate}
With these symmetries, we may also identify the \emph{unique} non-vanishing contraction of the curvature tensor: 
\begin{equation}
	\label{eq:ch3:Ric}
	R_{\mu\nu} := R\indices{^\alpha_\mu_\alpha_\nu}.
\end{equation}
(Note: The choice of indices to contract is a \emph{convention}. Due to symmetry, any other choice yields the same tensor up to an overall sign so we refer to the choice of convention by whether or not there is a minus sign in \eqref{eq:ch3:Ric}). This is known as the \textbf{Ricci Tensor}\index{Ricci Tensor}, and notice that the symmetries above imply it is symmetric: $R_{\mu\nu} = R_{\nu\mu}$ (always nice to have). Finally, we may also identify the one and only non-vanishing scalar that can be constructed from (one copy of) the curvature tensor and metric alone, and that is the \textbf{Ricci Scalar}\index{Ricci Scalar}:
\begin{equation}
	\label{eq:ch3:RicS}
	R := R\indices{^\mu_\mu}.
\end{equation}

One quick thing before moving on: how many degrees of freedom does this bad boy have? At first glance, the Riemann curvature tensor is a generic rank-4 tensor, so should have a whopping {\color{blue} $n^4$} degrees of freedom in $n$ spacetime dimensions. Fortunately for us (and therefore the world), the symmetries above cut this down significantly. Anti-symmetry in pairs means the two pairs of those $n$s are elements of anti-symmetric tensors, so only encode $\frac 12 n(n-1)$ unique degrees of freedom, cutting us down to  {\color{blue} $(\frac 12 n(n-1))^2$} degrees of freedom total. Next, symmetry across the two pairs relieves us of more degrees of freedom. Treating the two pairs as independent multi-indices of a higher-dimensional matrix (i.e., $R_{mn} = R_{\{\mu\lambda\}\{\nu\rho\}}$), we use the well-known formula for the unique degrees of freedom in a symmetric $m \times m$ matrix as $\frac 12 m(m+1)$ to write our total degrees of freedom as {\color{blue} $\frac 12[\frac 12 n(n-1)][\frac 12 n(n - 1) + 1]$}. Finally, the algebraic bianchi identity imposes a great many constraints, but not quite as many as it might seem at first. The previous constraints in fact make redundant all expressions of \eqref{eq:ch3:Rsymm3} in which any two of the indices are the same (it is a good exercise to show this). As a result, the cyclic permutations remove $\binom{n}{4}$ degrees of freedom from our system, leaving an algebraic mess which miraculously simplifies to a grand total of {\color{blue} $\frac {1} {12} n^2(n^2 - 1)$} degrees of freedom. In the usual $n = 4$ of the real world, this is a nice safe 20 degrees of freedom---still enough to cause headaches, but not too many for us to call it quits.

\begin{aside}[The Weyl Tensor]
    For our purposes, the Riemann curvature tensor and the Ricci tensor and scalar are all that will be necessary to understand the basics of General Relativity. In fact we will see that it is only the Ricci tensor and scalar that do the bulk of the work, but it is worth noting that these quantities have far fewer degrees of freedom---and so contain much less information about local curvature---than the full Riemann curvature tensor. Phrased another way, physics only couples directly to \emph{part} of the curvature of spacetime so it is sometimes convenient to decompose the Riemann curvature tensor into its traces (that speak directly to local physics), and the traceless remainder. This traceless remainder goes by the name of the \textbf{Weyl tensor}\index{Weyl Tensor} and is defined by simply subtracting away all possible traces from the Riemann tensor:
\begin{equation}
    \label{eq:ch3:Weyl}
    C_{\mu\nu\rho\sigma} := R_{\mu\nu\rho\sigma} - \frac 2{n - 2}\left( g_{\mu[\rho}R_{\sigma]\nu} + g_{\nu[\sigma}R_{\rho]\mu} \right) - \frac 2 {(n - 1)(n - 2)} g_{\mu[\sigma}g_{\rho]\nu}R.
\end{equation}
The subtracted terms on the RHS are all needed to account for the symmetries of the indices on the Riemann curvature tensor, so at the end of the day any contraction of indices on $C_{\mu\nu\rho\sigma}$ vanishes (you should try some for yourself). The Weyl tensor is a very interesting object geometrically, and in a sense encodes information about the part of curvature that twists and distorts without stretching or squeezing, but a closer look at its properties is well outside the scope of this work.
\end{aside}

For the reader in a hurry, this is sufficient to move on to the next chapter. For those with a little more time on their hands, it is a good exercise to prove the symmetries above, particularly because the easiest way to prove them involves a very helpful tool, the \emph{locally inertial} or \emph{Riemann normal} coordinates.

\subsubsection*{Locally Inertial (aka Riemann Normal) Coordinates}

\textbf{Locally inertial coordinates}\index{Locally Intertial Coordinates} are the formal notion of us as rulers of our manifold being kind to Tiny Alice and choosing our coordinates such that they align with her local frame. For any single point on the manifold, we are allowed to establish a system of coordinates such that our cardinal directions align with the coordinate system established by the tiny person that lives there, and the tangent space to the manifold at that point has an inner product given by the Minkowski metric $g_{\mu\nu}(p) = \eta_{\mu\nu}$. 

What's surprising is that this notion actually \emph{extends} just a little ways away from Tiny Alice's room. Not only can we establish coordinates such that Tiny Alice's metric looks like $\eta$, but we can also arrange for the \emph{first derivatives} of the metric to vanish at her location. One way to see why this should be the case is that as far as Tiny Alice can tell, inertial objects follow straight lines, so the geodesic equation from her perspective reads:
 \begin{equation}
	\label{eq:ch3:geodAtAlice}
	\frac{\dd^2 x^\mu(\tau)}{\dd \tau^2} + \Gamma^\mu_{\sigma\rho}\frac{\dd x^\sigma(\tau)}{\dd \tau}\frac{\dd x^\rho(\tau)}{\dd \tau} = \frac{\dd^2 x^\mu(\tau)}{\dd \tau^2} = 0, \qquad \text{(at $p$)}
\end{equation}
so we also have $\Gamma\vert_p = 0$ which is certainly evidence for the derivative of the metric vanishing at that point. Notice however that the vanishing of a derivative at a point \emph{does not imply the vanishing of a second derivative} at that point. This goes back to the idea of second derivatives sampling more of an underlying distribution---just because things look flat at a single point, and even a shmidge around that point, that doesn't mean you're actually looking at something that's flat overall (think about a local minimum of some curvy function). 

There are more thorough, formal ways of showing the existence of this coordinate system, but the easiest way to convince yourself they exist is through a parameter-counting argument. If these coordinates exist on the manifold, then they can be related to any other coordinates on the manifold by a coordinate transformation: $x(\hat{x})$, where $\hat{x}$ are our putative locally inertial coordinates (I am suppressing indices here because this is a schematic argument). In the vicinity of $p$, we may write this as a Taylor series: $x(\hat{x}^{-1}) = \hat{x}(p) + \frac{\dd \hat{x}}{\dd x}\big\vert_p\left( \hat{x} - \hat{x}(p) \right) + \ldots$. More importantly, the Jacobian of the transformation may similarly be expanded in a Taylor series:
\begin{equation}
	\label{eq:ch3:JacTay}
	\frac{\partial x^\mu}{\partial \hat{x}^\nu} = \frac{\partial x^\mu}{\partial \hat{x}^\nu}\big\vert_p + \frac{\partial^2 x^\mu}{\partial \hat{x}^\alpha \partial \hat{x}^\nu}\big\vert_p \left( \hat{x}^\alpha - \hat{x}^\alpha(p) \right) + \ldots,
\end{equation}
and so may be the components of the metric (individually as functions, not components of a tensor):
\begin{equation}
	\label{eq:ch3:MetTay}
        g_{\mu\nu}(x(\hat{x}^{-1})) = g_{\mu\nu}(\hat{x}(p)) + \frac{\partial g_{\mu\nu}}{\partial \hat{x}^\alpha}\big\vert_p \left( \hat{x}^\alpha - \hat{x}^\alpha(p) \right) + \ldots \qquad \text{(as functions)}.
\end{equation}
In this way, we can perturbatively change coordinates for the metric near our point of interest and see how far our freedom to choose the components of the coordinate transformation will take us.

So crunch away, we have (now as components of a tensor):
\begin{align}
	\label{eq:ch3:expMet}
	g_{\mu\nu}(\hat{x}) &= \frac{\partial x^\alpha}{\partial \hat{x}^\mu}\frac{\partial x^\beta}{\partial \hat{x}^\nu}g_{\alpha\beta}(x), \notag \\
						&= \frac{\partial x^\alpha}{\partial \hat{x}^\mu}\frac{\partial x^\beta}{\partial \hat{x}^\nu}\Big\vert_p g_{\alpha\beta}(\hat{x}(p)) \notag \\
						&\quad {} + \left( \hat{x}^\gamma - \hat{x}^\gamma(p) \right) \left\{ \frac{\partial x^\alpha}{\partial \hat{x}^\mu}\frac{\partial x^\beta}{\partial \hat{x}^\nu}\frac{\partial g_{\alpha\beta}(\hat{x})}{\partial \hat{x}^\gamma} + \frac{\partial^2 x^\alpha}{\partial \hat{x}^\gamma\partial \hat{x}^\mu}\frac{\partial x^\beta}{\partial \hat{x}^\nu}g_{\alpha\beta}(\hat{x}) + \frac{\partial x^\alpha}{\partial \hat{x}^\mu}\frac{\partial^2 x^\beta}{\partial \hat{x}^\gamma\partial \hat{x}^\nu}g_{\alpha\beta}(\hat{x}) \right\}\Big\vert_p \\
						&\quad{} + \ldots \notag \\
						& \stackrel{?}{=} \eta_{\mu\nu}. \notag
\end{align}
Now, obviously we have enough degrees of freedom to rotate the metric at a single point into the Minkowski form, so it's not hard to see there's a solution to $\frac{\partial x^\alpha}{\partial \hat{x}^\mu}\frac{\partial x^\beta}{\partial \hat{x}^\nu}\Big\vert_p g_{\alpha\beta}(\hat{x}(p)) = \eta_{\mu\nu}$. In fact, it is more than possible, the system is overdetermined. In 3+1 dimensions, we have an excess freedom of 6 parameters, which corresponds to the 6 generators of the Lorentz group, so not only can we choose the components of the Jacobian to take us to the Minkowski metric at a point, but we have enough degrees of freedom left over to also rotate it by any Lorentz transformation\footnote{In fact, in $n$ dimensions there are $\frac 12 n (n-1)$ generators of the Lorentz Lie algebra, and  $n^2 - \frac 12 n(n+1) = \frac 12 n(n-1)$ left-over degrees of freedom from fixing the metric to $\eta$ at a point.}. The question then is, is it technically possible to solve the linear system of equations $\{\ldots\} = 0$ for the terms in braces in \eqref{eq:ch3:expMet}? Without working too hard, what we can do is count parameters and see if the system of equations is under, over, or identically determined. 

The idea here is that the original metric and its derivatives are generic, so we want to pick the components of our coordinate transformation (and its derivatives) so as to eliminate the original metric and its derivatives. The Hessian matrix (the ``Jacobian of the Jacobian,'' if you will) looks at first like it has $n^3$ components, but it's actually smaller than that. Since partial derivatives commute, the Hessian has only $\frac 12 n(n+1)\times n$ degrees of freedom (a symmetric $n\times n$ matrix has $\frac 12 n(n+1)$ components, so think of the gradient of the Jacobian as $n$ symmetric $n\times n$ matrices). Meanwhile, the metric is a symmetric $n\times n$ matrix, so it has $\frac 12 n(n+1)$ degrees of freedom, and so its gradient has $n$ times that many components (one for each partial derivative). Altogether then, we want to kill off  $\frac 12 n^2(n+1)$ components, and we have $\frac 12 n^2(n+1)$ degrees of freedom to do it with, so the system is \emph{exactly} determined, and we can indeed choose coordinates such that the first derivative of the metric vanishes at a point.  

Finally, we have to ask if we can go further, can we set the second derivative to vanish? Here again we count. The next order up involves a gradient of the gradient of the metric, so \\$\frac 12 n(n+1)\times \frac 12 n(n+1) = \frac 14 n^2(n+1)^2$ components (again, partial derivatives commute, cutting the degrees of freedom down a bit), and the Jacobian of the Jacobian of the Jacobian, with \\$\frac 1{3!}n^2(n+1)(n+2)$ degrees of freedom (symmetry of all 3 partial derivatives leads to $n$ copies of $\mqty(n + 3 - 1 \\ 3)$ independent terms). In 3+1 dimensions, these are 100 and 80 degrees of freedom respectively, which means we're 20 degrees of freedom shy of being able to use a coordinate transformation to flatten a metric beyond its first derivative. Those 20 degrees of freedom are important, by the way, they are exactly the 20 degrees of freedom of the Riemann curvature tensor! With that out of the way, we can move on to proving the symmetries of $R\indices{^\mu_\nu_\rho_\sigma}$.

\subsection*{Proofs of Riemann Tensor Symmetries}

The symmetries above are most easily seen by direct computation in the Riemann normal coordinates. In those coordinates, at the point $p$ where the metric is Minkowski, we have $\hat{g}_{\mu\nu}(p) = \eta_{\mu\nu}$, $\partial_\rho \hat{g}_{\mu\nu}(p) = 0$ and so $\widehat\Gamma^\mu_{\nu\lambda}(p) = 0$. The Riemann tensor then boils down to (suppressing the evaluation at $p$):
\begin{align}
	\label{eq:ch3:compRInert}
	\hat{R}_{\mu\lambda\nu\rho} &= \hat{g}_{\mu\sigma}\hat{R}\indices{^\sigma_\lambda_\nu_\rho}, \notag \\
								&= \hat{g}_{\mu\sigma}\left( \widehat\Gamma^\sigma_{\rho\lambda,\nu} -  \widehat\Gamma^\sigma_{\nu\lambda,\rho}  \right), \notag \\
								&= -\frac 12 \hat{g}_{\mu\sigma}\hat{g}^{\sigma\alpha}\left(\hat{g}_{\rho\lambda,\alpha\nu} - \hat{g}_{\lambda\alpha,\rho\nu} - \hat{g}_{\alpha\rho,\lambda\nu} \right. \notag \\
								&\left. \qquad {} - \hat{g}_{\nu\lambda,\alpha\rho} + \hat{g}_{\lambda\alpha,\nu\rho} + \hat{g}_{\alpha\nu,\lambda\rho} \right), \notag \\
								&= -\frac 12 \left(\hat{g}_{\rho\lambda,\mu\nu} - \hat{g}_{\lambda\mu,\rho\nu} - \hat{g}_{\mu\rho,\lambda\nu} \right. \notag \\
								&\left. \qquad {} - \hat{g}_{\nu\lambda,\mu\rho} + \hat{g}_{\lambda\mu,\nu\rho} + \hat{g}_{\mu\nu,\lambda\rho} \right), \notag \\
								&= -\frac 12 \left(\hat{g}_{\rho\lambda,\mu\nu} - \hat{g}_{\mu\rho,\lambda\nu} - \hat{g}_{\nu\lambda,\mu\rho} + \hat{g}_{\mu\nu,\lambda\rho} \right).
\end{align}
(Here we used that $\partial_\rho \hat{g}_{\mu\nu} = 0$ to pull the overall factor of $ \hat{g}^{\sigma\alpha}$ out front, and we have used that partial derivatives commute). 

Normally if we were trying to derive a tensor relation from this, it would be fairly tedious to try and find another tensor that is equal to the RHS in this particular coordinate system at that particular point, but since our aim is only to permute indices, the job is already done. We know the tensor it will be equal to, it's just the (covariant) Riemann tensor with its indices permuted. So for instance, it is easy to see that if we swap $\mu$ and $\lambda$, the first pair and second pair of terms on the RHS swap themselves around inducing an overall minus sign, which immediately yields $ \hat{R}_{\mu\lambda\nu\rho} = - \hat{R}_{\lambda\mu\nu\rho}$, but since this is a \emph{tensor} equation, we can drop the hats and observe that it holds in any coordinate system. Furthermore, although this was only shown for a single point, a corresponding Riemann normal coordinate system can be established at \emph{any} point on the manifold, and will result in the same expression, so the point-wise derivation can be extended to apply to the Riemann tensor as a tensor \emph{field} over the whole manifold. In this way, \eqref{eq:ch3:compRInert} contains all the information needed to directly compute all of the symmetries listed above. It is a good exercise to pick one or two of them and follow the calculation through.

\section{Recap}

\begin{itemize}
	\item Physics needs derivatives, but the ordinary partial derivative is \emph{not} a good operator on tensors (it does not take tensors to tensors).
	\item Define a new kind of derivative, the \emph{Covariant Derivative} (or \emph{connection}) $\nabla$ to carry out the duties of a derivative in a manner consistent with tensor equations. It has the form of the ordinary derivative plus a correction: $ \nabla = \partial + \Gamma$, the correction being called the ``connection coefficients.''
	\item The choice of a covariant derivative is an \emph{additional structure} on a manifold, so points+coordinates $\to $ (points+coordinates)+metric $\to $ (points+coordinates+metric)+connection.
	\item For simplicity (among other reasons), we choose the \emph{Levi-Civita} connection, defined by having zero torsion ($\Gamma^\sigma_{\mu\nu} = \Gamma^\sigma_{\nu\mu}$) and being metric-compatible ($ \nabla g = 0$). For the Levi-Civita connection, the connection coefficients are called \emph{Christoffel symbols} and take on the simple form \eqref{eq:ch3:christofs}. 
	\item Generalizing equations from flat to curved space typically amounts to replacing $\partial \to \nabla $.
	\item Straight lines are generalized on manifolds to \emph{geodesics}, given by curves that solve the geodesic equation \eqref{eq:ch3:defGeod2}. 
	\item Intrinsic curvature of manifolds deduced by geodesic deviation, the failure of initially parallel lines to remain parallel. This failure is the geometric manifestation of tidal forces, and is entirely encoded in the \emph{Riemann Curvature tensor}, given in components by \eqref{eq:ch3:defRiem}.
	\item A very useful tool in calculating tensor relations is the \emph{locally inertial coordinate system}, where the metric at a single point takes the form of the Minkowski metric, and at which point the first derivative of the metric (and hence the Christoffel symbols) vanishes. Note that the \emph{second} derivative of the metric (and hence the derivative of the Christoffel symbols) does \emph{not} usually vanish there, which is thanks to the curvature of the manifold. 	
\end{itemize}




\begin{subappendices}

\section{An Example of a Geodesic}
\label{ch:curve:app:geodEx}

We've done a lot of work with generic geodesics, but it can be illuminating to solve for one explicitly. In general, solving for a geodesic is a daunting task---integrating a series of coupled, non-linear differential equations---and in practice it is usually done numerically. In some cases though, there are known analytic solutions; besides the trivial example of geodesics in flat space (i.e., straight lines), probably the most common analytically solvable geodesics are those on the 2-sphere. 

The standard angular metric on the 2-sphere is:
\begin{equation}
    \label{eq:geodEx:ds2}
    \dd s^2 = \dd\theta^2 + \sin[2](\theta) \dd\phi^2,
\end{equation}
(using the correct physicist conventions in which $\theta$ measures latitude and $\phi$ measures longitude). The non-vanishing Christoffel symbols derived from this metric are:
\begin{equation}
    \label{eq:geodEx:gammas}
    \begin{gathered}
        \Gamma^{\theta}_{\phi\phi} = -\sin\theta\cos\theta, \qquad \Gamma^{\phi}_{\theta\phi} = \Gamma^\phi_{\phi\theta} = \cot\theta.
    \end{gathered}
\end{equation}
(And for fun, the Ricci curvature tensor is diagonal with $R_{\theta\theta} = 1,\;\; R_{\phi\phi} = \sin[2](\theta)$, and Ricci scalar $R = 2$). The geodesic equation is $U^\nu \nabla_\nu U^\mu = 0$ , or in more calculationally useful form:
\begin{equation}
    \label{eq:geodEx:geodEq}
    \dot U^\mu + \Gamma^\mu_{\rho\sigma}U^\rho U^\sigma = 0, 
\end{equation}
in general, and specifically
\begin{align}
    \dot U^\theta - \sin\theta\cos\theta\, U^\theta U^\phi &= 0, \quad \text{and} \label{eq:geodEx:geodEq1}\\
    \dot U^\phi + 2\cot\theta\, U^\theta U^\phi &= 0 \label{eq:geodEx:geodEq2}
\end{align}
for us. Also important is the constraint equation (that the length of a geodesic is fixed), so we also have
\begin{equation}
    \label{eq:geodEx:const}
    (U^\theta)^2 + \sin[2](\theta) (U^\phi)^2 = K^2
\end{equation}
for some constant $K$ (note that $K^2 > 0$ since the metric is Riemannian, so positive definite for non-trivial vectors).

Solving the system of equations \eqref{eq:geodEx:geodEq1} to \eqref{eq:geodEx:const} in all generality is no simple task, but we need not be so adventurous. The 2-sphere is a highly symmetric creature, so any initial conditions $\{x_0^\mu, \, U_0^\mu\}$ can easily be used to define a coordinate system of exactly the same form as \eqref{eq:geodEx:ds2}. It is easiest therefore to start from coordinates adapted to $\{x_0^\mu,\, U_0^\mu\}$, so let us say in those coordinates the initial conditions are
\begin{align}
    x^\mu_0 &= (\pi/2, 0), \quad \text{and}  \label{eq:geodEx:IC1} \\
    U_0^\mu &= (K, 0). 
\end{align}
Then we get to integrating. Both of the geodesic equations are separable, but the second \eqref{eq:geodEx:geodEq2} is a little cleaner:
\begin{align}
    \label{eq:geodEx:solvePhi}
    \frac {\dd U^\phi}{U^\phi} &= -2\cot\theta\,\dd U^\theta, \notag \\
    \implies \quad U^\phi &= \frac{\sin[2](\theta_0)}{\sin[2](\theta)}U^\phi_0.
\end{align}
With the initial condition that $U_0^\phi = 0$, this sets $U^\phi = 0$ for all time, which is always nice to see. With the second coordinate out of the equations altogether now, it is almost trivial to solve for the first using the constraint equation (the remaining geodesic equation is redundant; the constraint equation always makes redundant one of the geodesic equations and is usually a little easier to use). From \eqref{eq:geodEx:const}, we have 
\begin{align}
    \label{eq:geodEx:solveTheta}
    (U^\theta)^2 &= K^2, \notag \\
    \implies U^\theta &= K
\end{align}
using the initial condition to select the positive root. The resultant geodesic is parameterized as 
\begin{equation}
    \label{eq:geodEx:solveGeod}
    \begin{aligned}
        \theta(\tau) &= K\tau + \frac \pi 2, \\ 
        \phi(\tau) &= 0
    \end{aligned}.
\end{equation}

In other words, a particle kicked along the equator of a sphere will just happily follow the equator until the end of time (in its happy frictionless world---we are theoretical physicists after all). More generally, one will always find the geodesics on spheres to be arcs of ``great circles,'' (like the equator or lines of longitude, the largest circles that can be sliced from the sphere). This may be our first example of a non-Euclidean geodesic, but it is far from a revelation for any of us lucky enough to have grown up on the surface of the Earth. It is well known that the shortest path between any two points on the Earth is not a straight line on any geographical map even though it is very much the case that the shortest path to physically get from one point to another is to point at it and go straight there (``as the crow flies,'' so the expression goes). The disparity arises because geographical maps are contrived coordinate systems designed to preserve our faith in local Cartesian coordinate systems, ensuring relative orientations and separations of landmarks are as familiar as possible (though the remarkable size of Greenland on most world maps attests to the increasing failure of this procedure near the edges of the projection).

\section{Integration on Manifolds}
\label{ch:curve:app:int}

Warning: this is a long appendix. So far, I would say we've done a great job taking derivatives of things in curved spaces, and even setting up systems of differential equations. That will take us a long way, but differentiation is only really half of calculus\footnote{Well okay, we actually have touched a bit on integration; solutions to the differential equations we've established are essentially anti-derivatives, but they take place in the land of real analysis, so they don't feel the wrath of curvy geometry.}, so if we leave it there, our tool kit will be missing an enormous amount of important material with which to study physics. To remedy this, we need to think about how integrals are going to work.

\subsection{Integrals in One Dimension}
When we started to look at derivatives we had to be very clear about what types of objects we were working with, what kind of operation a derivative is, and eventually we had to invent a whole new covariant derivative in order to handle curved spaces. It should come as no surprise then that we have to treat integration with a similar degree of care. Let us first consider the simplest manifold imaginable\footnote{Okay, second simplest, the actual simplest is probably just a point.}: $L$, a subset of $\mathbb{R}^1$ (i.e., a line segment). This is an example of a \emph{manifold with boundary}, the main manifold is the line segment \emph{without} the endpoints, and the ``boundary''---named $\partial L$---is the set of endpoints (thought of as a 0-dimensional manifold). A scalar field on $L$ is really just an ordinary one-dimensional function $F(x)$, and as long as it's a nice non-perverse function, we can apply to it all the usual operations of first-year calculus. In particular, for a lovely differentiable function $F(x)$ with derivative $f(x)$, we can apply the old $\frac{x}{x} = 1$ trick and write it in the more complicated form:
\begin{equation}
    \label{eq:int:oldInt}
    F(b) - F(a) = \int_{a}^b f(x) dx. 
\end{equation}
What is happening here? Well first we've taken the gradient of $F(x)$ at each point $x \in (a,b)$, then multiplied it by an infinitessimal displacement along the line (i.e., in the one coordinate direction available), and summed up the contributions at every point. Now let's find a way to write this in the language of differential geometry.

In \eqref{eq:int:oldInt}, we're thinking of $dx$ as an infinitessimal displacement (which is why I have typeset it with the $d$ in italics), but the keen reader will recall that this notation is of course almost identical to how we denoted cotangent basis vectors, so that same reader would be in good form to wonder if there's a connection---and indeed there is. Recall the differential geometric version of the gradient of a scalar field is a map $\dd: C^\infty(M) \to T^*\!M$, where $C^\infty(M)$ means smooth functions on the manifold and $T^*\!M$ is the cotangent bundle. This map went along the lines of $\dd F = \partial_i F \dd x^i$, which in one dimension is simply $\dd F = \partial_x F \dd x$. While we're using almost the same notation, keep in mind that the $\dd x$ here is a basis \emph{covector}, not some infinitessimal displacement. With that said, covectors are functions that eat vectors and spit out a number, so there's hope we can use the covector $\dd x$ to return a number that would serve the same purpose as the $dx$ in \eqref{eq:int:oldInt}. Indeed, in one dimension it's frightfully simple, we only really have one vector to eat, $\partial_x$! So we can imagine defining a new operation:
\begin{equation}
    \label{eq:int:newInt1d}
    \int_{U_i} \dd F := \int_{\phi(U_i)} f(x) dx,
\end{equation}
where the left-hand side \emph{defines} the operation of integration of covectors $\int_{U_i} : T^*\!M \to \mathbb{R}$ over a (1-dimensional) coordinate patch $U_i$ (with coordinate system $\phi$) as the usual integral on $\mathbb{R}$, identifying the measure\footnote{Could be either Riemann or Lebesgue.} on the right-hand side with the result of $dx := \dd x(\partial_x)$ at each point\footnote{Well, we're being a bit creative, in fact. Recall we had defined the covector basis by the condition $\dd x^i(\partial_j) = \delta^i_j$. That was in the context of tangent and cotangent spaces though, so we imagine the concept of ``unit length'' \emph{in a tangent space} equates to an infinitessimal length on the scale of the manifold.}. We'll stick with a single coordinate patch for the moment, but the result can easily be extended to the full manifold (which we will show soon enough).

\subsection{Integrals in Higher Dimensions}

Great, we've integrated integration into differential geometry! Well, in one dimension. Can we build on this to extend the result to higher dimensions? Of course we can, but some interesting complications arise. First, let us follow our noses and try naively extending the previous program to higher dimensions. We consider a 2-dimensional manifold that is a subset of $\mathbb{R}^2$ and a scalar field $\psi(x,y)$. If we wish to write this as the integral of its derivative, we could take the partial derivative across only a single direction and integrate that, but we've essentially already done this. The new thing is to write this as the integral across \emph{both} the  $x$ and $y$ directions at the same time, but to do this we need to take the derivative twice:
\begin{equation}
    \label{eq:int:oldInt2d}
    F(a_2, b_2) - F(a_1, b_2) - F(a_2, b_1) + F(a_1, b_1) = \int_{[a_1,b_1]}^{[a_2,b_2]} \partial_x\partial_y \psi(x,y) dx dy. 
\end{equation}
And so we hit our first snag: second-order differentiation. 

\begin{figure}[htp!]
    \centering
    \includegraphics[width=\textwidth]{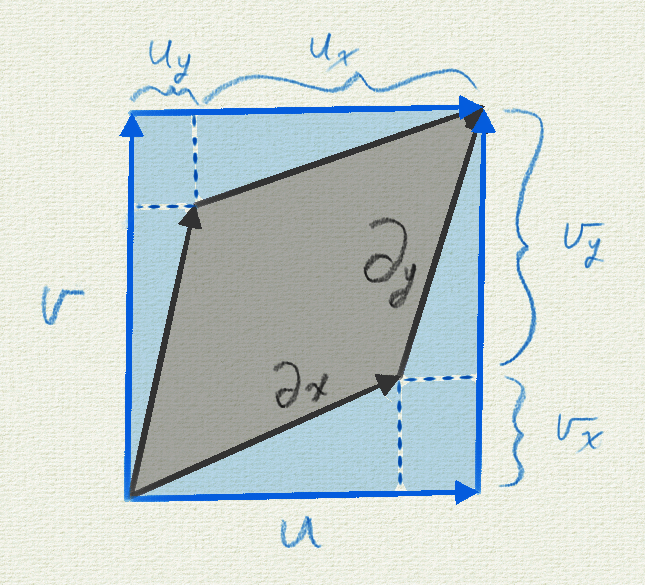}
    \caption[Calculating areas in non-orthogonal coordinate bases.]{To get an area (the grey shaded region) out of non-orthogonal basis vectors $\partial_x$ and $\partial_y$, use their projections ($u_x, u_y, v_x,$ and $v_y$) on orthogonal vectors $u$ and $v$ and carry out some elementary geometry calculations.}
    \label{fig:basis_areas}
\end{figure}

Recall that for curvy spaces (and even simply non-orthogonal coordinates) the simple partial derivative of a tensor (such as $\partial_i (\omega_j \dd x^j)$ for some covector $\omega$) is \emph{not} a tensor, so is not the sort of object we have any interest in. Instead, we had to define a whole new type of derivative, the covariant derivative. Could this be the generalization we need? Well, in \eqref{eq:int:oldInt2d}, the measure $dx dy$ is basically the product of infinitessimal displacements in the (orthogonal) $x$ and $y$ directions, so at first glance, the tensor $\nabla_x \partial_y \psi \, \dd x \otimes \dd y$ seems like it should be what we're looking for, but there is a subtle flaw to this attempt. This subtle error has to do with what we meant to calculate back in the day when we first wrote down the measure $dx dy$. In the old Riemann-style sums, we had intended this product to represent the \emph{area} of an infinitessimal box spanned by infinitessimal displacements in the $x$ and $y$ directions. For Cartesian coordinates, the tensor $\dd x \otimes \dd y$ does the trick, but as soon as we choose (or are forced) to use non-Cartesian coordinates, the coordinate basis vectors lose their orthogonality (i.e., as soon as we introduce a non-trivial metric tensor), so we're no longer measuring a sensible area everywhere: $\dd x\otimes \dd y (u, v) = u_x v_y$, where $u = u_x \partial_x + u_y \partial_y$, $v = v_x \partial_x + v_y \partial_y$ are orthogonal unit vectors $g(u,v) = 0$. This is easier to picture graphically; figure \ref{fig:basis_areas} shows the area spanned by non-orthogonal basis vectors $\partial_x$ and $\partial_y$, as well as the area spanned by orthogonal vectors $u$ and $v$, and the trick is to get the first from the second. To get the correct area, compute the total area $uv = (u_x + u_y)(v_x + v_y)$ and subtract off all the purple shaded regions, $2\times \frac12 \times u_xv_x$, $2\times \frac12 \times v_yu_y$, and $2\times v_x u_y$. Altogether, the desired region has area $A = u_x v_y - u_yv_x $. This is surprisingly close to the naive attempt using $\dd x\otimes \dd y$, all that's missing is to anti-symmetrize it to $\dd x \otimes \dd y - \dd y \otimes \dd x$ and we have a good measure of area! As it turns out, there is a deeper reason for this---let's explore.

\subsection{Differential Forms and the Exterior Derivative}
\label{ch:curve:app:int:forms}

Recall our big complaint about the ordinary partial derivative as an operator on tensors: that it fails to transform correctly with a coordinate transformation since it product-rules the factor of the Jacobian (or inverse) that comes from transforming the tensor it's attempting to differentiate. To solve the problem in all generality, we invented a new derivative (the covariant derivative) and used the wild freedom that came with such a definition to give it a whole bunch of lovely properties that made our lives much easier. Well for general tensors, that's still the only way to differentiate them (and so the only choice to build physical equations from), but for \emph{covectors}, there exists a happy coincidence that allows us to define an \emph{alternative} type of differentiation. 

First, calculate the transformation law for the partial derivative acting on a covector $\omega = \omega_\mu \dd x^\mu$:
\begin{align}
    \label{eq:partialCovector}
    \partial_\nu(\omega_\mu) \to& \frac{\partial x^\alpha}{\partial y^\nu}\partial_\alpha\left(\frac{\partial x^\beta}{\partial y_\mu}\omega_\beta\right), \notag \\
    =& \frac{\partial x^\alpha}{\partial y^\nu}\frac{\partial x^\beta}{\partial y^\mu}\partial_\alpha\omega_\beta + \omega_\beta\frac{\partial x^\alpha}{\partial y^\nu}\frac{\partial^2 x^\beta}{\partial x^\alpha\partial y^\mu}, \notag \\
    =& \frac{\partial x^\alpha}{\partial y^\nu}\frac{\partial x^\beta}{\partial y^\mu}\partial_\alpha\omega_\beta + \omega_\beta\frac{\partial^2 x^\beta}{\partial y^\nu\partial y^\mu}.
\end{align}
Now if this had been a vector, or a tensor with \emph{any} contravariant indices, we would have ended up with the second term having mixed $x$ and $y$ derivatives, but in the special case here where we are only operating on covariant indices, the ugly second term happens to be completely symmetric in $\nu$ and $\mu$. This affords us the opportunity to construct a new type of proper tensorial derivative out of \emph{just the ordinary partial derivative} by simply anti-symmetrizing the operation: $\partial_\nu(\omega_\mu) \to \dd \omega := \partial_\nu\omega_\mu - \partial_\mu\omega_\nu$. This operation is so important it gets its own name, the \textbf{exterior derivative}. The exterior derivative is a unique derivative operator that only takes purely covariant and anti-symmetric tensors to purely covariant and anti-symmetric tensors of one higher rank. Just before moving on, this is a good spot to remark upon one of the most powerful properties of this operation: that it is nilpotent, i.e., $\dd (\dd \omega) = 0$. This is easiest to see for the case $\omega = \dd \phi = \partial_\mu \phi \dd x^\mu$, in which case the exterior derivative is $\dd^2\omega = (\partial_\mu\partial_\nu - \partial_\nu\partial_\mu)\phi = 0$, but the result is completely general.

We've seen already that the alternating (anti-symmetric) purely covariant tensor of rank two in two dimensions (i.e., $\dd x\otimes \dd y - \dd y \otimes \dd x$) is an operator that computes the area spanned by the basis vectors at any point. In fact, one can show that the one and only purely covariant alternating tensor of rank $n$ in $n$ dimensions performs the same function for those $n$ dimensions. Now we've also seen that alternating purely covariant tensors are also naturally equipped with an extremely simple derivative operation that doesn't care about the underlying geometry of the manifold at all (i.e., the exterior derivative), so it's worth examining these tensors a little more closely. 

First of all, in more than two dimensions it can get unwieldy to explicitly write out all of the anti-symmetrized terms in an anti-symmetric tensor, so instead we define an operator to take care of that particular book-keeping. We define the \textbf{wedge product}\footnote{I am legally obliged here to tell the story of my experience learning about this operation. I learned about differential geometry from a wonderful math professor in Canada. This professor was, however, thoroughly Australian, and when he drew the $\wedge$ symbol on the board he described it as the ``widge product'' which my North American ears heard as the ``witch product''. After many fruitless internet searches while struggling to complete my assignment, my desperate search for ``anti-symmetric tensor product'' finally answered a great many of my questions.} as the anti-symmetrized tensor product: $\wedge : V\times W \to V\wedge W$ by
\begin{equation}
    \label{eq:defWedge}
    (v\wedge w)_{a_1\ldots a_pb_1\ldots b_q} = \frac{(p+q)!}{p!q!}v_{[a_1\ldots a_p}w_{b_1 \ldots b_q]},
\end{equation}
for $v \in V$, $w \in W$, and $\dim(V) = p$, $\dim(W) = q$. (The pre-factor on the right-hand side takes care of the inverse of that same factor that's built into the definition of the anti-symmetrization of the indices). Notice that by definition, we have $v\wedge w = -w \wedge v$, and so as an instant corollary,  $v\wedge v = 0$ (in Abelian land). When $V = W = T^*\!M$ as is the case of our interest, then we define $T^*\!M\wedge\ldots\wedge T^*\!M =: \Lambda^n (T^*\!M)$ the space of \textbf{differential $n$-forms} on the manifold $M$. These $n$-forms are evidently the type of object we can integrate on manifolds.

Clearly we have already seen the one-forms, those are just covectors. To make things consistent, we define 0-forms to simply be smooth functions on the manifold (we have already seen the partial derivative acting on functions yields covectors, and since a single index is already as anti-symmetric as it's going to get, the action of the partial derivative on functions must be exactly the action of the exterior derivative). Everything else follows as you would expect\footnote{In what follows, we will always explicitly write the wedges, but you should know that it is common practice to drop them when talking about differential forms---i.e., most texts will write $\dd x\wedge \dd y \to \dd x \dd y$. We would do this if we were going to say much more about them.}; using that $\dd x^\mu \wedge \dd x^\mu = 0$ and $\dd x^\mu \wedge \dd x^\nu = - \dd x^\nu \wedge \dd x^\mu$, there can clearly only be ${\binom{n}{m}} = \frac{(n+m)!}{n!m!}$ linearly independent basis vectors in  $\Lambda^m (T^*\!M)$ for an $n$-dimensional manifold. Importantly, it follows that there are no $m$-forms at all for $m > n$! (This is easy to show explicitly, it is a good exercise to do so). Moreover, the binomial coefficient is symmetric about $n/2$, so the space of $p$-forms happens to have the same dimension as the space of $(n-p)$-forms. These spaces are in fact so closely related that one can define an bijective operation to go back and forth between them---an operation given the name of the \textbf{Hodge Dual} operator $\star : \Lambda^p \to \Lambda^{n-p}$. 

There is \emph{much} more that can and should be said about differential forms and all the beautiful mathematical structures associated with them, but for present purposes we will settle on two more definitions. As mentioned, in any $n$-dimensional manifold there is only a single linearly-independent $n$-form, sometimes called the ``top form'' $\omega_\text{top} \propto \dd x^0 \wedge \dd x^1 \wedge \ldots \wedge \dd x^n$. On some manifolds, it is possible to find an everwhere-defined smooth field of these $\omega_\text{top} = \omega(x) \dd x^0 \wedge \dd x^1 \wedge \ldots \wedge \dd x^n$ where the function $\omega(x)$ has the same sign everywhere on the manifold and is nowhere vanishing. When this is the case, the form $\omega_{\text{top}}$ is called an \emph{orientation}, and the manifold is called \emph{orientable}. This is the case for all reasonable manifolds, and certainly all spaces we'll be looking at (the only counterexamples tend to be preposterous things like the M\"obius strip). If a manifold is orientable and the sign on $\omega(x)$ is positive then it is called \emph{right handed}, and if it is negative is called \emph{left handed} (these line up with the orientations we all learned when first we were introduced to three-dimensional vectors). An orientation is important for integration because in a sense it give a consistent direction of summation (imagine if you were adding the things on the intervals $[0,1] + [1,2] + [2,3] + [3,2] + [2,1] + [1,2] + \ldots$, it would be chaos!).

The second definition is that of the \emph{volume form}. So far we have the right objects to integrate, but they're not quite as intuitive as we were looking for. Ideally we would like to take a generic function on a manifold and ``integrate'' it over the whole manifold. At the moment though, we can only integrate differential forms, specifically ``top'' or $n$-forms. Component-wise, these are geometric objects with components that are indeed generic functions, but which must transform when asked to do so by a coordinate transformation. In particular, for an $n$-form we have under a coordinate transformation:
\begin{align}
    \label{eq:transformTopForm}
    \omega(x) \dd x^1 \wedge \ldots \dd x^n \to&\, \omega^\prime(x) J^1_\mu \ldots J^n_{\nu}\dd x^\mu\wedge \ldots\wedge \dd x^\nu, \notag \\   
    =&\, \frac{1}{n!}\omega^\prime(x) \epsilon_{\alpha\ldots\beta}(J^{-1})^\alpha_\mu \ldots (J^{-1})^\beta_{\nu}\dd x^\mu\wedge \ldots\wedge \dd x^\nu, \notag \\
    =&\, \omega^\prime(x) \det(J^{-1}) \epsilon_{\mu\ldots\nu}\dd x^\mu\wedge \ldots\wedge \dd x^\nu.
\end{align}
Here we define the inverse Jacobian matrix $(J^{-1})^a_b := \frac{\partial y^a}{\partial x^b}$, we use component notation to write $\dd x^1 \wedge \ldots \wedge \dd x^n = \epsilon_{\mu \ldots \nu} \dd x^\mu \wedge \ldots \wedge \dd x^\nu$, and in the last line we use the identity $\frac 1 {n!}\epsilon_{ab\ldots c} \det(M) = \epsilon_{mn\ldots o}M^{mn\ldots o}_{ab\ldots c}$. The takeaway from \eqref{eq:transformTopForm} is that either we have to live with transforming our generic functions any time we feel a change of coordinates coming on, or we find one function that already transforms like this and use that as a base to integrate anything we want. Of course the latter is preferable, and in fact there is a very convenient function to use! Consider how the metric transforms under a change of coordinates: $g_{\mu\nu} \to J^\alpha_\mu\,g_{\alpha\beta}\,J^\beta_\nu$ (since the metric is covariant). Then using basic linear algebra, $\det(g_{\mu\nu}) := g \to \det(J)^2 g$, so $\sqrt{-g} \to \det(J) \sqrt{-g} $ (accounting for the sign of a Lorentzian metric; if in Euclidean space simply use $\sqrt{g}$ ). Then since $\det(J)\det(J^{-1}) = 1$, we find the $n$-form $\text{vol} := \sqrt{-g}\dd x^1 \wedge \ldots \wedge \dd x^n$ does all the coordinate transforming we could hope for, so in general if we want to integrate a generic function $f(x)$ on an $n$-dimensional Lorentzian manifold (as one does), we simply integrate the $n$-form:
\begin{equation}
    \label{eq:whatToIntegrate}
    f(x) \to f(x) \text{vol} = f(x) \sqrt{-g} \dd x^1 \wedge \ldots \wedge \dd x^n.
\end{equation}

\subsection{Summing it Up: Partitions of Unity}
We now have almost everything we need to define a basic tool of integration on manifolds. In any coordinate patch $U_i$ with coordinate functions $\phi_i$, we define the integral over the coordinate patch of a scalar function $f(p)$ to be:
 \begin{equation}
    \label{eq:defIntInPatch}
    \int_{U_i} f(p) \text{vol} \to \int_{\phi_i(U_i)} f(x)\, \sqrt{-g}\, dx^1 \ldots dx^n.
\end{equation}
This expression is great, but we know a manifold can often (even usually) require more than one coordinate patch to be completely described, so if we want to integrate a function over an \emph{entire} manifold, we need to figure out how to transition from one coordinate patch to another. The solution to this conundrum lies in something called a \emph{partition of unity}.

A partition of unity is a fancy mathematical term for a trick that is very familiar to physicists: writing the number $1$ in a complicated way. In statistics, for example, we write the number $1$ as $1 = \sum_i p_i$, where the $p_i$ are a bunch of probabilities, and then we use this decomposition (one might even say ``partition'') to compute averages and variances, and all those lovely things. In quantum mechanics, when one has a complete basis of states $\{\ket{i}\}$, it is common to employ the identity operator written as\footnote{Notice this is just the tensor $\hat{e}_i\otimes \hat{e}^i$, which has components $\delta^i_j$.} $\mathbb{I} = \sum_i \ket{i}\bra{i}$. And now it should be fairly clear what the path forward here is.

On a manifold with $m$ coordinate patches, we define a set of $m$ smooth functions on the manifold $\{\psi_i\}$ by the conditions that $\psi_i(p) = 0$ for all points outside the chart $U_i$, that $\psi_i(p) \ge 0$ everywhere, and that $\sum_i \psi_i(p) = 1$ everywhere. This set of functions is called a \emph{partition of unity} and outside of the constraints above, the functions are completely arbitrary. Using this glue, we can define integration over the \emph{entire} manifold as:
\begin{equation}
    \label{eq:intEntireM}
    \int_M f(p)\text{vol} := \sum_i \int_{\phi_i(U_i)} \psi_i(x)f(x)\,\sqrt{-g} \, dx^1\ldots dx^n.
\end{equation}
It can be shown (although I won't) that the integral does not depend on the choice of partition or the covering of coordinate charts. This makes sense; coordinate systems are designed to mean the same thing on the overlap of their domains, so the integral of a function will be the same in that region regardless of the coordinate system used. Then to avoid over-counting the integral in the overlap of coordinate systems, we simply have to scale down the contribution from each when we sum them all up. To be precise, the functions $\psi_i$ must be smooth and so can be quite complicated, but for the purposes of the average physicist it often suffices to make them simple box-type functions, for example just halving the contribution from two different coordinates systems on their overlap.

\subsection{Stoke's Theorem}

Just before leaving integration, we must discuss an extremely powerful tool, and possibly the most elegant equation in mathematics: Stoke's Theorem. Our discussion of one-dimensional integration was relatively straightforward, and the expression in \eqref{eq:int:oldInt} is pristine, but our attempt to generalize it to higher dimensions in \eqref{eq:int:oldInt2d} is less so. In particular, while the one-dimensional integral translates cleanly to differential form notation---$\int \dd F$ for the 0-form $F(p)$---the two-dimensional integral did not have such a clean analogy. In fact, the reason our second integral translates poorly to differential form notation is because the premise is fundamentally flawed; we tried to have a meaningful discussion about a second derivative in higher dimensions, but the manifold-y gradient is the exterior derivative, and its anti-symmetry ensures all second (and higher) applications of the operator vanish! Fortunately, it turns out that if we're less ambitious and stick to just integrating a single derivative, we stumble across an extremely powerful relation. 

I won't prove it here\footnote{Textbooks rarely do as the proof is easiest to perform in an alternative (equivalent) formulation of integration on manifolds, integration over chains. See for example \cite{spivak1965calculus}}, but one can show the generalized version of Stoke's Theorem:
\begin{equation}
    \label{eq:Stokes}
    \int_{M} \dd \omega = \int_{\partial M} \omega.
\end{equation}
Simple as can be. Here, $M$ is an $n$-dimensional manifold with boundary, $\partial M$ is the $(n-1)$-dimensional boundary of that manifold, and $\omega$ is an $n$-form on $M$. In some sense, this equation is a more universal formulation of the Fundamental Theorem of Calculus (i.e., equation \eqref{eq:int:oldInt}).

Much the same way as the Einstein Equations encode a great deal of information in a compact form, the generalized Stoke's theorem contains \emph{four} vector calculus identities, as well as their as-yet un-named counterparts in higher dimensions. Let's start with the simplest case, one dimension. When  $M$ is a one-dimensional manifold, any top form can be written $\omega = \dd\phi$ for some scalar field  $\phi(x)$. Then Stoke's theorem reads $\int_M \dd \phi = \int_{\{b,a\}}\phi(x) = \phi(b) - \phi(a)$, where the boundary of a one-dimensional manifold is the zero-dimensional manifold consisting of the endpoints (or the empty set if the one-dimension forms a loop). This is of course just the fundamental theorem of calculus! Now let's get a bit more complicated.

In three dimensions, represent a traditional vectorfield $\vec f = (f_x, f_y, f_z)$ by a co-vectorfield $f = f_x \dd x + f_y \dd y + f_z \dd z$. It is not hard to show that the traditional divergence is just the exterior derivative acting on the dual of the co-vector: $\nabla \cdot \vec f \to \dd\star f$. Then plugging into \eqref{eq:Stokes},
\begin{equation}
    \label{eq:divThm}
    \int_V \nabla \cdot \vec{f} \to \int_M d\star f = \int_{\partial M} \star f \to \int_\Sigma \vec{f}\cdot \dd\vec{\Sigma},
\end{equation}
we find the Divergence theorem of ordinary vector calculus (here $\Sigma$ is the 2-d boundary of the 3-d volume of integration, and $\dd\vec\Sigma$ is the normal area element on the surface of the boundary). 

Next, as is tradition we will look at two-dimensional problems as subsets of three-dimensional problems. Consider again the traditional vectorfield $\vec{f}$. Now it is a simple calculation to show $ \nabla \times \vec{f} \to \star \dd f$, so for a two-dimensional sub-manifold of $\mathbb{R}^3$ (with associated boundary), we find:
\begin{equation}
    \label{eq:stokesThm}
    \int_{A} (\nabla \times \vec{f})\cdot \dd^2\!\vec{A} \to \int_\Sigma \dd f = \int_{\partial \Sigma} f \to \oint \vec{f} \cdot \dd\vec{\ell},
\end{equation}
which is just the ordinary vector calculus version of Stoke's theorem (such a limited view!). 

Finally, if we restrict our attention to just the $x-y$ plane, consider the traditional vectorfield $\vec{f_g} = f_x \dd x + f_y \dd y$. Then again we have:
\begin{equation}
    \label{eq:greenThm}
    \int_{A} (\partial_x f_y - \partial_y f_x) dxdy \to \int_\Sigma \dd f_g = \int_{\partial \Sigma} f_g \to \oint \vec{f_g} \cdot \dd\vec{\ell},
\end{equation}
which is commonly referred to as Green's theorem. (It is also fairly easy to see that this falls out of \eqref{eq:stokesThm} by restriction to the $x-y$ plane). 

\vspace{1em}

Beautiful, isn't it?

\end{subappendices}

\chapter{The Einstein Field Equations and General Relativity}
\label{ch:EFEs}

Finally we have all the pieces we need to construct the general theory of relativity. We have a working understanding of tensors on general manifolds, how to take their derivatives in a tensorial way, how to describe matter on a manifold, and a local, intrinsic measure of curvature. The question now is how to piece them together. It turns out there is actually no way to \emph{derive} the fundamental equation that relates the curvature of a manifold to its matter content, that has to be a postulate\footnote{But this shouldn't come as a surprise---after all, Newton's law of gravity is also an axiom, a postulated relation between matter and its gravitational attraction.} (the last postulate) of the relativistic theory of gravity
\begin{postulate}[Einstein Field Equations]
Spacetime can be modelled as a pseudo-Riemannian manifold and the curvature of spacetime relates to its energy-momentum content through the \textbf{Einstein Field Equations}\index{Einstein Field Equations}:   
\begin{equation}
	\label{eq:ch4:defEFE}
	R_{\mu\nu} - \frac 12 R\,g_{\mu\nu} = \kappa T_{\mu\nu},
\end{equation}
where $\kappa = 8\pi G$ in terms of the Newtonian gravitational constant $G$. 
\end{postulate}

Two quick notes about \eqref{eq:ch4:defEFE} just before moving on: 1) the choice of $+\kappa$ on the RHS instead of $-\kappa$ is a \emph{convention}, just like the sign on the Ricci tensor, so be mindful of that when reading the literature, and 2) the LHS is sometimes also written with a so-called ``cosmological constant'' term $\Lambda g_{\mu\nu}$. While the cosmological constant can be thought of as a geometrical property of spacetime, it is just as easy (and probably better) to think of it as a globally constant energy density (with equation of state $w = -1$), so it really belongs in the stress-energy tensor, which is the approach we will take here (see section \ref{ch:AppsII:CM:CC}). 

There are two main ways to go about justifying this postulate, a ``bottom-up'' way (by construction) and a ``top-down'' way (through an action principle). 

\section{The EFEs By Construction}
\label{ch:EFEs:sec:direct}

The first way to come at the Einstein field equations is to think ``gravity is geometry, and matter sources gravity, so how do I get matter to source geometry?'' Obviously we only have the one way to describe matter, it's the stress-energy tensor $T_{\mu\nu}$, and fundamentally the geometry just boils down to the metric, but there are lots of important constructions that involve the metric and its derivatives, like the connection and the various curvature tensors. What we need is to narrow things down a bit, and to do so, we'll use the correspondence theorem---i.e., the principle that we should be able to recover Newtonian gravity in an appropriate limit. In particular, we recall the Newtonian Poisson equation:
\begin{equation}
	\label{eq:ch4:poisson}
	\vec{\nabla}^2 \phi = 4\pi G \rho,
\end{equation}
where $\vec{\nabla}^2$ is the old 3+0-dimensional Laplacian (the notation not being far off, as it can indeed be represented by $\nabla_i \nabla^i$), $\phi$ is the gravitational potential, and $\rho$ is the mass density. Well mass density is just non-relativistic energy density (hence the same variable), which is just $T^{00}$, and we are hypothesizing that the gravitational potential is basically the metric, so we'll try looking for an equation that has two derivatives of the metric set proportional to the stress-energy tensor. A good try would be 
\begin{equation}
	\label{eq:ch4:TryEFE}
	R_{\mu\nu} \stackrel{?}{\propto} T_{\mu\nu}.
\end{equation}
This would be nice because we know $T_{\mu\nu}$ is symmetric, and so is the Ricci tensor, so it's a good start but we're missing something important with this equation: conservation of stress-energy. Recall we had imposed $\nabla^\nu T_{\mu\nu} = 0$ (I've flipped the superscripts and subscripts, but that's okay because the connection is metric-compatible). If \eqref{eq:ch4:TryEFE} is going to be a tensor equation, then the same must be true on the other side, but $\nabla^\nu R_{\mu\nu} \neq 0$ in general. 

What we need is something like $R_{\mu\nu}$ but that satisfies its covariant gradient vanishes. As it turns out, the Bianchi identities conveniently furnish exactly the tensor we need. Take \eqref{eq:ch3:Rsymm4} and contract on the first and third indices, as we did to define the Ricci tensor:
\begin{align}
	\label{eq:ch4:derivG}
	g^{\mu\nu}\left( R_{\nu\rho[\mu\lambda;\sigma]} \right) &= g^{\mu\nu}\left( R_{\nu\rho\mu\lambda;\sigma} + R_{\nu\rho\lambda\sigma;\mu} + R_{\nu\rho\sigma\mu;\lambda} \right), \notag \\
															&= R_{\rho\lambda;\sigma} + R\indices{^\mu_{\rho\lambda\sigma;\mu}} - R_{\rho\sigma;\lambda}, \notag \\
															&= 0.
\end{align}
Metric compatibility is a beautiful thing, isn't it? Since the metric can pass through the covariant derivative willy-nilly, we can raise and lower indices inside and outside of the derivative with impunity. Equation \eqref{eq:ch4:derivG} is nice and all, but it's not quite what we're looking for yet. To finally get there, contract again, this time on $\rho$ and $\lambda$:
\begin{align}
	\label{eq:ch4:derivG2}
	g^{\rho\lambda}g^{\mu\nu}\left( R_{\nu\rho[\mu\lambda;\sigma]} \right) &= g^{\rho\lambda} \left(R_{\rho\lambda;\sigma} + R\indices{^\mu_{\rho\lambda\sigma;\mu}} - R_{\rho\sigma;\lambda}\right), \notag \\
																		   &= R_{;\sigma} - R\indices{^\mu_{\sigma;\mu}} - R\indices{^\mu_{\sigma;\mu}}, \notag \\
																		   &= \left( \delta^\mu_\sigma R - 2R\indices{^\mu_\sigma} \right)_{;\mu}, \notag \\
																		   &= 0.
\end{align}
(The second equality uses $R\indices{^\mu_{\rho\lambda\sigma;\mu}} = R\indices{_{\mu\rho\lambda\sigma}^{;\mu}}$ and the symmetries of the curvature tensor). Finally, raising an index, we find:
\begin{equation}
	\label{eq:ch4:defG}
	\left( R^{\mu\nu} - \frac 12 g^{\mu\nu}R \right)_{;\mu} =: G\indices{^{\mu\nu}_{;\mu}} = 0, 
\end{equation}
which defines the \textbf{Einstein tensor}\index{Einstein Tensor} $G_{\mu\nu} := R_{\mu\nu} - \frac 12 g_{\mu\nu}R$. The (aptly named) Einstein tensor is exactly the symmetric, conserved tensor composed only of the metric and up to two derivatives that we were looking for. GR is then the theory of physics defined by the postulate 
 \begin{equation}
	\label{eq:ch4:efeAgain}
	G_{\mu\nu} = \kappa T_{\mu\nu},
\end{equation}
for some proportionality constant $\kappa$ (which we will elaborate more on later).

\section{The EFEs from an Action Principle}
\label{ch:EFEs:sec:action}

The correspondence argument is a nice enough way of deriving the field equations, but a more modern line of thinking involves the bewilderingly impressive action formalism. In a nutshell, classical mechanics can be formulated in a way such that the physical equations of motion followed by all forms of matter boil down to a single number, the action $S$. It is written in terms of an integral of a function called the Lagrangian $L$, which calculates the total kinetic energy $T$ minus the total potential energy $V$ of a system based on the matter content of the system. The action is related to the Lagrangian by $S = \int \dd t \, L$, and the simple principle that makes physics tick is the \emph{principle of stationary action}, that the configuration of matter that is actually seen in Nature is the one that extremizes the action. Mathematically, we say the variation of the action $\delta S = S - S_0$ vanishes. 

All of the fundamental laws of physics can be arranged in a way that they are the equations of motion of some Lagrangian\footnote{As far as we know.}. The geodesic equation \eqref{eq:ch3:defGeod2}, for example, is the equation of motion associated with the action $S = \int \dd\tau \, \sqrt{-U_\mu U^\mu}$ for a particle's four-velocity. The Maxwell equations of electromagnetism derive from the action $S = -\int \dd^4 x\, \frac 14 F_{\mu\nu}F^{\mu\nu}$, where the Lagrangian is $L = \int \dd^3x\, \mathcal{L} := -\int \dd^3 x \, \frac 14 F_{\mu\nu}F^{\mu\nu}$, which defines the Lagrangian density $\mathcal{L} := -\frac 14 F_{\mu\nu}F^{\mu\nu}$. One big advantage of the Lagrangian formulation is that it lends itself readily to quantization. Through the Feynman path integral, almost any action can become a quantum theory at the drop of a hat. Much of modern particle and condensed matter physics has to do with finding the right Lagrangian to describe the quantum physics of interest.

What does all of this have to do with general relativity? Well one can play the same game and try to find a Lagrangian whose variation yields the Einstein equations as its equation of motion. The usual strategy is to think of the simplest scalar thing you can that is made of the fields you're interested in, and complexify from there, if you have to. So what's the simplest scalar thing we have in gravity? Why the Ricci scalar, of course. So we simply give the following action a go:
\begin{equation}
	\label{eq:ch4:hilbert}
	S_{H} = \int \dd^4x\, \sqrt{-g} R.
\end{equation}
This is known as the \textbf{Einstein-Hilbert action}\index{Einstein-Hilbert Action}. Here, $R$ is just the Ricci scalar. The factor of $\sqrt{-g}$ (the square root of the absolute value of the metric determinant) is there because otherwise the volume factor $\dd^4x$ is not an invariant tensor (see appendix \ref{ch:curve:app:int}). Here the Ricci scalar encodes all of the purely gravitational physics; if we wanted to include other physics (such as some matter content) we would simply add another action $S_M = \int \dd^4x \mathcal{L}_M$. As it turns out, variation of this total action with respect to changes in the metric yields \emph{both} the Einstein field equations \emph{and} a general definition of the stress-energy tensor in terms of the Lagrangian density of matter. 

We can do the variation in pieces (we'll follow Carroll \cite[\S 4.3]{carroll_2019} quite closely). First, it's helpful to note that there is an easy relation between the variation of the metric $\delta g_{\mu\nu}$ and the variation of its inverse, $\delta g^{\mu\nu}$. Simply note that the identity matrix is an invariant, and write $\delta(\delta^\mu_\nu) = \delta(g^{\mu\sigma}g_{\sigma\nu}) = \delta g^{\mu\sigma} g_{\sigma\nu} + g^{\mu\sigma}\delta g_{\sigma\nu} = 0 \implies \delta g_{\rho\nu} = -g_{\mu\rho}g_{\sigma\nu}\delta g^{\mu\sigma}$. Then a stationary point of the action with respect to the metric will also be a stationary point with respect to the metric inverse, and it turns out to be a little more convenient to use the latter, so let's compute $\frac{\delta S_H}{\delta g^{\mu\nu}}$ (just the gravitational part first). 

A variation is much like a Taylor series; the goal is to take $g^{\mu\nu} \to g^{\mu\nu} + \delta g^{\mu\nu}$, calculate the result to first order in $\delta g^{\mu\nu}$, and rearrange terms until we have an expression of the form $\{\ldots\}\delta g^{\mu\nu}$. Once that's done, we can insist the term whole variation must vanish for \emph{any} variation $\delta g^{\mu\nu}$, and so the term in braces must vanish. Piecewise then, consider: 
\begin{align}
    \label{eq:calcVar}
    \delta S &= \int \dd^4 x \, \delta\left( \sqrt{-g} g^{\mu\nu}R_{\mu\nu} \right), \\ \notag
             &= \int \dd^4 x \, \left\{\delta\left( \sqrt{-g}\right) g^{\mu\nu}R_{\mu\nu} + \sqrt{-g} \,\delta \left(g^{\mu\nu}\right)R_{\mu\nu} + \sqrt{-g} g^{\mu\nu}\,\delta \left(R_{\mu\nu}\right)\right\}, \\ \notag
        &=: \delta S_1 + \delta S_2 + \delta S_3.
\end{align}
Evidently, $\delta S_2$ is already in precisely the correct form: $\delta S_2 = (\sqrt{-g} R_{\mu\nu}) \delta g^{\mu\nu}$. One down, two to go!

Next, we'll quickly tackle $\delta S_3$. Evaluating this term is a little more sneaky than it is elucidating, so we'll skim it. The key here is to use an old matrix identity:  $\ln(\det M) = \Tr(\ln M)$ (you might have seen this at some point in quantum mechanics, or statistical mechanics). Varying $M$ is akin to taking its derivative, which is a linear process that commutes with the trace, so generally we have $(\det M)^{-1}\delta(\det M) = \Tr(M^{-1}\delta M) \to \delta(\det M) = (\det M)\Tr(M^{-1}\delta M) $. Now specializing to the case that $\det M = g$, we have $\delta g = g g^{\mu\nu}\delta g_{\mu\nu}$ or using our trick from above, $\delta g = -g g_{\mu\nu}\delta g^{\mu\nu}$. Finally, we have:
\begin{align}
    \label{eq:compVar1}
    \delta\sqrt{-g} &= \frac{-1}{2\sqrt{-g}}\delta g \notag \\
                    &= \frac{g}{2\sqrt{-g}}g_{\mu\nu}\delta g^{\mu\nu}, \notag \\
                    &= -\frac 12\sqrt{-g}g_{\mu\nu}\delta g^{\mu\nu}, \notag \\
\end{align}
and so
\begin{equation}
    \delta S_1 = -\frac 12 \sqrt{-g} g^{\alpha\beta}R_{\alpha\beta} g_{\mu\nu}\delta g^{\mu\nu}. 
\end{equation}

Finally it remains to calculate $\delta S_3$. This term takes a little more effort to crunch through; we're effectively looking for the variation of the full Riemann tensor $\delta R\indices{^\mu_\nu_\rho_\sigma}$, and then contracting two of the indices of the variation. Since the curvature tensor is built up of Christoffel symbols, the most direct procedure is to compute the variation of $\Gamma^{\mu}_{\nu\rho}$ and chug through the tedious algebra. Once again though, there is a handy-dandy trick we can employ to greatly simplify the problem: recall that even though connection coefficients are not themselves tensors, the difference between two choices of connection coefficients \emph{is} a tensor ($\widehat\Gamma - \Gamma$ is a tensor). Well, the variation of the Christoffel symbols $\delta \Gamma$ is the difference $\Gamma(g + \delta g) - \Gamma(g)$, and Christoffel symbols evaluated at different metrics constitute \emph{different} connections, so their difference must be a tensor. Why is this relevant? Because if $\delta  \Gamma$ is a tensor, then we are well within our rights to compute its covariant derivative, and if we do that, then we are stumble upon a wonderful coincidence. So first compute:
\begin{align}
    \label{eq:compDelGamma}
    \nabla_\lambda(\delta \Gamma^{\rho}_{\nu\mu}) &= \partial_\lambda\left( \delta\Gamma^\rho_{\nu\mu} \right)  + \Gamma^\rho_{\lambda\sigma}\delta\Gamma^\sigma_{\nu\mu} - \Gamma^\sigma_{\lambda\nu}\delta\Gamma^\rho_{\sigma\mu} - \Gamma^\sigma_{\lambda\mu}\delta\Gamma^\rho_{\nu\sigma}.
\end{align}
Then anti-symmetrizing on $\nu$ and $\lambda$, $\nabla_\lambda(\delta \Gamma^\rho_{\nu\mu}) - \nabla_\nu(\delta\Gamma^\rho_{\lambda\mu})$, we find:
\begin{align}
    \label{eq:compDiffDelGamma}
    \nabla_\lambda(\delta \Gamma^{\rho}_{\nu\mu}) - \nabla_\nu(\delta \Gamma^{\rho}_{\lambda\mu}) &= \partial_\lambda\left( \delta\Gamma^\rho_{\nu\mu} \right)  + \Gamma^\rho_{\lambda\sigma}\delta\Gamma^\sigma_{\nu\mu} - \Gamma^\sigma_{\lambda\nu}\delta\Gamma^\rho_{\sigma\mu} - \Gamma^\sigma_{\lambda\mu}\delta\Gamma^\rho_{\nu\sigma} \notag \\
                  &\quad - \partial_\nu\left( \delta\Gamma^\rho_{\lambda\mu} \right)  - \Gamma^\rho_{\nu\sigma}\delta\Gamma^\sigma_{\lambda\mu} + \Gamma^\sigma_{\nu\lambda}\delta\Gamma^\rho_{\sigma\mu} + \Gamma^\sigma_{\nu\mu}\delta\Gamma^\rho_{\lambda\sigma}, \notag \\ 
                  &= \partial_\lambda\left(\delta\Gamma^\rho_{\nu\mu}\right) - \partial_\nu\left( \delta\Gamma^\rho_{\lambda\mu} \right) + \Gamma^\rho_{\lambda\sigma}\delta\Gamma^\sigma_{\nu\mu} + \Gamma^\sigma_{\nu\mu}\delta\Gamma^\rho_{\lambda\sigma}, \notag \\
                  &\quad - \Gamma^\sigma_{\lambda\mu}\delta\Gamma^\rho_{\nu\sigma} - \Gamma^\rho_{\nu\sigma}\delta\Gamma^\sigma_{\lambda\mu}, \notag \\
                  &= \partial_\lambda\left(\delta\Gamma^\rho_{\nu\mu}\right) - \partial_\nu\left( \delta\Gamma^\rho_{\lambda\mu} \right) + \delta\left(\Gamma^\rho_{\lambda\sigma}\Gamma^\sigma_{\nu\mu}\right) - \delta\left( \Gamma^\sigma_{\lambda\mu}\Gamma^\rho_{\nu\sigma} \right), \notag \\ 
                  &= \delta\left( R\indices{^\rho_\mu_\lambda_\nu} \right) 
\end{align}
Now simply contract $\rho$ and $\lambda$ and we have our variation! What's that? We still don't have anything remotely resembling $\{\ldots\}\delta g^{\mu\nu}$? No matter, we have something much better: a total derivative. Recall that this entire procedure is taking place under the auspices of an integral over all of spacetime, so any quantity that is a total derivative can be immediately integrated to yield the argument evaluated on the boundary. However, something I neglected to mention earlier is that when varying quantities in an action, it is customary to assume the variations all vanish at the boundaries of the problem (imagine trying to find the most efficient path between two points by trial-and-error: you might try any of a thousand different paths, but you'll always choose test paths that begin and end at the same points since that's a non-negotiable property of the path you're searching for). Given that we now have an expression composed entirely of functions of $\delta g^{\mu\nu}$ and we are to evaluate that expression at $\delta g^{\mu\nu} = 0$, we are very happy to be left with the statement $\delta S_3 = 0$. 

All that remains is to directly compute:
\begin{align}
    \label{eq:compVarSTot}
    \delta S &= \delta S_1 + \delta S_2, \notag \\
             &= \int \dd^4 x \left\{ -\frac 12 \sqrt{-g} R\, g_{\mu\nu} + \sqrt{-g} R_{\mu\nu} \right\}\delta g^{\mu\nu}. 
\end{align}
Finally, requiring the action to be stationary under variations of the metric amounts to demanding the contents of the braces in \eqref{eq:compVarSTot} vanish. Cancelling through the common factors, this returns precisely the vacuum Einstein field equations:
\begin{equation}
    \label{eq:varVacEFE}
    R_{\mu\nu} - \frac 12 R g_{\mu\nu} = 0. 
\end{equation}

Clever, eh? While it's a neat party trick to derive the vacuum equations, you could be forgiven for complaining that general relativity only really picks up when we have matter to play with. However, this formalism is more than a fun way to brag to your friends about your skills in variational calculus, it is also the single best tool to use to define a \emph{general} stress-energy tensor. Recall in section \ref{ch:geomRel:stressEnergy} we hacked together some ad-hoc\footnote{Ad-hack?} descriptions of matter-energy into some definitions of specific stress-energy tensors? Well, consider what happens when we add another piece to the action describing solely non-gravitational physics:
\begin{equation}
    \label{eq:totalAction}
    S = S_H + S_M,
\end{equation}
where now we conveniently re-scale\footnote{Go ahead, try and stop me.} $S_H = \frac{1}{2\kappa}\int \dd^4x \sqrt{-g} R$, then the variation of the total action yields:
\begin{align}
    \label{eq:varTotAction}
    \frac{\delta S}{\delta g^{\mu\nu}} = -\frac{1}{4\kappa} \sqrt{-g} R\, g_{\mu\nu} + \frac{1}{2\kappa}\sqrt{-g} R_{\mu\nu} + \frac{\delta S_M}{\delta g_{\mu\nu}} = 0,
\end{align}
or, with a little rearranging,
\begin{equation}
    \label{eq:varEFE}
    R_{\mu\nu} - \frac12 R\, g_{\mu\nu} = -\frac{2\kappa}{\sqrt{-g}}\frac{\delta S_M}{\delta g^{\mu\nu}}.
\end{equation}
In other words, for any non-gravitational matter-energy that can be described by an action, we simply \emph{define} the stress-energy tensor associated with it to be:
\begin{equation}
    \label{eq:goodDefStressEnergy}
    T_{\mu\nu} := -\frac{2}{\sqrt{-g}} \frac{\delta S_M}{\delta g^{\mu\nu}}.
\end{equation}
Excellent.

\section{The Newtonian Limit}
\label{ch:EFEs:newt}

Finally, we cannot construct a whole new theory to replace an old one without first checking in with the Correspondence theorem. The Correspondence theorem is a semi-formal way of saying we require any theory that supplants an old one to reduce to the old one under the conditions under which the old one was a functional description of Nature---for general relativity, this means we must find a way to recover Newton's theory of gravity under the right circumstances. 

Formally, we would expect the Newtonian approximation to general relativity to look like a small perturbation to the Minkowski metric (the Minkowski metric being non-gravitating spacetime): $g_{\mu\nu} \approx \eta_{\mu\nu} + h_{\mu\nu}$. In fact, we will explore general perturbations of this form later when we study the source-free Einstein field equations, but for the Newtonian approximation we can simplify further with some intuition. First, we'd better not have any mixing between the time and space dimensions---that would be entirely out of place in Newton's theories---so we can immediately set $h_{0i} = h_{i0} = 0$. Next, the shape of space in Newton's theory is plainly Euclidean, so we'd also better have $h_{ij} \propto \delta_{ij}$. Finally, in Newtonian mechanics temporal gradients better be much smaller than spatial gradients (a system ill-suited to the beautiful natural units), so we'll require the perturbation to be static, $\partial_t h = 0$. This essentially leaves us with one single degree of freedom, which we will knowingly write: $h_{\mu\nu} = -2\phi\delta_{\mu\nu}$ (note that this does include a contribution to the time-time component). Hence, we take as our Newtonian ansatz:
\begin{equation}
    \label{eq:ch4:newt:guessNewton}
    ds^2 \approx -(1 + 2\phi)\dd t^2 + (1 - 2\phi)\left( \dd x^2 + \dd y^2 + \dd z^2 \right).
\end{equation}
With this guess, we simply plug-n-chug into the fundamental equations of general relativity, judiciously apply some Taylor series', and watch Einstein's hard work pay off.

First up, consider the geodesic equation. Before doing any calculations, let's list the Christoffel symbols, taking them just to first order in $\phi$:
\begin{align}
    \label{eq:ch4:newt:newtGammas}
    \Gamma^0_{00} \sim 0 \quad && \Gamma^i_{00} \sim \delta^{ij}\partial_j \phi \quad && \Gamma^0_{0j} \sim  \partial_j \phi \notag \\
    \Gamma^0_{ij} \sim 0 \quad && \Gamma^i_{0j} \sim 0 \quad && \Gamma^i_{ij} \sim -\partial_j \phi \quad && \Gamma^i_{jj} \sim \partial_i \phi \quad (i \neq j)
\end{align}
(note that there are no sums here, $\Gamma^i_{ij}$ simply refers to the component with $i$ in the top and bottom). With those in hand, let's first calculate the time-component of the geodesic equation (and re-scale the equation by $m$ ):
\begin{align}
    \label{eq:ch4:newt:timeGeodNewt}
    m\frac{\dd p^0}{\dd \tau} + \Gamma^0_{\alpha\beta}p^\alpha p^\beta &= 0, \notag \\
    \implies \quad m\frac{\dd p^0}{\dd \tau} + \Gamma^0_{00}p^0 p^0 &= 0, \notag \\
    \implies \quad \frac{\dd p^0}{\dd t} = 0&,
\end{align}
which uses first that any Newtonian particle is non-relativistic so $p^i \ll p^0$, second that $p^0 \sim m$, and third that $t(\tau) \sim \tau$. Equation \eqref{eq:ch4:newt:timeGeodNewt} is simply Newtonian energy conservation. Good to have, for sure, but we're looking for some more iconic equations. 

Next, we need the spatial components of the geodesic equation:
\begin{align}
    \label{eq:ch4:newt:spaceGeodNewt}
    m\frac{\dd p^i}{\dd\tau} + \Gamma^i_{\alpha\beta}p^\alpha p^\beta &= 0, \notag \\
    \implies \quad m\frac{\dd p^i}{\dd\tau} + \Gamma^i_{00}p^0p^0 &\sim 0, \notag \\
    \implies \quad \frac{\dd p^i}{\dd t} \sim -m \partial_i \phi&.
\end{align}
Much more interesting---equation \eqref{eq:ch4:newt:spaceGeodNewt} is Newton's second law for a particle influenced by a gravitational potential. This provides much stronger evidence that we can identify $\phi$ with the Newtonian gravitational potential, but we've got one more check to really clinch it, the Einstein equations.

While we could just plug in the Christoffels and calculate everything directly, we need to start at the Riemann curvature tensor, and that's a rank-4 beast, so it would be better if we could find an easier way to calculate it in the approximations above. Fortunately, a direct calculation turns out to be surprisingly easy given that $R\indices{^\mu_\nu_\rho_\sigma}$ is composed of a lot of products of the metric, but we are only taking things to first order in $\phi$. First, let's approximate the Christoffel symbols:
\begin{align}
    \label{eq:ch4:newt:newtApproxGamma}
    \Gamma^\mu_{\nu\rho} &= -\frac12 g^{\mu\alpha} \left( g_{\nu\rho,\alpha} - g_{\nu\alpha,\rho} - g_{\rho\alpha,\nu}\right), \notag \\
                         &\sim (\eta^{\mu\alpha} + 2\phi\delta^{\mu\alpha})\left( \phi_{,\alpha}\delta_{\nu\rho} - \phi_{,\rho}\delta_{\nu\alpha} - \phi_{,\nu}\delta_{\rho\alpha} \right), \notag \\
                         &\sim \eta^{\mu\alpha}\left( \phi_{,\alpha}\delta_{\nu\rho} - \phi_{,\rho}\delta_{\nu\alpha} - \phi_{,\nu}\delta_{\rho\alpha}  \right), 
\end{align}
and then quite easily the derivative is:
\begin{equation}
    \label{eq:ch4:newt:newtApproxDGamma}
    \partial_\lambda \Gamma^\mu_{\nu\rho} \sim \eta^{\mu\alpha}\left( \phi_{,\alpha\lambda}\delta_{\nu\rho} - \phi_{,\rho\lambda}\delta_{\nu\alpha} - \phi_{,\nu\lambda}\delta_{\rho\alpha}  \right). 
\end{equation}
Now we can see why this direct approach turns out to be fairly easy: $\Gamma$ is already of order $\phi$, so the $\Gamma\Gamma$ terms in the curvature tensor simply vanish at leading order! Thus we are left with:
\begin{align}
    \label{eq:ch4:newt:newtApproxCurv}
R\indices{^\mu_\rho_\lambda_\nu} &\sim \partial_\lambda \Gamma^\mu_{\nu\rho} - \partial_\nu \Gamma^\mu_{\lambda\rho}, \notag \\
                                     &\sim \eta^{\mu\alpha}\left( \phi_{,\alpha\lambda}\delta_{\nu\rho} - \phi_{,\alpha\nu}\delta_{\lambda\rho} - \phi_{,\rho\lambda}\delta_{\nu\alpha} + \phi_{,\rho\nu}\delta_{\lambda\alpha} \right). 
\end{align}

And that's all there is to the full Riemann curvature tensor. Next up: contract a couple of indices to find the Ricci tensor:
\begin{align}
    \label{eq:ch4:newt:newtApproxRCurve}
    R_{\rho\nu} &\sim \eta^{\mu\alpha} \left( \phi_{,\alpha\mu}\delta_{\nu\rho} - \phi_{,\alpha\nu}\delta_{\mu\rho} - \phi_{,\rho\mu}\delta_{\nu\alpha} + \phi_{,\rho\nu}\delta_{\mu\alpha} \right), \notag \\
                &\sim  \delta_{\rho\nu}\partial^\alpha\partial_\alpha \phi - \eta^{\mu\alpha}\delta_{\mu\rho}\partial_\alpha\partial_\nu\phi - \eta^{\mu\alpha}\delta_{\nu\alpha}\partial_\mu\partial_\rho\phi + 2\partial_\rho\partial_\nu\phi, \notag \\ 
&\sim  \delta_{\rho\nu} \Box \phi - \eta^{\mu\alpha}\delta_{\mu\rho}\partial_\alpha\partial_\nu\phi - \eta^{\mu\alpha}\delta_{\nu\alpha}\partial_\mu\partial_\rho\phi + 2\partial_\rho\partial_\nu\phi. 
\end{align}
(This defines the d'Alembertian operator $\Box := \partial_\mu \partial^\mu$). 
\begin{aside}[Kronecker Tensor]
    We were able to sneak by without confronting this, but it is important to pay attention to the symbol $\delta_{\mu\nu}$. We've been casually throwing this around as if it were a tensor, and while it is a tensor, it is also a little notationally dangerous. For instance, the contraction $\eta^{\mu\nu}\delta_{\nu\rho}$ should according to the rules of contractions be $\delta\indices{^\mu_\rho}$. However, component-wise this symbol is numerically the same as the components of $\eta$, so maybe we should define it as $\eta\indices{^\mu_\rho}$ just as a special case? Absolutely not! The symbol $\eta\indices{^\mu_\rho} = \eta^{\mu\nu}\eta_{\nu\rho} = \delta^\mu_\rho$ has the same components as the Kronecker delta. In such a sticky situation, it's best to leave things as explicit as possible and leave both the metric and Kronecker delta as they were.
\end{aside}
Before proceeding further, it will be beneficial to use the static property of $\phi$ to clean up the expression \eqref{eq:ch4:newt:newtApproxRCurve} a little bit. First, notice that if either of $\rho$ or $\nu$ are 0, all terms but the first involve $\partial_0 \phi$ and so vanish. Next, scrutinizing the middle two terms, for purely spatial $\rho$ and $\nu$, the expressions $\eta^{\mu\alpha}\delta_{\mu\rho}$ and $\eta^{\mu\alpha}\delta_{\nu\alpha}$ evaluate numerically to $\delta^\alpha_\rho$ and $\delta^\mu_\nu$, and since the partial derivatives commute, these two terms sum to exactly $-2\partial_\rho\partial_\nu \phi$ and so cancel the last term. Finally, without temporal derivatives, the d'Alembertian simply reduces to the spatial Laplacian $\Box \to \vec\nabla^2$, so altogether the Ricci tensor reduces to:
\begin{equation}
    \label{eq:ch4:newt:newtApproxRCurve2}
    R_{\rho\nu} \sim \delta_{\rho\nu} \vec\nabla^2\phi.
\end{equation}

All that's left to compute for the Einstein tensor is now the Ricci scalar, which is profoundly simple to compute from \eqref{eq:ch4:newt:newtApproxRCurve2}:
\begin{align}
    \label{eq:ch4:newt:newtApproxR}
    R = g^{\nu\rho}R_{\rho\nu} &\sim \eta^{\nu\rho}R_{\rho\nu}, \notag \\
                               &\sim  \eta^{\nu\rho}\delta_{\rho\nu} \vec\nabla^2 \phi, \notag \\
                               &\sim  2 \vec\nabla^2 \phi.
\end{align}
Putting it all together, we have the approximation to the Einstein Field Equations:
\begin{equation}
    \label{eq:ch4:newt:newtApproxEFE}
    R_{\mu\nu} - \frac12 g_{\mu\nu} R \sim (\delta_{\mu\nu} - \eta_{\mu\nu})\Box\phi = \kappa T_{\mu\nu}. 
\end{equation}
Now the expression in parentheses $\delta_{\mu\nu} - \eta_{\mu\nu}$ is easy to evaluate, it is simply $2$ for the 00-component, and 0 otherwise, so in the Newtonian limit, the LHS of the Einstein equations is only non-trivial for the time-time component, which suggests the only component of the stress-energy tensor that is non-trivial for Newtonian physics is also the time-time component. This of course makes perfect sense; $T^{00} = \rho$ is the energy density, which for non-relativistic matter is strongly dominated by the rest mass $m$ of the system, and which also dominates all other stresses and pressures. Thus the only non-trivial equation to survive from the Einstein Field Equations is:
\begin{equation}
    \label{eq:ch4:newt:newtApproxEFE00}
    2\vec{\nabla}^2\phi = \kappa\rho,
\end{equation}
describing the response of the Newtonian gravitational field to the presence of Newtonian mass density. Comparing \eqref{eq:ch4:newt:newtApproxEFE00} to the Poisson equation for Newtonian gravity that was already known to describe the same thing:
\begin{equation}
    \label{eq:ch4:newt:newtPoisson}
    \vec{\nabla}^2\phi = 4\pi G\rho_m,
\end{equation}
and using that the non-relativistic limit of the full relativistic energy density is simply the mass density $\rho \sim \rho_m$, we may identify $\kappa = 8\pi G$.

\section{Recap}

\begin{itemize}
    \item General Relativity is born by postulating a relation between curvature and matter. This relation takes the form of the Einstein Field Equations, \eqref{eq:ch4:defEFE}.
    \item While the Einstein Field Equations cannot be derived, they certainly can be justified. Two good justifications are 1) finding the simplest rank-2 tensor constructed entirely of the Riemann curvature tensor that is also conserved, and 2) varying the simplest action that can be constructed from the Riemann curvature tensor.
    \item The Newtonian limit is achieved by trying as an ansatz for the metric the perturbation $g_{\mu\nu} \sim \eta_{\mu\nu} - 2\phi\delta_{\mu\nu}$, which upon substitution into the geodesic equation yields energy conservation and Newton's second law in a gravitational potential, and upon substitution into the Einstein Field Equations yields the Poisson equation for Newtonian gravity.
    \item The Newtonian approximation allows an identification of the gravitational coupling $\kappa$ with the Newtonian gravitational constant $G$ as $\kappa = 8\pi G$.
\end{itemize}

\begin{subappendices}

\section{Energy Conditions}
\label{ch:EFEs:app:energy}

Without any other input, the Einstein Field Equations are not more than a re-labelling of a specific combination of geometric tensors: $R_{\mu\nu} - \frac 12 g_{\mu\nu}R \to \kappa T_{\mu\nu}$. Identifying $T_{\mu\nu}$ as a wholly separate entity representing the matter-energy content of the universe is what turns the field equations from a triviality to one of the foundational postulates of a physical theory. When one has a physically-motivated stress-energy tensor ready to go the process is automatic. This will always be the case for us in this work, but some very theoretically-minded theoretical physicists like to try to understand at the most abstract level the physical content of theories like general relativity, and for them it is helpful to wonder what it means for a general stress-energy tensor to be physically sensible. To that end, a number of \textbf{energy conditions}\index{Energy Conditions} have been proposed to be added to the theory to put physical constraints of possible stress-energy tensors in the way that direct observations are used in addition to Newton's theory to require mass be positive. Here we will briefly describe the three most popular conditions, but for a thorough treatment of the topic see \cite{curiel_primer_2017}.

The first and probably most common condition is the \emph{null energy condition} which requires $T_{\mu\nu}k^\mu k^\nu \ge 0$ for any null vectorfield $k^\mu$. This can be thought of as saying lightlike things only feel positive stress-energy---a very weak generalization of the positive mass postulate indeed. Interestingly, this condition can be shown to have as a consequence the very natural notion that gravity is seen to be attractive acting on lightlike particles (this is done with a more detailed calculation of the sort we used to derive the Riemann curvature tensor, looking at more general distributions of geodesics and the different ways their shapes are distorted by local curvature). 

Next is the \emph{weak energy condition}, defined by $T_{\mu\nu}\xi^\mu \xi^\nu \ge 0$ for any timelike vectorfield $\xi^\mu$. This condition is a bit more personal to us; recall that timelike vectors represent the four-velocities of massive bodies and we can always find coordinates such that at any point in spacetime, any given timelike vectorfield only has support in the timelike direction (i.e., such that $\xi^i = 0$). That means the weak energy condition can always be phrased as $T^{00} \ge 0$, which is quite directly the statement that energy density is positive. The weak energy condition also actually implies\footnote{This can be seen by a limiting procedure. If we can always locally write a timelike vector as aligned with the time axis, we can also always locally Lorentz boost it. The weak energy condition is locally Lorentz invariant so if it holds for locally ``stationary'' timelike vectors, then it holds for arbitrarily strongly boosted timelike vectors. Take the limit that the boost approaches the speed of light and you arrive at the null energy condition.} the null energy condition, so it is widely considered one of the most reasonable things to expect from a physically meaningful stress-energy tensor.

Finally, one also often hears tell of the \emph{strong energy condition}, which is applied to the trace-reversed stress-energy tensor, $(T_{\mu\nu} - \frac 12 g_{\mu\nu}T)\xi^\mu \xi^\nu \ge 0$ with $\xi^\mu$ again a timelike vectorfield. This is motivated by extending the ``attractive gravity'' outcome of the null energy condition to all causal observers (i.e.~also to timelike observers). Local geodesic expansion/contraction is related directly to the Ricci tensor (just as local geodesic deviation is related directly to the Riemann curvature tensor), and from the Einstein field equations the Ricci tensor is given by the trace-reversed stress-energy tensor, which is why that is the quantity addressed by this condition. Note that this condition also implies the null energy condition, but does not imply the weak energy condition. 

All of these conditions can also be phrased in an averaged way to appease the statistally-minded, or the quantumly-minded, or just the cautiously-minded critics. At the end of the day though, these are just guidelines based on physical intuition, and even the weakest condition above (the null energy condition) can be found wanting for otherwise non-pathological matter-energy distributions \cite{rubakov_null_2014}. Keep this in mind when reading about theorems that make use of these assumptions.

\end{subappendices}

\chapter{Applications Part I: Source-Free Solutions}
\label{ch:AppsI}

At long last, we can finally employ all of these tools we have derived towards the study of actual physics. In this short course we will take a quick pass over some of the most important physical applications of general relativity. We'll follow a path of increasingly complicated matter content in a way that exactly parallels a path of increasingly complicated charge distributions in electromagnetism. We begin with the free vacuum equation; in electromagnetism, the source-free Maxwell equations give rise to electromagnetic waves, and in exactly the same way, the source-free Einstein field equations give rise to gravitational waves. Next, the Maxwell equations sourced by a point-charge distribution give rise to the ubiquitous Coulomb force, and in exactly the same way, the Einstein field equations sourced by a point-mass give rise to the Schwarzschild metric. Finally, the electromagnetic field in media is first studied in the simplest example of homogeneous and isotropic dielectric materials, and so too is the cosmology of the universe studied by modelling all matter, radiation, and dark energy in the universe by a simple perfect fluid (with varying choices for the equation of state). 

In this chapter, we consider the source-free gravitational wave solutions, and as a lead-up to sources, we also consider the point-of-view of a uniformly accelerating observer. This latter discussion is somewhat unphysical in that it does not describe an observer being accelerated by any real force (and so is just a somewhat distorted view of Minkowski space), but furnishes an excellent example of adapting coordinates to suit the situation, and as we will see later, is highly relevant for realistic physics.

\section{Source-Free EFEs: Gravitational Waves}
\label{ch:AppsI:sec:GW}

As promised, we begin to explore the consequences of Einstein's field equations in a simple vacuum, $T_{\mu\nu} = 0$. Unfortunately---and in a major deviation from electromagnetism---the Einstein field equations in a vacuum are still highly complex and non-linear, unlike Maxwell's equations. In fact, almost all of our study of black holes in \ref{ch:AppsII:BH} will take place where the stress-energy vanishes as well. So rather than try to unpack the entire solution space of the source-free field equations, we'll instead start by looking at the \textbf{weak field} solutions, i.e., solutions to the \emph{linearized} field equations. We touched on these above in \ref{ch:EFEs:newt}, but there we were very restrictive, taking the weak field and non-relativistic limits with the aim of reconstructing Newton's law of gravity. Here we will be more open to new physics, while still only looking for small deviations from the Minkowski metric. This is more useful than it sounds; in many cases we are interested in gravitational dynamics that are localized in spacetime, and so expect the geometry far away from the fun to be approximately flat, and only slightly perturbed. And even in situations where the large-scale asymptotic background isn't Minkowski (in a de Sitter universe, for example), life will always look flat on small enough scales, and on those scales perturbations from the rest of the universe will still fit the linearized Einstein field equations.

So we begin again with a metric $g_{\mu\nu} \sim \eta_{\mu\nu} + h_{\mu\nu}$, where the perturbation $h_{\mu\nu}$ is still required to be small, $ \abs{h_{\mu\nu}} \ll 1$, but is otherwise unconstrained. Only keeping terms to linear order in the perturbation implies all indices can be raised and lowered with impunity using only the Minkowski metric, and calculation of the Christoffel symbols is straightforward:
\begin{equation}
    \label{eq:ch5:GW:christofs}
    \Gamma^\mu_{\nu\rho} \approx -\frac 12 \eta^{\mu\alpha}\left( h_{\nu\rho,\alpha} - h_{\nu\alpha,\rho} - h_{\rho\alpha,\nu} \right),
\end{equation}
since $\eta$ is constant. Similarly, the Ricci curvature tensor is easily computed:
\begin{align}
    \label{eq:ch5:GW:Ricci1}
    R_{\rho\nu} = R\indices{^\lambda_\rho_\lambda_\nu} &\sim \partial_\lambda \Gamma^\lambda_{\nu\rho} - \partial_\nu \Gamma^\lambda_{\lambda\rho}, \notag \\
                                     &= -\frac 12 \eta^{\lambda \alpha}\left( h_{\nu\rho, \alpha \lambda} - h_{\nu\alpha, \rho \lambda} - h_{\lambda\rho, \alpha \nu} + h_{\lambda\alpha, \rho\nu} \right), \notag \\
                                     &= -\frac 12 \left( \Box_0 h_{\nu\rho} - h\indices{^\lambda_\nu_,_\rho_\lambda} - h\indices{^\alpha_\rho_,_\alpha_\nu} + h_{,\rho\nu} \right), 
\end{align}
using that $\Gamma^2 \sim \mathcal{O}(h^2)$, and defining $\Box_0 := \eta^{\alpha\beta}\partial_\alpha\partial_\beta$ and $h := h\indices{^\alpha_\alpha}$. Rather than proceed to compute the curvature scalar and chug through some algebra, it is more convenient to re-write Einstein's field equations in the so-called ``trace-reversed'' form: first, contract both sides of the Einstein equations \eqref{eq:ch4:defEFE} with the metric and take the trace to find $R - 2R = -R = \kappa T$ (which defines the trace of the stress-energy tensor as  $T := g^{\mu\nu}T_{\mu\nu}$). Then substitute this expression for the curvature tensor back into the original field equations, and rearrange to find:
\begin{equation}
    \label{eq:ch5:GW:EFEtraceRev}
    R_{\mu\nu} = \kappa \left(T_{\mu\nu} - \frac 12 g_{\mu\nu} T\right). \qquad \text{(Trace-Reversed EFEs)}.
\end{equation}
Then for a source-free configuration, it is sufficient to solve 
\begin{equation}
    \label{eq:ch5:GW:TraceRevSourceFree}
    R_{\mu\nu} = 0,
\end{equation}
or in our case,
\begin{equation}
    \label{eq:ch5:GW:undetermWeakField}
    \Box_0 h_{\nu\rho} - h\indices{^\alpha_\nu_,_\rho_\alpha} - h\indices{^\alpha_\rho_,_\nu_\alpha} + h_{,\rho\nu} = 0. 
\end{equation}
So far so good, but we can make our lives easier still by counting parameters.

\subsection{Gauge-Fixing the Linearized EFEs}

At face value, the Einstein field equations are perfectly well-determined: the $\frac 12 n(n+1)$ degrees of freedom in the symmetric rank-2 tensor $g_{\mu\nu}$ should be solvable in terms of the $\frac 12 n(n+1)$ unique components of the symmetric rank-2 stress-energy tensor $T_{\mu\nu}$. Unfortunately, the last part of that sentence was actually a lie, there are only $\frac 12 n(n-1)$ independent components of the stress-energy tensor on account of the law of its conservation, $ \nabla_\nu T^{\mu\nu} = 0$ ($n$ independent constraints). On the left-hand side of the field equations this fact was encoded in conservation of the Einstein tensor $\nabla_\nu G^{\mu\nu} = 0$ ensured by the Bianchi identities. Fundamentally, this is deeply tied to the same issue in electrodynamics---written in terms of the scalar and vector potentials, Maxwell's equations appear at face value to be four equations for four unknowns, but conservation of electromagnetic four-current makes one of those equations redundant. At the end of the day, this leaves a degree of freedom (a ``gauge symmetry'') in the solution to the electromagnetic potentials which can (and should) always be exploited to make life simpler. In electromagnetism, this symmetry speaks to a freedom to choose coordinates in an ``internal'' space traversed by all electromagnetically charged creatures; in general relativity the analogous symmetry is similarly associated with the freedom to choose the coordinates of spacetime itself (or more concretely, the freedom to choose the tangent vector bases).

All of this is true for the metric and Einstein field equations in general, but the consequences for our current calculation are that we are free to impose a set of four external conditions on the metric perturbation, and just as in the case of Maxwell's theory, we should be inclined to choose constraints that make life as simple as possible. In principle, the form of the gauge transformation we can employ derives from the Bianchi identity, but in this case we can see it more directly from our requirement that the metric perturbation remain small. 

We have written our metric as ``close to'' the Minkowski metric $g_{\mu\nu} \sim \eta_{\mu\nu} + h_{\mu\nu}$ in some particular coordinate system $x^\mu$. Of course we're always free to change coordinates, but in order to do so while maintaining the same ``closeness'' to the Minkowski metric, the inverse Jacobian of the transformation must take the form $(J^{-1})^\alpha_\beta \sim \delta^\alpha_\beta + \partial \xi^\alpha/\partial x^\beta + \ldots$ for some small functions $\xi^\alpha$. Under this coordinate transformation, and to leading order in the small quantities, we have the inverse metric:
\begin{equation}
    \label{eq:ch5:GW:deriveGaugeGW}
(g^\prime)^{\mu\nu} \sim  \left(  \delta_\alpha^{\mu} + \frac{\partial \xi^\mu}{\partial x^\alpha} \right) \left(\eta^{\alpha\beta} - h^{\alpha\beta}\right) \left( \delta_{\beta}^{\nu} + \frac{\partial \xi^\nu}{\partial x^\beta} \right) \sim \eta^{\mu\nu} - h^{\mu\nu} + \xi^{\mu,\nu} + \xi^{\nu,\mu}.
\end{equation}
(The negative sign on $h$ arises from requiring the inverse of the perturbed metric to satisfy $g^{\mu\alpha}g_{\alpha\nu} \approx \delta^\mu_\nu$. It is a good exercise to show this). Consequently, any 
\begin{equation}
    \label{eq:ch5:GW:gaugeGW}
    h^\prime_{\mu\nu} = h_{\mu\nu} - \xi_{\mu,\nu} - \xi_{\nu,\mu} 
\end{equation}
will describe the same perturbative physics, and we may use the $n$ degrees of freedom in $\xi^\mu$ to paramterize our remaining freedom in specifying the metric.

Now let us use this freedom to simplify the field equations \eqref{eq:ch5:GW:undetermWeakField}. The Laplacian is a very well-understood differential operator, it would be wonderful if we could just get rid of the other terms, wouldn't it? While it may appear that we don't have enough degrees of freedom to do this, in fact it turns out to be eminently doable. Re-write the ugly terms as:
\begin{equation}
    \label{eq:ch5:GW:uglyTerms}
    h\indices{^\alpha_\nu_,_\rho_\alpha} + h\indices{^\alpha_\rho_,_\nu_\alpha} - h_{,\rho\nu} = \partial_\rho \left( h\indices{^\alpha_\nu_,_\alpha} - \frac 12 h_{,\nu} \right) + \partial_\nu \left( h\indices{^\alpha_\rho_,_\alpha} - \frac 12 h_{,\rho} \right),
\end{equation}
using that the partial derivatives commute. The terms in brackets are the same, so if that expression vanishes then both of its appearances in the field equations vanish as well. That is, we will choose a gauge such that
\begin{equation}
    \label{eq:ch5:GW:LorenzGauge}
    h\indices{^\alpha_\nu_,_\alpha} - \frac 12 h_{,\nu} = 0.
\end{equation}
This is known as the \textbf{Lorenz Gauge}\footnote{Named for Ludvig Lorenz, not Hendrik Lorentz.} or \textbf{Harmonic Gauge}\index{Harmonic Gauge}, in analogy with electromagnetism. 

The gauge fixing condition \eqref{eq:ch5:GW:LorenzGauge} is an $n$-dimensional vector equation, and so neatly fixes our $n$ coordinate degrees of freedom. There is however a subtle but important point to be made here about degrees of freedom, one that is easiest to discuss by first finding the explicit form of the gauge transformation that ensures this gauge condition is satisfied. Under the change of coordinates \eqref{eq:ch5:GW:gaugeGW}, the quantity\footnote{Which I pronounce as ``bar h'' so as to avoid confusion with the reduced Plank constant ``h bar'' $\hbar$.} $\overline{h}_{\mu\nu} := h_{\mu\nu} - \frac 12 \eta_{\mu\nu} h$ transforms as:
\begin{equation}
    \label{eq:ch5:GW:gaugeBarh}
    \overline{h}_{\mu\nu} \to \overline{h}_{\mu\nu} - \xi_{\mu,\nu} - \xi_{\nu,\mu} + \eta_{\mu\nu}\xi\indices{^\alpha_,_\alpha}.
\end{equation}
A general metric perturbation can therefore be brought to the form \eqref{eq:ch5:GW:LorenzGauge} by finding $\xi^\mu$ such that:
\begin{align}
    \label{eq:ch5:GW:fixGauge}
    \overline{h}\indices{^\alpha_\nu_,_\alpha} - \xi\indices{^\alpha_,_\nu_\alpha} - &\eta^{\alpha\beta}\xi\indices{_\nu_,_\beta_\alpha} + \xi\indices{^\alpha_,_\alpha_\nu} = 0, \notag \\
\implies \qquad \Box_0 \xi_\nu &= \overline{h}\indices{^\alpha_\nu_,_\alpha}.
\end{align}

\begin{aside}[Local vs.~Boundary Degrees of Freedom]
    The condition \eqref{eq:ch5:GW:fixGauge} is a system of $n$ partial differential equations---specifically the inhomogeneous wave equation. Thought of as constraints, differential equations strictly-speaking only uniquely fix (certain orders of) derivatives of undetermined functions, not the entire functions themselves. Through the magic of calculus\footnote{Actually, this is imprecise. The analysis of partial differential equations is almost shockingly more complicated than the analysis of ordinary differential equations, and in the general case it is still simply unknown whether a system of PDEs has any solutions at all, let alone a fixed, finite number of them. However, for our purposes, it will almost always be the case that we care about Cauchy (i.e.~boundary/initial value) problems with analytic coefficients (i.e.~coefficients representable as Taylor series'), in which case the Cauchy-Kowalevski theorem guarantees what is said above.}, these constraints can be used to infer general characteristics of the desired functions, but cannot pin them down globally without the additional information of boundary conditions. This has an important impact on how we count degrees of freedom versus constraints, turning it into a two-step process.

    Suppose we have a collection of $n$ variables we wish to constrain. Algebraically, we simply need $n$ independent, consistent equations to solve for them. If those variables are instead $n$ unknown functions of some parameters, the counting is the same only the constraints must now be $n$ equations between functions (in effect, each functional equation is equivalent to an infinite number of algebraic equations, but counting infinities is a dangerous road to go down). So far so simple: $n$ equations for $n$ unknowns (algebraic or functional, as required). 

    Now suppose instead we have $n$ unknown functions, but only have $n$ first-order differential equations to constrain them. It is certainly true that these equations will uniquely fix the derivatives of our unknown functions, but from those derivatives we can only infer so much about the entire desired functions. In effect, these differential equations only fix \emph{local} information about the unknown functions, just relations between values of the functions at nearby points. We still absolutely need this information, so it's important to count them as \textbf{local degrees of freedom}\index{Degrees of Freedom!Local}, but the job is not done until we also contribute global information about the function through boundary conditions---i.e., by fixing the remaining \textbf{boundary degrees of freedom}\index{Degrees of Freedom!Boundary}. Furthermore, the higher the order of the differential equations, the more local the information they fix and the more boundary data is needed in the end---i.e., $N^\text{th}$ order differential equations require $N$ sets of boundary conditions.

    The takeaway is the following. When looking to fix $n$ undetermined functions with $N^\text{th}$-order differential equations, we need $n$ such equations (i.e., to fix $n$ local degrees of freedom), and $Nn$ additional constraints to fix the remaining $Nn$ boundary degrees of freedom.
\end{aside}

\subsection{Gravitational Wave Solutions}

Choosing the Lorenz gauge, the Einstein field equations for the weak field now reduce to
\begin{equation}
    \label{eq:ch5:GW:fixedEFE}
    \Box_0 h_{\mu\nu} = 0,
\end{equation}
which is simply the homogeneous wave equation with solutions 
\begin{equation}
    \label{eq:ch5:GW:genGWs}
    h_{\mu\nu} = \real(A_{\mu\nu}(k) e^{i k_\rho x^\rho}),
\end{equation}
with $n^2$ complex-valued integration constants $A_{\mu\nu}(k)$ (so $2n^2$ real \emph{boundary} degrees of freedom) and light-like Minkowski vector wavenumber $k_\mu$ (corresponding to the $n-1$ coordinates on the boundary of spacetime, albeit written in momentum-space). The solutions \eqref{eq:ch5:GW:genGWs} are \textbf{gravitational waves}\index{Gravitational Waves}, propagating wave-like perturbations to the background Minkowski metric. Note that strictly speaking the full solution is an integral over all possible modes, but we will restrict our attention to a single mode for simplicity (effectively a physical boundary condition setting $A_{\mu\nu}(k) \sim \delta^{(3)}(\vec{k} - \vec{k}_0)$).

At this point, we have used up all our local degrees of freedom, so now it is time to use some boundary conditions. Typically, boundary conditions speak to physics, to the specific physical scenario we wish to describe. In this case however, some of our boundary degrees of freedom are spurious, which we can see from the gauge condition \eqref{eq:ch5:GW:fixGauge} only fixing the local degrees of freedom in our gauge functions $\xi^\mu$, leaving us with $2n$ boundary degrees of freedom available to be fixed arbitrarily. 

Let's get counting: first of all, $h_{\mu\nu}$ is a metric perturbation, and therefore is necessarily symmetric. We therefore only really have $n(n+1)$ boundary degrees of freedom in $A_{\mu\nu}$. Moreover, the gauge condition \eqref{eq:ch5:GW:LorenzGauge} is actually an independent equation from the Einstein equation we've just solved, and must separately be satisfied as well. Direct substitution of the general solution \eqref{eq:ch5:GW:genGWs} into that condition returns:
\begin{equation}
    \label{eq:ch5:GW:transverse1}
    k_\alpha A\indices{^\alpha_\nu} = \frac 12 k_\nu A, 
\end{equation}
defining the trace $A := A\indices{^\alpha_\alpha}$. This is an $n$-dimensional complex equation, so we are left with $n(n-1)$ boundary degrees of freedom. This uses up our mathematical constraints, so now we employ our gauge freedom, which it will turn out to be convenient to choose as follows. 

First, for some generic constant timelike vectorfield $U^\mu$ (suspiciously named like a four-velocity), we'll require 
\begin{equation}
    \label{eq:ch5:GW:transverse2}
    A_{\alpha\beta}U^\beta = 0.
\end{equation}
This is almost all of our degrees of freedom, but in fact one of these complex equations is redundant since we already have \eqref{eq:ch5:GW:transverse1}, meaning it was already known that  $k^\alpha A_{\alpha\beta} U^\beta = 0$ (since the timelike $U^\mu$ is always orthogonal to any lightlike $k^\mu$, the RHS of \eqref{eq:ch5:GW:transverse1} necessarily vanishes when contracted with $U^\nu$). With our last remaining complex degree of freedom then, let us impose the simplest, cleanest condition of tracelessness:
\begin{equation}
    \label{eq:ch5:GW:traceless}
    A\indices{^\alpha_\alpha} = 0. 
\end{equation}
With the traceless condition, we arrive at a grand total of $n(n-3)$ real physical degrees of freedom in our gravitational wave solutions. 

The conditions \eqref{eq:ch5:GW:transverse1} and \eqref{eq:ch5:GW:transverse2} are known as ``transverse'' conditions, and together with \eqref{eq:ch5:GW:traceless} they make up the so-called \textbf{transverse-traceless} or \textbf{TT} gauge\index{Transverse-Traceless Gauge}. These conditions suffice to describe a superposition of propagating gravitational waves; in most applications it is only important to consider waves propagating in a single direction (i.e., emanating from a single source), in which case we only need the outgoing  $+ikx$ solution, and not the incoming $-ikx$ solution, reducing the degrees of freedom again by half (aligning our coordinates with the direction of propagation, this amounts to fixing $A^*_{\mu\nu} = A_{\mu\nu}$). Note that this constitutes the imposition of a physical boundary condition.

That was a lot of counting and counting is hard\footnote{Especially the way we do it here. For a cleaner derivation that uses the representation theory of the Lorenz group, see for example \cite[\S 7.2]{carroll_2019}.}, so let's recap. We started with $2n^2$ real boundary degrees of freedom in the complex coefficients $A_{\mu\nu}(k)$ of the gravitational wave solutions. We then found:
\begin{itemize}
    \item Symmetry in $h _{\mu\nu}$ takes one factor of $n$ to  $\frac 12 (n + 1)$: {\color{blue}  $n(n+1)$}
    \item Compatibility with the Lorenz gauge \eqref{eq:ch5:GW:LorenzGauge} imposes \eqref{eq:ch5:GW:transverse1}, which is $n$ complex equations: {\color{blue} $n(n-1)$}
    \item Next, we use our remaining $2n$ real gauge boundary degrees of freedom. We choose the integration constants of $\xi^\mu$ to satisfy $A_{\alpha\beta}U^\beta = 0$ for some fixed timelike vector $U^\beta$ (recalling that one complex equation is redundant thanks to the previous condition): {\color{blue}  $n(n-3) + 2$}
    \item And we polish off the remaining 2 real degrees of freedom from our gauge parameter by simply requiring the trace of $A$ to vanish, $A\indices{^\alpha_\alpha} = 0$: {\color{blue} $n(n-3)$}
    \item The remaining degrees of freedom are physical, but in practice we wish to study uni-directional monochromatic propagating waves, so we choose boundary conditions to make that happen: {\color{blue} $\frac 12 n(n-3)$}
\end{itemize}
To be clear, we will write $h_{\mu\nu}^\text{TT}$ to represent the gravitational wave solution \eqref{eq:ch5:GW:genGWs} with the transverse-traceless gauge restrictions above. Note however that the degree-of-freedom \emph{counting} is generic, so any other gauge would similarly leave $\frac 12 n(n-3)$ physical degrees of freedom, which in our four spacetime dimensions evaluates to a nice clean two independent components. 

\subsection{A Gravitational Wave Just Passing Through}

Now consider how these waves appear from the perspective of an inertial four-dimensional Minkowski observer. Choose coordinates such that $\partial_t$ aligns with the four-velocity of our inertial observer---i.e., choose coordinates such that $U^\mu(\tau_0) = \delta^\mu_0$, as always. Now choose this $U^\mu$ to be the same as the constant general timelike four-vectorfield $U^\mu$ that we used above to fix (almost all) the gauge degrees of freedom in $h^\text{TT}_{\mu\nu}$. In these coordinates, we therefore have $A_{\alpha 0} = A_{0 \alpha} = 0$ for all $\alpha$. Furthermore, it is advantageous to align one coordinate axis (of our background Minkowski spacetime) with the spatial components of $k^\mu$. In other words: choose coordinates such that $k^i \partial_i = k^z \partial_z$. Then using that the momentum four-vector is lightlike, define the gravitational wave's frequency  $\omega := k^0$, so in these coordinates we simply have $k^\mu = (\omega, 0, 0, \omega)^\text{T}$. Again using the transverse gauge condition, this results in $A_{\alpha 3} = A_{3 \alpha} = 0$ for all $\alpha$. What's left? Clearly we only still have component combinations with coordinates 1 and 2, so $A_{11}, A_{22}$, and $A_{12} = A_{21}$. The traceless condition however finally relates the diagonals, so $A_{22} = - A_{11}$, leaving the only two independent, physical components of the gravitational wave perturbation, $A_{11}$ and $A_{12}$.

Okay, now we have a choice of coordinates and a metric tensor---the next step, as ever, is to ask what happens to things in this geometry. First up, what are the geodesics? The connection coefficients for our metric are (going back to \eqref{eq:ch5:GW:christofs}):
\begin{gather}
    \label{eq:ch5:GW:TTChristoffels}
    \Gamma^0_{ij} = \omega A_{ij} \sin(kx) = \Gamma^3_{ij}, \qquad \Gamma^i_{j0} = \omega A_{ij} \sin(kx) = - \Gamma^i_{j3},
\end{gather}
with $i,j \in [1,2]$ (the 0 and 3 directions obviously being special), and the rest vanishing. The geodesic equations are then:
\begin{align}
    \label{eq:ch5:GW:TTGeods}
    \dot U^0 + \omega A_{ij} \sin(kx) U^iU^j &= 0, \notag \\
    \dot U^i + \omega A_{ij}U^j\left( U^0 - U^3 \right) &= 0,  \\ 
    \dot U^3 + \omega A_{ij} \sin(kx) U^iU^j &= 0. \notag
\end{align}
In general, the equations \eqref{eq:ch5:GW:TTGeods} are difficult to solve\footnote{Although the situation can be simplified quite a bit with the Killing vectors we will see in \ref{ch:AppsII:BH}.}, but there is a class of solutions that are readily available. In an ideal, acceleration-free universe, what is the comfortable worldline we would like ourselves to follow? Of course simply $x^\mu(\tau) = (\tau, 0, 0, 0)^\text{T}$, and as if by magic, taking $U^\mu = (1, 0, 0, 0)^\text{T}$ as an ansatz easily solves the geodesic equations, so with this metric, in these coordinates, an inertial particle sitting still at the origin stays where it is. What's more, the particle doesn't need to be sitting at the origin; in fact any $U^\mu = (a, 0, 0, b)^\text{T}$ solves the geodesic equation, so particles sitting at \emph{any} point in spacetime stay where they are \emph{in these coordinates}. What we are seeing is the power of the ``transverse'' condition, motion confined to the $t-z$ plane is really entirely unaffected by a lone gravitational wave propagating in the $z$-direction, the wave's effects are entirely confined to the \emph{transverse} (perpendicular) $x-y$ plane. 

The geodesic equations tell us that transverse gravitational waves propagating in the $z$-direction do not affect the \emph{coordinates} of paths followed by individual inertial observers at fixed locations in the $x-y$ plane. Does this mean that those observers do not feel any physical effects of the gravitational waves as they pass through? Well yes and no: they are \emph{inertial} observers, so individually they won't feel any gravitational disturbance no matter what the metric is (i.e.,~the equivalence principle), but if they compare their lives to their neighbours, they will surely notice a non-Minkowski metric tensor (through the Levi-Civita \emph{connection} it induces in the manifold). In other words, the fact that the coordinates of inertial observers' worldines are constant in the transverse plane is a statement about the coordinates themselves, that they are \textbf{adapted} to these observers (we will have much more to say about adapting coordinates to observers in the following sections), but how these observers see \emph{each other} is not so obvious, and the distinction between these two speaks to the fundamental lack of physicality in coordinates\footnote{``Coordinates'' as in ``maps from the manifold to subsets of $\mathbb{R}^n$,'' whose arbitrariness should always have precluded physicality. Having said that, the physical reality of the points on the manifold themselves is an open question, which I will not try to solve here.}.

\subsection{What is physical?}

Certainly not coordinates, we just spent several subsections making use of our total freedom to choose coordinates which must make them a mathematical formality, not a physical reality. Instead, physical things must be represented by those geometric objects we just spent several chapters formally constructing. In particular, distances in space and time are intimately tied to spacelike and timelike vectors respectively. Go back once again to Special Relativity---how did we used to define ``proper time''? Proper time was the name Einstein gave to the time elapsed in an inertial observer's rest frame, and was the only physically meaningful value of elapsed time, all other measures of the same time interval were only relative (cf time \emph{dilation}, recovering the proper time interval from a Lorentz transformed coordinate system required a contribution from some \emph{contracted} length as well). In the affine space that is Minkowski space, any finite proper time interval is the length of a \emph{displacement} vector corresponding to some stretch of an inertial observer's life, but when the affine structure of spacetime is broken by a non-Minkoski metric tensor (such as our gravitational wave metric), all but infinitessimal displacement vectors are abolished. That is to say, proper time displacements can only be defined as the sum over infinitessimal displacements\footnote{In fact, the geodesic equation can be derived as the variation of \eqref{eq:ch5:GW:properTime} with respect to the metric---an exercise you should perform for yourself.}:
\begin{equation}
    \label{eq:ch5:GW:properTime}
    \Delta\tau := \int_{\tau_i}^{\tau_f} \sqrt{-U(\tau)^2} \dd\tau,
\end{equation}
where $U(\tau)^2 := g(U(\tau),U(\tau))$ as usual, with $U(\tau)$ the (timelike) four-velocity of an inertial observer\footnote{Actually, equation \eqref{eq:ch5:GW:properTime} looks wrong at first glance and trivial at second, but is just annoying in practice. It looks wrong because $\Delta \tau = \tau_f - \tau_i$, so the integral should just be over $\dd\tau$, but in fact the four-velocity of an inertial observer satisfies $U(\tau)^2 = -1$ so the prefactor is always unity---hence seemingly trivial. In practice, one tends to fix $x_f$ and solve for $\tau_f$ (hence $\Delta \tau$) which is just a pain.}.

\begin{figure}[ht]
    \centering
    \includegraphics[width=1.\textwidth]{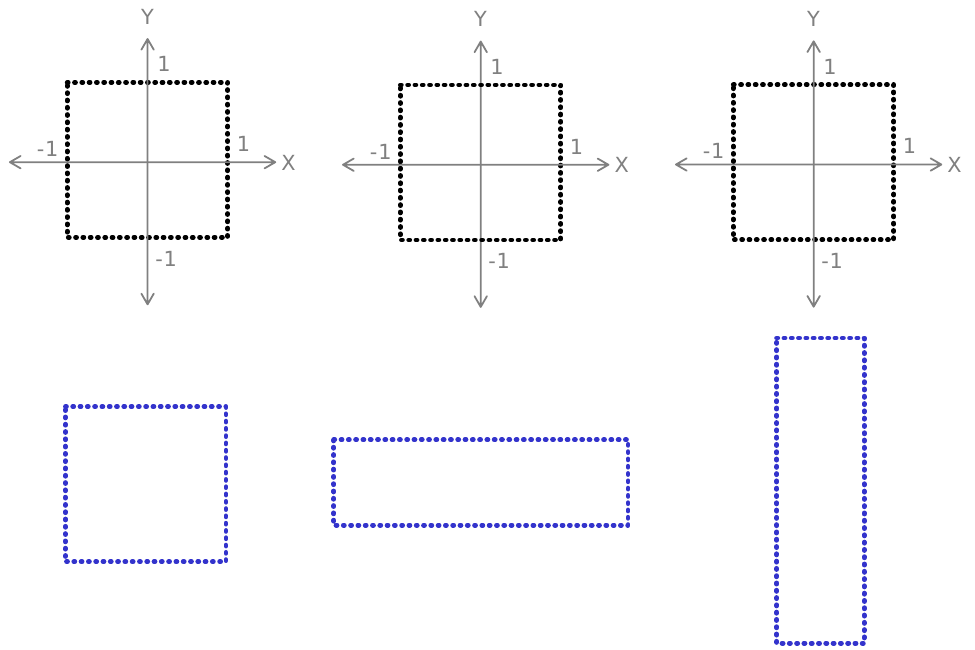}
    \caption[``Plus'' type linear polarization of gravitational waves.]{``Plus'' type linear polarization of gravitational waves at proper times $\tau_0 = -3\pi/(2\omega),\, \tau_1 = 0,$ and $\tau_2 = \pi/\omega$. Notice the \emph{coordinates} of the slab never change, but the \emph{physical} proper lengths of its sides do.}
    \label{fig:ch5:GW:plus}
\end{figure}

We could in fact already probe the effects of gravitational waves by using \eqref{eq:ch5:GW:properTime} from the perspective of an observer moving towards or away from another in the $x-y$ plane, but it is even more direct to do so for a proper \emph{distance} calculation instead. Imagine constructing some structure $\gamma^\mu(\tau, u, v)$ in the $x-y$ plane, say a disk of sand, with each grain constituting an inertial observer at a fixed $(x,y)$, and for convenience align their proper times such that each geodesic is parameterized by the same $\tau$. It is a continuous distribution, so we can parameterize the $x$ and $y$ directions respectively by some continuous parameters $u$ and $v$. Then the \emph{physical} distance across the disk in the $x$ direction\footnote{We're being a little imprecise here by only studying one-dimensional cuts through our slab, but the result is okay because gravitational waves only induce shear. If we wished to study more complicated effects though, we would need the formalism of geodesic congruences---see \cite[Appendix F]{carroll_2019} for details.} (say) is the sum over the infinitessimal geodesic separation between grains of sand in the $x$ direction, $X^\mu := \frac {\dd}{\dd u} \gamma^\mu(\tau, u, v)$:
\begin{equation}
    \label{eq:ch5:GW:properDelX}
    \Delta X(\tau, v) = \int_{u_i}^{u_f} \sqrt{X(\tau, u, v)^2}\dd u.
\end{equation}
(Note the lack of negative sign---$X^\mu$ is a spacelike vector, so its magnitude is positive and we can take its square root directly). Try this for some uniformly distributed sand: take $\gamma^\mu(\tau, u, v) = (\tau, u, v, 0)$:
 \begin{equation}
    \label{eq:ch5:GW:calcDelX}
    \Delta X(\tau, v) = \int_{u_i}^{u_f} \sqrt{(1 + 2A_{11}\cos(\omega \tau))}\dd u \sim (1 + A_{11}\cos(\omega \tau)) (u_f - u_i),
\end{equation}
using that $\abs{h} \ll 1$. So physically, the actual width of this distribution in the $x$ direction will vary sinusoidally along the lifetime of the elements of the disk---as long as $A_{11} \neq 0$. A similar calculation in the $y$ direction yields:
\begin{equation}
    \label{eq:ch5:GW:calcDelY}
    \Delta Y(\tau, u) \sim (1 - A_{11}\cos(\omega\tau))(v_f - v_i). 
\end{equation}
If $A_{11}$ is the only non-zero component of gravitational wave polarization, these are the only effects: physical lengths along the $x$ and $y$ directions deform oppositely, even though the coordinate locations never change. This is visualized in figure \ref{fig:ch5:GW:plus}; for obvious reasons this polarization is called the ``plus'' polarization, and in order to be more clear notationally we define the magnitude in this direction as $A_{+} := A_{11}$.

\begin{figure}[ht]
    \centering
    \includegraphics[width=1.\textwidth]{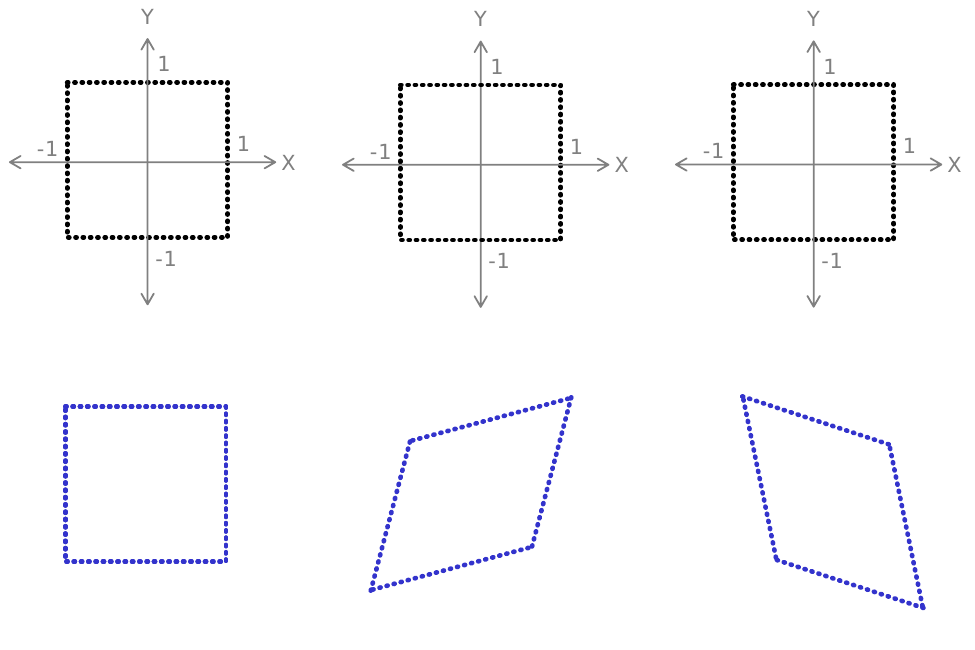}
    \caption[``Cross'' type linear polarization of gravitational waves.]{``Cross'' type linear polarization of gravitational waves at proper times $\tau_0 = -3\pi/(2\omega),\, \tau_1 = 0,$ and $\tau_2 = \pi/\omega$. Notice the \emph{coordinates} of the slab never change, but the \emph{physical} proper lengths of its diagonals do.}
    \label{fig:ch5:GW:cross}
\end{figure}

What about the off-diagonal polarization $A_{12}$? Evidently it doesn't change either \eqref{eq:ch5:GW:calcDelX} or \eqref{eq:ch5:GW:calcDelY}, so it does not affect proper lengths directly along either transverse axis, but the off-diagonal terms of the metric mix $x$ and $y$ components of vectors, so surely a gravitational wave with off-diagonal polarization must affect off-axis vectors. Consider the most off-axis vector possible, one running right across a diagonal of our slab of geodesics: $D_+^\mu = \frac 1 {\sqrt{2}} (X^\mu + Y^\mu)$. If the gravitational wave has only off-diagonal polarization (i.e., $A_+ = 0$), this vector has length $D_+^2 = g(D_+, D_+) = 1 + 2A_{12}\cos(\omega\tau)$, so the proper length of one diagonal goes as:
\begin{equation}
    \label{eq:ch5:GW:calcDplus}
    \Delta D_+ = \int_{s_i}^{s_f} \sqrt{D_+^2} \dd s \sim (1 + A_{12}\cos(\omega\tau))(s_f - s_i), 
\end{equation}
where $s \in [0, \sqrt{(u_f - u_i)^2 + (v_f - v_i)^2}]$ (the diagonal is essentially a parameterized curve in the field of geodesics, so we can define it by a parameter $s$ and label the path through the field of geodesics by $u(s) = v(s) = s$). Of course, we could just as easily have defined a vector along the opposite diagonal, $D_-^\mu = \frac 1 {\sqrt{2}}(X^\mu - Y^\mu)$ with corresponding length $D_-^2 = g(D_-, D_-) = 1 - 2A_{12}\cos(\omega\tau)$, and so with proper length
\begin{equation}
    \label{eq:ch5:GW:calcDminus}
    \Delta D_- = \int_{s_i}^{s_f} \sqrt{D_-^2} \dd s \sim (1 - A_{12}\cos(\omega\tau))(s_f - s_i).
\end{equation}
Clearly this is the same motion as the plus polarization only rotated 45 degrees, and so is named the ``cross'' polarization, and given the descriptive notation $A_\times := A_{12}$. The cross polarization is shown schematically in figure \ref{fig:ch5:GW:cross}.

\begin{aside}[Circular Polarizations]
    The polarizations $+$ and $\times$ are the gravitational analogue of linear polarizations in electromagnetism, where the $E$ field oscillates along either the $x$ or $y$ axes (transverse to its direction of propagation $z$). Again exactly as in the theory of electromagnetic waves, it can be useful at times to consider complex linear combinations of linear polarizations, the left- and right-handed circular polarizations, $A_R := \frac 1{ \sqrt{2} }( A_+ + iA_\times )$ and $A_L := \frac 1{ \sqrt{2} }( A_+ - iA_\times )$. These have exactly the effect you'd imagine, the ``squish'' distortions of the linear modes are aligned in just such a way as to propagate either clockwise or anti-clockwise around the distribution of geodesics. 
\end{aside}

A gravitational wave's effects on proper distances is the heart of how interferometric detectors work. Gravitational wave detectors like LIGO use null geodesics (light rays) to very precisely measure proper distances along perpendicular directions in local coordinates, and look for the characteristic sinusoidal disruptions calculated (simplistically) above. However, this is not the only way to go about hunting these delicate creatures. The exercise we have just performed is akin to an electromagnetic computation using the integral forms of Maxwell's equations, but we may also make similar use of their differential forms. 

Consider again the slab of geodesics we designed above, $\gamma^\mu(\tau, u, v)$---where else have we seen a similar distribution of geodesics? Of course this is precisely what we used to find a \emph{local} measure of non-uniform gravitational fields, i.e.~local curvature. The local separation between geodesics in our slab along the $x$ direction at some initial proper time $\tau_0$ is again $X^\mu = (0, 1, 0, 0)^\text{T}$ as above, but now let us ask what happens to this particular along the lifetime of the geodesics. Following the discussion in \ref{ch:curve:sec:curve}, we can use the geodesic deviation equations \eqref{eq:ch3:geodAcc4}:
\begin{equation}
    \label{eq:ch5:GW:geodAcc}
    \frac{D^2}{d \tau^2}\Big\vert_{\tau_0} X^\mu = R\indices{^\mu_{\lambda\nu\rho}}U^\lambda U^\nu X^\rho.
\end{equation}
Obviously $U^\mu = (1, 0, 0, 0)^\text{T}$ in these coordinates with this geodesic congruence parameterization, so the acceleration boils down to:
\begin{equation}
    \label{eq:ch5:GW:geodAcc2}
    \frac{D^2}{d \tau^2}\Big\vert_{\tau_0} X^\mu = R\indices{^\mu_{001}}.
\end{equation}
Of course now we need a component of the Riemann curvature tensor, but in fact this is not so hard. The linearized curvature tensor takes the form:
\begin{equation}
    \label{eq:ch5:GW:linRiem}
    R\indices{^\mu_{\lambda\nu\rho}} \sim \partial_\nu \Gamma^\mu_{\lambda\rho} - \partial_\rho \Gamma^\mu_{\lambda\nu} = -\frac 12 \eta^{\mu\alpha}\left( h_{\lambda\rho,\alpha\nu} - h_{\rho\alpha,\lambda\nu} - h_{\lambda\nu,\alpha\rho} + h_{\nu\alpha, \lambda\rho} \right).
\end{equation}
The only components we need are:
\begin{equation}
    \label{eq:ch5:GW:linRiem2}
    R\indices{^\mu_{001}} \sim -\frac 12 \eta^{\mu\mu}\left( h_{01,\mu 0} - h_{1\mu, 00} - h_{00,\mu 1} + h_{0\mu, 01} \right) \qquad \text{(not summed)}. 
\end{equation}
We already have $h_{0\nu} = 0$ for all $\nu$, so the first, third, and fourth terms all disappear immediately, leaving only the second, $R\indices{^\mu_{001}} \sim \frac 12 h_{1\mu,00}$. The geodesic acceleration is therefore
\begin{equation}
    \label{eq:ch5:GW:geodAccX}
    \frac{D^2}{d \tau^2}\Big\vert_{\tau_0} X^1 = -\omega^2 A_{+}\cos(\omega\tau),
    \quad \text{and} \quad
    \frac{D^2}{d \tau^2}\Big\vert_{\tau_0} X^2 = -\omega^2 A_{\times}\cos(\omega\tau).
\end{equation}
Similarly, geodesic separations in the $y$ direction will accelerate according to:
\begin{equation}
    \label{eq:ch5:GW:geodAccY}
    \frac{D^2}{d \tau^2}\Big\vert_{\tau_0} Y^1 = -\omega^2 A_{\times}\cos(\omega\tau), \quad \text{and} \quad
    \frac{D^2}{d \tau^2}\Big\vert_{\tau_0} Y^2 = \omega^2 A_{+}\cos(\omega\tau),
\end{equation}
In other words, as a gravitational wave passes through a lab, very sensitive detectors could in principle measure a small \emph{force} parallel or perpendicular (depending on the polarization) to some linear structure. This approach is the aim of resonance and bar-type gravitational wave detectors---notably less successful so far than their large-scale proper distance-based competitors. 

\subsection{Measuring Curvature}

One last note before we move on. In the above, we used knowledge of our metric ansatz to describe geodesics and the curvature tensor, and arrange physical predictions to be made with respect to a specific few parameters labelling our solution. However, in principle we could flip this around and use geodesic deviation to directly measure local curvature without a preconceived notion of what it should be. In electromagnetism, for instance, we can use the Lorentz force law to directly measure an electric field based on how it affects the lives of electrically charged particles---this is fundamentally no different, the metric tensor is roughly our equivalent of the electromagnetic four-potential, and the Riemann curvature tensor our analog of the electric and magnetic fields.

\section{A prelude to sources: Uniform Acceleration}
\label{ch:AppsI:sec:Rind}

Why did we go through all of the effort to phrase things in the language of differential geometry? The equivalence principle: a universe with more than one massive object is one in which \emph{no} observer is unaccelerated, but at least everyone accelerates the same way, and differential geometry is the tool we can use to understand the structure of this effectively curved spacetime. But something we've neglected to do is look at the perspective of the realistic observer, one who is now necessarily accelerated. All of Special Relativity was derived by looking at the lives of different inertial observers, why have we been so cruel to real scientists! Time to rectify that omission\footnote{Relativistic uniform acceleration is not detailed very thoroughly in most textbooks (when it's even mentioned at all), but an excellent discussion can be found in \cite[\S 12.2]{Gourgoulhon:2013gua} and references therein.}.

Let's try to make life simple by studying a single, uniformly accelerated observer. Realistically, a perfect uniformly accelerated particle would require something like a charged particle in a universe permeated by a uniform electric field, but since we only care about the experience of the particle, it suffices to look at Minkowski space through the lens of the world-line of a fictious observer associated with the solution to Newton's second law for a constant acceleration. What is meant by ``constant'' acceleration in relativity, however, is not quite trivial. Restrict life to 1+1 dimensions for a moment; intuitively, constant acceleration should mean the acceleration four-vector for a particle is given by $a^\mu = (0, a)^\text{T}$ for some constant parameter $a$. Let's try that:
\begin{equation}
    \label{eq:ch5:Rind:newton2RindNaive}
    \frac{\dd U^\mu}{\dd \tau} = a^\mu,
\end{equation}
trivially yields $U^\mu(\tau) = ( 1, a\tau )^\text{T}$, where we set the initial condition at $U(\tau = 0) = (1, 0)^\text{T}$. But now recall that for a \emph{physical} observer, we can at any time find an inertial reference frame the instantaneously coincides with the particle's rest frame, and in which the particle's instantaneous four-velocity is $(1, 0)^\text{T}$, so it must always have the same norm,  $U^2 = -1$. Applying that to the solution to \eqref{eq:ch5:Rind:newton2RindNaive} though, we find $-1 + a^2\tau^2 = -1$ for all $\tau$, which immediately implies $a = 0$. 

So a constant four-acceleration actually isn't anything reasonable---what \emph{is} the relativistic version of constant acceleration? Thinking more physically, the instantaneous 3-velocity of a particle is the parameter $\vec\beta$ of the boost required to bring it to its instantaneous rest frame from some reference inertial frame. As a function of proper time, $\vec\beta(\tau)$ certainly can't be linear since its magnitude $\abs{\beta}$ is restricted to be less than 1 and linear growth would always violate that bound in finite proper time. Instead, we need to work with some notion of a ``relativistic'' velocity, that has infinite range and can grow without bound. This is easy enough to construct, just define the \textbf{rapidity}\index{Rapidity} $\phi$ such that $\beta = \tanh{\phi}$. Now $\phi$ can cover all real numbers, so we can propose a definition of constant acceleration as linear growth in rapidity: $\phi(\tau) = a\tau$. Then the Lorentz-boosted four-velocity of a ``uniformly accelerated'' particle has the convenient form (again truncated to 1+1 dimensions for ease)
\begin{equation}
    \label{eq:ch5:Rind:uniformU}
    U(\tau) = \mqty( \cosh(a\tau) \\ \sinh(a\tau) ),
\end{equation}
and so has four-acceleration
\begin{equation}
    \label{eq:ch5:Rind:uniformA}
    a(\tau) = a\mqty( \sinh(a\tau) \\ \cosh(a\tau) ).
\end{equation}
One can easily verify that these satisfy $U^2 = -1$ and  $a^2(\tau) = a^2$, leading us to a clean definition of relativistic uniform acceleration as a system for which the four-acceleration has constant \emph{magnitude}, but not constant components\footnote{Indeed, this condition---along with the condition on the magnitude of the four-velocity and the relativistic version of Newton's second law---is sufficient to derive the form \eqref{eq:ch5:Rind:uniformU}.}. It is also easy to see that in the limit of small relativistic acceleration, $\phi \sim \beta$ and $a^\mu \sim (0, a)^\text{T}$, as we would like to see in the non-relativistic regime. 

\subsection{Life in a Perfect Rocket}

\begin{figure}[htp!]
    \centering
    \includegraphics[width=\textwidth]{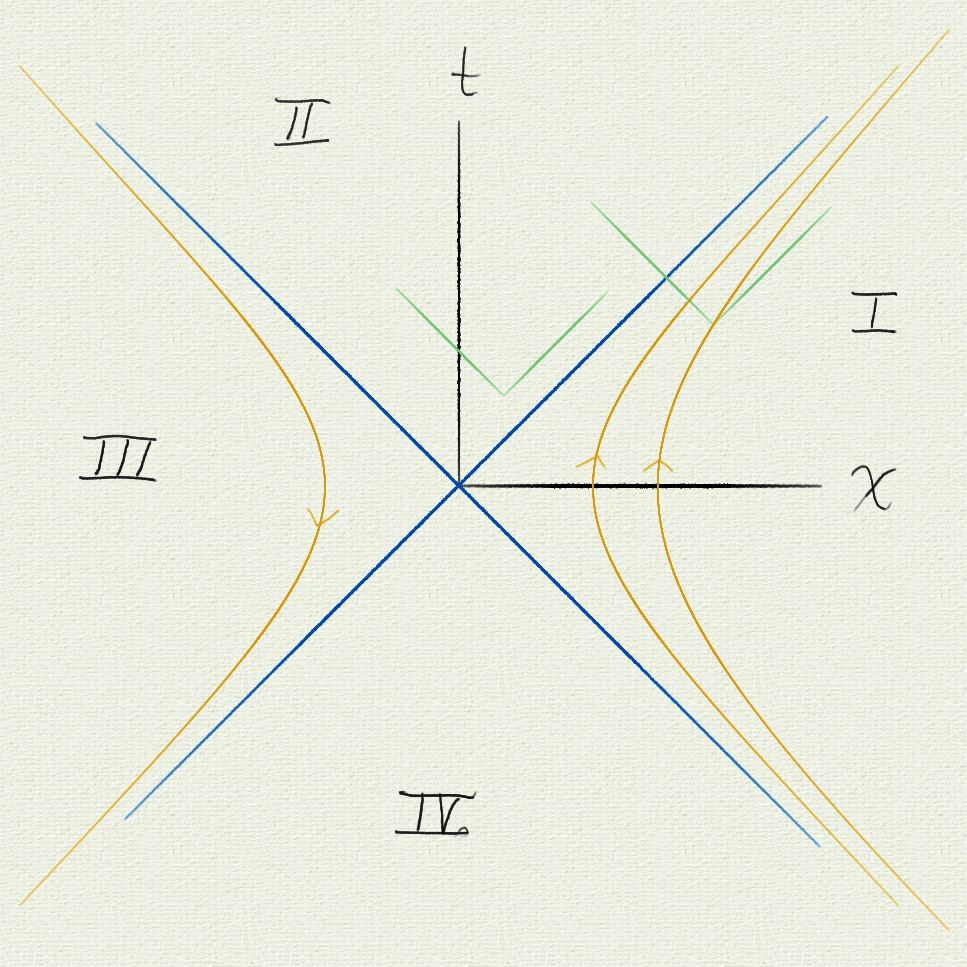}
    \caption[Accelerated observer worldlines in Minkowski (Cartesian) coordinates.]{Accelerated observer worldlines in Minkowski (Cartesian) coordinates (yellow). The Rindler horizons (i.e., light-cones at $x = 0$) are in blue, and example light cones inside and outside the horizon are in green. The disconnected Rindler charts are labelled I-IV. }
    \label{fig:ch5:Rind:rindler_mink}
\end{figure}

So in Cartesian Minkowski coordinates, a uniformly accelerated particle traces out a hyperbolic world-line (see figure \ref{fig:ch5:Rind:rindler_mink}). What about from the point of view of the accelerated particle? The particle is surely not inertial, but that's no trouble for us now with all of the might of differential geometry on our side. Let $\gamma(\tau)$ be the world-line of our uniformly-accelerated particle. The worldline $\gamma$ runs through the manifold of spacetime and is independent of our choice of coordinates. We know in Cartesian coordinates that  $x^\mu(\gamma(\tau)) = a^{-1}( \sinh(a\tau), \cosh(a\tau))^\text{T}$ (easily integrating \eqref{eq:ch5:Rind:uniformU}), and we seek coordinates $X^\mu$ such that  $X^\mu(\gamma(\tau)) = ( \tau + A, B)^\text{T}$ for some constant offsets $A$ and $B$. There is a lot of freedom in the problem at the moment, but if we impose the coincidence that these coordinates overlap for $\tau = 0$ (i.e., so that $U^\mu(\tau = 0) = ( 1, 0)^\text{T}$ for both observers), then we have $A = 0$ and  $B = 1/a$, and a compatible coordinate transformation\footnote{Actually, the problem is under-determined, we're trying to pin down two degrees of freedom with only one, so a choice still has to be made at the end.} can easily be seen to be:
 \begin{equation}
    \label{eq:ch5:Rind:RindToCart}
    x^{\mu}(X^{-1}) = \mqty( X\sinh(a T) \\ X\cosh(a T)),
\end{equation}
from accelerated coordinates to Cartesian (defining $x^0 = t,\, x^1 = x,\, X^0 = T,$ and $X^1 = X$), and
\begin{equation}
    \label{eq:ch5:Rind:CartToRind}
    X^{\mu}(x^{-1}) = \mqty( a^{-1} \tanh[-1]( t/x ) \\ \sqrt{x^2 - t^2}  ),
\end{equation}
for the inverse. The coordinates \eqref{eq:ch5:Rind:CartToRind} are known as \textbf{Rindler}\index{Rindler Coordinates} coordinates\footnote{Named for Wolfgang Rindler who did a deep dive into this space in 1960, although the coordinates and concept had been studied by almost all the big names since the early days of Special Relativity.}, and have some very interesting things to say about relativistic physics. 

While we could explore the life of the individual accelerated observer, more important for what follows is an analysis of the global structure of these coordinates, and Minkowski space through the lens of this \emph{class} of observer. To that end, first note the range of the coordinates defined by \eqref{eq:ch5:Rind:CartToRind}. Certainly the time coordinate $T$ is free to be anything it likes, but the space coordinate $X$ is restricted to be positive. Tracking back, positive semi-definite $X$ and unconstrained $T$ lead to the Rindler chart covering only half of $\mathbb{R}^2$, and only a quarter of Minkowski space (the positive-$x$ spacelike quadrant of the lightcone at the origin, region I in figure \ref{fig:ch5:Rind:rindler_mink}). The boundary at $X = 0$ corresponds to the path of a light ray emitted at $x = 0$ in the Cartesian Minkowski chart, and constitutes an \textbf{event horizon}\index{Event Horizon}---no signal from the left side of this line will ever reach the observer on the accelerated trajectory. This is easy to see in the Cartesian chart (and in fact impossible to see in the Rindler chart, since the coordinates do not have sufficient range), all light rays in Cartesian Minkowski coordinates propagate along lines of slope 1, so any light ray emitted at $x < t$ will never make it to a point at $x > t$. On the outside of the event horizon, however, \emph{any} signal except for a light ray directed away from the horizon will eventually reach it in finite proper time. This is graphically depicted in figure \ref{fig:ch5:Rind:rindler_mink}, the green light cone behind the horizon will always be enclosed by the horizon, while the green light cone outside the horizon will easily intersect it.

\begin{figure}[ht]
    \centering
    \includegraphics[width=\textwidth]{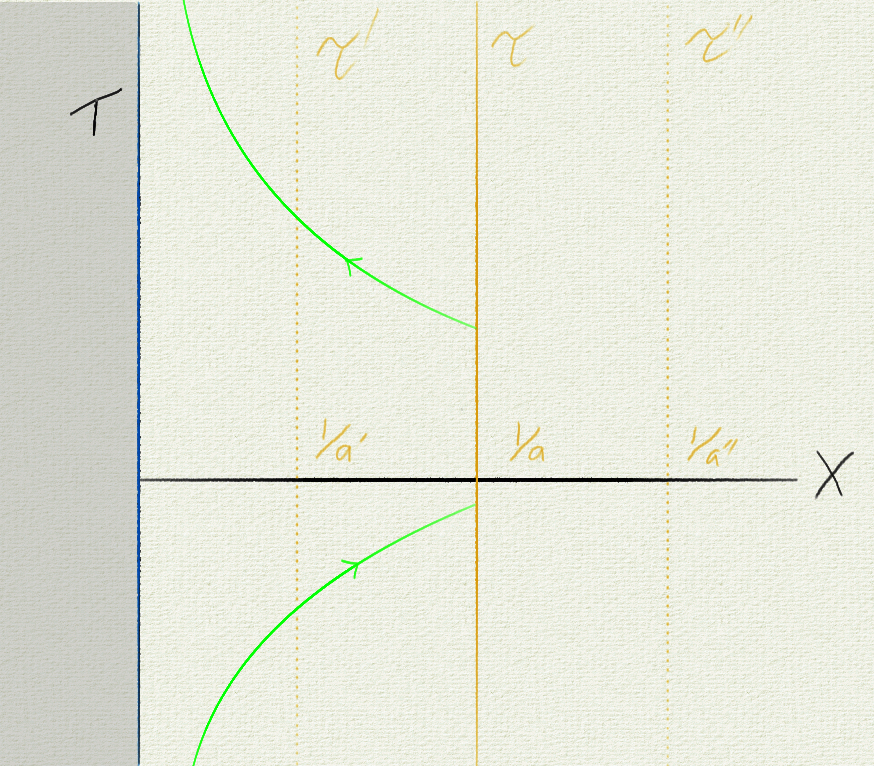}
    \caption[Accelerated observer worldlines in Rindler coordinates.]{Rindler coordinates (region I in figure \ref{fig:ch5:Rind:rindler_mink}). The accelerated observer's worldline is in gold, and the Rindler horizon is blue. Light rays received by and sent by the observer are in green. Other accelerating observers' worldlines are in dotted gold lines, displaying relative time dilation $\tau^\prime = a^\prime/a \tau$ for  $a^\prime < a$ and time contraction for  $\tau^{\prime\prime} = a^{\prime\prime}/a \tau$ for  $a^{\prime\prime} > a$.}
    \label{fig:ch5:Rind:rindler_rind}
\end{figure}

What's even more interesting is what this event horizon looks like from the perspective of the accelerated observer, from within the Rindler coordinates. It is easy enough to do this by simply changing coordinates for paths we know and love, but since Rindler coordinates only cover a quarter of spacetime it is an illuminating exercise to solve for the geodesics using the metric in those coordinates. A simple coordinate transformation shows that the Minkowski metric (in region I, the open set of Minkowski space serving as the domain of the Rindler coordinate chart) in Rindler coordinates takes the simple form:
\begin{equation}
    \label{eq:ch5:Rind:RindlerMetric}
    \dd s^2 = -a^2 X^2 \dd T^2 + \dd X^2.
\end{equation}
For a general path, we would need to solve the geodesic equations in full, but we can learn the salient details of causal trajectories in these coordinates from only the light-like paths, which are easier to solve for. Tangents to geodesics satisfy that their lengths do not change along the curve (this is essentially how we derived the geodesic equation, but can easily be shown explicitly by evaluating $\dv[]{}{s}U^2 = U^\alpha \nabla_\alpha (U^2)$ and using the geodesic equation), and for null vectors in a (1+1)-dimensional space with a diagonal metric, this leads to a very simple relation:
\begin{equation}
    \label{eq:ch5:Rind:solveNullRind}
    U^0 = \pm \frac{1}{aX} U^1,
\end{equation}
which can be integrated to find
\begin{equation}
    \label{eq:ch5:Rind:solveNullRind2}
    T(s) = \pm \ln(aX(s)),
\end{equation}
for some affine parameter $s$. Clearly $X$ is constrained to be greater than 0, as required, and $T$ is unconstrained. The key takeaway is the logarithmic relationship---an accelerated observer moving along a vertical line in Rindler space will see photons emitted towards the event horizon approach closer and closer, but never actually reach the horizon (see the green curves in figure \ref{fig:ch5:Rind:rindler_rind}. Massive timelike observers move slower than light, so their curves will be similarly logarithmic but approach the horizon even slower). 

How can it be that signals seemingly reach and pass the horizon in Minkowski coordinates, but fail to ever reach the horizon from the point of view of an accelerated observer? The answer is that \emph{coordinate} time and \emph{proper} time are \emph{different} things---time is a \emph{place} not a \emph{parameter}. Rindler coordinates are adapted to the Rindler observer such that the coordinate $T$ coincides with that observer's proper time in their reference frame, so paths through $T$ are paths through time \emph{as seen by the Rindler observer}. The same thing goes in Minkowski coordinates; any $t$ coordinate is adapted to the worldline of some inertial observer, and inertial observers in different Lorentz frames perceive each others' time as dilated with respect to their own. The accelerated Rindler observer is just constantly being Lorentz transformed, so is seeing other observers' time being more and more dilated, effectively watching the rest of the universe grind to a halt. What we've just seen is a stark example of the lack of physical meaning in coordinates. The only real physical variables at play here are the proper times of the observers involved, and while the emitted light ray's Rindler coordinate position doesn't reach the horizon for finite Rindler time, it easily reaches the horizon in finite \emph{affine parameter} $s$ (this can be seen explicitly either by solving the geodesic equation for the null ray, or by applying the change of coordinates \eqref{eq:ch5:Rind:CartToRind} on the path $(s, 1-s)$), the light ray's affine parameter serving as its version of proper time. 

Now inertial observers and accelerated observers are fundamentally quite different, as we've seen here comparing the two from each other's perspectives. Actually, long ago we started this whole course discussing how inertial observers see each other (related via Lorentz transformations), so we should finish the program in this section by comparing accelerating observers to each other. Take our reference accelerating observer to have uniform acceleration $a$ as above, and consider another observer with uniform acceleration $a^\prime$. In our reference Rindler observer's coordinates, the worldline of our second observer reads:
\begin{equation}
    \label{eq:ch5:Rind:obs2}
    X_{a^\prime}^\mu(x^{-1}) = \mqty( \frac {a^\prime} a \tau \\ 1/a^\prime ).
\end{equation}
So other uniformly accelerating observers will also sit on vertical lines in Rindler coordinates (see figure \ref{fig:ch5:Rind:rindler_rind}), fixed at $X = 1/a^\prime$, but their clocks will run at a different rate from our reference observer, either dilated or contracted depending on the ratio of $a^\prime$ to $a$. One caveat about this result: we are specializing to accelerating observers lined up at $x_0 = X_0 = 1/a$ (or $a^\prime$, etc.). This is the hyperbolic equivalent of studying a series of concentric circles, but the point is that placing the hyperbola (or circles) at generic locations mixes up the inertial and accelerating effects, so by only studying the concentric orbits we cleanly study only the effects of relative acceleration without having to convolve the result with additional linear motion.

Of course, in a world with Cartesian coordinates, this is just a fun exercise in thinking physically with coordinates, but in a world with gravity (the next sections) it is \emph{mandatory}. Without the safety net of a preferred coordinate system (the affine/Cartesian coordinates), the next best thing is coordinates that are \emph{adapted} to certain physical observers. So far we have discussed two sets of coordinates adapted to different observers: Cartesian coordinates adapted to inertial observers, and Rindler coordinates adapted to uniformly accelerating observers. We will discuss this more in section \ref{ch:AppsII:BH}, but first we have a little more book-keeping to do with Rindler coordinates. 

\subsection{Thinking Globally}

The analysis above covers the life and times of an observer accelerating in the positive $x$ direction. As we've seen, that observer (and their personal Rindler coordinates) will only ever see one quarter of Minkowski space, region I in figure \ref{fig:ch5:Rind:rindler_mink}. Nevertheless, that observer's uniformly accelerating world could conceivably give rise to its own uniformly accelerating Einstein (UA-Einstein), who would discover the much cleaner Minkowski coordinate system. UA-Einstein would then be impossibly tempted to consider the full potential range of these Minkowski coordinates, and speculate about the existence of parallel, disconnected universes populated by similar yet distinct classes of observers, none of whom can ever interact with each other. 

UA-Einstein would now have at his disposal a single non-singular metric and enormous coordinate patch of which his world only spans a quarter the volume. Having been born and raised a uniformly accelerating being, he would then seek to interpret the other quadrants in terms of physics familiar to him---hyperbolic trajectories. The easiest quadrant to interpret is region III, which can be obtained from his Rindler world by simply extending the range of his $X$ coordinate to cover negative values. Uniformly accelerating observers on either side of $X = 0$ would never hear from each other, but UA-Einstein could very reasonably argue that in principle there could exist a mirror universe of reflected accelerating observers described by the same $(T,X)$ coordinates and metric as his own world. Worldlines of fixed positive $X$ translate to hyperbola in Minkowski space with positive spatial coordinate, and in the same way, negative $X$ describe hyperbola with negative spatial coordinate in Minkowski space, so quadrants I and III in \ref{fig:ch5:Rind:rindler_mink} can be understood through the coordinate transformation:
 \begin{equation}
    \label{eq:ch5:Rind:RindToCart13}
    x_{I/III}^{\mu}(X^{-1}) = \mqty( X\sinh(a T) \\ X\cosh(a T)).
\end{equation}
where now both coordinates run over all real values (except $X = 0$), and each positive $X$ coordinate is associated with the worldline of an accelerated observer with acceleration $a = 1/X$, while each negative $X$ coordinate corresponds to observers accelerating in the negative $x$ direction. The only catch is that the worldlines of observers with negative $X$ are now parameterized by $\tau \to -\tau$, but this could also be avoided by taking $T \to -T$ in quadrant III instead.

Regions I and III are the only ones that describe the trajectories of real physical accelerated observers. UA-Einstein would still however be determined to interpret the remaining volume in a similar way, through the lens of a uniformly accelerating observer. With a judicious application of the Tilted Head Theorem, this theorist would surely notice that the hyperbolic trajectories of fixed $X$ in the regions I and III could instead populate regions II and IV if they were at fixed $T$, living life by moving through the $X$ direction. This would be adapted to a world in which the roles of space and time were reversed, populated by what we would perceive as imaginary some spacelike observers who flow through what we consider proper distance instead of proper time. This observer would live on some worldline $X^\mu(\gamma(s)) = (1/a, s)^\text{T}$ in their perverse world, with its bizarro metric
\begin{equation}
    \label{eq:ch5:Rind:bizzRind}
    \dd s^2_{II/IV} = a^2T^2 \dd X^2 - \dd T^2,
\end{equation}
and correspondingly reversed coordinate transformation to Minkowski space
\begin{equation}
    \label{eq:ch5:Rind:RindToCart24}
    x_{II/IV}^\mu(X^{-1}) = \mqty( T\cosh(a X) \\ T\sinh(a X)).
\end{equation}
In the global picture, these worldlines would correspond to the spacelike hyperbolic worldlines $x^\mu(\gamma(s)) = a^{-1}(\cosh(as), \sinh(as))^\text{T}$. Obviously in Minkowski space, and from the perspective of the uniformly accelerating Rindler observers, these regions are physically meaningless---physical meaning to these regions could only be found in some bizarre world where timelike and spacelike coordinates somehow magically trade places in certain regions of spacetime. 

This exercise may have seemed somewhat frivolous, but it is an excellent warmup for the exploratory character of work in general relativity. In a general gravitating world, we do not have privileged inertial observers who authoratatively define the global coordinates and geometry, we have to start from the perspective of some representative observer, explore their world as fully as possible, then take seriously the possibility of a world beyond their purvue, investigating as much physics as can be imagined. This process of coordinate transforming your way through an event horizon to a bigger world is called \emph{extending coordinates}, and pursuing the extension to its greatest extent as we have just done is called a \textbf{maximal extension}\index{Maximal Extension}. 

\section{Recap}
\label{ch:AppsI:sec:recap}

\begin{itemize}
    \item The source-free Einstein equations support wave-like perturbative solutions akin to electromagnetic waves
    \item Solving for these \textbf{gravitational wave} solutions involves a great deal of choices to fix a great many degrees of freedom (both of a local and a boundary nature)
    \item Gauge symmetry associated with the non-uniqueness (and non-physicality) of coordinates provides the majority of the simplification, and the \textbf{Transverse-Traceless gauge} in particular reduces the metric perturbation to two (complex, momentum-dependent) degrees of freedom, while physical boundary conditions (a monochromatic wave propagating in the positive-$z$ direction) fix this further to only two degrees of freedom, or \emph{polarizations} ``plus'' $A_+$ and ``cross'' $A_\times$ 
    \item The transverse nature of the TT gauge leads to an entire class of geodesics determined by fixed $x-y$ coordinates and linear motion in the $t-z$ plane
    \item Despite their fixed relative coordinates, geodesics in this plane will \emph{physically} wibble-wobble as a gravitational wave passes through, since the magnitude of vectors is determined by the metric tensor, and physical (i.e.,~proper) distances are integrals over lengths of tangent vectors
    \item The last points bear repeating: \emph{coordinates are not physical}, the only physical distances are \emph{proper distances/times}
    \item As coordinates are only mathematical tools, it is very important to tailor our choice of coordinates to the physics under study
    \item For gravitational waves in the TT gauge, coordinates are adapted to follow inertial observers in the transverse plane
    \item In Minkowski space, coordinates can be adapted to the life and times of a uniformly accelerated observer---these are \textbf{Rindler coordinates} 
    \item A Rindler observer sees an \textbf{event horizon}. Signals from across this horizon will never reach the Rindler observer
    \item The Rindler horizon can be thought of as a consequence of length contraction and time dilation, that the accelerated observer's relativistic velocity (rapidity) relative to inertial observers increases without bound, so it sees inertial observers' clocks as running slower and slower
    \item Rindler coordinates only cover a quarter of Minkowski spacetime. They can however be \emph{extended} to the rest of Minkowski space by considering \emph{similar} coordinate charts in the other 3 quadrants (one simply the time-reversed image of the original chart, and the others representing a analogous spacelike ``accelerated'' observer)
\end{itemize}

\begin{subappendices}

\section{Sourcing Gravitational Waves}
\label{ch:AppsI:app:GWSources}

In section \ref{ch:AppsI:sec:GW} we solved for linear perturbations to the vacuum solution of general relativity with vanishing local stress-energy\footnote{In fact, wavelike local vacuum solutions can be found in full non-linear generality as well, though with a little extra effort. See e.g.~\cite[\S 17]{griffiths_exact_2009}.}. Even though we found our gravitational wave solutions with \\$T_{\mu\nu} = 0$, they do not represent \emph{global} vacuum solutions (except when their amplitudes vanish) because that is uniquely\footnote{Unless there is a non-zero cosmological constant in which case it is uniquely (anti)-de Sitter space---see \ref{ch:AppsII:Cosmo}.} Minkowski space, so non-trivial physics must be encoded in their boundary conditions. Here we will quickly run through the process of injecting this physics for a somewhat idealized astrophysical example (which derivation will mostly follow the arguments in \cite{schutz_2009}). 

Imagine a universe completely empty except for a single spatially-localized astrophysical system that will serve as the source of stress-energy for the universe. For ease, we will preserve the linearity of the field equations everywhere, so assume $\abs{\kappa T_{\mu\nu}} \sim \mathcal{O}(h_{\mu\nu})$ over a region of linear spatial dimension $\sim \mathcal{O}(\epsilon)$ and vanishing elsewhere. Since the stress-energy is no longer vanishing over the whole domain we will have to work with the non-trace-reversed field equations, but this is easy enough to accommodate by absorbing the trace-reversal into a re-definition of the metric perturbation. That is, we will work with $\overline{h}_{\mu\nu} = h_{\mu\nu} - \frac 12 \eta_{\mu\nu}h$, in terms of which the Lorenz gauge Einstein field equations become
\begin{equation}
    \label{eq:ch5:srcs:EFEs}
    \Box_0 \overline{h}_{\mu\nu} = -2\kappa T_{\mu\nu}. 
\end{equation}
Taking it a bit further, use that the stress-energy is unconstrained in time to decompose it and the metric perturbation in a Fourier series, $T_{\mu\nu} = \int \dd \omega\, e^{i\omega t}\, \tilde T_{\mu\nu}(\omega, \vec{x})$, and $h_{\mu\nu} = \int \dd \omega\, e^{i\omega t}\,\tilde h_{\mu\nu}(\omega, \vec{x})$. For simplicity, assume the Fourier-space distributions are sharply peaked about some fixed $\omega_0 = \Omega$, then drop the tildes and find
\begin{equation}
    \label{eq:ch5:srcs:EFEft}
    \left( \Omega^2 + \nabla_0^2 \right) \overline{h}_{\mu\nu} = -2\kappa T_{\mu\nu}. 
\end{equation}
where now it is understood that the stress-energy and metric perturbations are functions of $\Omega$ and not $t$. 

We do not seek to accurately model the internal gravitational physics of some fun astrophysical lab, just to detect its consequences on spacetime far away, so start by looking outside the source, where $T_{\mu\nu} = 0$. The solutions to \eqref{eq:ch5:srcs:EFEft} are just the plane waves we found earlier, but in this case they are a poor choice of basis. We will have to apply boundary conditions both at infinity and at a small local Gaussian surface around the source, so the basis better suited to the problem is one associated with spherical-polar coordinates instead (with origin somewhere inside the source). These can be found most conveniently straight from the vacuum field equation by writing the spatial Laplacian in spherical-polar coordinates, using separation of variables, and identifying the angular solutions as the spherical harmonics $Y_{\ell m}(\theta, \phi)$ and the radial functions as spherical Bessel functions $j_{\ell}(\Omega r), y_{\ell}(\Omega r)$ (see e.g.~\cite[\S 10.47]{NIST:DLMF}). Actually, with some foresight it will be more convenient to use the spherical Hankel function basis for the radial functions, so we'll write the general metric perturbation solution exterior to the source as
\begin{equation}
    \label{eq:ch5:srcs:genSol}
    \overline{h}_{\mu\nu}(\Omega, \vec{x}) = \sum_{\ell, m} \left( A^{(\ell,m)}_{\mu\nu} h^{(1)}_{\ell}(\Omega r) + B^{(\ell,m)}_{\mu\nu}h^{(2)}_{\ell}(\Omega r) \right)Y_{\ell m},
\end{equation}
where $h_{\ell}^{(1,2)}$ are the spherical Hankel functions of the first and second kind, $A$ and $B$ are integration constants, and $\ell \in \mathbb{Z}_0^+$ (the non-negative integers), and $m \in [-\ell, \ell]$. All of these functions and integration constants are complex at the moment, so we will remind ourselves to take the real part at the end of the caluclation.  

The expression \eqref{eq:ch5:srcs:genSol} is the general solution for our exterior gravitational waves in spherical-polar coordinates, so now we have to use gauge conditions and physical boundary conditions to fix the integration constants. Since we have them this time, it's actually more convenient to fix the physical conditions first. Asymptotically far from the source, for instance, there ought not be gravitational waves emanating from \emph{further} away, so we must impose that the large-$r$ behaviour is that of a radially outgoing wave, which selects $B = 0$. Easy enough. The near-source boundary condition is a little trickier with the form of $\overline{h}_{\mu\nu}$  not being known within the source, but can be handled neatly using the fact that we know the solutions to the homogeneous equation $(\Omega^2 + \nabla^2)f = 0$  (i.e., the spherical Bessel functions and spherical harmonics). The only spherical Bessel function that is finite at the origin is the first kind $j_{\ell}(\Omega r)$ , so we can focus in on the behaviour of the metric perturbation and stress-energy near the source by projecting out the $j_{\ell}(\Omega r)$  component and integrating over the source:
\begin{equation}
    \label{eq:ch5:srcs:proj}
    \int_{\mathcal{B}_{\epsilon}} \dd^3 x \,j_{\ell}(\Omega r)Y^*_{\ell m}\left(\Omega^2 + \nabla^2 \right)\overline{h}_{\mu\nu} = -2\kappa \int_{\mathcal{B}_{\epsilon}} \dd^3 x \, j_{\ell}(\Omega r) Y^*_{\ell m} T_{\mu\nu},
\end{equation}
where the integral\footnote{For those who have read the appendix on integration, \ref{ch:curve:app:int}, note that the leading order metric is Minkowski, so $\sqrt{-g} \sim 1$.} is over a ball of radius $\epsilon$ centred at the origin. Integrating by parts twice on the LHS and using that $(\Omega^2 + \nabla^2)j_{\ell}(\Omega r)Y^*_{\ell m} = 0$, the LHS reduces to an expression on the boundary of the source which can be evaluated using the \emph{exterior} form for $\overline{h}_{\mu\nu}$ , allowing us to pin down the second integration constant. Substituting \eqref{eq:ch5:srcs:genSol} (with $B_{\mu\nu} = 0$ ), the LHS evaluates exactly to $iA_{\mu\nu}^{(\ell, m)}/\Omega$ , so the amplitude of the emitted gravitational wave is given by
\begin{equation}
    \label{eq:ch5:srcs:solveA}
    A_{\mu\nu}^{(\ell, m)} = 2i\kappa J_{\mu\nu}^{(\ell, m)},
\end{equation}
where the source term is defined as
\begin{equation}
    \label{eq:ch5:srcs:defJ}
    J_{\mu\nu}^{(\ell, m)} := \int \dd^3 x\, j_{\ell}(\Omega r) Y^*_{\ell m}T_{\mu\nu}.
\end{equation}
This is the essential result, that a simple oscillatory source generates spherical gravitational waves at the same frequency with a magnitude that depends on the spatial shape of the source stress-energy. Obviously one should make all of this more precise and realistic, but there are libraries full of these treatments already, and the aim of this work was just to set you up to go explore them.

We'll just wrap up with one quick observation. Most realistic astrophysical distributions are highly non-relativistic and very far away, so the limit $\Omega \epsilon \ll 1$  is very appropriate, in which case the $s$-wave ($\ell = 0$) mode of the gravitational wave strongly dominates the amplitude. In that limit, the source term $J_{\mu\nu}^0$  only has spatial components to leading order, which can (using conservation of stress-energy) be written as
\begin{equation}
    \label{eq:ch5:srcs:quadSrc}
    J_{ij} \approx -\frac {\Omega^2} 2 I_{ij},
\end{equation}
where $I_{ij}$ is the quadrupole moment of $T^{00}$ (i.e., roughly the quadrupole moment of the mass density distribution of the source). That the \emph{leading} contributor to the generation of gravitational waves in common gravitational systems is the quadrupole moment of the system helps explain why they are so weak and hard to spot; it is not sufficient for a large gravitationally-active system to be dynamic to generate appreciable gravitational radiation, but it also has to be rather rapidly changing shape in a typically subdominant way. In the universe we can readily observe, that pretty much leaves us with rapidly orbiting, tightly bound systems of heavy objects like systems of black holes and neutron stars, and even then only in the wild closing acts of their dances. These are exactly the systems that have started to be observed with the new gravitational wave observatories. 


\end{subappendices}

\chapter{Applications Part II: Sources}
\label{ch:AppsII}

Finally we approach the two most prominent examples of general relativity with sources: black holes, and cosmology. Again in analogy with electromagnetism, these are the solutions to point-source and continuous medium sources (respectively), though only the black hole solutions maintain a tight correspondence in interpretation (i.e., mapping cleanly to the Coulomb potential, while the cosmological connection to fields in media is perhaps a bit murkier). 

\section{Point-Source EFEs: Black Holes}
\label{ch:AppsII:BH}

Point sources are some of the most important geometries to study in General Relativity. On large enough scales, any astrophysical object is approximately a point-like source of energy density, so the solutions will generically describe almost every gravitationally-relevant individual object in the universe to a good approximation. This is exactly the case in Newton's theory; the Newtonian potential generated by a point-source (i.e., the Coulomb potential for gravity) is the starting point of virtually any Newtonian calculation in astrophysics. 
In principle, the procedure we need to follow should be as straightforward as writing\footnote{Note that this form of the stress energy means we have chosen coordinates aligned with the centre of mass of the source.} $T\indices{_{0}^0} = -\rho = -\frac{M}{\sqrt{-g}} \delta^{(3)}(\vec{x})$ with the rest of $T\indices{_\mu^\nu} = 0$ and solving the corresponding Einstein field equations. The Dirac delta function is less of a ``function,'' however, and more of a ``boundary condition.'' Look to quantum mechanics and the Schr\"odinger equation with a delta function potential. The procedure there is to solve the \emph{free} Schr\"odinger equation on either side of the delta function's argument (typically $x = 0$), then patch the solutions on either side together with continuity and a boundary condition determined by integrating the equation of motion across the delta function's argument. We will follow a similar path here.

What does it mean to solve the Einstein field equations in vacuum everywhere \emph{except} the specific location of the point mass? We already know the vacuum solution to the Einstein field equations, it is the \emph{unique} Minkowski metric (possibly perturbed by gravitational waves, but here we will seek the lowest-energy configuration). But hidden in that statement is an assumption about boundary conditions, that the metric and its derivatives are all smooth at every point. Removing a specific point and applying a unique boundary condition at that favoured point is akin to selling off the spatial translation invariance---or affine structure---of Minkowski space, as well as the Lorentz boost invariance (there is a special point in space, it ought not mix with time). 

So fine, we still know what Minkowski space looks like with only the remaining rotation invariance and time-translation, surely it is just the Minkowski metric in spherical coordinates, as those are adapted to spatial rotation-invariance, and only defined on Minkowski space with a single point removed:
\begin{equation}
    \label{eq:ch6:BH:polMink}
    \dd s^2 = -\dd t^2 + \dd r^2 + r^2 (\dd\theta^2 + \sin[2](\theta) \dd\phi^2).
\end{equation}
But this metric \eqref{eq:ch6:BH:polMink} still \emph{is} the Minkowski metric, even though it's written in a different coordinate system, so it must still satisfy all of the symmetries of Minkowski space, even if these coordinates obscure some of them. Phrased another way, the symmetries of Minkowski space are properties of the \emph{geometry} of the space, not any particular coordinate system. How do we see this explicitly? Is there a way to quantify a symmetry of a metric? Indeed there is: \emph{Killing vectors}\index{Killing Vectors}.

\subsection{Symmetries Made Manifest: Killing Vectors}
\label{ch:AppsII:BH:killingVec}

A symmetry of spacetime is something that when done \emph{actively} to the points on the manifold leaves the metric tensor invariant; roughly speaking if $\phi: M \to M$ and $g(\phi(p)) = g(p)$, then $\phi$ is a symmetry. This is very abstract, but through the magic of coordinate charts can be made more tangible by adapting a set of coordinates to the symmetry in question. Symmetric coordinates can be ``aligned'' with a symmetry such that the composition $X_\phi^\mu(\phi(p)) = X_\phi^\mu(p) + C^\mu$, where  $C^\mu \propto \delta^\mu_\sigma$ with $\sigma$ a specific fixed direction. The most obvious example of how this works is translations in Cartesian coordinates. In that case, the effect of the map $\phi$ is exactly what we imagined the symmetry to be; in Cartesian coordinates, a translation in the $\nu$ direction is $X_C^\mu(p) \to X_C^\mu(p) + C\delta^\mu_\nu$. The Jacobian of this transformation is the identity, so as long as components of the metric tensor don't depend on the coordinate $X_C^\nu$ itself, the form of the metric tensor is exactly the same in the translated coordinates, hence clearly ``invariant'' under the symmetry. The inverse holds as well in fact, if the components of the metric tensor in some coordinate system are independent of one of those coordinates, then translations in that direction correspond to a symmetry of spacetime. 

This is an important property of symmetries, but it can't be \emph{enough}. Sure we can always adapt a coordinate system to a single symmetry, but we cannot in general find a single coordinate system simultaneously adapted to \emph{all} the symmetries of a spacetime---the most obvious counter-example being Minkowski space. In Minkowski space, we know we have the symmetries of translations along the spatial and time directions (and indeed Cartesian coordinates are impressively adapted to all four of those symmetries) but we also know the space to be symmetric with respect to the 3 independent Lorentz boosts and the additional 3 independent spatial rotations, and certainly we can't use 10 coordinates to represent a 4-dimensional space! So we must somehow find a quantity that can characterize a symmetry no matter the coordinates used, and just such a solution can be found with a little intuition from classical mechanics. In classical mechanics, we know from Noether's theorem that continuous symmetries of the action correspond to conserved currents and charges. The same is in fact true in general relativity, and a complete discussion of the topic would start with Noether's theorem on a general differentiable manifold, but it suffices for our purposes to be a little bit sneaky and use our advance knowledge that continuous symmetries of spacetime correspond to conserved \emph{momenta}, and hence should have something to do with the geodesic equation (the GR analogue of Newton's second law, remember).

So consider again the geodesic equation, this time written from the get-go in terms of the four-momentum instead of the four-velocity\footnote{This calculation follows closely the calculation in \cite[\S 3.8]{carroll_2019}} (just pass a factor or two of $m$ through, the right-hand side is 0, no one will notice):
\begin{equation}
    \label{eq:ch6:BH:geod1}
    m\frac{D}{d\tau}p^\mu = p^\nu \nabla_\nu p^\mu = 0.
\end{equation}
Using metric compatibility, we may lower the $\mu$ index, and write
\begin{equation}
    \label{eq:ch6:BH:geod2}
    m\frac{\dd}{\dd\tau}p_\mu - \Gamma^\alpha_{\mu\nu}p^\nu p_\alpha = 0,
\end{equation}
(recalling that the covariant derivative acting on covectors takes a negative on the connection coefficients, and the definition of the non-covariant directional derivative $\frac{\dd}{\dd\tau} V_\mu = U^\nu\partial_\nu V_\mu$). The important part comes with the second term:
\begin{equation}
    \label{eq:ch6:BH:geod3}
    \Gamma^\alpha_{\mu\nu}p^\nu p_\alpha = -\frac 12 g^{\alpha\rho}\left( g_{\mu\nu,\rho} - g_{\mu\rho,\nu} - g_{\nu\rho,\mu} \right)p^\nu p_\alpha,
\end{equation}
but the overall inverse metric contracted with the covariant four-momentum just raises the index $g^{\alpha\rho}p_\alpha \to p^\rho$, and then symmetry in $\nu$ and $\rho$ (i.e., $p^\nu p^\rho = p^\rho p^\nu$) eliminates the first two (anti-symmetric) terms, leaving only the third:
 \begin{equation}
    \label{eq:ch6:BH:geod4}
    \Gamma^\alpha_{\mu\nu}p^\nu p_\alpha = \frac 12g_{\nu\rho,\mu} p^\nu p^\rho.
\end{equation}

Putting it altogether, we find
\begin{equation}
    \label{eq:ch6:BH:geod5}
    m\dv[]{}{\tau} p_\mu = \frac 12 g_{\nu\rho,\mu}p^\nu p^\rho.
\end{equation}
In other words, in a coordinate system adapted to some symmetry such that the metric components are independent of the coordinate $x^*$, the component of covariant momentum along that direction  $p_*$ is \textbf{conserved} along the particle's worldline. Notice that \eqref{eq:ch6:BH:geod5} is not tensorial (the derivatives are not covariant), but is still valid in any coordinate system since both sides transform in the same non-tensorial way (as the Christoffel symbols were explicitly designed to do). However, the conservation equation
\begin{equation}
    \label{eq:ch6:BH:cons1}
    \dv[]{}{\tau} p_* = 0,
\end{equation}
is explicitly only valid in the coordinate system for which $\partial_* g_{\nu\rho} = 0$. Notice that the Minkowski metric in Cartesian coordinates is independent of all four coordinates, so \eqref{eq:ch6:BH:cons1} is the statement that energy and linear momentum are all independently conserved along the life of every inertial observer, as they should be. 

The conservation equation \eqref{eq:ch6:BH:cons1} is very close to what we're looking for, but its coordinate-dependence is stubbornly impeding our progress---we need to generalize the statement. The reason this conservation equation is coordinate-dependent is because we started with the geodesic equation and in a sense ``discovered'' this result in a particular coordinate system. Suppose instead we go the other way and ask ``is there a direction along which momentum is conserved?'' In that case, consider a scalar quantity $p_\star$ defined as the \emph{projection} of the four-momentum along some direction $K^\mu$, i.e.: $p_\star := g(p, K) = g_{\mu\nu}p^\mu K^\nu = K^\nu p_\nu$. In a coordinate system $y^\mu$ adapted to a symmetry such that $\partial_{*}g_{\mu\nu} = 0$ (with $\partial_* := \partial_{y^*}$) for all metric components, and $K = K^\mu \partial_\mu = \delta^\mu_* \partial_\mu$ we have exactly $p_\star = p_*$. If the coordinate-dependent $p_*$ is conserved along a particle's worldline then so too must be $p_\star$, but its conservation is an explicitly coordinate-\emph{invariant} statement. Mathematically:
\begin{align}
    \label{eq:ch6:BH:cons2}
    \dv[]{}{\tau} p_\star &= \dv[]{}{\tau} \left( K^\nu p_\nu \right), \notag \\
                      &= p^\nu \partial_\nu \left( K^\nu p_\nu \right), \notag \\
                      &= p^\mu p_\nu \nabla_\mu K^\nu + p^\mu K^\nu \nabla_\mu p_\nu, \notag \\
                      &= p^\mu p_\nu \nabla_\mu K^\nu = 0, 
\end{align}
where we used the Levi-Civita connection's compatibility with contractions in the second line, and the geodesic equation in the third. It is worth stressing that \eqref{eq:ch6:BH:cons2} is entirely independent of the geodesic equation, even though that's what we used to motivate it. Conservation of $p_*$ falls out of the vectorial geodesic equation evaluated in a specific ($y^\mu$) coordinate system, but conservation of $p_\star$ is a consequence of properties of the metric and the projection vector $K^\mu$---properties which are coordinate-\emph{in}dependent, as they must be for geometric objects. These two statements overlap \emph{only} in the specific coordinate system adapted to the symmetry in question, but the latter statement is much more powerful since it can be made with any choice of coordinates.

Equation \eqref{eq:ch6:BH:cons2} is essentially what we're after, but we should tidy it up a little bit first. For one, we may again use metric-compatibility to swap the position of the $\nu$s and put the four-momenta on the same level: $p^\mu p^\nu \nabla_\mu K_\nu = 0$. For another, notice that the rank-two tensor $p^\mu p^\nu$ is explicitly symmetric, so contracting it with the anti-symmetric part of the second factor $\nabla_\mu K_\nu$ automatically vanishes, so the equation \eqref{eq:ch6:BH:cons2} is specifically a statement about the symmetric piece of the second factor. In other words, the conservation of a component of momentum due to a symmetry of spacetime is encoded in the equation
 \begin{equation}
    \label{eq:ch6:BH:killingEq}
    \nabla_{(\mu} K_{\nu)} = \frac 12 (\nabla_\mu K_\nu + \nabla_\nu K_\mu) = 0
\end{equation}
known as \textbf{Killing's equation}\index{Killing's Equation}\index{Killing's Equation}. This equation allows us to go backwards and say that any vector $K^\mu$ that satisfies Killing's equation is the vector associated with a symmetry of spacetime, and even though we may be writing it in some other coordinate system, there does indeed exist \emph{some} coordinate system in which it is a coordinate basis vector and the metric components are independent of the coordinate direction with which it is associated. Such a vector $K^\mu$ is known as a \textbf{Killing vector}\index{Killing Vector}\footnote{These are named after the mathematician Wilhelm Killing, which I for one find marvelously serendipitous on account of how the vectors essentially kill resistance to motion in their chosen directions.}. Back to Minkowski space.

\subsection{Hidden Minkowski Symmetries}
\label{ch:AppsII:BH:minkSym}

I have asserted that the spherical coordinate rendition \eqref{eq:ch6:BH:polMink} of the Minkowski metric still retains all ten symmetries of Minkowski space including translations---now we have the tool to see it. The Killing vectors associated with spatial translations are $X, Y, \text{and } Z$, with Cartesian components $\delta^\mu_1, \delta^\mu_2, \text{and } \delta^\mu_3$, respectively. In spherical coordinates, these components transform to\footnote{In covariant form---Killing's equation can't be made mixed by raising the index on the Killing vector because the indices are symmetrized, the mixed form would involve $\nabla_\mu V^\nu + \nabla^\mu V_\nu$.}:
\begin{align}
    \label{eq:ch6:BH:polSpatialK}
    X_\mu\dd x^\mu &= \dd x = \sin(\theta)\cos(\phi)\dd r + r \cos(\theta)\cos(\phi) \dd \theta - r \sin(\theta)\sin(\phi)\dd \phi , \notag \\
    Y_\mu\dd x^\mu &= \dd y = \sin(\theta)\sin(\phi)\dd r + r \cos(\theta)\sin(\phi) \dd \theta + r \sin(\theta)\cos(\phi)\dd \phi, \\
    Z_\mu\dd x^\mu &= \dd z = \cos(\theta)\dd r - r \sin(\theta) \dd \theta. \notag 
\end{align}
Making use of Killing's equation will then require the Christoffel symbols, which for the metric \eqref{eq:ch6:BH:polMink} vanish except for:
\begin{equation}
    \label{eq:ch6:BH:polChris}
    \begin{gathered}
        \Gamma^r_{\theta\theta} = -r \qquad \Gamma^r_{\phi\phi} = -r \sin[2](\theta) \qquad \Gamma^\theta_{r\theta} = \Gamma^\phi_{r\phi} = \frac 1r \\
        \Gamma^\theta_{\phi\phi} = -\sin(\theta)\cos(\theta) \qquad \Gamma^\phi_{\theta\phi} = \cot(\theta).
    \end{gathered}
\end{equation}

As a proof-of-concept, let's verify that $Z$ indeed satisfies Killing's equation in spherical coordinates. Clearly all elements with a timelike index vanish, so it remains to check the 6 independent combinations for the spatial indices. We have:
\begin{equation}
    \label{eq:ch6:BH:killZ}
    \begin{aligned}
        \nabla_r Z_r &= -\Gamma^\mu_{rr}Z_\mu &= 0,  \\
        \nabla_r Z_\theta + \nabla_\theta Z_r &= -\sin(\theta) - \sin(\theta) - 2\Gamma^\mu_{r\theta}Z_\mu &= 0, \\
        \nabla_r Z_\phi + \nabla_\phi Z_r &= -2\Gamma^\mu_{r \phi} Z_\mu &= 0, \\
        \nabla_\theta Z_\theta &= -r \cos(\theta) - \Gamma^\mu_{\theta \theta}Z_\mu &= 0,  \\
        \nabla_\theta Z_\phi + \nabla_\phi Z_\theta &= -2\Gamma^\mu_{\theta \phi}Z_\mu &= 0, \\
        \nabla_\phi Z_\phi &= -2(\Gamma^r_{\phi \phi}Z_r + \Gamma^\theta_{\phi \phi}Z_\theta) &= 0,  \\
    \end{aligned}
\end{equation}
as expected, and the same calculation for $X$ and $Y$ yields the same result.

\subsection{Breaking Symmetries}
\label{ch:AppsII:BH:breakSym}

Now that we've verified spherical coordinates preserve all of the symmetries of Minkowski space, let's get to breaking the ones we no longer want: spatial translations and Lorentz boosts. Actually, breaking Lorentz boosts is trivial. Boosts have the form $t X_\mu + x T_\mu$ (and similar for $y$ and $z$), so if $X_\mu, Y_\mu,$ and $Z_\mu$ fail to satisfy Killing's equation while $T_\mu$ still \emph{does} satisfy it, the boosts will automatically be broken (show this explicitly---note that this is the same story with breaking global rotations down to cylindrical symmetry).

We need to preserve $T_\mu$ and rotations, so here they are for reference:
\begin{equation}
    \label{eq:ch6:BH:polPreservedK}
    \begin{aligned}
        T &= -\dd t, \\
        \Phi &= r^2 \sin[2](\theta) \dd \phi, \\
        \Psi &= r^2 \cos(\phi) \dd \theta - r^2 \sin(\theta) \cos(\theta) \sin(\phi) \dd \phi, \\
        \Omega &= -r^2 \sin(\phi) \dd \theta - r^2 \sin(\theta) \cos(\theta) \cos(\phi) \dd \phi, 
    \end{aligned} 
\end{equation}
where $\Phi, \Psi,$ and $\Omega$ are rotations about the  $z$-, $y$-, and $x$-axes, respectively. All of the covectors in \eqref{eq:ch6:BH:polSpatialK} and \eqref{eq:ch6:BH:polPreservedK} (as well as the corresponding boosts) describe symmetries of the Minkowski metric \eqref{eq:ch6:BH:polMink}, just written in spherical coordinates; \emph{breaking} any of those symmetries can only be done by constructing a \emph{different} metric, but preserving the rest of the symmetries gives us clear guidance on how to do that. Case in point: notice that none of the Killing vectors we seek to preserve have any components in the $\dd r$ direction, while \emph{all} of the ones we seek to break do. Take the $rr$ component of Killing's equation for $Z_\mu$, for example, the first line in \eqref{eq:ch6:BH:killZ}. The $\Gamma^r_{rr}$ Christoffel symbol will never survive in any of the preserved Killing equations anyway, so if we make it non-zero, only the translations and boosts will ever know. If we also keep the metric diagonal for simplicity, we have $\Gamma^r_{rr} = -\frac 12 g^{rr}g_{rr,r}$, so let $g_{rr} = f(r)$ and we should be good!
\begin{equation}
    \label{eq:ch6:BH:putMetric}
    \dd s^2_\text{put} = -\dd t^2 + f(r) \dd r^2 + r^2\left( \dd \theta^2 + \sin[2](\theta) \dd \phi^2 \right). \qquad \text{(Putative point-source metric)}
\end{equation}
But sadly this is not quite enough. Indeed, while a general $f(r)$ does break the Killing equations for the translation vectors, it can not do so while retaining vanishing Ricci curvature (except when $f(r) = 1$), hence cannot describe a \emph{vacuum} solution to the Einstein field equations. 

Thanks to a little foresight, we will build on this putative metric because the factor $f(r)$ will come in handy shortly. It is now inevitable that we interfere with the preserved vectors \eqref{eq:ch6:BH:polPreservedK}, but to interfere as little as possible, we should look carefully at the time components. Only $T_\mu$ has a $\dd t$ component, so if we change the metric in a way that tampers with $\Gamma^r_{t \mu}$ as well as $\Gamma^r_{ij}$, we can be cautious about the time-like Killing vector while feeling safe that the rotational Killing vectors remain intact. Going back to \eqref{eq:ch6:BH:killZ}, a non-zero $\Gamma^r_{tt}$ means:
\begin{equation}
    \label{eq:ch6:BH:killZwT}
    \nabla_{(t} Z_{t)} = -\Gamma^r_{tt} Z_r = \frac 12 g^{rr}g_{tt,r}Z_r,
\end{equation}
which uses that we're still hoping to keep the metric diagonal, and that keeping $T_\mu$ as a Killing vector means the metric components must still be independent of $t$ (not to mention keeping the angles all the way out of the discussion). A non-zero result for \eqref{eq:ch6:BH:killZwT} is easy enough to achieve with a simple $g_{tt} = -q(r)$, so that the simplest metric we can construct that maintains time-translation and rotational symmetries while \emph{not} being symmetric under spatial translations or boosts is:
\begin{equation}
    \label{eq:ch6:BH:genSchw}
    \dd s^2 = -q(r) \dd t^2 + f(r) \dd r^2 + r^2 \left( \dd \theta^2 + \sin[2](\theta) \dd\phi^2 \right),
\end{equation}
with at the very least $q(r) \neq 1 \neq f(r)$ (it's a bit more stringent than that, but there are better ways to construct the metric if formality is your goal anyway). This is occasionally referred to as the \textbf{standard form}\index{Standard Form} of the general static spherically symmetric spacetime metric.

\subsection{Pinning it Down}
\label{ch:AppsII:BH:pin}

While we won't go into it here, it can be shown that highly symmetric spaces like Minkowski space or the almost-Minkowski space determined by \eqref{eq:ch6:BH:putMetric} are highly constrained by their symmetries, so there only a few distinct metrics not connected by coordinate transformations. After breaking six of those symmetries though, the metric \eqref{eq:ch6:BH:genSchw} describes a great many inequivalent spacetimes, so physics and boundary conditions must be used to narrow it down. Most importantly, we need to ensure the Einstein equations are satisfied---a general choice of $q(r),f(r)$ won't necessarily be flat, so we need to enforce $R_{\mu\nu} = 0$ everywhere except the origin. The non-zero Christoffel symbols are now the slightly modified ones from \eqref{eq:ch6:BH:polChris}:
\begin{equation}
    \label{eq:ch6:BH:fullChris1}
    \begin{gathered}
        \Gamma^r_{\theta\theta} = -\frac{r}{f(r)} \qquad \Gamma^r_{\phi\phi} = -\frac{r}{f(r)} \sin[2](\theta) \qquad \Gamma^\theta_{r\theta} = \Gamma^\phi_{r\phi} = \frac 1r \\
        \Gamma^\theta_{\phi\phi} = -\sin(\theta)\cos(\theta) \qquad \Gamma^\phi_{\theta\phi} = \cot(\theta),
    \end{gathered}
\end{equation}
and the now non-vanishing:
\begin{equation}
    \label{eq:ch6:BH:fullChris2}
    \Gamma^t_{tr} = \frac 12 \frac{q^\prime(r)}{q(r)}, \qquad \Gamma^r_{tt} = \frac 12 \frac{q^\prime(r)}{f(r)} \qquad \Gamma^r_{rr} = \frac 12 \frac {f^\prime(r)}{f(r)}.
\end{equation}
The Ricci curvature tensor turns out to be diagonal (this is thanks to symmetries, but ultimately the beast still has to be computed directly):
\begin{equation}
    \label{eq:ch6:BH:ricciSph}
    \begin{aligned}
        R_{tt} &= \frac {q^{\prime\prime}}{2f} - \frac 1 4 \frac {q^\prime}{f} \left( \frac {f^\prime} f + \frac {q^\prime} q \right) + \frac 1 r \left( \frac {q^\prime} f \right)   \\
        R_{rr} &= -\frac {q^{\prime\prime}}{2q} + \frac 1 4 \frac {q^\prime}{q} \left( \frac {f^\prime} f + \frac {q^\prime} q \right) + \frac 1 r \left( \frac {f^\prime} f \right) \\
        R_{\theta\theta} &= 1 - \partial_r \left( \frac r f \right) - \frac 12 \frac r f \left(\frac {f^\prime}{f} + \frac {q^\prime}{q}  \right) \\
        R_{\phi\phi} &= \sin[2](\theta)R_{\theta\theta}
    \end{aligned},
\end{equation}
where primes denote differentiation with respect to $r$. The Einstein equations then amount to setting each of the components in \eqref{eq:ch6:BH:ricciSph} to 0.

In fact, of the four non-vanishing components of the Ricci tensor, only two are linearly independent---this is as it should be since we only have the two functions $f(r)$ and $q(r)$ free. It is convenient to start with a linear combination:
 \begin{equation}
    \label{eq:ch6:BH:evalRicci1a}
    \frac {R_{tt}} q + \frac {R_{rr}} f = \frac 1 {rf}\left( \frac {f^\prime} f + \frac {q^\prime} q \right) = 0.
\end{equation}
For general $r$ and non-vanishing $f(r)$, this is only satisfied if the expression in brackets is 0. Assuming $f,q \neq 0$, multiply through by the product of the two, then find the term in parentheses reduces to:
 \begin{equation}
    \label{eq:ch6:BH:Ricci1b}
    \partial_r(fq) = 0, 
\end{equation}
which implies $fq =: C$ with $C$ a real constant. Notice that we started with two local degrees of freedom and four boundary degrees of freedom (since the EFEs are second order differential equations), but this solution directly relates $f$ and $q$ in their entirety, so all that remains is one local degree of freedom (say $q$) and two boundary degrees of freedom. In fact we can already pin down the value of one of the boundary degrees of freedom $C$ with a physical boundary condition; way out at the edge of space, we'd better not still be suffering from the effects of a point mass on the other side of the universe, so the metric must reduce to Minkowski. For large $r$ then, we must have $f\to 1$ and independently $q\to 1$ so that $fq \to 1$. But since $C$ is a constant, this must be its value \emph{everywhere} (except possibly exactly at the source).

Using that $f = 1/q$, the $\theta\theta$ equation cleans up very nicely:
\begin{equation}
    \label{eq:ch6:BH:Ricci1c}
    R_{\theta\theta} = 1 - \partial_r(rq) = 0,
\end{equation}
so that
\begin{equation}
    \label{eq:ch6:BH:solveQ}
    q(r) = 1 + \frac B r 
\end{equation}
for some real constant $B$. This $B$ is the one remaining integration constant and it requires some physical boundary condition to be fixed. The easiest way to do this is to note that for any finite constant $B$, there always exist large enough values of the radial coordinate such that the inverse-$r$ decay reduces the metric coefficients to the perturbative $q(r) \sim 1 + 2\phi(r)$, $f(r) \sim 1 - 2\phi(r)$ with $2\phi(r) = B/r$. As we saw in \ref{ch:EFEs:newt}, the weak field limit of general relativity reduces to Newtonian gravity to leading order, so the metric perturbation $\phi(r)$ equates to the Newtonian potential for the same point mass, which is known to be $\phi(r) = -GM/r$, hence the integration constant must be $B = -2GM$. If it feels a bit shakey relying on a Newtonian analogy to fix this integration constant instead of directly involving the source stress-energy tensor, see \ref{ch:AppsII:app:nsBC} for a more proper derivation. 

At the end of the day, the final expression for this metric is:
\begin{equation}
    \label{eq:ch6:BH:schwMetric}
    \dd s^2 = -\left( 1 - \frac {2GM} r \right) \dd t^2 + \left( 1 - \frac {2GM} r \right)^{-1} \dd r^2 + r^2 \dd\Omega^2,
\end{equation}
where $\dd\Omega^2 = \dd \theta^2 + \sin[2](\theta) \dd\phi^2$ is shorthand for the solid angle element. This metric was discovered almost immediately after Einstein published the theory in late 1915 by Karl Schwarzschild, a German physicist serving on the Eastern front in World War I. For his troubles, Karl died the following year, but this elegant solution still bears his name: the \textbf{Schwarzschild metric}\index{Schwarzschild Metric}. One final note here: although our derivation was pseudo-formal, a precise calculation reveals the Schwarzschild metric is the \emph{unique} static, spherically symmetric (but not translationally-invariant) vacuum solution to Einstein's field equations. This proof is known as \textbf{Birkhoff's theorem}\index{Birkhoff's Theorem} (and somewhat related to the \textbf{No-Hair theorem}).

\subsection{Investigating the Schwarzschild Metric}
\label{ch:AppsII:BH:Schw}
We have accomplished our first goal, \eqref{eq:ch6:BH:schwMetric} is the metric tensor of spacetime sourced by a point mass; the task now is to study our solution. Clearly something bizarre happens at $r = r_s := 2GM$. This radius $r_s$ is the \textbf{Schwarzschild radius}\index{Schwarzschild Radius}, and not only does the time component of the metric vanish at this point (as did the same component in the Rindler metric), but the radial component even diverges. The least we can take away from that observation is that these coordinates are not able to describe that radius specifically, but they should be acceptable everywhere else\footnote{Strictly speaking, the derivation in the previous section only fixed the metric outside $r_s$, but it is trivial to extend the radial coordinate to everywhere outside $r = 0$ except of course exactly at $r_s$.}, so let us explore the properties of the metric away from this point first, then find some better coordinates. For the most part, we will focus on the region exterior to the Schwarzschild radius which equally well describes the region outside \emph{any} static spherically symmetric mass (Birkhoff's theorem), but we will take seriously the region inside the Schwarzschild radius too, and when we do so the geometry can only correspond to a point-source energy density and is known as a \textbf{black hole}\index{Black Hole} (for reasons that will become apparent).

As ever, we begin with a quick look at geodesics. A general geodesic can be quite complicated, but we can simplify the problem with some symmetry arguments. First, we know that $\partial_\phi$ is a Killing vector (our coordinates are \emph{adapted} to the rotational symmetry in the $\phi$ direction), so any geodesic will have a (covariantly) constant component $U^\phi = A/r^{2} \sin[2](\theta(\tau))$ for some constant angular velocity $A$. Let's make our lives simple, fix $A = 0$ so we may fix the $\phi$ coordinate to our choosing. Looking back to \eqref{eq:ch6:BH:polPreservedK}, if we fix $\phi = 0$, then the $\theta$ direction is also a Killing vector (along that fixed $\phi = 0$ hypersurface only, mind), so we can similarly fix $U^\theta = C/r^2$. Again, let's look at the simplest case, so set $C = 0$ and fix $\theta = \pi/2$ so we're just sitting in the equatorial plane. Then the problem is reduced to the $t-r$ plane, and if we only look at null geodesics the problem can be entirely solved without ever lowering ourselves to suffering through the geodesic equation, we may be entirely satisfied with the constant magnitude $0 = -(1 - r_s/r)(U^t)^2 + (1 - r_s/r)^{-1}(U^r)^2$, or:
 \begin{equation}
    \label{eq:ch6:BH:setupNullGeod}
    U^t = \pm \frac{U^r}{1 - r_s/r}, 
\end{equation}
which has solutions 
\begin{equation}
    \label{eq:ch6:BH:nullGeods}
    t(s) - t_0 = \pm\left( r(s) - r_0 + r_s\ln( \frac{r(s) - r_s}{r_0 - r_s} ) \right),
\end{equation}
for some affine parameter $s$. Note the similarity to \eqref{eq:ch5:Rind:solveNullRind2}, the null geodesics in Rindler space; here again $t(s)$ depends logarithmically on $x(s)$, only there is a linear piece and some offsets involved as well. Take for example initial conditions such that $U^r_0 = -1$,  $r(s_0) = r_0 \gg 1$ so that $U^t_0 \to 1$ (i.e., from very far away, shine a light directly at the point-source). Initially we are very far from the gravitational source, so its influence is negligible which we see in the linear term dominating the logarithm for large arguments leading to approximately Minkowski trajectories. On approach to the Schwarzschild radius, the logarithmic term picks up the pace and thoroughly outperforms the linear term, replicating exactly the form of null rays in Rindler space and leading us to suspect $r_s$ as an event horizon. 

With the null geodesics being so similar to those of Minkowski space in Rindler coordinates, it is worth also investigating the non-geodesic paths of observers at fixed radii. In Rindler coordinates, those paths corresponded to the worldlines of uniformly accelerating observers, so could the same be true here? The four-velocity of an observer at some fixed radius $r_a$ is the unit timelike vector with support only in the time direction. That is, the four-velocity must satisfy $g_{\mu\nu}U^\mu U^\nu = g_{tt}(U^t)^2 = -1$, or 
\begin{equation}
    \label{eq:ch6:BH:schFourVel}
    U^t = \left( 1 - \frac {r_s} {r_a} \right)^{-1/2},
\end{equation}
choosing the positive sign on the square root to represent life moving forward in time. An observer with this four-velocity will experience a four-acceleration $a^\mu = U^\nu \nabla_\nu U^\mu = \delta^\mu_r \Gamma^r_{tt} = \delta^\mu_r \frac {r_s}{2r_a^2}$, using the Christoffels from \eqref{eq:ch6:BH:fullChris2}. Then the magnitude of the four-acceleration is $a = \sqrt{g_{\mu\nu}a^\mu a^\nu} = \sqrt{g_{rr}a^r a^r}$ or
\begin{equation}
    \label{eq:ch6:BH:accel}
    a = \frac {r_s}{2r_a^2} \left( 1 - \frac {r_s} {r_a} \right)^{-1/2},
\end{equation}
so as one might expect, the gravitational acceleration required to remain at a fixed radius\footnote{Hence the acceleration is outward, hence positive.} approaches the Newtonian result $r_s/2r_a^2 = GM/r_a^2$ for large radii, but possibly surprisingly soars to infinity near the Schwarzschild radius. 

In fact, the four-velocity alone already tells us something interesting about Schwarzschild geometry---\textbf{gravitational time dilation}\index{Gravitational Time Dilation}. The worldline found by integrating the four-velocity \eqref{eq:ch6:BH:schFourVel} is
\begin{equation}
    \label{eq:ch6:BH:fixedWL}
    x^\mu(\tau) = \mqty( \frac \tau {(1 - r_s/r_a)^{1/2}} \\ r_a ).
\end{equation}
As $r_a \to \infty$, the time coordinate approaches the proper time parameter so these coordinates are adapted both to observers at fixed radii, and in particular to observers infinitely far from the source, for whom the local geometry is approximately Minkowski. To asymptotic observers, those closer to a gravitational source will have clocks that run at a reduced speed---their times will appear dilated. Thinking about this in terms of acceleration, the effect is purely kinematic, it is the same time dilation associated with observers at fixed (hyperbolically concentric) relative accelerations, as depicted in figure \ref{fig:ch5:Rind:rindler_rind}, only where the reference accelerating observer is the asymptotically Minkowski observer, so there are no observers with smaller accelerations (hence no time contraction). 

Observers at fixed Schwarzschild radii are uniformly accelerating, but this is not very obvious from the form of the Schwarzschild metric. In a small enough region around any given radius however, the accelerating geometry can be teased out as follows. Define a region about some fixed radius $r_a$ in terms of $r \approx r_a + r_s\delta$, for a small perturbation  $\delta \ll r_a/r_s$ (scaling with $r_s$ is mandatory as it is the only physical scale in the problem). In this region, the Schwarzschild metric is approximately:
 \begin{equation}
    \label{eq:ch6:BH:approxSchw}
    \dd s^2 \approx -\left( 1 - \frac {r_s}{r_a} + \frac {r_s^2}{r_a^2}\delta \right)\dd t^2 + \left( 1 - \frac {r_s}{r_a} + \frac {r_s^2}{r_a^2}\delta \right)^{-1} \dd r^2 + r^2 \dd^2\Omega. 
\end{equation}
Defining a new coordinate 
\begin{equation}
    \label{eq:ch6:BH:hatr}
    \hat{r} := \frac {2r_a^2}{r_s}\sqrt{1 - \frac{r_s}{r_a} + \frac{r_s^2}{r_a^2}\delta},  
\end{equation}
the metric takes the form
\begin{equation}
    \label{eq:ch6:BH:localRindSchw}
    \dd s^2 \approx -(a_N \hat{r})^2 \dd t^2 + \dd \hat{r}^2 + r^2(\hat{r}) \dd^2\Omega,
\end{equation}
where $a_N := r_s/2r_a^2$ is the Newtonian gravitational acceleration, and the factor of $r^2$ appearing in the angular component now has to be thought of as a function of the new radial coordinate $\hat{r}$ (one could spell it out, but it would be an ugly mess for no benefit). 

In fact, the existence of this approximation is not entirely surprising. Recall the Newtonian approximation \eqref{eq:ch4:newt:guessNewton}; perturbing about a fixed radius safely inside the weak-field regime, we can approximate the Newtonian potential $\phi(r_a + \delta) \sim \phi_a + \phi^\prime_a\delta$ , in which case the metric takes the form:
\begin{equation}
    \label{eq:ch6:BH:approxNewt}
    \dd s^2 \approx -(1 + 2\phi_a + 2\phi^\prime_a\delta)\dd t^2 + (1 + 2\phi_a + 2\phi^\prime_a\delta)^{-1} \left( \dd r^2 + r^2 \dd^2\Omega \right), 
\end{equation}
and a coordinate transformation analogous to \eqref{eq:ch6:BH:localRindSchw} effects a local Rindler approximation of the metric. All this is saying is that in the weak gravitational field regime, the Newtonian acceleration appears uniform in a small enough coordinate patch. What makes this approximation interesting in the Schwarzschild geometry is that it is valid \emph{everywhere}, even near the horizon where the gravitational field is decidedly non-Newtonian. For this reason, we will take a moment to explore this Rindler structure a bit deeper, with some bias towards the stronger field regime.

\subsection*{Exploring the Local Rindler Metric}

The perturbative metric \eqref{eq:ch6:BH:localRindSchw} is defined in a small neighbourhood about an observer located at $r = r_a$, $\delta = 0$, or $\hat{r}_0 = \frac {2 r_s}{r_a^2} \sqrt{1 - \frac{r_s}{r_a}}$. This observer experiences a physical acceleration given by \eqref{eq:ch6:BH:accel}, but the acceleration parameter in the metric at the moment is only the Newtonian acceleration. The mismatch arises because we have only adapted the radial coordinate to our little motionless accelerating observer, the time coordinate is still adapted to the Minkowskian observers at infinity. To really see the Schwarzschildian world through the eyes of a fixed local observer, we have to build the time dilation right into the local time coordinate (picture some crazy sci-fi alien somehow stuck on a giant rigid surface surrounding the centre of mass of some big, round, astronomical body). Writing  $ \hat{t} := (1 - r_s/r_a)^{1/2} t$, the local perturbative metric takes the form
\begin{equation}
    \label{eq:ch6:BH:localRindSchwDilated}
    \dd s^2 \approx -(a \hat{r})^2 \dd \hat{t}^2 + \dd \hat{r}^2 + r^2(\hat{r})\dd^2\Omega,
\end{equation}
where now the acceleration parameter $a$ exactly lines up with the relativistic acceleration \eqref{eq:ch6:BH:accel}, and moreover, tracks with our reference observer following the simple worldline:
\begin{equation}
    \label{eq:ch6:BH:localRindWorldline}
    \hat{x}^\mu = \mqty( \tau \\ a^{-1}).
\end{equation}
(Note we restrict our attention to the $\hat{t}$-$\hat{r}$ plane).

\begin{figure}[htp!]
    \centering
    \includegraphics[width=0.8\textwidth]{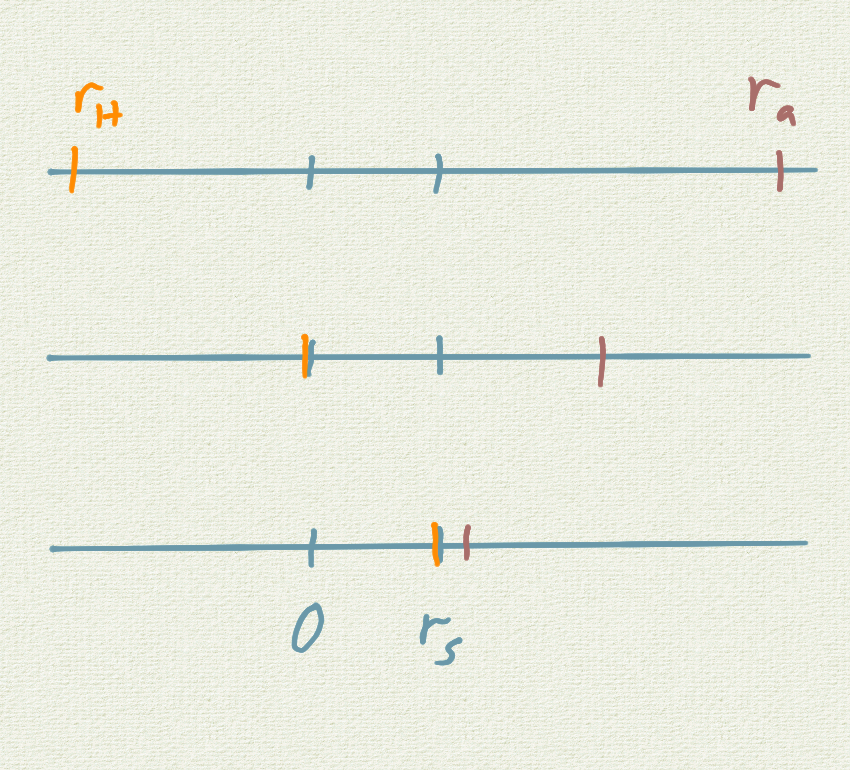}
    \caption[The Schwarzschild-Rindler horizon as a function of distance from the Schwarzschild horizon.]{In local Rindler coordinates about a point $r_a$, the Rindler horizon at $r_H$ starts impossibly distant and very fictional (top) and ``catches up'' to the real Schwarzschild horizon as $r_a \to r_s$ (bottom). The point at which the Rindler horizon becomes a real radius and can begin to be taken seriously is when $r_a = 2r_s$ (middle). }
    \label{fig:ch6:BH:RindHorizons}
\end{figure}

The $(\hat{t},\hat{r})$ coordinates in which the local Rindler metric is defined are very restricted, subject to $\delta \ll r_a/r_s$ (large enough at times, but still always small in the grand scheme of things), but it is still worth looking at the hypothetical horizon and null rays. The Rindler horizon in these coordinates would be found where $\hat{r} = 0$ (if it were in the coordinate domain), that is, where $\delta_H = \frac {r_a}{r_s}(1 - \frac {r_a}{r_s})$ or $r_H = r_a(2 - \frac {r_a}{r_s})$. From $r_a \ge 2r_s$, this results in a very distant and fictional negative global radius, but as $r_a$ approaches $r_s$, that fictional radius increases, becoming more and more reasonable until it eventually aligns exactly with the Schwarzschild radius when $r_a = r_s$. In other words, a tiny local scientist will extrapolate out a Rindler horizon that is much further away than the real Schwarzschild horizon, but the projection becomes more and more realistic the closer the observer gets to $r_s$ (see \ref{fig:ch6:BH:RindHorizons}). In all cases, the local observer will see light rays moving through their space logarithmically, seemingly chasing that distant horizon.

This is all good and physical locally, but it takes a chunk of work to get to see these details for any particular observer starting from Schwarzschild coordinates. To study the life and times of observers at different radii, one must start from Schwarzschild and perform a completely different coordinate transformation each time (see the transformation \eqref{eq:ch6:BH:hatr} and change $r_a$ to  $r_b, r_c,$ etc.---each reference radius defines an entirely new coordinate system). But since we know this structure exists locally about any point in space, it would be much more revealing if we could find a coordinate system that was defined once-and-for-all globally but still demonstrated this underlying local Rindler structure.

\subsection*{Coordinate Hunting: Rindler-Tortoise Coordinates}
Searching for a useful coordinate transformation is a bit more of an art than a science. Generally speaking, we are trying to find a global coordinate system that makes clear at least some of the properties of Rindler space that are seen locally in the Schwarzschild metric (of course it can't be all them, otherwise it would be globally Rindler space, which is decidedly not a point-source geometry). One of the most common ways to do something like this is to focus on null geodesics, since their constant vanishing norm cleanly relates one metric component to the others, and through the notion of lightcones they define one of the most important characteristics of any spacetime---its causal structure. To that end, we'll look for a new radial coordinate that encodes radial null Schwarzschild geodesics $\dd t = (1 - r_s/r)^{-1} \dd r$ in a Rindler form $\dd t \propto R^{-1} \dd R$:  
\begin{align}
    \label{eq:ch6:BH:deriveR}
    \left( 1 - \frac {r_s}{r} \right)^{-1} \dd r &= \frac {\alpha r_s}{R} \dd R, \notag \\
    \quad r + r_s \ln( \frac r {r_s} - 1) &= \alpha r_s \ln(R/r_s) + \alpha r_s \ln K, \notag \\
    \implies \quad R(r) &= K r_s e^{r/\alpha r_s} \left( \frac{r}{r_s} - 1 \right)^{1/\alpha}.
\end{align}
The scale $r_s$ is mandatory as it is the only physical scale in the problem, but the proportionality constants $\alpha$ and $K$ are otherwise free, and a convenient choice turns out to be $\alpha = K = 2$, so we define the \textbf{Rindler-Tortoise}\index{Rindler-Tortoise Coordinate} coordinate (see aside \ref{aside:tortoise} for an explanation of the name):
\begin{equation}
    \label{eq:ch6:BH:defRindTort}
    R(r) := 2r_s e^{r/2r_s}\left( \frac{r}{r_s} - 1 \right)^{1/2},
\end{equation}
in terms of which, the Schwarzschild metric is:
\begin{equation}
    \label{eq:ch6:BH:schwRindTort}
    \dd s^2 = e^{-r/r_s} \frac{r_s}{r} \left\{ -\left( \frac{R}{2r_s} \right)^2 \dd t^2 + \dd R^2 \right\} + r^2(R) \dd^2\Omega.
\end{equation}

The Schwarzschild metric in Rindler-Tortoise coordinates \eqref{eq:ch6:BH:schwRindTort} is \textbf{conformal}\index{Conformal Metric} to the Rindler metric in the $t-r$ plane, meaning the metrics differ only by an overall function of spacetime. Conformally related metrics are (generically) \textit{distinct} metrics not related by coordinate transformations, and therefore represent distinct physics. Nevertheless, conformally related metrics are still similar in the sense that \textit{local} physics looks the same (if scaled a bit). In our case, this is exactly what we were after, a globally-defined coordinate system (at least exterior to the Schwarzschild radius) that looks locally like the perturbative form of the metric in Schwarzschild coordinates. Exactly how locally is ``locally'' depends on how much the conformal factor $f(r) = e^{-r/r_s}\frac{r_s}{r}$ varies compared to the rest of the metric components. In this case, the $g_{rr}$ term always goes as the perturbation, so as long as the perturbation remains a perturbation, the leading term in the conformal factor will suffice. The $g_{tt}$ component is where the usefulness of approximating the metric finds its limits. Starting from $R$, we think of the original radial coordinate in terms of the new one, $r(R)$ implicitly defined by the equation \eqref{eq:ch6:BH:defRindTort}, so that a perturbation $\frac {\partial f}{\partial r}\Delta r = \frac {\partial f}{\partial r} \frac {\partial r}{\partial R} \Delta R$. Then the question is, how do the leading perturbations compare? 

Define the ratio
\begin{align}
    \label{eq:ch6:BH:comparePerts}
    \delta &:= \frac{\frac{R^2_a}{4r^2_s}(\pdv{f}{r}\pdv{r}{R})\vert_{R_a}\Delta R}{f(r_a)\frac{R_a}{2r^2_s}\Delta R} \notag \\ 
           &= \frac {R_a}{2}\frac {1}{f(r_a)}\left(\pdv{f}{r}\pdv{r}{R}\right)\Big\vert_{R_a}, \notag \\
           &= -\left( 1 - \frac{r^2_s}{r^2_a} \right).
\end{align}
The numerator of $\delta$ is the leading order correction from perturbations to the conformal factor, while the denominator is the leading order correction from perturbations to $(R/4r_s)^2$, the $g_{tt}$ component of the Rindler part of the metric. While the magnitude of this ratio remains small, the metric is well approximated by a rindler metric with radial coordinate $\hat{R} := \sqrt{f(r_a)}(R + \Delta R)$, and we can see that this is easily the case as long as $r_a \sim r_s$. In that regime, 
\begin{align}
    \label{eq:ch6:BH:findVarR}
    \hat{R} &\approx 2r_s e^{-r_s/2r_s}e^{(r_s + r_s\epsilon)/2r_s} \left( \frac {r_s}{r_s} + \epsilon -1 \right)^{1/2}, \notag \\
                   &\approx 2r_s \sqrt{\epsilon},
\end{align}
which is exactly $\hat{r}$ above when $r_a = r_s$. So near the Schwarzschild horizon these coordinates are exactly the Rindler approximation of the Schwarzschild coordinates, but further away where the gravitational acceleration becomes very weak, the coordinates need to rapidly compress the Schwarzschild coordinates to keep the accelerating aspect of the metric relevant\footnote{Another way to see this: in Rindler-Tortoise coordinates, the horizon is always at $R = 0$, but in perturbed Schwarzschild coordinates it is deep inside an imaginary parallel universe, so reconciling the two takes drastic measures.}. So job done, we have found global (exterior) coordinates that make obvious the hidden Rindler structure of the Schwarzschild metric. 
\begin{aside}[Tortoise Coordinates]
\label{aside:tortoise} 
Although the Schwarzschild coordinates are adapted to uniformly accelerating observers at fixed radii, that acceleration is very weak through most of space, and asymptotically reaches an observer who feels nothing at all (this is reflected in the Schwarzschild null geodesic \eqref{eq:ch6:BH:nullGeods} which has a logarithmic portion associated with uniform acceleration, but also a linear portion associated with free motion). If one so desired, one could perform this same exercise to extract the Minkowski structure hidden in the Schwarzschild coordinates, seeking a radial coordinate that takes  $\dd t = (1 - r_s/r)^{-1} \dd r$  to $\dd t = \dd r^*$ . This is simply the second line in \eqref{eq:ch6:BH:deriveR},
\begin{equation}
    \label{eq:ch6:BH:deriveRstar}
    r^* := r + r_s \ln(\frac r {r_s} - 1),    
\end{equation}
defines the \textbf{tortoise coordinate}\index{Tortoise Coordinate} $r^*$ in terms of which the metric becomes:
\begin{equation}
    \label{eq:ch6:BH:tortoiseMetric}
    \dd s^2 = \left( 1 - \frac {r_s} r \right)\left( -\dd t^2 + (\dd r^*)^2 \right) + r^2(r^*)\dd^2\Omega.
\end{equation}

We again find the metric is conformal in the $t$-$r$ plane, but this time to the Minkowski metric, and this time it is only approximately Minkowski in the regime $r_a \gg r_s$, so asymptotically far from the event horizon. In fact, in these coordinates the horizon is at $r^* \to \infty$, so everyone is infinitely far from the horizon and they are of limited use in investigating the interesting properties of the Schwarzschild geometry. This is incidentally where the name of the coordinate is derived from; the infinitely distant yet finite-radius horizon evidently reminded someone (possibly MTW \cite{Misner:1973prb}) of the tortoise that raced Achilles and won by asymptotic confusion. 
\end{aside}

\subsection*{Decelerating Schwarzschild Coordinates}

Where there is Rindler, there is Minkowski. The Rindler-Tortoise coordinates are a natural transformation from Schwarzschild coordinates that looks at the world through the eyes of uniformly accelerating (i.e.~fixed-radii) observers near the horizon. But with the metric in this form, it is impossibly tempting to wonder what the world looks like from the perspective of a Minkowski observer in the same place. In other words, let our observer at the horizon be UA-Einstein, let him transform to his trademark Minkowski coordinates and see what that means in the Schwarzschild geometry. Apply the inverse Rindler transformation \eqref{eq:ch5:Rind:RindToCart13},
\begin{equation}
    \label{eq:ch6:BH:RindToCart13}
    \begin{aligned}
        \tau &:= R\sinh(t/2r_s), \\
        \rho &:= R\cosh(t/2r_s)
    \end{aligned}
\end{equation}
and find the metric
\begin{equation}
    \label{eq:ch6:BH:KSMetric}
    \dd s^2 = \frac{4r_s}{r}e^{-r/r_s}\left( -\dd \tau^2 + \dd \rho^2 \right) + r^2(\rho) \dd^2\Omega.
\end{equation}
These coordinates are called \textbf{Kruskal-Szekeres}\index{Kruskal-Szekeres Coordinates}, and they are a triumph in the theoretical analysis of Schwarzschild geometry, as we will now see.

\begin{figure}[htp!]
    \centering
    \includegraphics[width=\textwidth]{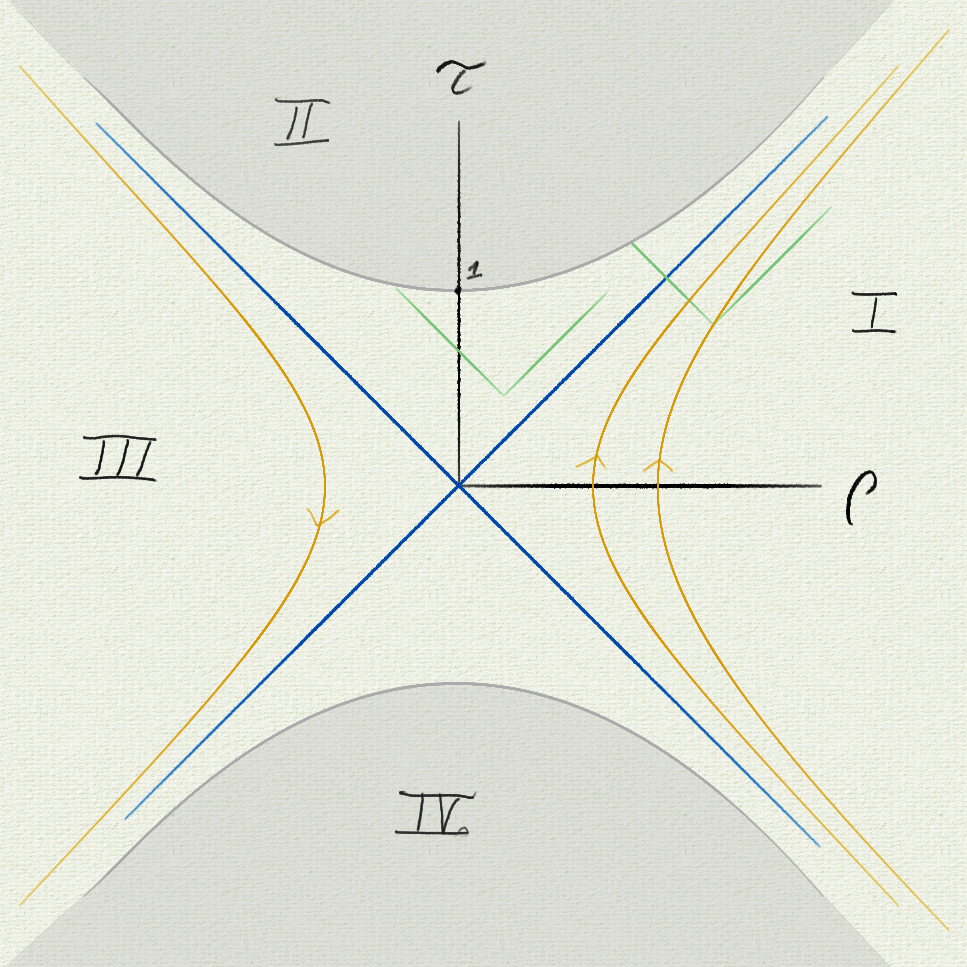}
    \caption[The maximally extended Kruskal-Szekeres coordinates.]{The maximally extended Kruskal-Szekeres coordinates. The diagram is analogous to Minkowski space divided into Rindler wedges (figure \ref{fig:ch5:Rind:rindler_mink}), except now regions II and IV have physical interpretations in terms of the interior regions of (respectively) black hole and white hole geometries (``interior'' being defined by the lines $\tau = \pm \rho$ for the white hole since it has no event horizon). The shaded regions are excluded by the finite range of the Schwarzschild radial coordinate. }
    \label{fig:ch6:BH:KSdiagram}
\end{figure}

Exactly as for Rindler coordinates proper, the original coordinates cover only a quarter of the domain of the new Minkowski coordinates, region I in figure \ref{fig:ch6:BH:KSdiagram}. In this region, null and massive geodesics easily reach the horizon (the line $\rho = \tau$) and it is only natural to suspect they cross over just as easily as they do in Minkowski space. Unlike a true Rindler observer however, we no longer have to imagine a fictional world to extend the Kruskal-Szekeres coordinates to region II to follow these geodesics, there is a perfectly real world that covers that domain already. Inside the horizon, $r_s/r < 1$ and time and radius reverse their roles in the Schwarzschild metric:
 \begin{equation}
    \label{eq:ch6:BH:schwInt}
    \dd s^2 = \left( \frac {r_s} r - 1 \right) \dd t^2 - \left( \frac {r_s} r - 1 \right)^{-1} \dd r^2 + r^2 \dd^2\Omega.
\end{equation}
In this region, we can perform an analogous Rindler-Tortoise transformation to the region II Rindler metric, defining a timelike radial coordinate
\begin{equation}
    \label{eq:ch6:BH:intR}
    R_{<} := 2r_s e^{r/2r_s}\left( 1 - \frac r {r_s} \right)^{1/2},
\end{equation}
in terms of which the metric is the interior Rindler-Tortoise form
\begin{equation}
    \label{eq:ch6:BH:intMetric}
    \dd s^2 = e^{-r/r_s}\frac {r_s}r \left\{ \left( \frac{R_{<}}{2r_s} \right)^2 \dd t^2 - \dd R^2_{<} \right\} + r^2 \dd^2\Omega.
\end{equation}
Then the region II transformation from Rindler \eqref{eq:ch5:Rind:RindToCart24} can be used, only this time we don't need to swap the Rindler radial and time coordinates since they've already done that themselves:
\begin{equation}
    \label{eq:ch6:BH:RindToCart24}
    \begin{aligned}
        \tau &= R_{<}\cosh(t/2r_s), \\ 
        \rho &= R_{<}\sinh(t/2r_s)
    \end{aligned}.
\end{equation}
So what we see is that regions I and II of the Kruskal-Szekeres coordinate space completely map out the Schwarzschild geometry for all values of $r$ without any issue at the event horizon, null geodesics are just straight lines that don't feel a thing through all of space---with one exception. Unlike the fictional time-space-exchanged Rindler coordinates covering region II of Minkowski space, the real interior coordinate \eqref{eq:ch6:BH:intR} has a finite range to it, it is confined to $R_{<} \in (0,1)$, corresponding to $r \in (r_s, 0)$. So the Kruskal-Szekeres coordinates (and by extension all geodesics) actually terminate along an hyperbola $\tau^2 - \rho^2 = 1$ corresponding to $r = 0$. This speaks to real, meaningful physics at the origin that can't be ignored, but fundamentally we knew that when setting up the problem from the very beginning. In no other area of physics do we expect a true point source---an infinite density---to actually exist, so we should always have expected to need to modify the model close to the origin.

Finally, nothing is stopping us from extending the Kruskal-Szekeres coordinates as far as they will go and trying to interpret regions III and IV. From the perspective of us Schwarzschild observers, those regions correspond to all the same physics taking place in reverse, but entirely causally disconnected from us. So if we take the extension of these coordinates seriously as the correct manifold of a point-mass-sourced geometry, then for every black hole there exists a disconnected parallel universe containing a mirror image \emph{white hole}. Although a fun theoretical exercise to think about, impossiblility of communicating with this hypothetical parallel universe makes it a moot point scientifically. 

\subsection{The Proof of the Pudding is in the Tasting}
\label{ch:AppsII:BH:obs}
All of the above is an inescapable conclusion of the experimental facts of the principle of relativity, the invariance of the speed of light, and the equivalence principle. Nevertheless, the bread and butter of science is in the production of predictions, and the execution of experiments to test them. It is therefore crucial to test the finished product regardless of how solid its foundations, so we will wrap up this section by covering a couple of the classic tests of the Schwarzschild metric---at least the exterior region which is applicable to the spacetime outside of massive bodies like the Earth and the Sun.

Originally, the first tests of general relativity were the calculation of the precession of Mercury's orbit (an observation which confounded Keplerian mechanics), and the deflection of light by the sun's Schwarzschild exterior. Before covering complicated examples like those however, we already have at hand a testable prediction that has become so commonplace it's hard to imagine twenty-first century life without it. We saw in \eqref{eq:ch6:BH:fixedWL} that observers at different fixed Schwarzschild radii will notice a certain relative time dilation. Although small, it can be relevant if extreme precision is required, such as for satellites---especially geolocation satellites. Global Positioning System (GPS) and other satellites need to include corrections due to the time dilation they experience both from their high orbital velocities\footnote{They move in the $\phi$ direction at a high velocity relative to us on the surface, and at fixed values of $r$ and $\theta$ the $t-\phi$ plane is simply Minkowski space, so inertial observers are related by the usual Lorentz transformations.} as well as their radial separation from the Earth's surface. With the easiest test out of the way, let's move on to the most visible effects of Schwarzschild geoemtries: orbits. 

\subsubsection{Orbits Existence and Classifications}
We are back to looking at exterior geodesics, but this time we have to open ourselves up to paths not confined to the radial direction. Fortunately we still have plenty of symmetries to work with and the problem can be simplified quite a bit. Recall the Schwarzschild geometry has four Killing vectors, time $T$ and three rotations $\Phi, \Psi,$ and $\Omega$, each with a corresponding conserved (covariant) momentum. The Schwarzschild coordinates are adapted to $T$ and $\Phi$, but rotations need an additional word, so we'll just take care of the time component first. A conserved component of momentum in the time direction means geodesics (written in terms of momentum) satisfy 
 \begin{equation}
    \label{eq:ch6:BH:defE}
    p^t = \frac E {\sqrt{1 - \frac {r_s}{r}}}. 
\end{equation}
with $E := p_t$ identified as the conserved energy of the particle.

Now to rotations. In principle there are three rotational Killing vectors so three generic angular momenta (though only one aligns with a component of four-momentum). These three momenta form a single three-dimensional vector (the total angular momentum vector), and spherical symmetry means we can align one of our coordinate axes along this vector\footnote{Note that we can only do this maneouver once, so if we wish to track two particles at once, their \emph{relative} angular momentum will be generically non-zero and need to be tracked in full.}. Typically one chooses to align the $z$-axis with this vector so that in these coordinates the only non-trivial angular momentum is the one that aligns with geodesics, $p_\phi$. At the same time, fixing vanishing momentum in the $\Psi$ and $\Omega$ directions implies purely equitorial motion $\theta = \pi/2$ (see \eqref{eq:ch6:BH:polPreservedK}), so we have 
\begin{equation}
    \label{eq:ch6:BH:fixTheta}
    p^\theta = 0, 
\end{equation}
and the $\phi$ component
\begin{equation}
    \label{eq:ch6:BH:defL}
    p^\phi = \frac L {r^2},
\end{equation}
which identifies $L$ as the particle's conserved angular momentum. This takes care of three components of four-momentum, so the remaining component is easiest to solve for through the constraint equation $p^2 = -m^2$:
\begin{equation}
    \label{eq:ch6:BH:constLength}
    -E^2\left( 1 - \frac {r_s}{r} \right)^{-1} + m^2 \left( 1 - \frac {r_s}{r} \right)^{-1}\left( \frac {\dd r}{\dd s} \right)^2 + \frac {L^2}{r^2} = -m^2,
\end{equation}
which can be neatly normalized and arranged for a phase-space treatment into
\begin{equation}
    \label{eq:ch6:BH:orbits}
    \dot{r}^2 = \hat{E}^2 - V(r), 
\end{equation}
where $\hat{E} := E/m$ for massive particles and simply $E$ for massless particles, and the effective potential term $V(r)$ is
\begin{align}
    V(r) &= \left( 1 - \frac {r_s} r \right)\left( 1 + \frac {\hat{L}^2}{r^2} \right), \qquad m \neq 0, \label{eq:ch6:BH:VeffMass} \\
    V(r) &= \left( 1 - \frac {r_s} r \right)\frac {\hat{L}^2}{r^2} , \qquad m = 0, \label{eq:ch6:BH:VeffLight} 
\end{align}
with again $\hat{L} := L/m$ for massive particles and simply $L$ otherwise.

\begin{figure}[ht!]
    \centering
    \includegraphics[width=0.76\textwidth]{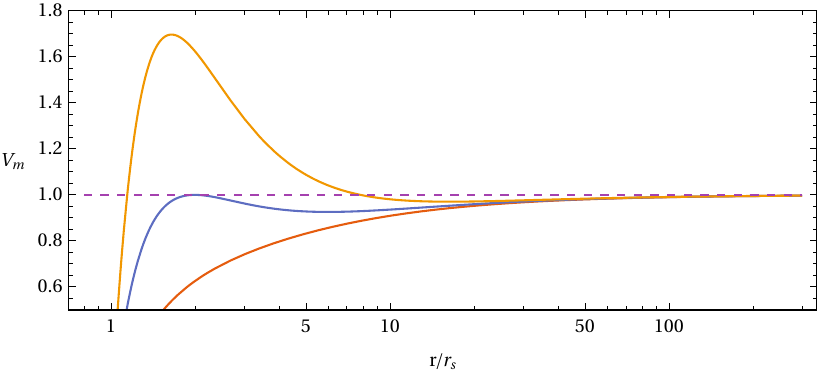}
    \includegraphics[width=0.76\textwidth]{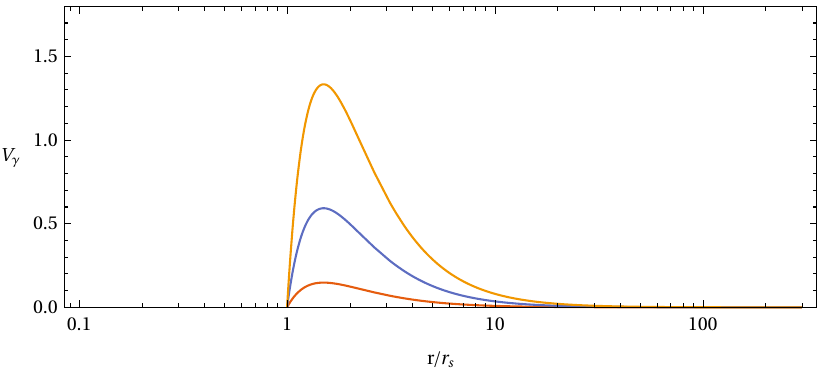}
    \includegraphics[width=0.76\textwidth]{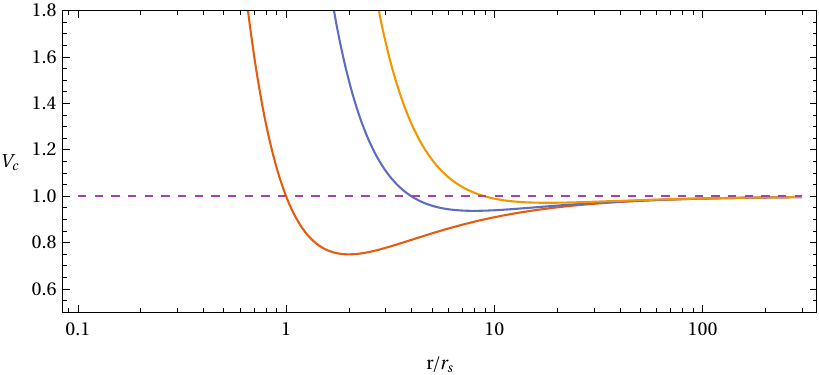}
    \caption[The effective potentials of orbital motion for massive, massless, and non-relativistic particles in Schwarzschild spacetime.]{The effective potential for massive particles (top), massless particles (middle), and the Newtonian limit (bottom), for $\hat{L}/r_s = $ 1 (red), 2 (blue), and 3 (orange). The massive and Newtonian cases both asymptote to 1 so the dip below that value demonstrates the existence of a stable minimum (i.e.~orbit) at finite $r$. The positive asymptote at the origin in the Newtonian potential demonstrates can never reach the origin, while the negative asymptote at the origin in the general relativistic cases demonstrates that a sufficient ratio of $E/L$ will lead to a particle \emph{always} reaching the origin.}
    \label{fig:ch6:BH:potentials}
\end{figure}

The general solutions are still too hard to solve for analytically, but some analysis of the potentials (and physical intuition) enables an excellent qualitative view of the situation. The potentials are plotted for a few values of $\hat{L}/r_s$ in figure \ref{fig:ch6:BH:potentials}, as well along with the Newtonian analogue $V(r) = 1 - \frac{r_s}{r} + \frac{\hat{L}^2}{r^2}$ for reference. Possible physical trajectories are determined by comparing fixed values of total energy $E$ to the plot of the potential for a fixed angular momentum---physical trajectories must keep $E > V$ (otherwise $\dot{R}$ would become imaginary). 

Start with the Newtonian example. Here it is always possible to find a stable minimum of $V$ such that there exists a regime of energies $V_\text{min} < E < 1$ where a particle's radius will oscillate about a fixed value, i.e.~be confined to an elliptical orbit (or ``closed'' elliptical paths). For higher particle energies, there is always a regime close to the source where $V > E$ such that incoming particles will always reach a minimum and return to infinity, i.e.~flyby trajectories (or ``open'' elliptical paths).

Relativistic dynamics are more intricate. There still usually exist regimes for both massive and massless particles in which there is a point near the source where $V > E$ and particles may return to where they came from, and for sufficient values of $E/L$ massive particles still exhibit a local minimum some distance from the source that enables the existence of stable orbits, but there are now two completely new phenomena. First, we see both relativistic potentials asymptote to a negative value at the origin, meaning any particle with enough energy or too little angular momentum will take a nosedive into the source. This is as we should expect, since we know already that any particle that makes it to the event horizon will be irressistably compelled to the origin. What might come as a surprise is the second new phenomenon, the unstable equilibrium near the horizon. This turning point represents an unstable orbit and is impressively seen in both the massive and massless cases, so it is termed the \textbf{photon sphere}\index{Photon Sphere}, since in principle there could exist a ring of photons trapped at this radius around a sufficiently compact object (though again, the orbit is unstable so it is unlikely it would be well populated in reality). We can make the statement of the existence of the orbits more precise by calculating the turning points $V^\prime(r_0) = 0$,
 \begin{equation}
    \label{eq:ch6:BH:turnM}
    V^{\prime}(r) = \frac 1 {r^2} \left( r_s - 2\frac {\hat{L}^2}{r} + \frac {3r_s \hat{L}^2}{r^2} \right), 
\end{equation}
for massive particles with zeroes at
\begin{equation}
    \label{eq:ch6:BH:minM}
    r_0 = \frac{\hat{L}^2}{r_s}\left( 1 \pm \sqrt{1 - 3r_s^2/\hat{L}^2}  \right).
\end{equation}
and 
\begin{equation}
    \label{eq:ch6:BH:turnL}
    V^\prime(r) =  -\frac {2 \hat{L}^2}{r^3} + \frac{3r_s \hat{L}^2}{r_4}   
\end{equation}
with the single zero at
\begin{equation}
    \label{eq:ch6:BH:minL}
    r_0 = \frac 32 r_s,
\end{equation}
irrespective of angular momentum.

\subsubsection{Open Null Orbits: Deflection of Light}
The existence of an unstable orbit for photons and the inescapability of the event horizon, while both novel aspects of the general relativistic solution, are both outside the realm of easy observables (especially for early twentieth-century observers). Small deviations from the orbits of massive particles can be computed and lead to a successful prediction for the precession of the perihelion of Mercury's orbit, but a slightly easier and perhaps more novel observation is the deviation of the trajectories of massless particles. In Newtonian mechanics, massless particles do not feel gravity, so feel an effective potential of only $V(r) = \frac{L^2}{r^2}$ (this is just a funny way of writing a straight line, the solution for no acceleration). But in the relativistic model the effective potential for null rays is strongly modified by the presence of mass (i.e., the term with $r_s$), and even those paths that don't fall into the black hole or get temporarily confined to the photon sphere will see a deviation from the simple Newtonian limit, so let us see how that happens.

First, we should parameterize the trajectory in terms of $\dd \phi/\dd r$ , since we want to track the deviation of a photon's path as it travels from $r \to \infty$  to a point of closest approach, and then back again to $r \to \infty$  (a funny but useful way of describing a flyby). This quantity can be calculated as
\begin{equation}
    \label{eq:ch6:BH:dphidr}
    \dv[]{\phi}{r} = \frac{\dd \phi/\dd s}{\dd r/ \dd s} = \frac {p^\phi}{p^r} = \pm \frac {1}{r^2} \left( \frac 1 {b^2} - \frac 1 {r^2}\left( 1 - \frac {r_s}{r} \right)  \right)^{-1/2},
\end{equation}
where we define the non-relativistic impact parameter (radius of closest approach) as $b := L/E$. This is a little easier to treat in terms of $u := 1/r$,
\begin{equation}
    \label{eq:ch6:BH:dphidu}
    \dv[]{\phi}{u} = \pm\left( \frac 1 b^2 - u^2 + r_s u^3 \right)^{-1/2}.
\end{equation}
This is looking a little unfriendly, so perhaps a good time for a reality check. The Newtonian limit is $r_s \to 0$ (for a photon), in which case we can use a nice simple trigonometric integral $\dv[]{}{x} \asin(x) = 1/ \sqrt{1 - x^2}$ to see the reference solution
\begin{equation}
    \label{eq:ch6:BH:refSolPhi}
    \phi(u) = \asin(bu),
\end{equation}
or $r\sin(\phi) = b$, the equation for a line of constant  $x = b$. 

\begin{figure}[htp!]
    \centering
    \includegraphics[width=0.8\textwidth]{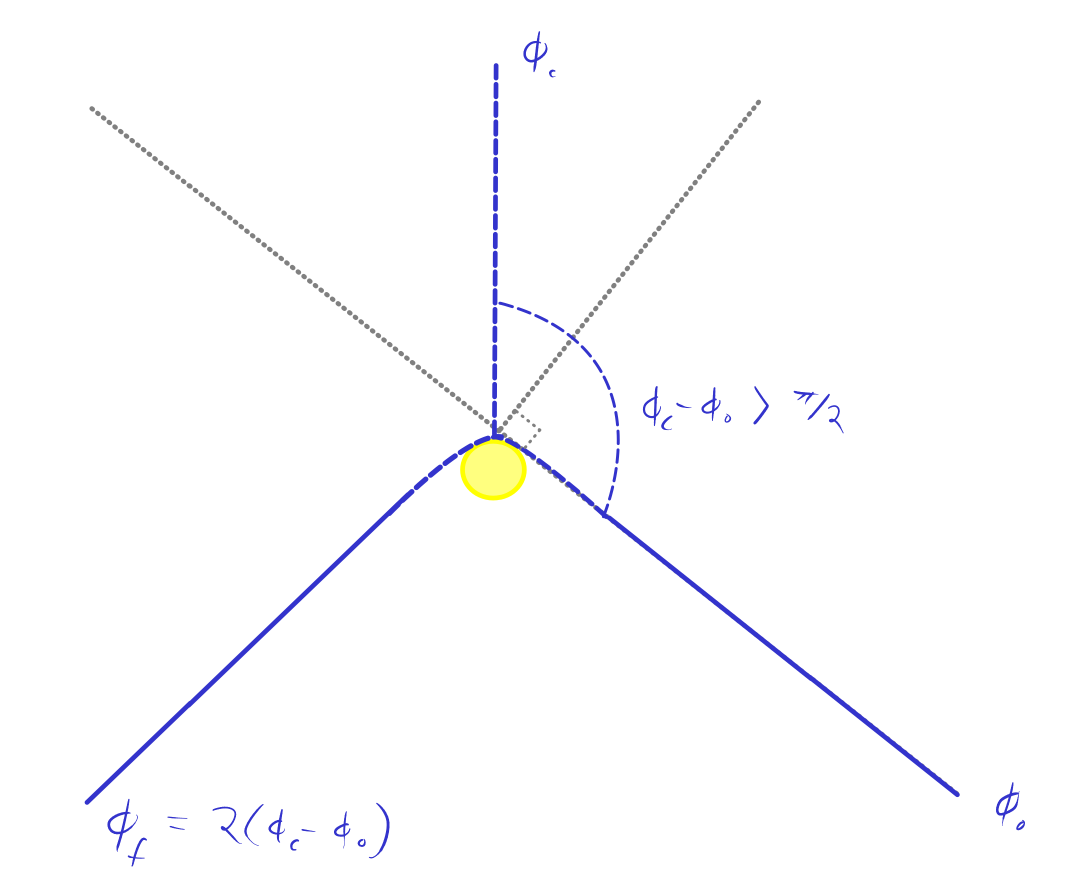}
    \caption[Deflection of massless particles by a massive body in a Schwarzschild geometry.]{Deflection of massless particles by a massive body in a Schwarzschild geometry (exaggerated for clarity). The grey path is the undeflected Newtonian path, a simple straight line.}
    \label{fig:ch6:BH:lightDef}
\end{figure}

Recovering the Newtonian solution is a good sign, but \eqref{eq:ch6:BH:dphidu} is still generally intractible, so an approximation would help. Rather than try to solve all possible flyby photon trajectories, we can look for those that do not get very close to an object's Schwarzschild radius, $b \gg r_s$. In addition to affording us an opportunity to approximate, this is also the most relevant scenario for obsrvations, since the deflection of light by massive objects is something we only expect to observe around geometrically large, massive objects like stars and galaxies (there are not many black holes we get to observe on a daily basis). With another handy reparameterization (courtesy of \cite{schutz_2009}), $y := u(1 - \frac 12 r_s u)$, one finds
 \begin{equation}
    \label{eq:ch6:BH:dphidy}
    \dv[]{\phi}{y} \approx \frac {1 + r_s y}{\sqrt{1/b^2 - y^2}} + \mathcal{O} (r_s^2/b^2),
\end{equation}
with solution
 \begin{equation}
    \label{eq:ch6:BH:solvePhi}
    \phi \sim \phi_0 + \frac{r_s}{b} + \arcsin(by) - r_s\left( \frac 1 {b^2} - y^2 \right)^{1/2}.
\end{equation}
Under our approximation, $y_\text{min} \sim 1/b$, so $\phi_\text{min} \sim \phi_0 + r_s/b + \pi/2$ . So a photon traces out an angle $\phi_\text{min} - \phi_0 = r_s/b + \pi/2$  getting to its closest point, and must symmetrically trace out the same angle going back out to infinity, so in total traces out $\Delta \phi = 2(r_s/b + \pi/2)$, which leaves a deviation from the Newtonian result ($\pi$) of  $\Delta\phi_\text{gr} = 2r_s/b$  (see figure \ref{fig:ch6:BH:lightDef}). For light rays passing right by the surface of the sun, this amounts to a deflection of just $1.74''$---small but detectable!  


\section{Perfect Fluid EFEs: Cosmology}
\label{ch:AppsII:Cosmo}

Finally, we can't run from it any longer, we have to try to solve the Einstein field equations for an everywhere non-empty spacetime. Nevertheless, we can still try to do this as simply as possible, so we consider spacetime to be filled with the most symmetric non-trivial energy-momentum that we can, which turns out to be that of an homogeneous and isotropic (though time-dependent) perfect fluid. There are not a lot of astrophysical systems that would satisfy quite this degree of simplicity, but fortunately there is one very important system that is well described by such a stress-energy tensor: the entire universe.

\subsection{Modelling the Universe}
\label{ch:AppsII:CM:FRW}
\begin{figure}[ht!]
    \centering
    \includegraphics[width=0.7\textwidth]{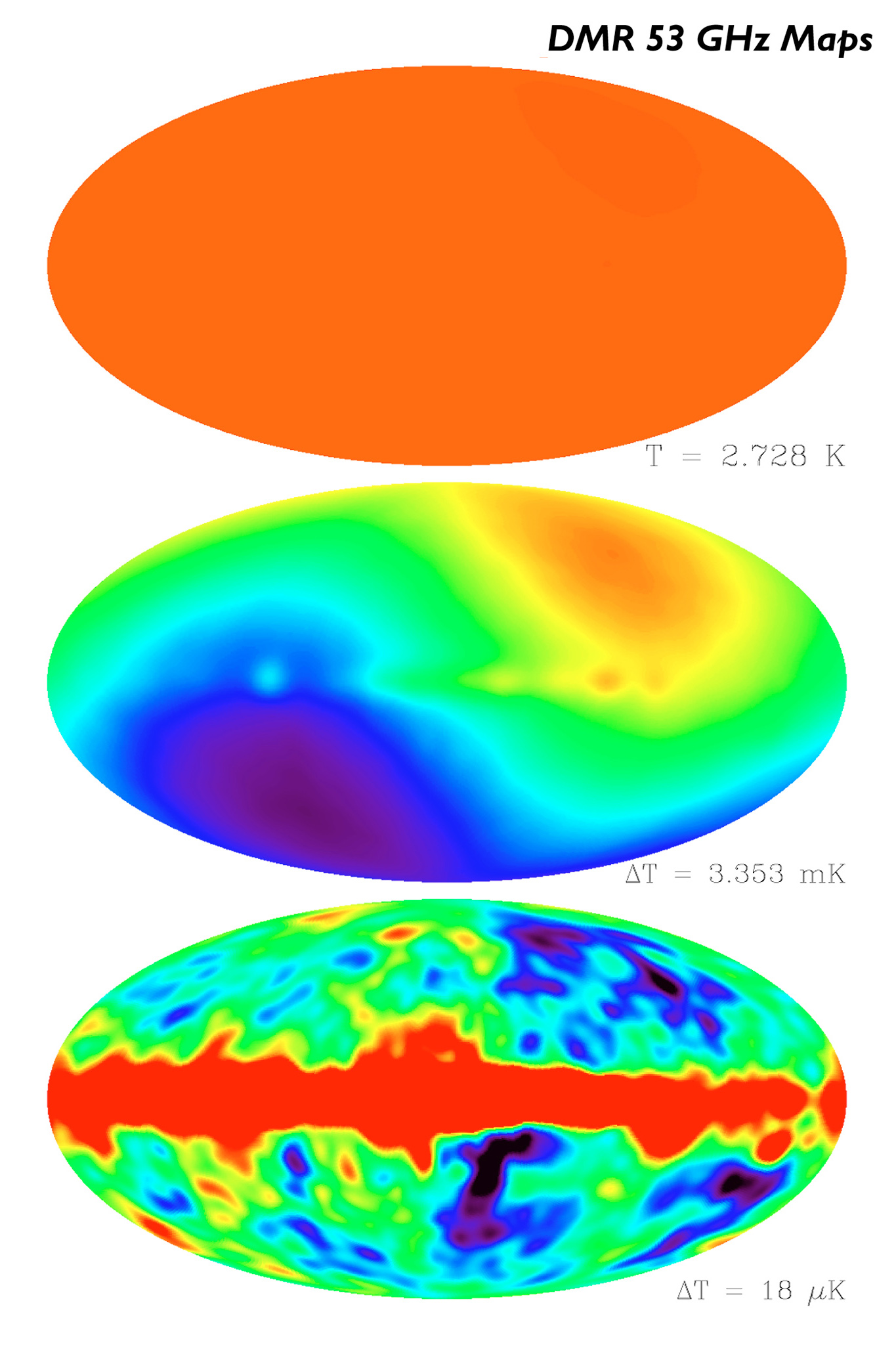}
    \caption[The scales of the CMB, as presented by the COBE project.]{The scales of the CMB, as presented by the COBE project \cite{cobe}. Top: the complete CMB picture is almost indistinguishable from a single blackbody temperature. Middle: subtracting the background temperature reveals only a $0.1\%$ deviation attributable to the sun's peculiar motion relative to the cosmic background (our motion through the galaxy). Bottom: subtracting the sun's dipole motion reveals the remaining inhomogeneities in the CMB appear only at the level of $\delta T/\ev{T} \sim \mathcal{O}(10^{-6})$.}
    \label{fig:ch6:CM:CMB}
\end{figure}

On the largest scales, the universe is very smooth. Look up at the night sky from anywhere on earth; if you focus in on individual constellations, you can figure out where you are, but if you just quint it all looks the same. More quantitatively, on the largest scales we can detect there is a lingering background of radiation, the \textbf{Cosmic Microwave Background Radiation (CMB)}\index{Cosmic Microwave Background Radiation}. The spectrum of the CMB radiation is a nearly perfect blackbody, to leading order it is well described by a single number (the corresponding blackbody temperature). At the $0.1\%$ level, there is a dipole variation corresponding to our relative motion through the Milky Way galaxy. Subtracting that dipole, variations in the colour of light we see in the universal background radiation across the sky appear only at the level of one part in ten thousand (see figure \ref{fig:ch6:CM:CMB})---a very uniform distribution indeed! 

Of course, our cosmological observations are all taken from only a single laboratory (the Earth), so the CMB uniformity technically only speaks to isotropy (i.e., spherical symmetry). It would however be a very chance circumstance if we just happened to develop in the one point in an exciting, wildly varying universe from which vantage it happened to appear mundane. Until strong evidence to the contrary can be found, it is best to also adopt the \textbf{Copernican Principle}\index{Copernican Principle}, that we do not exist in a privileged point in spacetime, and instead that all parts of the universe see the same isotropy. Mathematically, that is homogeneity, so before even writing down a stress-energy tensor, we know we seek a spacetime geometry that is both spatially homogeneous and isotropic. 

\subsection{A Maximally Symmetric Moment in Time}
\label{ch:AppsII:CM:maxSym}
The symmetries we have described have been observed only at a moment in cosmic time. As beautiful and convenient as it would be if we could also assume the universe to be static (i.e., time-translationally invariant), that is simply an unfounded stretch. I emphasize this point because although we now have ample evidence to the contrary, it was for a long time assumed the universe must be static, and the result we will soon derive as an inescapable consequence of the theory of general relativity was rather uncomfortable to some of its authors. What we \emph{can} assume though, is that the spatial symmetries we observe now are consistent throughout the universe's history. This is actually directly supported by some observations, but it can also be considered a sort of temporal Copernican principle, that it would be quite the coincidence if it just so happened that we evolved to observe the universe at the one point in its entire existence when it just happened to appear as spatially symmetric as possible. So with these assumptions in mind, we propose an ansatz for a cosmological metric as:
\begin{equation}
    \label{eq:ch6:CM:preFRW}
    \dd s^2 = -\dd t^2 + a^2(t)\dd s_3^2,
\end{equation}
where $a(t)$ is called the \textbf{scale factor}\index{Scale Factor}, and $\dd s_3^2$ is the homogeneous and isotropic metric of purely spatial slices of spacetime.

The highly symmetrical spatial metric $\dd s_3^2$ is not necessarily quite as trivial as it may seem. It is tempting to think we already know the unique spherically symmetric and homogeneous three-dimensional spatial geometry---Euclidean space. But while spatial translations are easy to see experimentally in local Cartesian frames, on the scale of the universe we only have experimental data from one specific point in space, and can only speculate about how exactly the Copernican principle manifests on larger scales. Phrased another way, there's no way to know just from isotropy and our lack of specialness whether or not the spacial geometry of the universe is satisfactorily described by a Cartesian system of coordinates, or some other equally but differently symmetric coordinates. 

So for the sake of generality, it is best to start with a metric we found while pursuing the Schwarzschild solution:
\begin{equation}
    \label{eq:ch6:CM:genS3}
    \dd s_3^2 = f(r) \dd r^2 + r^2 \dd^2\Omega.
\end{equation}
This metric we know to be isotropic (cf section \ref{ch:AppsII:BH:breakSym}), and for $f(r) = 1$ it preserves translations as symmetries. For $f(r) \neq 1$, we know it fails to satisfy the vacuum field equations so we didn't pursue it further. But now that we are no longer in a vacuum, it may still be of value to consider non-trivial $f(r)$. Although we know it breaks translations as Killing vectors, we do not actually know if a non-trivial $f(r)$ could not support another set of Killing vectors instead. What we do know is that the symmetric space we seek must have three more symmetric directions in addition to the spherically symmetric directions for a total of six symmetries. This number is important because it allows us to explore the possible symmetries of \eqref{eq:ch6:CM:genS3} without explicitly hunting down Killing vectors using a sneaky trick: maximal symmetry.

Maximally symmetric spaces should rightfully be explored in detail with a study of Lie algebras and representation theory, but it is sufficient for us here to give an heuristic description and pull a theorem from the literature. In principle, a manifold's geometry is completely specified by its metric tensor $g_{\mu\nu}$ (and chosen connection, but for our Levi-Civita connection this too is specified by the metric), which as a symmetric tensor is comprised of $\frac 12 n(n+1)$ degrees of freedom for an $n$-dimensional manifold. But for each symmetry the manifold possesses, we know there exist coordinates such that $\partial_* g_{\mu\nu} = 0$ in the direction of the symmetry, effectively reducing the degrees of freedom in the metric by one. In other words, if a space has $\frac 12 n(n+1)$ symmetries, there is no room for individuality in the metric, it must be specified entirely by the symmetries of the manifold. If one stubbornly tried to impose more symmetries on the manifold, one would be out of luck as the metric is already fully specified, so a new constraint would either be dependent on the others, or mathematically inconsistent. Thus a manifold with $\frac 12 n(n+1)$ symmetries is \textbf{maxmially symmetric}\index{Maximal Symmetry}, and is completely characterized by a unique rank-2 symmetric tensor constructed entirely from the symmetries of the space. What's more, symmetries of the metric translate to symmetries of the Ricci curvature tensor (you'll have to trust me or check the literature), so in a maximally symmetric space, $R_{\mu\nu} \propto g_{\mu\nu}$. 

Back to our search for a highly symmetric spatial geometry; we seek a three-dimensional space with six symmetries, which is exactly a maximally symmetric space. The ansatz \eqref{eq:ch6:CM:genS3} narrows down the possibilities by fixing the three rotational symmetries, and it turns out we can classify all possible remaining geometries by simply imposing the maximal symmetry condition $R_{\mu\nu} = 2kg_{\mu\nu}$ (normalizing the proportionality with a bit of foresight). From section \ref{ch:AppsII:BH:pin}, we find we have to solve:
\begin{align}
    \label{eq:ch6:CM:Ricci}
    R_{rr} &= \frac 1 r \frac {f^\prime}{f} = 2kf, \\
    R_{\theta\theta} &= 1 - \partial_r\left( \frac{r}{f} \right) - \frac 12 \frac{rf^\prime}{f^2} = 2kr^2, \\
    R_{\phi\phi} &= \sin[2](\theta) R_{\theta\theta} = 2kr^2\sin[2](\theta).
\end{align}
This time we only have one degree of freedom to solve for, so only one equation is linearly independent and we are free to use the simplest one, which is clearly the $rr$ equation. A simple separation of variables integration yields
 \begin{equation}
    \label{eq:ch6:CM:solveF}
    f(r) = \frac 1 {1 - kr^2},
\end{equation}
where the initial condition for $f$ is taken to be 1 (to align with the $k = 0$ solution for Euclidean space), but any other choice can be made with a suitable re-scaling of $r$. 

It is a matter of convention if one chooses to consider $k$ to be real valued (in which case $k^2$ has dimensions of inverse length), or to absorb the length scale of $k$ into $r$, which transfers that dimensionality to the scale-factor $a$. For the moment we will absorb the dimensionality of $k$ into the radial coordinate, in which case all that matters is its sign. Obviously if $k = 0$ we recover ordinary flat Euclidean space, but if $k = \pm 1$, we find equally valid maxmially symmetric non-Euclidean geometries, both of which can be supported thanks to a non-vanishing (though highly symmetric) cosmological stress-energy tensor. When $k = -1$ the geometry is hyperbolic, the radial coordinate is unbounded but $f(r)$ is confined to the range $(0,1)$ (this is sometimes called an \textbf{open}\index{Cosmological Curvature!Open} universe). When $k = +1$ the geometry is spherical, $f(r)$ is now bounded only from below by 1, but the radial coordinate is now stuck to the range $(0,1)$ (this is sometimes called a \textbf{closed}\index{Cosmological Curvature!Closed} universe). These three geometries are a little clearer to see with a little change of coordinate: 
\begin{equation}
    \label{eq:ch6:CM:hyperSph}
    \dd s^2_3 = \dd \hat{r}^2 + S_k^2(\hat{r})\dd^2\Omega
\end{equation}
defines $\hat{r}$ through $r^2 = S_k^2(\hat{r})$ with
\begin{equation}
    \label{eq:ch6:CM:defSk}
    S_k(\hat{r}) := \begin{cases}
        \sin(\hat{r}),  \qquad &k = +1, \\
        \hat{r}, \qquad &k = 0,\\
        \sinh(\hat{r}), \qquad &k = -1
    \end{cases}.
\end{equation}
These \emph{hyperspherical} coordinates emphasize the distinct nature of the three different symmetric geometries. The spherical geometry is that of a 3-sphere, a universe where every direction is periodic, not just the rotational directions. The flat geometry is as expected, and the hyperbolic geometry is explosive as ever, with a radial direction that exponentially disperses. Altogether, our ansatz for a cosmological metric takes the form
\begin{equation}
    \label{eq:ch6:CM:defFRW}
    \dd s^2 = -\dd t^2 + a^2(t) \left\{ \dd \hat{r}^2 + S_k^2(\hat{r}) \dd^2\Omega \right\},
\end{equation}
known as the \textbf{Friedmann-Robertson-Walker (FRW) metric}\index{Friedmann-Robertson-Walker Metric}.

Not that it needs to be said at this point, but of course remember the distinction between coordinates and physical distances. The spatial coordinates in the FRW metric are sometimes called \textbf{comoving coordinates}\index{Comoving Coordinates} since they are adapted to inertial cosmological observers in exactly the same way that the transverse-traceless gauge coordinates were adapted to inertial observers in a gravitational wave geoemtry. Physical separations between objects across the universe will vary with the scale factor through cosmological time (the time coordinate is also not the proper time of any individual observer in the universe but is close to it, being adapted to the averaged proper time of the near uniform mass-energy density of the universe). The reason this distinction is typically stressed in cosmology is because when the spatial geometry of the universe is flat ($k=0$) it is common to use a pseudo-Cartesian coordinate system to label points in space, which naively leads one to think about comoving ``distances,'' but we know better. Just like with distributions in a gravitational wave background, we plot out the distribution once in comoving coordinates, then follow the physical separation of particles through time by measuring proper distances.


\subsection{The Fluid Universe}
\label{ch:AppsII:CM:fluidUni}
With the metric ansatz sorted, it remains to connect the scale factor (the only dynamic quantity remaining associated with the geometry of spacetime) to the matter-energy content of the universe. We began this whole discussion with an observation of how incredibly symmetric the stress-energy of the universe appears to be, so that symmetry needs to be built into the mathematical representation of that matter-energy. Fortunately, we have already done this---in section \ref{ch:geomRel:stressEnergy}, we constructed the stress-energy tensor by describing a macroscopic distribution of matter that was as dynamically simple as possible. The perfect fluid stress-energy reflected a macroscopic collection of matter (possibly relativistic) that was free of complicated dynamics, which translated to being spatially isotropic and having no mixing between space and time components (in the fluid's rest frame). So all we need to do to use this to model the stress-energy of the universe is to simplify it one step further, to impose homogeneity by insisting the energy density and pressure be spatially uniform. In other words, a good model for the energy-density of the universe turns out to be:
\begin{equation}
    \label{eq:ch6:CM:uniSET}
    (T_\text{uni})\indices{^\mu^\nu} = (\rho(t) + p(t))u^\mu u^\nu + p(t) g^{\mu\nu},
\end{equation}
which identifies our chosen coordinates as the rest frame of the cosmos. 

Just before moving on to the field equations, we can already learn something valuable just from the conservation of stress-energy. To evaluate this we of course need the Christoffel symbols, which are as follows (excluding vanishing symbols and those connected by symmetry of the lower indices, and using a dot for differentiation with respect to $t$):
\begin{equation}
    \label{eq:ch6:CM:FRWChris}
    \begin{gathered}
        \Gamma^t_{rr} = \frac {a\dot a}{1 - kr^2}, \qquad \Gamma^t_{\theta\theta} = a \dot a r^2, \qquad \Gamma^t_{\phi\phi} = a \dot a r^2 \sin[2](\theta) , \\
        \Gamma^r_{tr} = \Gamma^\theta_{t\theta} = \Gamma^\phi_{t\phi} = \frac {\dot a}{a} , \qquad \Gamma^\theta_{r\theta} = \Gamma^\phi_{r \phi} = \frac 1 r, \\ 
        \qquad \Gamma^r_{rr} = 2kr(1 - kr^2), \qquad \Gamma^r_{\theta\theta} = (\sin\theta)^{-1} \Gamma^r_{\phi\phi} = -r(1 - kr^2), \\
\Gamma^\theta_{\phi\phi} = -\cos(\theta)\sin(\theta), \qquad \Gamma^\phi_{\theta\phi} = \cot(\theta).
    \end{gathered}
\end{equation}
From here, it's easiest to use the mixed form of the stress-energy tensor $T\indices{_\mu^\nu} = \text{diag}(-\rho, p, p, p)$ , in which case the conservation of stress energy $\nabla_\nu T\indices{_\mu^\nu} = 0$ yields  
\begin{equation}
    \label{eq:ch6:CM:consT}
    \dot \rho + 3 \frac {\dot a}{a}\left( \rho + p \right) = 0,
\end{equation}
for the timelike component, with all others being trivial. This is a very powerful conservation law on its own, but thinking about the universe in thermodynamic equilibrium, we should expect there to be a relation between $\rho$ and $p$, an \textbf{equation of state}\index{Equation of State}. For all of the types of cosmological matter-energy we consider, it is sufficient to consider equations of state of the form:
\begin{equation}
    \label{eq:ch6:CM:eos}
    p = w\rho
\end{equation}
for some constant $w$. In that case, \eqref{eq:ch6:CM:consT} can be simplified and solved to find:
\begin{equation}
    \label{eq:ch6:CM:evolveRho}
    \rho = \rho_0 a^{-3(1+w)}.
\end{equation}
In other words, even without knowing exactly how the scale factor evolves, we already know that the energy density of the universe must shrink or grow (if we know the equation of state).

Better still, we can easily construct the equation of state for two of the types of cosmological materials we are interested in. The simplest of all is dust, non-relativistic matter. In the case that the relative motions of cosmological matter are non-relativistic, the energy density is dominated by the matter density, and we have 
\begin{equation}
    \label{eq:ch6:CM:wDust}
    w_\text{dust} \approx 0,
\end{equation}
so that the energy density dilutes as $\rho_\text{dust} \sim 1/a^3$, exactly as it should (put the same mass in progressively larger boxes, the mass density will shrink as the inverse of the volume). Relativistic matter is almost as easy, but takes a second more thought. In that case the pressure is no longer trivial, but if the matter is \emph{highly} relativistic, the pressure in any given direction will be on the order of the energy density. With the same energy isotropically driving pressure in three independent directions, the pressure in each direction can only feel about a third of the total available energy density, so the equation of state for a highly relativistic perfect fluid is
\begin{equation}
    \label{eq:ch6:CM:wRad}
    w_\text{rad} \approx 1/3,
\end{equation}
leading to a dilution of $\rho_\text{rad} \sim 1/a^4$ (another way to think of this: highly relativistic particles are more wave-like than particle-like, and their energies go as their wavelengths $E \sim 1/\lambda$, so their energy densities redshift as well as dilute with volume $E/V \sim 1/(\lambda V^3)$).

Finally, we must gravitate\footnote{You're lucky I waited this long to say that.} towards the field equations. Here again with such a simple stress-energy tensor, it is easiest to use the trace-reversed form of the Einstein equations. From the mixed form of the stress-energy, we have the trace $T = -\rho + 3p$, so the trace-reverse is $T_{\mu\nu} - \frac 12 g_{\mu\nu}T = (\rho + p)u_\mu u_\nu + \frac 12(\rho - p)g_{\mu\nu}$. The only non-trivial and unique equations of motion are then the $tt$ and $rr$ components, respectively:
\begin{align}
    \label{eq:ch6:CM:EFEs}
    -3 \frac {\ddot a}{a} &= \frac 12 \kappa \left( \rho + 3 p \right), \\
    \frac {\ddot a}{a} + 2\left( \frac {\dot a}{a} \right) + 2\frac k {a^2} &= \frac 12 \kappa \left( \rho - p \right).
\end{align}
For convenience, these are typically arranged into the Einstein Constraint
\begin{equation}
    \label{eq:ch6:CM:EConst}
    H^2 = \frac \kappa 3 \rho - \frac k {a^2},
\end{equation}
(which defines the misleadingly named \textbf{Hubble constant}\index{Hubble Constant} $H := \dot a / a$), and the \textbf{Friedmann Equation}
\begin{equation}
    \label{eq:ch6:CM:Friedmann}
    \frac {\ddot a}{a} = -\frac \kappa 6 \left( \rho + 3 p \right).
\end{equation}
Although the Hubble parameter is certainly not a constant, it is still valuable as a geometrical scale the same way the Schwarzschild radius furnished a geometric interpretation of the matter-energy content of a black hole ($r_s \leftrightarrow M$ and similarly $H \leftrightarrow \rho$). In particular, a measurement of the Hubble parameter in our cosmological neighbourhood today yields $H_0 \sim 70 \text{ km/s/Mpc}$, which is a silly unit, but inverted turns into a time scale $t_H \sim 10^{10}$y, setting the scale for the age of the universe as around 10 billion years. 

Just before plugging in the dust and radiation energy densities and pressures, we will temporarily make our lives simpler with a little conceptual diversion and a single measurement. The constraint equation \eqref{eq:ch6:CM:EConst} is an excellent visual representation of the fundamental principle of general relativity, that matter-energy and geometry are two sides of the same coin. In this case, we can use the Hubble parameter to define an energy scale, 
\begin{equation}
    \label{eq:ch6:CM:defRhoCrit}
    \rho_\text{crit} := \frac {6H^2}{\kappa},
\end{equation}
with respect to which, one finds
\begin{equation}
    \label{eq:ch6:CM:curvByRho}
    \frac k{a^2H^2} = \frac \rho {\rho_\text{crit}} - 1.
\end{equation}
In other words, the extent of spatial curvature of the universe\footnote{Here we haven't incorporated $\abs{k}$ into the scale factor, but if we did we can simply scale it out again for this calculation.} is dictated by the scale of the energy density of the universe with respect to some (dynamic) critical energy density scale. Having said that, observations indicate that the energy density of the universe today is almost exactly $\rho = \rho_\text{crit}$, so the spatial geometry of the universe is comfortably flat, $k = 0$. We will briefly restore $k$ in the next section for an historical discussion, but for the rest of our work we will be content to work in a Euclidean spatial geometry.

Finally, with $k = 0$ and two explicit examples of  $p(\rho)$ and  $\rho(a)$, it is almost trivial to solve for the (cosmological) time evolution of the scale factor in different universal epochs, namely non-relativistic and relativistic matter domination. The first satisfies  $\rho \sim a^{-3}$ , and using the constraint equation \eqref{eq:ch6:CM:EConst} leads to
\begin{equation}
    \label{eq:ch6:CM:aDust}
    a_\text{dust}(t) = a_0 t^{2/3},
\end{equation}
while the second satisfies $\rho \sim a^{-4}$ , leading to
\begin{equation}
    \label{eq:ch6:CM:aRad}
    a_\text{rad}(t) = a_0 t^{1/2}.
\end{equation}
In both cases, we see positive (if sub-linear) growth of the universe in time, and in both cases there is an issue going too far backwards in time, as both scale factors approach $0$ as $t\to 0$, which leads to the metric becoming spatially degenerate at that point---a problem which deserves attention, but not from us. Of course, realistically the universe is neither composed of pure dust or pure radiation, but at least a mixture of the two (and possibly something else we will see soon), so the time-dependence of the scale factor is not quite so clean as either \eqref{eq:ch6:CM:aDust} or \eqref{eq:ch6:CM:aRad}, and a proper investigation of the history of the universe requires properly treating this mixture. Nevertheless, the power-law difference between the two spread over the time scales involved in the lifetime of the universe means there exist long matter (dust) dominated epochs and seperately long radiation dominated epochs within which it is appropriate to approximate the scale factor by either of those pure expressions, and much of practical cosmological calculations involves doing just that.

\subsection{An Historical Insight Into the Present}
\label{ch:AppsII:CM:staticUni}

The application of general relativity to cosmology happened quite quickly. Friedmann's metric appeared as early as 1922 \cite{friedman_uber_1922}, but Einstein was already hard at work trying to describe the universe with his new theory right from the get-go, publishing a seminal paper on the topic in 1917 \cite{1917SPAW.......142E}. Einstein's solution was crafted in a clever but ad hoc way. At the time he had two concerns to reconcile: first the community consensus that the universe should be eternal (i.e., not time-dependent), and second the ambiguity in choosing boundary conditions. The first concern was easy enough, you just don't even think about time dependence in your solution, but the second concern was a real puzzle. With any local gravitational source, you can always impose asymptotic boundary conditions that reduce to Minkowski space, that's the point of locality. But the gravitation of the universe itself presents the problem that there's no asymptotic limit \emph{away} from the universe, nowhere for it to reduce to Minkowski space. His solution was to consider a closed spatial geometry for the universe, in which case the periodic boundary conditions essentially reduce to physics being local everywhere. 

The obvious test case of Einstein's solution was to model the matter-energy of the universe as a non-relativistic perfect fluid (a universe of stars and galaxies does not quite resemble a photon gas), but this alone turns out to be incompatible with the theory of general relativity. In fact, we already have all of the tools we need to see this; consider the FRW metric with $k > 0$,  $T^{\mu\nu} = T_\text{dust}^{\mu\nu}$ , and importantly, $a(t) = 1$. Then from the Friedmann equation \eqref{eq:ch6:CM:Friedmann}, we must have:
 \begin{equation}
    \label{eq:ch6:CM:badEOS}
    0 = \frac {-\kappa}{6}\left( \rho + 3p \right),
\end{equation}
which for dust with $p = 0$ implies vanishing energy density. Worse still, no type of physical matter can satisfy this constraint since it would require either the energy or pressure densities to be negative, which is not how any relativistic or non-relativistic matter that we know behaves. 

With just the field equations and the stress-energy of normal matter, there is really no way out of this predicament, so Einstein proposed an alteration of the field equations instead---something that would turn the left-hand side of \eqref{eq:ch6:CM:badEOS} into $-\Lambda$ instead of 0. In fact, the proposed alteration is not far-fetched, it is rather fixing an assumption we tacitly made when proposing the relation between curvature tensors and the matter-energy tensor. Recall from chapter \ref{ch:EFEs} that we sought a rank-2 symmetric tensor composed only of the metric and up to second-order derivatives, which was also covariantly conserved. This is almost uniquely satisfied by the Einstein tensor $G_{\mu\nu} = R_{\mu\nu} - \frac 12 R g_{\mu\nu}$, but what we neglected to mention is that the same conditions are met by any tensor related to $G_{\mu\nu}$ by a constant offset proportional to the metric, $G_{\mu\nu} + \Lambda g_{\mu\nu}$ (from the action perspective, this is a constant offset in the Lagrangian density $\mathcal{L} \to R + \Lambda$). Einstein called this constant offset a \textbf{cosmological constant}\index{Cosmological Constant}, a sort of background cosmological geometry that just turns out not to be Minkowski. Including this constant augments the trace-reversed Einstein equations as
\begin{equation}
    \label{eq:ch6:CM:traceRevEFEsLambda}
    R_{\mu\nu} - \Lambda g_{\mu\nu} = \kappa\left( T_{\mu\nu} - \frac 12 T g_{\mu\nu} \right),
\end{equation}
and the Einstein constraint and Friedmann equation to be respectively:
\begin{equation}
    \label{eq:ch6:CM:EConstLambda}
    H^2 = \frac \kappa 3 \rho - \frac k {a^2} + \frac 13 \Lambda,
\end{equation}
and
\begin{equation}
    \label{eq:ch6:CM:FriedmannLambda}
    \frac {\ddot a}{a} - \frac 13 \Lambda = -\frac \kappa 6 \left( \rho + 3 p \right).
\end{equation}
A static, non-relativistic matter dominated universe can therefore exist if there is a cosmological constant related to the constant matter energy density by $\Lambda = \kappa\rho/2$. Such a universe would necessarily be closed, with radius of curvature set by $k/a^2 = \Lambda = \kappa\rho/2$.

\begin{figure}[htpb]
    \centering
    \includegraphics[width=\textwidth]{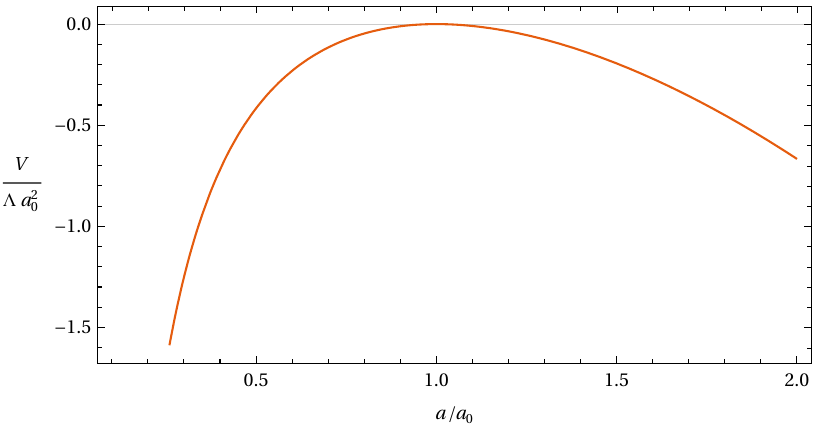}
    \caption[The potential term in the Einstein constraint for the static universe model.]{The potential term in the Einstein constraint for the static universe model, normalized by $\Lambda/a_0^2$. Einstein's static universe is the unstable equilibrium at $V/\Lambda a_0^2 = 1$.}
    \label{fig:ch6:CM:staticV}
\end{figure}

This static solution is nice and clean, and it satisfied all of Einstein's concerns, but it is fundamentally incompatible with general relativity as a dynamic theory, it relies on a metric ansatz that is unjustifiably finely tuned. The most general metric that satisfies the (justifiable and immutable) symmetries of the problem is the FRW metric with cosmological constant and positive spatial curvature. The equations of motion \eqref{eq:ch6:CM:EConstLambda} and \eqref{eq:ch6:CM:FriedmannLambda} are the actual equations of motion of the universe, and it takes only a moment to see that while the static solution indeed satisfies these equations, it is not \emph{stable}. Any deviation from perfect homogeneity and isotropy whether statistically, quantumly, or artificially generated, will instantly set the universe on a path towards either infinite outward expansion, or total collapse. There are two easy ways to see this, both using that $\rho = \rho_\text{dust} \sim a^{-3}$. The first is with the Friedmann equation; $\ddot a$  is at the tipping point between a positive term that scales as $a$  and a negative term that scales as $1/a^2$. For fluctuations that make the scale factor slightly smaller, the negative term wins and continues the negative trajectory, while for fluctuations that increase $a$, the second derivative is positive and encourages it to continue its growth. More visually, one can instead express the Einstein constraint in a phase-space form, with  $\dot{a}$ serving as a momentum term:
 \begin{equation}
    \label{eq:ch6:CM:phaseSpace}
    \dot{a}^2 + V(a) = 0,
\end{equation}
is an equilibrium expression akin to a Hamiltonian in classical mechanics, with the ``potential'' term
\begin{equation}
    \label{eq:ch6:CM:staticPotential}
    V(a) = k - \frac \kappa 3 \rho_0 a^{-1} - \frac 1 3 \Lambda a^2.
\end{equation}
Again, there is a delicate, finely tuned balance such that the potential vanishes at the critical value of Einstein's static universe, but becomes negative on either side of that value (see figure \ref{fig:ch6:CM:staticV}). 

So as beautiful as it is, this static universe could never be, the equations of general relativity are simply incompatible with such an existence. In principle, this is in fact a test of the theory; if measurements came out showing no discernable expansion in the cosmos, it would demand a new theory of gravity. Fortunately for those readers who have made it this far in this text, that is not the case. In 1929, Hubble produced conclusive evidence of cosmic expansion, and numerous experiments since then have not only reproduced that result but have dramatically improved upon it. According to the legends, upon learning of Hubble's result, Einstein discarded his cosmological constant and declared it his ``greatest blunder.'' As it turns out however, that dramatic disposal may have been a bit hasty. Ever more precise cosmological observations appear to be consistent with the presence of a very small, but non-vanishing positive cosmological constant after all. With the likely possibility that we do indeed need to take this term seriously, it is worth looking a bit closer at the cosmological constant.

\subsection{The Cosmological Constant}
\label{ch:AppsII:CM:CC}
Introducing the cosmological constant term to the Einstein field equations is mathematically trivial, but physically it introduces some degree of ambiguity in its interpretation. Constructing the field equations came with a very simple intepretation of the left-hand side as purely geometry, and the right-hand side as purely matter-energy, so introducing a new term means picking which side it belongs to---is the cosmological constant a geometric artifact, or a source of stress-energy? Of course the equality means these are equivalent interpretations, so it is useful to explore both ideas. 

\subsubsection{$\Lambda$ as Geometry}
A geometrical interpretation is a hair more natural given how we introduced the constant in our narrative above, by augmenting the Einstein tensor. In this case, the way to see what $\Lambda$ means is to look at the vacuum field equations (a pure, global vacuum, not a local perturbative situation like the gravitational waves). Here again it is useful to use the trace-reversed form $R_{\mu\nu} - \Lambda g_{\mu\nu} = 0$, which reveals 
\begin{equation}
    \label{eq:ch6:CM:maxSymLambda}
    R_{\mu\nu} = \Lambda g_{\mu\nu},
\end{equation}
exactly the equation satisfied by a maximally symmetric metric in 3+1 dimensions. This is as it should be, we expect purely empty space to have the highest degree of symmetry possible (how can ``nothing'' look different from any perspective?). But seeing the generic expression \eqref{eq:ch6:CM:maxSymLambda} shows that our stubborn insistence upon Minkowski space as the unique vacuum of general relativity was simply a mistake. The geometry of the vacuum is characterized by the curvature scale and signature of $\Lambda$, and it is an experimental question what that value is. 

Just as with the maximally symmetric 3-spaces uncovered in \ref{ch:AppsII:CM:maxSym}, the maximally symmetric Lorentzian 4-spaces can be grouped according to the sign of the curvature parameter. The space with $\Lambda = 0$ is of course simply Minkowski. The spaces with $\Lambda > 0$ are \textbf{de Sitter} spaces. These spaces can be written in the FRW coordinates with $a \sim e^{Ht}$ for constant $H^2 = \Lambda/3$, and it is approximately this cosmology that the evidence suggests we live in today. Finally, $\Lambda < 0$ is a peculiar beast indeed. A negative cosmological constant corresponds to an \textbf{anti-de Sitter} spacetime, which exhibits bizarre properties like a periodic time coordinate and consequent issues with causality. This space likely has very little cosmological relevance, but as it turns out it is of great interest in a number of very formal applications, particularly certain lines of inquiry in string theory.

\subsubsection{$\Lambda$ as Matter-Energy}

The geometric interpretation of $\Lambda$ is very natural when considering the vacuum equations of general relativity and studying maximally symmetric spacetimes, but in studying the universe we are not looking at empty space or (globally) maximally symmetric solutions. The universe we live in contains a non-vanishing, time-dependent source of stress-energy, so what $\Lambda$ represents in terms of curvature of the vacuum is moot. Solving for the scale-factor in a universe with a general mix of stress-energy densities (in our case some ratio of dust plus radiation), it is only natural to bring the highly symmetric cosmological constant term into the fold, and identify a stress-energy $T_{\Lambda}^{\mu\nu} = -\Lambda g^{\mu\nu}$ . In terms of mixed indices, this is
\begin{equation}
    \label{eq:ch6:CM:lambdaSET}
    (T_{\Lambda})\indices{_\mu^\nu} = -\Lambda \delta_\mu^\nu.
\end{equation}
Comparing to the general perfect fluid stress-energy $(T_\text{pf})\indices{_\mu^\nu} = \text{diag}(-\rho, p, p, p)$ identifies
\begin{align}
    \label{eq:ch6:CM:lambdaRhoP}
    \rho_\Lambda &:= \Lambda, \\
    p_\Lambda &:= -\Lambda,
\end{align}
and so an equation of state of $w = -1$, a peculiar form of matter-energy indeed. In fact there is one plausible source for this sort of background energy density, the ``zero-point'' vacuum energy density that comes from the quantum fields of matter. In quantum calculations, this background energy density is always subtracted away since the measurements of interest are always about deviations from the vacuum, but gravity couples to \emph{all} matter-energy, so even the background values should matter here. Unfortunately, quantum field theoretic calculations vastly over-shoot the measured value of the cosmological constant, so a satisfactory explanation for its physical origin is still very much an open question. 


\section{Recap}
\label{ch:Apps2:recap}

\begin{itemize}
    \item Non-trivial sources of stress-energy are best treated by first incorporating all symmetries of the stress-energy into a metric ansatz.
    \item For every continuous symmetry of the spacetime manifold, coordinates can be found such that the symmetry takes the form of translation along a specific coordinate direction, and the metric tensor is independent of that coordinate, $\partial_* g_{\mu\nu} = 0$.
    \item In any coordinate system, continuous symmetries of the spacetime manifold are identified with Killing vectors, vectors $K^\mu$ which satisfy the Killing equation $\nabla_{(\nu} K_{\mu)} = 0$.
    \item The component of momentum in a symmetric direction is given by the projection of (covariant) momentum along a Killing vector, and is conserved along the proper (or affine) time of a geodesic, $\dv[]{}{\tau}p_\star = \dv{}{\tau}{(K_\mu p^\mu)} = 0$.
    \item The metric outside a point-source of stress-energy is the Schwarzschild metric, \\$\dd s^2 = -(1 - r_s/r)\dd t^2 + (1 - r_s/r)^{-1}\dd r^2 + r^2 \dd^2\Omega$.
    \item The Schwarzschild metric is adapted to Minkowskian observers infinitely far away from the source, and slightly less strictly adapted to all observers at fixed radii outside the horizon (their time coordinate is dilated with respect to the observer at spatial infinity).
    \item Observers at fixed radii are uniformly accelerating, which can be seen locally by performing a Taylor expansion of the Schwarzschild metric to find local Rindler coordinates, and globally by performing a coordinate transformation to the Rindler-Tortoise coordinates in which the tempero-radial sector of the metric is conformal to the Rindler metric.
    \item Where there's Rindler there's Minkowski, and the standard inverse Rindler coordinate transformation applied to Rindler-Tortoise reveals the Kruskal-Szekeres coordinates, in which the Schwarzschild metric becomes smoothly conformal (in the $t$-$r$ plane) to Minkowski coordinates.
    \item Kruskal-Szekeres coordinates show no physical issues arise as geodesics cross the Schwarzschild event horizon, but that the singularity at the origin is indeed a problematic point. Furthermore, treating the full range of the coordinates reveals a parallel ``white hole'' sector, completely causally disconnected from the black hole sector, but exhibiting interesting conceptual phenomena nonetheless.
    \item The Schwarzschild solution can be used to make many testable predictions of physical observables, including gravitational time dilation, the existence of photon spheres, the deflection of light by gravitating sources, and the precession of perihelion of nearly elliptical orbits.
    \item The least complicated everwhere non-trivial source of stress-energy to use is one which is as spatially symmetric as possible, so is isotropic and homogeneous. This type of stress-energy well describes the universe as a whole, so it is used as a model of cosmology.
    \item A spatially isotropic and homogeneous stress-energy leads to a metric ansatz of the Friedmann-Robertson-Walker (FRW) form \\ $\dd s^2 = -\dd t^2 + a^2(t)\left( (1 - kr^2)^{-1} \dd r^2 + r^2 \dd^2\Omega \right) $
    \item In the purely spatial sector, isotropy and homogeneity together amount to the maximum number of symmetries available to a three-dimensional space, so the spatial part of the metric $(g_3)_{ij}$ satisfies the condition of maximal symmetry $(R_3)_{ij} \propto (g_3)_{ij}$, leading to three classes of spatial geometries depending on the sign of the spatial curvature $k$, with $k > 0$ open (hyperbolic), $k = 0$ flat (Euclidean), and $k < 0$ closed (spherical).
    \item The stress-energy that satisfies the symmetries of the universe is that of the perfect fluid, $(T_\text{pf})\indices{_\mu^\nu} = \text{diag}(-\rho, p, p, p)$. We consider fluids with a very simply equation of state $p = w\rho$, namely dust with $w = 0$ and radiation with $w = 1/3$. Conservation of stress-energy already enables calculation of the scaling of $\rho$ as $\rho \sim a^{-3(1 + w)}$, so $\rho_\text{dust} \sim a^{-3}$ and $\rho_\text{rad} \sim a^{-4}$.
    \item With the FRW metric and the perfect fluid stress-energy, the Einstein equations evaluate to the Einstein constraint $H^2 = \frac \kappa 3 \rho - \frac k {a^2}$, and the Friedmann equation $\ddot a/a = -\frac \kappa 6 \left( \rho + 3p \right) $. Time evolution of non-relativistic and relativistic matter is then found to go as $\rho_\text{dust} \sim t^{2/3}$ and $\rho_\text{rad} \sim t^{1/2}$. 
    \item If it were necessary for the universe to be static ($\dot a = \ddot a = 0$), this could only be effected with a positive spatial curvature and an additional cosmological constant term in the Einstein equations, $G_{\mu\nu} \to G_{\mu\nu} + \Lambda g_{\mu\nu}$. With only a constant density of non-relativistic matter, this is the Einstein Static Universe with $\Lambda = \kappa\rho_0/2$ and $k = \Lambda$. This solution is unstable and such a fine tuning of universal parameters would quickly fail as any perturbation to the size of the universe would quickly and consistently grow in magnitude.
    \item The cosmological constant still turns out to have value as recent measurements seem to indicate universal growth consistent with a very small but non-zero $\Lambda > 0$. 
    \item $\Lambda$ can be interpreted either geometrically or as matter-energy. 
    \item As a geometrical parameter, the cosmological constant is the curvature of the four-dimensionally maximally symmetric solution to the empty-space Einstein equations (directly analogous to $k$ for the maximally symmetric three-dimensional spatial metric). $\Lambda = 0$ corresponds to flat Minkowski space, $\Lambda > 0$ is the hyperbolically expanding de Sitter space, and $\Lambda < 0$ is the peculiar temporally periodic anti-de Sitter space.
    \item As a form of matter-energy, the cosmological constant appears as a constant energy density associated with empty space. As a fluid it has the form $\rho = \Lambda$ and $p = -\Lambda$ so has equation of state $w = -1$ and naturally scales as expected $\rho_\Lambda \sim a^0$.

\end{itemize}

\begin{subappendices}

\section{Schwarzschild Near-Source Boundary Condition}
\label{ch:AppsII:app:nsBC}
In the main text, the boundary condition used to solve for the second integration constant in the Schwarzschild metric (correspondence with the Newtonian limit) was a bit of a hack. In essence, we did supply a delta function potential to the Einstein field equations and integrate over a sphere about the origin, but we did so on such a large sphere that we approximated the field equations within the entire volume by the Newtonian limit. It is an important exercise to see how the same procedure works without such an approximation since general relativity is meant to \emph{supercede} Newtonian mechanics, not use the latter as a crutch. Unfortunately, this turns out to be difficult to do, and there is even still some debate in the literature about how to do it correctly \cite{vaidya_random_1985,balasin_energy-momentum_1993,kawai_distributional_1997,pantoja_distributional_2002, hayman_purely_2024}. There are three key issues that arise: first, since the worldline (geodesic) of the source point particle is coupled to the geometry it induces, ambiguity arises in exactly how to define the point-like stress-energy tensor. Second, since the Einstein field equations are non-linear, the usual integration procedure fails to evaluate in a simple way so some regularization scheme becomes necessary. Lastly, although the plan is to reinsert the location of the point source as a part of spacetime via distributional methods, the spherical polar coordinates themselves are still separately undefined at that point. We will address this last point by working almost exclusively in ``Cartesian'' coordinates\footnote{In quotes because the coordinates have the Cartesian form with respect to the spherical polar coordinates, but they are not Cartesian in that they do not describe the affine Euclidean space.} that are well-defined at the origin:
\begin{equation}
    \label{eq:ch6:nsBC:defCart}
    \hat{x}^1 := r\sin\theta\cos\phi, \quad \hat{x}^2 := r\sin\theta\sin\phi, \quad \text{and} \quad \hat{x}^3 := r\cos\theta.
\end{equation}
We take the other issues in turn. 

    In principle, the stress-energy due to the point source is given in terms of the source's action,
     \begin{equation}
        \label{eq:ch6:nsBC:ppAction}
        S_{\text{p}} = -M \int \dd\tau \sqrt{-U_\mu U^\mu} = -M \int \dd^4 x\int \dd \tau\,  \delta^{(4)}(x^\mu - x_{\text{p}}^\mu(\tau)) \sqrt{-g_{\mu\nu}U^\mu U^\nu},
    \end{equation}
    so the (covariant) stress-energy is
    %
    %
    \begin{equation}
        \label{eq:ch6:nsBC:ppSET}
    T_{\mu\nu} = -\frac 2 {\sqrt{-g}} \frac{\delta S_\text{p}}{\delta g^{\mu\nu}} = \frac M {\sqrt{-g}} \int \dd \tau\, \delta^{(4)}(x^\mu - x_\text{p}^\mu(\tau)) U_\mu U_\nu. 
    \end{equation}
    While this looks like a simple algorithm telling us unambiguously what the stress-energy must be, closer inspection reminds us that the worldline of our particle is \emph{a priori} unknown. Instead, it must be solved for self-consistently with the metric using the geodesic equation, but this cannot be done unambiguously. To see this, take an iterative approach; for any timelike geodesic $x^\mu_0(\tau)$, we can choose coordinates such that $t(\tau) = \tau$ and $x^i(\tau) = 0$. But in these coordinates, the Einstein field equations can be solved outside the source to find the Schwarzschild solution, which has a singularity at the source. This means that while $U^\mu$ may be well defined in these coordinates, $U_\mu$ cannot be defined as its dual $g_{\mu\nu}(\tau)U^\nu$ since $g_{00}(x_0(\tau))$ is degenerate everywhere on that path\footnote{And note that it is evaluated at exactly $r = 0$ so cannot be defined as a distribution through a limiting procedure either.} (and similarly, the inverse diverges the entire time so $U^\mu = g^{\mu\nu}(\tau)U_{\nu}$ cannot be defined). The only way out is to input the source four-velocity as an external parameter, presuming it not to be hampered by its own gravitational field:
    \begin{equation}
        \label{eq:ch6:nsBC:defU}
        U_{\mu} := \mqty(-1 & 0 & 0 & 0), \qquad \text{and} \qquad U^\mu = \mqty(1 \\ 0 \\ 0 \\ 0),
    \end{equation}
    but in doing so we must be exceedingly careful. Defining the covariant and contravariant forms of the four-velocity externally and separately means we get to pick one form of the stress-energy (one combination of indices) and then have to stick with it, since raising or lowering indices on the stress-energy tensor is ill-defined compared with the same activity on the other tensors in the theory. The standard choice is to define the mixed index version since that way one can also calculate the trace of the stress-energy tensor unambiguously:
    \begin{equation}
        \label{eq:ch6:nsBC:fixPPSET}
        \hat{T}\indices{_\mu^\nu} = -\frac {M}{\sqrt{-g}} \delta^0_\mu\delta^\nu_0 \delta^{(3)}(\vec{x}).
    \end{equation}
    (We'll use hats to indicate expressions in Cartesian coordinates). This expression for the stress-energy tensor is now a well defined tensor in any coordinate system that is itself well defined at the origin. 

    Since we can no longer tensorially use spherical polar coordinates, we have to re-visit section \ref{ch:AppsII:BH:pin} now in the Cartesian coordinates \eqref{eq:ch6:nsBC:defCart}. In mixed index form, the Ricci tensor is:
    \begin{align}
        \hat{R}\indices{_t^t} &= R\indices{_t^t}, \quad \text{and}  \label{eq:ch6:nsBC:cartRicT} \\
        \hat{R}\indices{_i^j} &= \frac{\hat{x}^i \hat{x}^j}{r^2}R\indices{_r^r} + \left( \delta^j_i - \frac{\hat{x}^i\hat{x}^j}{r^2} \right)R\indices{_\theta^\theta}. \label{eq:ch6:nsBC:cartRicSp} 
    \end{align}
    from which we robustly find $R\indices{_r^r} = R\indices{_\theta^\theta}$ and so $\hat{R}\indices{_i^j} = \delta^j_i R\indices{_r^r} = \delta^j_i R\indices{_\theta^\theta}$. Moreover, since\\ $R\indices{_t^t} = R\indices{_r^r} - \frac 1 {rf}\frac{(fq)^\prime}{fq} = R\indices{_\theta^\theta} - \frac 1 {rf}\frac{(fq)^\prime}{fq}$, we must have
     \begin{equation}
        \label{eq:ch6:nsBC:cartScalar}
        \hat{R} = 4R\indices{_\theta^\theta} - \frac 1 {rf}\frac{(fq)^\prime}{fq}, 
    \end{equation}
    and so the Einstein tensor is
    \begin{align}
        \label{eq:ch6:nsBC:cartG}
        \hat{R}\indices{_t^t} - \frac 12 \hat{R} &= -R\indices{_\theta^\theta} - \frac 12 \frac 1 {rf}\frac{(fq)^\prime}{fq}, \\
        \hat{R}\indices{_i^j} - \frac 12 \delta^j_i \hat{R} &= -R\indices{_\theta^\theta} + \frac 12 \frac 1 {rf} \frac{(fq)^\prime}{fq}.
    \end{align}
    Since the stress-energy only has a non-trivial time-time component, it follows that 
    \begin{equation}
        \label{eq:ch6:nsBC:exactR33}
        R\indices{_\theta^\theta} = \frac 1 {r^2}\left( 1 - \partial_r\left( \frac r f \right)  \right) - \frac 1 {2rf} \frac{(fq)^\prime}{fq} = \frac 12 \frac 1 {rf} \frac{(fq)^\prime}{fq},
    \end{equation}
    and so subsituting in either the time-time equation \eqref{eq:ch6:nsBC:cartRicT} or the scalar equation \eqref{eq:ch6:nsBC:cartScalar}, we obtain an expression that can be used to solve for the near-source boundary condition: 
    \begin{equation}
        \label{eq:ch6:nsBC:protoBC}
        \frac 1 {r^2} \partial_r\left( r\left(1 - \frac 1 f\right) \right)  = \frac {M\kappa} {\sqrt{-g}} \delta^{(3)}(\vec{x}). 
    \end{equation}

    It is easy enough to show that outside the source the same solutions are found as before, that is $fq = 1$ and  $f = (1 - B/r)^{-1}$. At the origin itself however, neither of these can be exact owing to the delta function in the field equations, so some means of regularization must be imposed \cite{pantoja_distributional_2002}---I will offer two very general options that yield the same result. For both, consider a regularizing function $h(r;\lambda)$ defined such that $\lim_{\lambda \to 0} h(r; \lambda) = 1$ and $h(0;\lambda) \to 0$ at least as fast as $r^2$, and attach it to the singular part of $f$, 
     \begin{equation}
        \label{eq:ch6:nsBC:regF}
        f_\lambda(r) = \frac 1 {1 - (B/r)h(r;\lambda)}.
    \end{equation}
    (For example, try $h(r;\lambda) = r^2/(r^2 + \lambda^2)$). Then integrating the boundary condition \eqref{eq:ch6:nsBC:protoBC} over a Euclidean\footnote{We can use a Euclidean integration since we are just manipulating a functional equation to find an algebraic relation, not using either side as a geometric object or measure on a curved space. Also note that if you are uneasy about the exchanged roles of $t$ and $r$ inside the horizon, this same procedure works identically for a particle of negative mass which has a naked singularity, so a perhaps slightly more rigorous approach would be to solve for the \emph{white hole} solution first, then infer the correct black hole solution by taking $-M \to +M$.} sphere of radius $\epsilon$:
    \begin{equation}
        \label{eq:ch6:nsBC:protoBC2}
        4\pi B \lim_{\lambda \to 0}  h(r;\lambda) \big\vert^\epsilon_0 = M\kappa.
    \end{equation}
    From here, we can either impose that $h(0;\lambda) = 0$ to smooth out the metric behaviour at the origin, or we can notice that  the metric is invariant under $r \to -r$  and $h \to -h$ , so if we impose $h(-r;\lambda) = -h(r; \lambda)$, then we can extend the integration range to $[-\epsilon, \epsilon]$  (and normalize by $1/2$ for consistency). In either case, we find
     \begin{equation}
        \label{eq:ch6:nsBC:BC}
        B = 2GM,
    \end{equation}
    as expected. This makes it clear that the Newtonian result does indeed derive from the relativistic one, and not somehow the other way around. 

    \subsection*{Postscript: Back to Polar}
    One final note; although the point-like stress-energy tensor only makes sense in coordinates that are well-defined at the origin, it is interesting to see what form this solution implies the stress-energy would take in spherical polar coordinates if they could be defined at the origin. It turns out to be quite easy to see that the correct stress-energy density in polar coordinates is in fact the same form as in Cartesian coordinates, \eqref{eq:ch6:nsBC:fixPPSET}. Curiously, this is not as trivial a result as it might sound. Our result is obtained by starting with a physically-motivated stress-energy tensor and solving for the most general spherically symmetric metric, finding two distinct distributional relationships for $f$ and $q$: \eqref{eq:ch6:nsBC:exactR33} and \eqref{eq:ch6:nsBC:protoBC}. If one instead assumed a \emph{less general} form for the metric in which the condition from the main text $fq = 1$ is exact even at the origin, this instead forces a more complicated and physically unmotivated form for the stress-energy tensor that contains non-trivial elements on all the diagonals (see e.g.~\cite{balasin_energy-momentum_1993, pantoja_distributional_2002}).



    
%
\end{subappendices}

\chapter*{Conclusion}
\addcontentsline{toc}{chapter}{Conclusion}

And so ends this brief introduction to the magnificent theory of general relativity. Of course there is still very much more to learn, and I highly encourage the novice reader to continue to explore the literature, particularly with the texts cited throughout (especially \cite{schutz_2009,carroll_2019} and \cite{wald2010general}). But hopefully now you have everything you need to make some sense of the modern literature in the modern language of the field, and to get up to speed quickly on any active line of inquiry. 

%
\bibliographystyle{unsrt} 
\bibliography{gr}
%
%
%
%

\printindex

\end{document}